\documentclass[twocolumn,english,aps,pra,eqsecnum]{revtex4-1}
\usepackage[T1]{fontenc}
\usepackage[latin9]{inputenc}
\usepackage{color}
\usepackage{float}
\usepackage{bm}
\usepackage{amsmath}
\usepackage{amssymb}
\usepackage{graphicx}

\makeatletter
\usepackage{varwidth}

\usepackage{babel}

\makeatother

\usepackage{babel}
\begin{document}
\title{Forward-backward stochastic simulations:  Q-based model for measurement
and Bell-nonlocality consistent with weak local realistic premises}
\author{M. D. Reid and P. D. Drummond}
\affiliation{Centre for Quantum Science and Technology Theory, Swinburne University
of Technology, Melbourne 3122, Australia}
\begin{abstract}
While Bell nonlocality can be explained through mechanisms such as
retrocausality and superluminal influences, it is not clear  why
such mechanisms do not manifest at an everyday macroscopic level.
In this paper, we show how measurement and nonlocality can be explained
consistently with macroscopic realism {}and no-signaling, and causal
relations for macroscopic quantities. Considering measurement of
a field amplitude $\hat{x}$, we derive  theorems that lead to
an equivalence between a quantum phase-space probability distribution
$Q(x,p,t)$  and stochastic trajectories for real amplitudes $x(t)$
and $p(t)$ propagating backwards and forwards in time, respectively.
We present forward-backward stochastic simulations  that motivate
a $Q$-based objective-field model of reality. Amplification plays
a key role in the measurement. With amplification, similar to decoherence,
contributions due to interference become unobservable, leading to
``branches'' in $x$ that correspond to distinct eigenvalues $\lambda$.
This elucidates how the system evolves from a superposition to an
eigenstate, from which Born's rule follows. For superposition states,
we find that the backward and forward trajectories are connected,
leading to ``hidden loops''.  We deduce a hybrid causal structure
involving causal deterministic relations for amplified variables,
along with microscopic noise inputs and hidden loops for unobservable
quantities. Causal consistency is confirmed, removing Grandfather-type
paradoxes.  The simulations allow evaluation of a state inferred
for the system, conditioned on a particular branch $\lambda$, from
which we deduce a model for projection and the collapse of the wave
function. The theory is extended to Einstein-Podolsky-Rosen correlations
and Bell nonlocality. We demonstrate consistency with three weak
local realistic premises: (1) the existence of real properties (defined
for the system after operations that fix measurement settings); (2)
a partial locality implying no-signaling; (3) ``elements of reality''
that apply to the predictions of a system by a meter, once meter-settings
are fixed. A Q-based hidden variable model  and mechanism for Bell
nonlocality is identified. Our work shows how forward-backward
stochastic simulations lead to a hybrid causal structure, involving
both deterministic causal relations and ``hidden'' stochastic loops,
explaining measurement and entanglement, with paradoxes typically
associated with retrocausality avoided.
\end{abstract}
\maketitle

\section{Introduction}

All local, strictly causal hidden variable theories are falsifiable,
since quantum predictions and experiments lead to a violation of Bell
inequalities \citep{Bell1964,brunner2014bell,clauser1978bell,bell2004speakable}.
However, it is known that violations of Bell inequalities can indeed
arise using the advanced solutions of classical electrodynamics \citep{pegg1980objective,cramer1980generalized,pegg1986absorber,cramer1986transactional}.
Classical electrodynamics with future boundary conditions has been
widely studied \citep{schwarzschild1903elektrodynamik,fokker1929invarianter,tetrode1922causal,dirac1945analogy,wheeler1945interaction,wheeler1949classical},
and is essential to the theory of radiating fields. Following from
Bohr \citep{bohr1987essays} who considered delayed-choice experiments
\citep{ma2016delayed}, Wheeler speculated that retrocausality due
to future boundary conditions may explain quantum paradoxes \citep{wheeler1978past,wheeler2014quantum}.
Bell's work also motivated the question of potential superluminal
disturbances, leading to no-signaling theorems \citep{eberhard1989quantum}.
These results have inspired analyses of causality in quantum physics
\citep{aharonov1964time,scully1982quantum,scully1991quantum,mohrhoff1996restoration,englert1999quantum,kim2000delayed,walborn2002double,jacques2007experimental,price2008toy,argaman2010bell,aharonov2008two,ionicioiu2011proposal,tang2012realization,peruzzo2012quantum,kaiser2012entanglement,ma2013quantum,ionicioiu2014wave,manning2015wheeler,zheng2015quantum,ma2016delayed,chaves2018causal,rossi2017restrictions,rab2017entanglement,qin2019proposal,polino2019device,kastner2019delayed,huang2019compatibility,yu2019realization,wharton2020colloquium,la2021classical,thenabadu2022macroscopic,herzog1995complementarity,oreshkov2012quantum,chaves2012entropic,wood2015lesson,cavalcanti2018classical,pearl2019classical,pearl2021classical,costa2016quantum,allen2017quantum,Shrapnel2018causationdoesnot,shrapnel2019discovering,barrett2021cyclic,pienaar2020quantum,Allen2021QuantumPhysRevX.7.031021,weilenmann2020analysing,chiribella2012perfect,chiribella2013quantum,pollock2018operational,araujo2014computational,araujo2015witnessing,oreshkov2016causal,giarmatzi2019quantum,gachechiladze2020quantifying,daley2022experimentally,araujo2017quantum,wharton2014quantum,weinstein2017learning,barrett2019quantum,guo2024experimental,chiribella2022quantum,goswami2018indefinite,Friederich_2025,price2024mechanism,dewitt2025forward}.

Studies of causal structure are closely related to studies of realism
in quantum mechanics. Classical models assume real properties that
describe the system at a given time; retrocausality supposes these
properties to be influenced by future events. Proposals to explain
Bell violations and wave-function collapse include superluminal causal
influences \citep{BohmPhysRev.85.166,struyve2010pilot,toner2003communication,scarani2014strong,Bancal2012quantum,struyve2007minimalist},
retrocausality \citep{pegg1980objective,cramer1986transactional,wharton2010novel,price2008toy,Wharton2018New,wharton2020colloquium,Donadi2022Toy,argaman2010bell,almada2016retrocausal,sutherland2008causally,price1996time}
cyclic causation \citep{vilasini2022general,vilasini2022impossibility}
and unobservable loops \citep{castagnoli2021unobservable,castagnoli2018completing,castagnoli2019relational},
or superdeterminism \citep{hossenfelder2020rethinking} $-$ but an
open question is why then are such mechanisms not apparent at a macroscopic
level? It has not been clear how to reconcile Bell nonlocality with
the absence of superluminal signaling. Causal models based on superluminal
or retrocausal mechanisms require a fine-tuning of parameters in order
to explain no-signaling, which seems undesirable \citep{wood2015lesson,pearl2021classical,cavalcanti2018classical}.
The fine-tuning may not always be a problem however, and no-signaling
does not rule out retrocausal or cyclic causation \citep{almada2016retrocausal,argaman2010bell,vilasini2022general,vilasini2022impossibility}.

As well as no-signaling, we argue that macroscopic realism is a natural
requirement for a model complying with causality at a macroscopic
level. Macroscopic realism (MR) is the assumption that a system in
a superposition of two macroscopically distinct states, ``will at
all times be in one or other of these states'' \citep{leggett1985quantum}
(something that cannot be changed by a future event).  Yet, Schrödinger
argued that MR is inconsistent with the notion that quantum mechanics
is a complete description of reality (``realdings'') \citep{schrodinger1935gegenwartige}:
Quantum causal models have been developed to address these problems
\citep{costa2016quantum,Allen2021QuantumPhysRevX.7.031021,shrapnel2019discovering,barrett2021cyclic,weilenmann2020analysing,allen2017quantum,pienaar2020quantum,barrett2019quantum}.
It is not clear however whether one can present a unified framework
combining causality and realism, and there remains the fundamental
question of the level (from microscopic to macroscopic) at which causal
concepts, or MR, will hold.

In this paper, we show how quantum stochastic phase-space theorems
lead to stochastic simulations of real amplitudes $x$ and $p$ 
propagating in a backward and forward time-direction respectively,
that may contribute towards a resolution of these questions. The amplitudes
are based on the well-known positive phase-space distribution, the
Q function in quantum optics \citep{Husimi1940}, and have been proposed
as a way to address the ``measurement problem'' \citep{drummond2020retrocausal,drummond2021objective,Friederich2021Introducing,drummond2021time,fulton2024alternative,Reid2023Short}.
The measurement problem is to understand how a system, when being
measured, evolves from a superposition to a single eigenstate (the
``collapse of the wave function'') \citep{wigner1963problem,bell1990against,bell2004speakable,weinberg2017trouble,born1955statistical}.
Questions arise, such as the timing of the ``collapse'' (when can
the system be considered to be in the eigenstate), and Born's question,
of ``what is the reality which our theory has been invented to describe?''
\citep{born1955statistical}. These questions are closely related
to the questions of cause and effect in Einstein-Podolsky-Rosen (EPR)
and Bell experiments \citep{wood2015lesson}. What hidden-variable
theories are possible that describe the EPR and Bell correlations,
and how would nonlocality be achieved, when there is no-signaling?
Bohm's pilot-wave hidden variable theory for quantum mechanics is
deterministic \citep{BohmPhysRev.85.166}, and the Pusey-Barrett-Rudolph
theorem excludes most epistemic models \citep{pusey2012reality}.

In this paper, we seek to resolve these questions in the context of
a model of reality motivated by the Q function. We analyze the measurement
of $\hat{x}$ on a superposition of eigenstates $|x_{j}\rangle$ of
$\hat{x}$, which we define as highly squeezed states in $\hat{x}$.
Measurement of $\hat{x}$ is treated as amplification of $\hat{x}$,
by a factor $G=e^{gt}$ ($g$ is real) modeled by a Hamiltonian $H_{amp}$
\citep{drummond2020retrocausal}. In the second part of the paper,
we apply the model of reality to  deduce how hidden variables satisfying
weak forms of local realism put forward in Ref. \citep{fulton2024alternative}
can explain EPR correlations and Bell nonlocality, consistently with
no-signaling. This gives insight into a mechanism for Bell nonlocality
which is consistent with causality at a macroscopic level.

The resolution we propose involves an equivalence between the quantum
state of the system at a fixed time $t$, which is given uniquely
by the Q function $Q(x,p,t)$, and a set of amplitudes $\{x(t),p(t)\}$
that are solutions (``trajectories'') of a pair of forward-backward
stochastic equations (Fig. 1). In this paper, we solve a pair of stochastic
differential equations that model the measurement as $H_{amp}$: 
one (for $p$) is solved in the forward-time direction with a boundary
condition at the initial time $t_{0}$; the other (for $x$) is solved
in the backward-time direction with a boundary condition (referred
to as a future boundary condition, FBC \citep{dirac1945analogy})
at the final time, $t_{f}$. Put simply, the initial-time boundary
condition is determined by the initial state of the system. The final-time
boundary condition is determined by the final state of the system.
However, the use of a FBC to resolve problems about reality seems
counter-intuitive. The ``paradox of the future boundary condition''
is introduced $-$ how can such a condition be consistent with our
understanding of cause-and-effect? How can we determine what that
FBC condition should be, if we do not know the solutions in advance?
A similar paradox applies to retrocausality, conventionally defined
as the past affecting the future.

It is often also speculated that a phase-space interpretation of quantum
mechanics is not possible, since the obvious candidate, the Wigner
function $W(x,p)$, can be negative \citep{Wigner1932,hillery1984distribution}.
The Q function, being a Gaussian convolution of $W(x,p)$,  has additional
noise and is hence always positive $-$ but then the $x$ and $p$
do not correspond to directly observable quantities. The Q function
of an eigenstate $|x_{j}\rangle$ of $\hat{x}$ is a Gaussian distribution
about a mean value of $x_{j}$. The fluctuations about the mean are
at the level of the quantum vacuum, but the physical significance
is unclear, since (for the eigenstate) these fluctuations must remain
undetected in the measurement process. In this paper, we clarify the
role of this ``hidden noise'' as part of the resolution of the FBC
paradox.

The solutions we put forward resolve these paradoxes, by avoiding
traditional notions of retrocausality, instead leading to a different
``hidden'' type of causation, originating from microscopic inputs
$\delta x$ at the future boundary. In the first part of this paper,
we analyze the measurement problem, by considering measurement of
$\hat{x}$ on a system prepared in a superposition $|\psi_{sup}\rangle=\sum_{j}c_{j}|x_{j}\rangle$
($c_{j}$ are probability amplitudes) of eigenstates $|x_{j}\rangle$
of $\hat{x}$. We find that the future boundary condition (FBC) for
$x$ has two parts: The first part originates from a deterministic
causal relation $x_{j}\rightarrow Gx_{j}$, from the initial time
$t_{0}$ to the final time $t_{f}$, that describes the amplification
of $\hat{x}$. The first part of the FBC is hence of a classical nature,
originating from the description of the state of the system at the
initial time. Moreover, we show (Result III.1 of this paper) that
the interference terms $\mathcal{I}nt$ that distinguish the superposition
$|\psi_{sup}\rangle$ from a mixture of $|x_{j}\rangle$ vanish in
the FBC. In other words, the amplification process plays a role similar
to decoherence: This works, because the preparation procedure has
implemented the particular choice of measurement setting $\hat{x}$,
and to fully describe the initial state it is necessary that $|\psi_{sup}\rangle$
be expanded in terms of the eigenstates of $|x_{j}\rangle$, referred
to as the measurement basis. This defines particular interference
terms $\mathcal{I}nt$ in $Q(x,p,t_{0})$, which vanish in the  FBC.

\begin{figure}
\begin{centering}
\includegraphics[width=1\columnwidth]{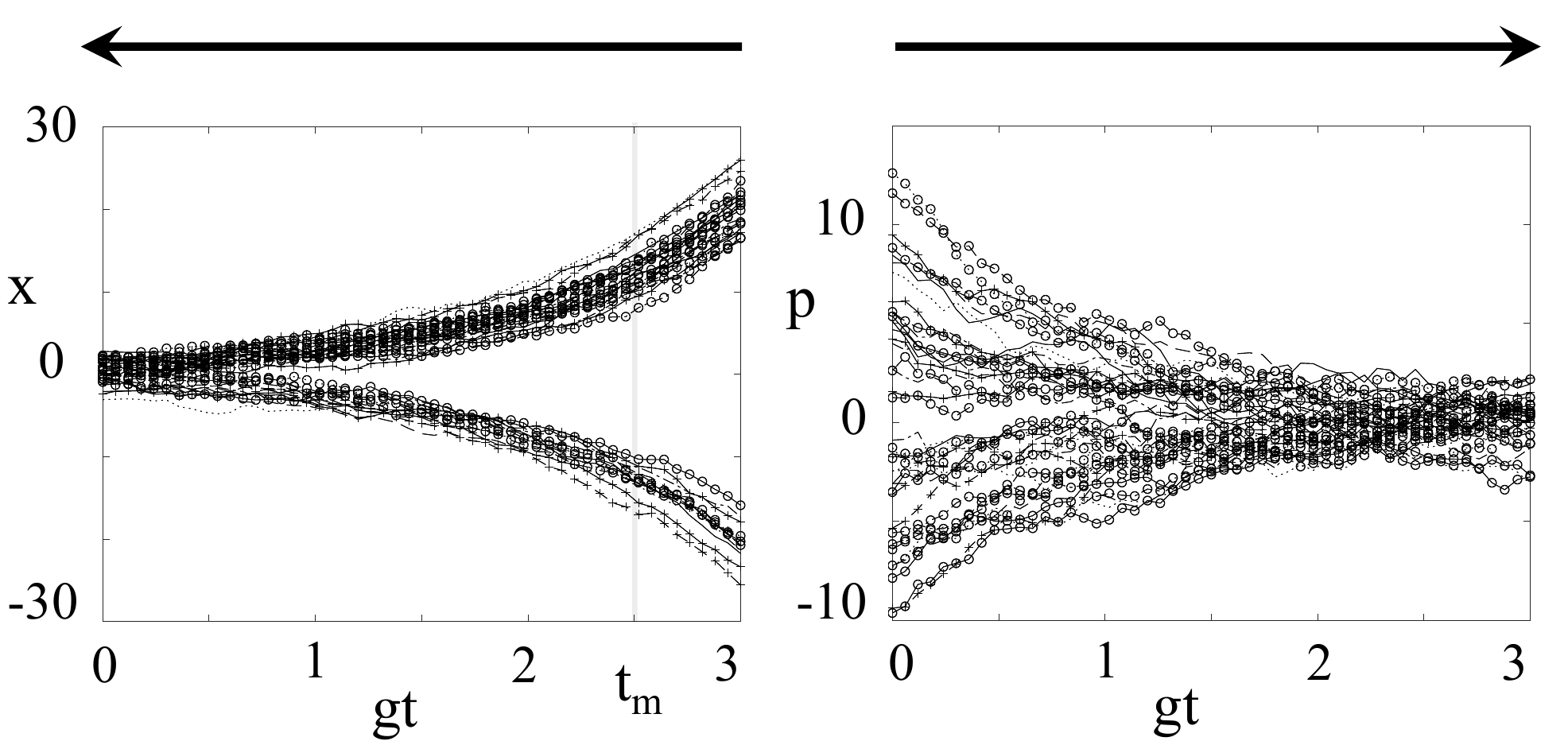}
\par\end{centering}
\caption{\textbf{\emph{Forward-backward stochastic simulation modeling quantum
measurement:}} Individual trajectories for $x$ and $p$ model a measurement
$\hat{x}$ on a system prepared in a superposition $|\psi_{sup}\rangle=\frac{1}{\sqrt{2}}(|x_{1}\rangle+i|-x_{1}\rangle)$
(Eq. (\ref{eq:sup-sq}) with $r=2$) of eigenstates of $\hat{x}$.
The system is amplified by $H_{amp}$ (Eq. (\ref{eq:ham-2-1})), where
$t_{0}=0$. The variable $x$ is amplified (left) and $p$ deamplified
(right). The trajectories for $x$ propagate in the negative-time
direction from a future boundary condition at time $t_{f}=3/g$; those
for $p$ propagate forward in time. Two branches for $x$ are evident
at time $t_{m}$ when the system is macroscopic.\label{fig:summary-pics}
Here, $x_{1}=0.8$, $G=e^{gt}$. (Refer Secs. II - \ref{sec:Forward-backward-stochastic-simu}.)}
\end{figure}

\begin{figure}
\begin{centering}
\par\end{centering}
\begin{centering}
\includegraphics[width=0.8\columnwidth]{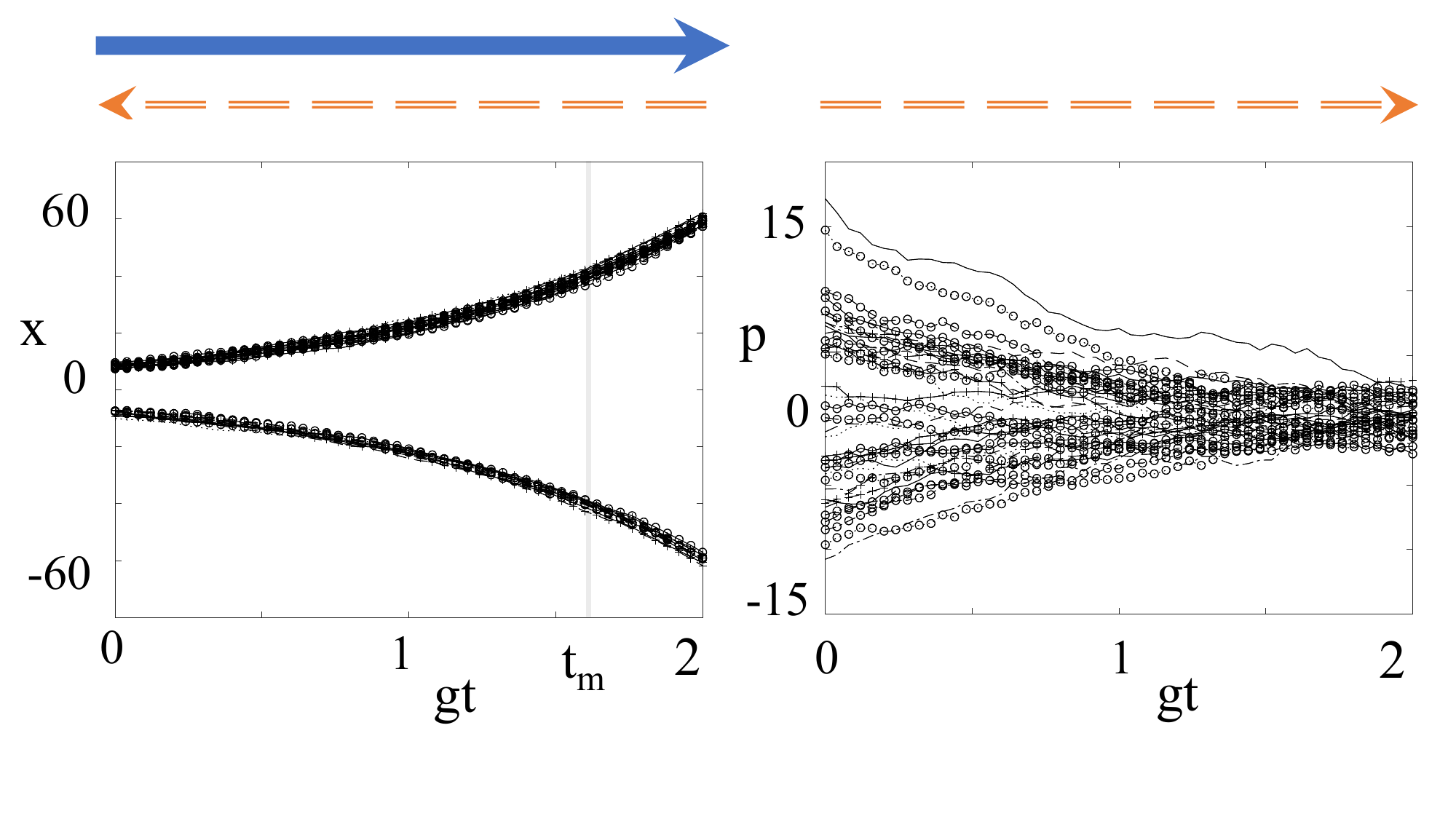}
\par\end{centering}
\begin{centering}
\includegraphics[width=0.9\columnwidth]{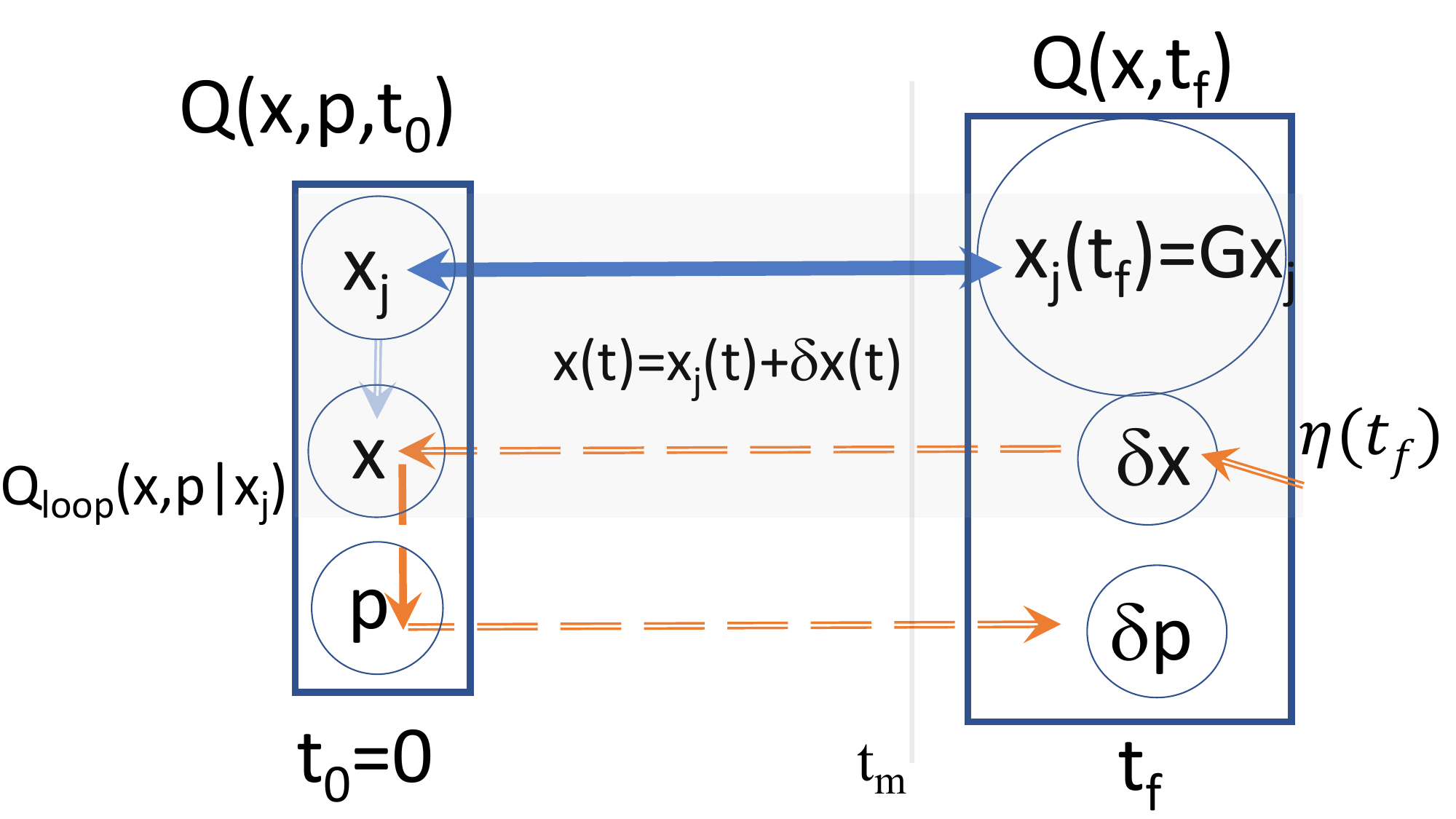}
\par\end{centering}
\caption{\textbf{\emph{Causal structure:}}\emph{ Top:} As for Fig. \ref{fig:summary-pics},
with $x_{1}=8$ and $t_{f}=2/g$.  Here, the system starts in a
superposition of macroscopically distinguishable states. Each branch
 follows the causal relation $x(0)\rightarrow Gx(0)$ (depicted
by the solid blue line). \emph{Lower:} Diagram of the causal relations
applied to the simulation. The Q function $Q(x,p,t_{0})$ describes
the system at time $t_{0}$.  A future boundary condition is determined
by the marginal $Q(x,t_{f})$ in $x$, after amplification for a time
$t_{f}$. A deterministic relation (solid two-way blue arrow) connects
each $x_{j}$ at time $t_{0}$ to the amplified value $Gx_{j}$ at
$t_{f}$. Gaussian noise $\delta x\equiv\eta(t_{f})$ enters at the
future boundary $t_{f}$.  A connection between forward and backward
trajectories $x(t)$ and $p(t)$ at time $t_{0}$ exists, which defines
a distribution $Q_{loop}(x,p|x_{j})$, giving rise to a ``hidden
loop'' (orange dashed lines). (Refer main text and Secs. \ref{sec:Forward-backward-stochastic-simu}
- \ref{sec:Causal-model-for} for details). \label{fig:The-causal-structure-1}}
\end{figure}

The second part of the future boundary condition for $x$ constitutes
a random microscopic input $\delta x\equiv\eta(t)$, the distribution
of which is a Gaussian $\mathcal{G}(0,\sigma_{vac}$), with a mean
of zero and a variance $\sigma_{vac}$ at the level of quantum vacuum.
 The fluctuation $\delta x$ has a constant average magnitude with
time $t$, which hence corresponds to and explains the ``hidden noise''
level in the Q function of the eigenstate $|x_{j}\rangle$. The result
is  the emergence in the simulations of macroscopic ``branches''
$\mathcal{B}_{j}$ $-$ distinct bundles of amplitudes $x(t)$, appearing
as ``lines'' with a width $\sim\delta x$ that correspond to a definite
eigenvalue $x_{j}$ (Figs. \ref{fig:summary-pics} and \ref{fig:The-causal-structure-1}).
Extending previous work \citep{drummond2020retrocausal,drummond2021objective,Friederich2021Introducing},
this leads to a model of reality to explain quantum measurement, from
which Born's rule follows.

The ``\emph{paradox of the future boundary condition}'' is hence
addressed by the hybrid nature of the boundary condition (Fig. 2).
There is no genuine backward causation, because the future inputs
$\delta x$ are random determined by the Gaussian $\mathcal{G}(0,\sigma_{vac})$,
being independent of the time $t_{f}$ at which the FBC is placed,
and not determined by any choice of setting or outcome $x_{j}$ at
the final time $t_{f}$. As we show in the second part of the paper,
this model can explain Bell nonlocality.

A main conclusion of this paper is that macroscopic realism (MR),
as well as no-signaling, follows naturally from the stochastic solutions
and the proposed model of reality. The deterministic relation $x_{j}\rightarrow Gx_{j}$
that forms part of the future boundary condition (FBC) dominates the
dynamics for the amplitudes $x(t)$, at the times $t_{m}$ when the
$x(t_{m})$ are macroscopic. The branches have emerged at time $t_{m}$
(Fig. 1), allowing definition of \emph{macroscopic variables} $\widetilde{\lambda}_{x}(t_{m})$,
the values of which represent the measurable property $x$ of a system
at a given time $t_{m}$, a property that can be detected. A main
conclusion of this paper is that there is \emph{no actual retrocausality},
because $\widetilde{\lambda}(t_{m})$ has a fixed value at the time
$t_{m}$. A change to its value \emph{requires a further local interaction},
which changes $\widetilde{\lambda}(t)$ at a \emph{future }time $t_{m2}>t_{m}$.
Macroscopic realism, and macroscopic causality (defined as causal
relations for the variables $\widetilde{\lambda}_{x}(t_{m})$, refer
Sec.I.B)), hold (Fig. 2). This is despite the first-glance apparent
``retrocausality'' of the model in the form of the backward-in-time
propagating solutions and the FBC. 

Consistent with these results, we are able to prove \emph{causal consistency},
as in absorber theory \citep{wheeler1945interaction,wheeler1949classical,cramer1986transactional}
and theories of closed time-like curves \citep{deutsch1991quantum,ringbauer2014experimental}.
Causal consistency is defined as the requirement that the joint density
of the amplitudes $\{x(t),p(t)\}$, when averaged over the ``branches'',
is given by the Q function $Q(x,p,t)$, at any time $t$ during the
evolution (Fig. 3). The proof is based on a conditional constraint
determined by $Q(x,p,t_{0})$ at the initial time. For a given branch
$\mathcal{B}_{j}$ (associated with eigenvalue $x_{j}$), the deterministic
relation $x_{j}\rightarrow Gx_{j}$ when inverted implies a set of
values around $x_{j}$ at time $t_{0}$.  A correlation between
the measured and complementary variables $x$ and $p$ at time $t_{0}$
allows a connection between backward and forward-propagating amplitudes,
that we refer to as a ``\emph{hidden loop}''.  The loop appears
for superpositions of $|x_{j}\rangle$, but not for mixtures, and
is hence a distinguishing quantum feature.

Following from this, we show in this paper how a probability distribution,
that we call $Q_{loop}(x,p,t_{0}|x_{j})$, for $x$ and $p$ at the
initial time $t_{0}$ conditioned on a given branch $\mathcal{B}_{j}$
(associated with an outcome $x_{j}$ for $\hat{x}$) can be defined
and calculated.  Perhaps surprisingly, (since the properties of the
inputs $\delta x$ at the FBC are unchanged with $t$) this \emph{postselected
distribution} $Q_{loop}(x,p,t_{0}|x_{j})$ is inferred by \emph{retrodiction}
(not retrocausality), by reversing the deterministic relation $x_{j}\rightarrow Gx_{j}$,
and using knowledge of the initial distribution $Q(x,p,t_{0})$. Calculation
shows that the amplitudes $x$ and $p$ of the distribution $Q_{loop}(x,p,t_{0}|x_{j})$
are too precisely defined to correspond to a quantum state. This leads
to a hidden-variable (HV) model for measurement on a superposition
state, and an analysis of the Schrödinger cat paradox (Fig. 4).

\begin{figure}
\begin{centering}
\includegraphics[width=1\columnwidth]{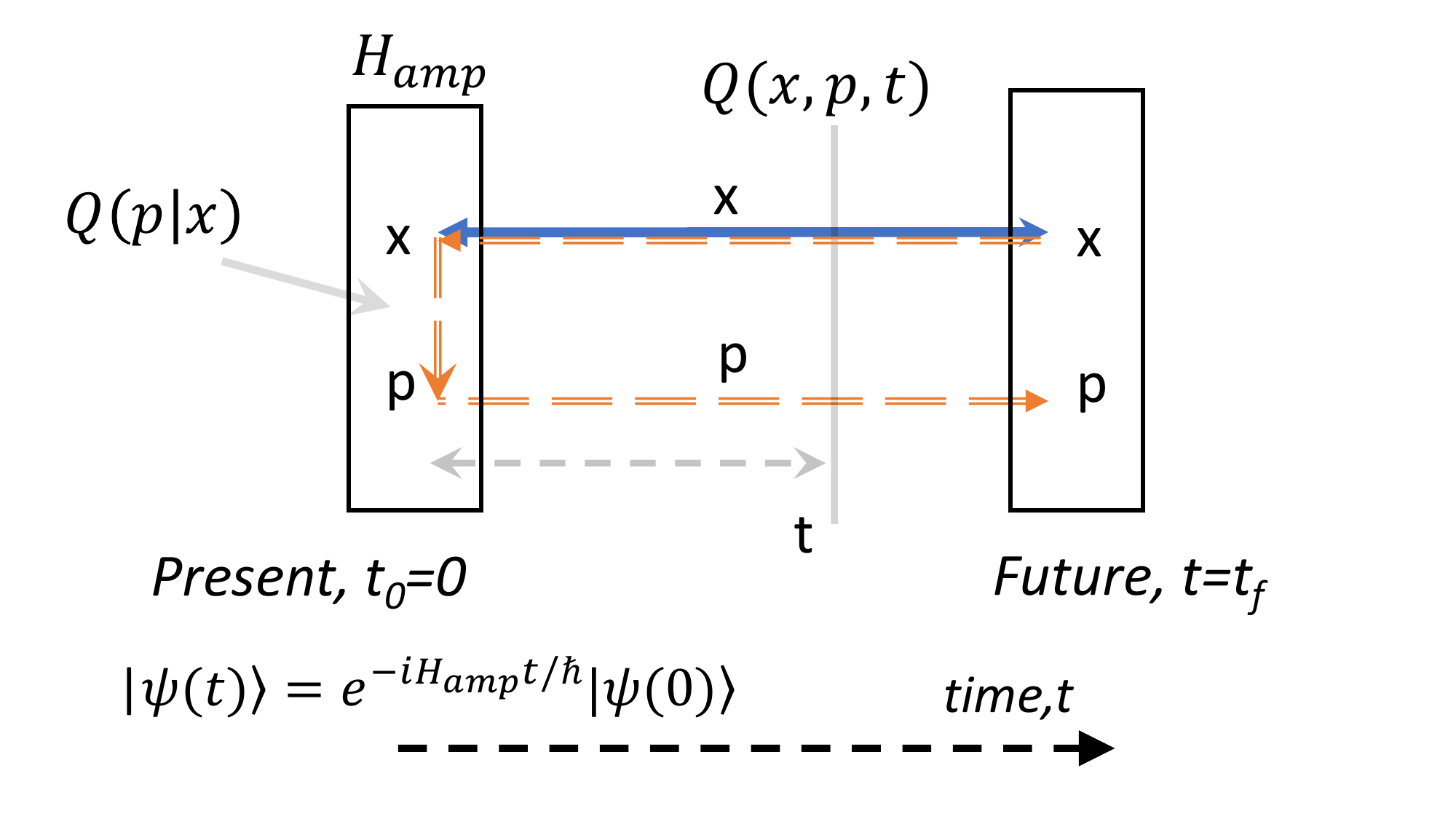}
\par\end{centering}
\caption{\textbf{\emph{Test of causal consistency:}} How can a future boundary
condition be consistent with the causal behavior of the Q function?
The diagram illustrates the equivalence of the probability density
of the forward- and backward-propagating $p(t)$ and $x(t)$ to the
Q function $Q(x,p,t)$, which defines the quantum state $|\psi(t)\rangle$.
The system evolves under amplification $H_{amp}$.  The equivalence
 is true for all $t_{f}$, even when $t_{f}$ is far into the future.
That the state at time $t$ does not depend on the future inputs is
explained by the causal structure of the simulation. An experimental
test is feasible. \label{fig:causal-consistency-2-1} (Refer Sec.
\ref{sec:Causal-model-for}.)}
\end{figure}

\begin{figure}
\begin{centering}
\includegraphics[width=0.8\columnwidth]{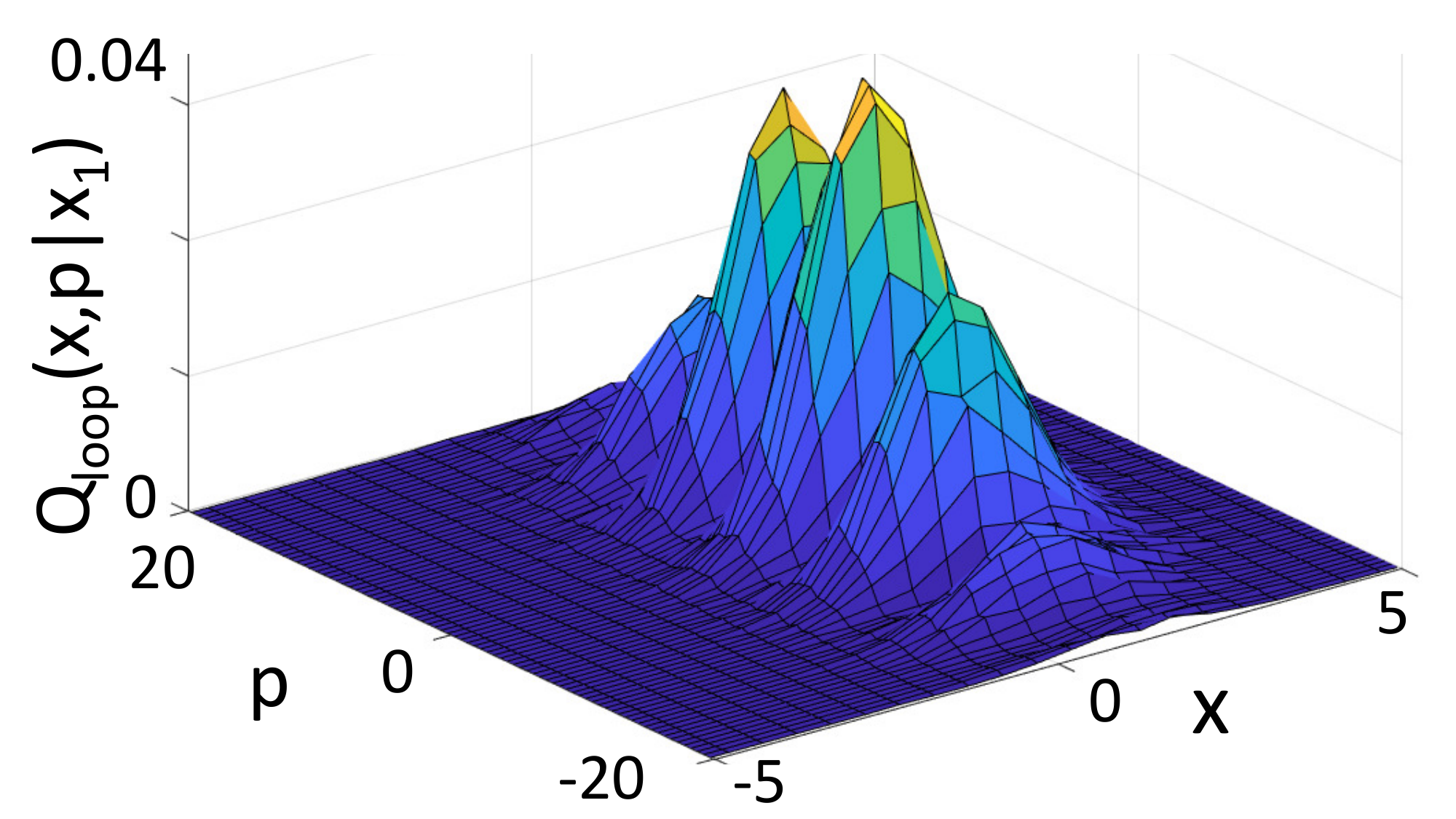}
\par\end{centering}
\caption{\textbf{\emph{Schrodinger's cat paradox:}} How can cat states be consistent
with macroscopic realism? A system is prepared in the superposition
$|\psi_{sup}\rangle=\frac{1}{\sqrt{2}}(|x_{1}\rangle+i|-x_{1}\rangle$
of eigenstates $|x_{j}\rangle$ of $\hat{x}$ . The measurement of
``which state the cat is in'' proceeds by amplification of $\hat{x}$,
via $H_{amp}$ (Fig. \ref{fig:The-causal-structure-1}). The simulation
enables evaluation of $Q_{loop}(x,p|x_{1})$, the postselected distribution
inferred for the ``cat'' at the initial time, given that the final
outcome is $x_{1}$. It can be shown that $Q_{loop}(x,p|x_{1})$
is not equivalent to a quantum state $|\psi\rangle$ (refer Sec. \ref{secQmodel-of}),
highlighting the argument for the incompleteness of quantum mechanics
put forward by Schrödinger \citep{schrodinger1935gegenwartige}. \label{fig:scat-1-1}Here,
$x_{1}=-x_{2}=1$.}
\end{figure}
\textbf{Entanglement:} In Part II of this paper, the results are extended
to treat Einstein-Podolsky-Rosen (EPR) entanglement and Bell nonlocality.
This enables construction of a Q-based hidden-variable (HV) model
for nonlocality, and also a full examination of quantum measurement.
In particular, we provide a model for \emph{projection} (``collapse
of the wave-function'') by analyzing the coupling of a system to
a meter. We demonstrate forward-backward stochastic simulations of
entangled systems, $A$ and $B$, described by a Q function $Q(\bm{\lambda},t_{0})$
where $\bm{\lambda}\equiv(x_{A},p_{A},x_{B},p_{B})$ are phase-space
amplitudes.

In order to examine entanglement, we allow for adjustable measurement
settings, $\theta$ and $\phi$, by considering measurement of quadrature
phase amplitudes $\hat{x}_{\theta A}=\hat{x}_{A}\cos\theta+\hat{p}_{A}\sin\theta$
and $\hat{x}_{\phi B}=\hat{x}_{B}\cos\phi+\hat{p}_{B}\sin\phi$ at
sites $A$ and $B$ respectively. The $\hat{x}_{K}$ and $\hat{p}_{K}$
are quadrature amplitudes for the system $K\in\{A,B\}$, in analogy
to position and momentum. The measurement settings are adjusted by
independent local interactions $H_{\theta}^{A}$ and $H_{\phi}^{B}$.
We show that these interactions correspond to\emph{ deterministic
causal transformations} of the local amplitudes $x_{K}$, $p_{K}$,
which precede the independent amplifications $H_{amp}^{A}$ and $H_{amp}^{B}$
that complete the measurements of $\hat{x}_{\theta A}$ and $\hat{x}_{\phi B}$
made on each system. For a Bell test, we consider binary outcomes
$+1$ and $-1$ defined by the sign of the quadrature phase amplitude
($\hat{x}_{\theta A}$ or $\hat{x}_{\phi B}$), measured at each of
site, with systems $A$ and $B$ spacelike-separated. We prove the
equivalence of the dynamics of the Q function $Q(\bm{\lambda},t)$
with the forward-backward dynamics of the amplitudes $\bm{\lambda}$:
 Causal consistency is demonstrated in the bipartite case (refer
Fig. 5 and \emph{Result IX.7} of this paper).

Since the forward-backward simulations involve a future boundary condition
(FBC), it might appear that the Bell violations are ``put in by hand'':
The ``put-in-by-hand'' argument is as follows: The final state at
time $t_{f}$ that determines the boundary condition of the backward
stochastic differential equation is a bipartite amplified state that
already possesses the correlations violating the Bell inequality.
Hence it would seem that nothing can be learned about the origin of
the Bell correlations.

\begin{figure}
\begin{centering}
\includegraphics[width=1\columnwidth]{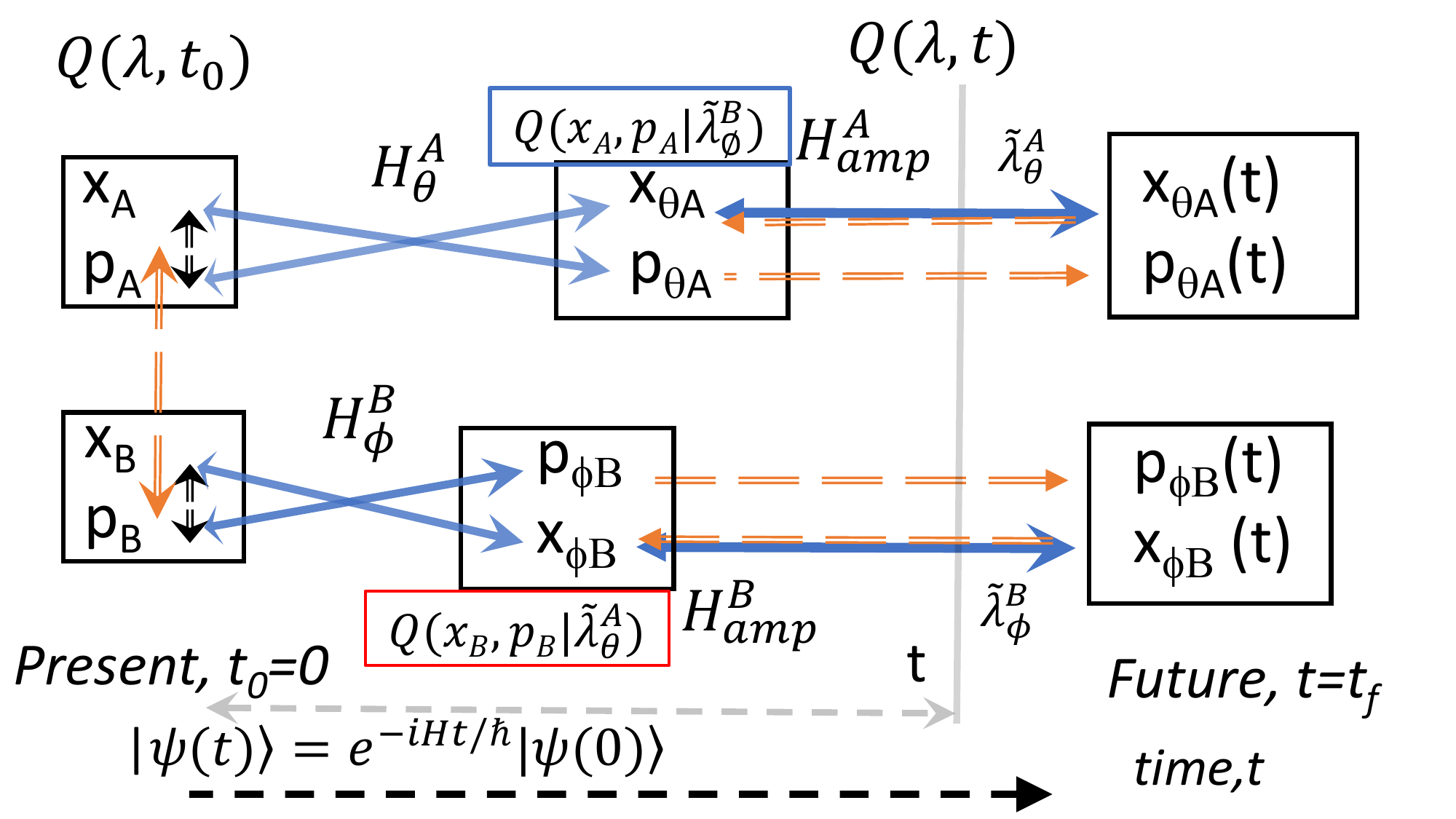}
\par\end{centering}
\caption{\textbf{\emph{Causal consistency for Bell-nonlocal systems $A$ and
}}\emph{$B$}\textbf{\emph{.}} The Q function $Q(\lambda,t)$ at time
$t$ is determined by $|\psi(t)\rangle$ which evolves causally in
the positive-time direction, in accordance with Hamiltonians $H_{\theta}^{A}$
and $H_{\phi}^{B}$ that determine the measurement settings, and $H_{amp}^{A}$
and $H_{amp}^{B}$ that amplify $\hat{x}_{\theta A}$ and $\hat{x}_{\phi B}$.
Here, $\mathbf{\lambda}=(x_{A},p_{A},x_{B},p_{B})$. The probability
density $Q(\lambda,t)$ is equivalent to that of the amplitudes $x_{A}(t)$,
$p_{A}(t)$, $x_{B}(t)$, $p_{B}(t)$ (and $x_{\theta K}(t),p_{\theta K}(t)$),
as defined by the forward-backward stochastic equations. The $x_{\theta K},p_{\theta K}$
are deterministic functions of $x_{K}$, $p_{K}$ for each $K\in\{A,B\}$
(depicted by two-way crossed blue arrows). The $Q(\lambda,t_{0})$
exhibits local correlation between $x_{K}$ and $p_{K}$ (short vertical
black dashed lines), as well as a correlation between $A$ and $B$
(vertical orange dashed line). \label{fig:epr-causal-consistency-1}
Branches denoted by $\widetilde{\lambda}_{\theta}^{A}$ and $\widetilde{\lambda}_{\phi}^{B}$
emerge on amplification, where here $\widetilde{\lambda}_{\theta}^{A}$
and $\widetilde{\lambda}_{\phi}^{B}$ are variables that assume the
values $x_{\theta j}^{A}$ and $x_{\phi k}^{B}$ representing the
outcomes of $\hat{x}_{\theta A}$ and $\hat{x}_{\phi B}$, respectively.
The distributions $Q(x_{A},p_{A}|\widetilde{\lambda}_{\phi}^{B})$
and $Q(x_{B},p_{B}|\widetilde{\lambda}_{\theta}^{A})$ determine the
probabilities of outcomes at $A$ given the outcome $\widetilde{\lambda}_{\phi}^{B}$
at $B$, and the probabilities of outcomes at $B$ given the outcome
$\widetilde{\lambda}_{\theta}^{A}$ at $A$. These are deduced from
the backward trajectories, using knowledge of $Q(\lambda,t_{0})$
and deterministic relations, and explain the Bell nonlocality. Being
mutually consistent, Lorentz-invariance paradoxes are avoided. (Refer
to Sec. IX for details.)}
\end{figure}

We explain in this paper that the violations (in the model of reality
proposed here) are\emph{ }not ``put in by hand'' at the time $t_{f}$,
but emerge at an earlier time $t_{Bell}$, through the dynamics of
the interactions (\emph{Results VII.5, IX.4 and IX.}8).  Similar
to the conclusions of Part I of the paper, the paradox of the future
boundary condition is avoided (Fig. \ref{fig:epr-causal-consistency-1})
and a genuine retrocausality, where the future impacts the measurable
past properties, does not occur. Moreover, there is no ``spooky action''
where the change of setting at one location ($B$) can change the
existing property at the other $A$. For entangled states, the setting
change at $B$ can lead to correlations that violate a Bell inequality
for \emph{future} outcomes at $A$,\emph{ if there is a further change
of setting} at $A$. Fine-tuning is explained consistently with a
hidden-variable model, which includes ``hidden'' noise and interference.

These conclusions are arrived at from three different perspectives,
which we summarize below for the reader's convenience. First, a hidden
variable model can be constructed based on the Q function (\emph{Result
VII.9}). The hidden variables (HV) are the phase-space amplitudes
$\bm{\lambda}\equiv(x_{A},p_{A},x_{B},p_{B})$; the Q function $Q(\bm{\lambda},t_{0})$
is the associated probability distribution. Using Bell's formalism,
the probability of an outcome $+1$ at both sites is 
\begin{equation}
P_{++}(\theta,\phi)=\int Q(\lambda,t_{0})P(++|\theta,\phi,\lambda)d\lambda\label{eq:bell-2}
\end{equation}
where $P(++|\theta,\phi,\lambda)$ is the probability of an outcome
$+1$ at both sites conditioned on the values $\lambda$. Since we
prove causal consistency, we can show that the distribution $Q(\bm{\lambda},t_{0})$
is independent of the future settings, $\theta$ and $\phi$ (Fig.
5). This demonstrates there is no retrocausality, where the future
affects the present ``state'' of the system, as defined by the Q
function. In this Q-based HV model, we identify that the Bell violations
arise from a breakdown of Bell's locality assumption (\emph{Result
IX.1}0).

The second perspective is to examine the macroscopic properties $\widetilde{\lambda}$
associated with the branches defined from the solutions (``trajectories'')
of the forward-backward stochastic equations. We conclude that the
Q-based HV model satisfies a \emph{partial locality}, which applies
to the systems after settings are established. Such weak forms of
local realism were proposed in Ref. \citep{fulton2024alternative,Fulton2024Weak}
and shown to be not contradicted by Bell's theorem. This leads to
a main result of this paper, that the Q-based model of reality is
consistent with \emph{weak local realism}.   We first demonstrate
consistency with the three premises of \emph{Weak Macroscopic Realism}
(wMR) \citep{fulton2024alternative,joseph2024macroscopic,Fulton2024Weak,hatharasinghe2023macroscopic,thenabadu2022bipartite}.
We consider both systems $A$ and $B$ to be at rest in a frame $F$.

\textbf{(1)} The first premise specifies macroscopic realism: Macroscopic
properties $\widetilde{\lambda}_{x}^{A}(t)$ for a system $A$ (say)
exist\emph{ once the local settings are fixed} and after amplification,
at a time $t=t_{m}$ (Fig. 1).

\textbf{(2)} The second premise specifies a partial locality: The
macroscopic value $\widetilde{\lambda}_{x}^{A}(t)$ cannot be changed
by a spacelike-separated event occurring after $t_{m}$, as in a subsequent
change of setting $\phi$ at $B$. Once the setting is fixed at $A$,
only a further change of setting at $A$ can change the value of $\widetilde{\lambda}_{x}^{A}(t)$,
which requires further dynamics.

\textbf{(3)} The third premise concerns projection: Once the value
$\widetilde{\lambda}_{x}^{A}(t_{mA})$ of the outcome (as in \textbf{(1)})
for system $A$ is determined, at a time $t_{mA}$,  consistency
with the initial state at time $t_{0}$ may imply restrictions on
outcomes at system $B$. Details are given below in Sec. I.B

Examination of the simulations reveals that the weak local realistic
premises can be generalized to apply to systems after the measurement
settings are implemented, but that are not necessarily macroscopic.
Hidden variables $\{\lambda_{K}\}$ can be identified  that define
probabilities for  outcomes at site $K\in\{A,B\}$. In this paper,
we demonstrate the consistency of the simulations with \emph{Weak
Local Realism} (wLR) \citep{fulton2024alternative}, defined below
in Sec. 1.B.

The weak local realistic Premises are useful for understanding EPR
correlations. We deduce a consequence for the EPR paradox: there
can be predetermined values (weak ``elements of reality'') for measurements
at $B$, prior to the choice of setting $\phi$ at $B$, provided
the settings are fixed at $A$.  In this paper, we demonstrate these
``elements of reality'' in a forward-backward stochastic simulation
of EPR correlations, which illustrates the emergence of macroscopic
correlated EPR ``branches''. The analysis may explain recent experiments,
which detect EPR correlations for macroscopic BEC systems \citep{colciaghi2023einstein}.

The third perspective given in this paper is to provide a model for
\emph{projection}. This allows us to understand the workings of a
meter.  Extending the analysis of the postselected distribution
defined in Part I of the paper, we evaluate the probability distribution
$Q(x_{B},p_{B}|x_{j}^{A})$ inferred for system $B$ conditioned on
a branch $\mathcal{B}_{j}$ (with outcome $x_{j}^{A}$) of a second
system $A$, correlated with $B$. The $Q(x_{B},p_{B}|x_{j}^{A})$
describes the ``projected state'' for system $B$, modeling the
``collapse of the wave function'' at $B$. We show that the ``collapse''
arises as information is lost about complementary variables of the
meter $A$. Limits on the timing of projection are deduced. 

We ask the following question about a meter. If the measurement at
$A$ allows inference of a state, or outcome, at $B$, then does the
measurement at $A$ \emph{cause} the outcome at $B$, through some
``spooky action-at-a-distance'', or is the predicted outcome at
$B$ due to correlations present at an earlier time? We show, by constructing
a local HV theory consistent with the simulations, that, provided
the settings of the meter $A$ are fixed, the latter is correct.

How then can a violation of a Bell inequality occur?  The mechanism
relates to \emph{hidden interference} that arises from the entangled
nature of the initial state. If the hidden-variable amplitudes $\lambda$
given in (\ref{eq:bell-2}) were directly measurable, there would
be no Bell violation, since the transformations $H_{\theta}^{K}$
and $H_{amp}^{K}$ are local. However, the amplitudes are not directly
measurable:  This is evident from the Q function $Q(\lambda,t_{0})$
which possesses ``hidden'' interference terms $\mathcal{I}nt_{AB}$.
These terms are defined with respect to the initial measurement basis,
determined by the preparation of the state $Q(\lambda,t_{0})$. We
choose the initial basis so that $\theta=\phi=0$ at time $t_{0}$.
Extending Result III.1 from Part I of the paper, we prove (\emph{Result
VII.2}) that the $\mathcal{I}nt_{AB}$ terms are not amplified by
$H_{amp}$  and are hence not detected by measurement of $\hat{x}_{A}$
and $\hat{x}_{B}$. Where changes of settings take place however,
the Q function changes: The $\mathcal{I}nt_{AB}$ terms change, and
can contribute to measurable probabilities, in a way that violates
the Bell inequalities (\emph{Result VII.5}). We show that this can
only happen if certain conditions are satisfied: The settings $\theta$
and $\phi$ must change at both locations $A$ and $B$, mixing $\hat{x}_{K}$
with $\hat{p}_{K}$ ($K\in\{A,B\}$) locally (\emph{Result VII.7}),
so that there is a ``loop'' between the backward and forward trajectories
of the same system $K$ (refer to Definitions below).

More detail is evident from the stochastic solutions (``trajectories''),
where we find the Bell nonlocality can be explained by the projection
mechanism.  A hidden interference effect comes into play if we change
the setting $\phi$ at $B$, to measure $\hat{x}_{\phi B}$, keeping
the setting at $A$ fixed. The projected distribution $Q(x_{A},p_{A}|x_{\phi k}^{B})$
at $A$ conditioned on a branch with outcome $x_{\phi k}^{B}$ at
$B$ can be inferred, by \emph{retrodiction}, at $B$, based solely
on a deterministic relation $x_{\phi k}^{B}\leftrightarrow Gx_{\phi k}^{B}$,
and a further set of deterministic relations $\mathcal{D}$ that specify
correlations between $\hat{x}_{\phi B}$ and $\hat{x}_{A}$, as well
as knowledge of the setting $\phi$ at $B$ and of the initial state
$Q(\lambda,t_{0})$. This is surprising because the projected state
for $A$ is sufficient to explain the Bell violations (\emph{Result
IX.4}). The $Q(x_{A},p_{A}|x_{\phi k}^{B})$ defines the state at
$A$, consistent with any future adjustment of setting $\theta$ at
$A$. For entangled states, the projected state $Q(x_{A},p_{A}|x_{\phi k}^{B}$)
contains extra ``hidden'' interference terms $\mathcal{I}_{A}$,
as in a quantum superposition, if $\phi$ satisfies $\phi\neq n\pi$
($n$ is an integer) $-$ but these terms are\emph{ }undetectable,
with the setting at $A$ is fixed ($\theta=0$). Hence, the predictions
for outcomes of $\hat{x}_{A}$ and $\hat{x}_{\phi B}$ are consistent
with a local HV model. We show there is consistency with \emph{no-signaling}:
With the setting fixed at $A$, a change of setting at $B$ does not
influence the outcome at $A$. The interference $\mathcal{I}_{A}$
of system $A$ can be detected however, if there is a \emph{further
change} of setting to $\theta\neq n\pi$ at $A$, and this leads to
correlations that can violate a Bell inequality.  The conclusion
is that Bell nonlocality emerges over the time span $t_{Bell}$, as
settings are changed at both sites, consistent with the premises of
weak local realism.

A final question examined in this paper is how to resolve an apparent
inconsistency with Lorentz-invariance, since for spacelike-separated
systems, the time-order of setting-changes may be reversed, in a different
frame. In one frame, the outcomes for $A$ for any setting are constrained
conditionally on the outcomes of $B$; in another frame, the outcomes
for $B$ are constrained by those of $A$. We suggest this is resolved
by the symmetry of the setting-changes, which must occur at both sites,
for Bell violations to occur. There are two projected states, one
for system $A$ and one for system $B$, which are each defined based
on one setting-change, but are unchanged by the time-order of the
setting-changes. The \emph{projected states} \emph{are} \emph{mutually
consistent}, with each other, and with the initial distribution $Q(\lambda,t_{0})$
(Fig. 5).

To conclude, we explain how experimental tests of the compatibility
of the forward-backward simulations with quantum predictions are possible,
elucidating paradoxes (Figs. 1-5). Below is a summary of the layout
of the paper, followed by a list of Definitions associated with the
main concepts.

\subsection{Layout of paper}

In Part I of the paper (Secs. II-V), we treat \emph{quantum measurement}.
In Sec. \ref{sec:Measurement-model}, we review the measurement Hamiltonian
$H_{amp}$, the Q function equations of motion and the equivalent
forward-backward stochastic dynamics. The path-integral and stochastic
equivalence theorems on which the paper is based are derived in Appendix
A, along with the numerical verification of the computed distribution
in Appendix B. In Sec. \ref{sec:Forward-backward-stochastic-simu},
the equations modeling a measurement $\hat{x}$ on a system in a superposition
$|\psi_{sup}\rangle$ of eigenstates of $\hat{x}$ are solved.  The
Q-based model of reality is explained in Sec. \ref{secQmodel-of}.
Born's rule is derived and the role of the Wigner function explained.
Justification of macroscopic realism and macroscopic causality is
presented in Sec. \ref{secQmodel-of}.C. The derivation of the postselected
distribution based on an outcome of a measurement, and explanation
of a ``hidden loop'' is given in Secs. \ref{secQmodel-of}.D and
E. Hidden-variable models $\mathcal{P}_{sup}$ and $\mathcal{P}_{det}$
for measurement on a superposition state and the justification of
weak local realism are presented in Secs. \ref{secQmodel-of}.F and
G. A causal model for measurement and a proof of causal consistency
are given in Sec. \ref{sec:Causal-model-for}.

In Part II of the paper, we treat \emph{nonlocality and entanglement}.
In Sec. \ref{sec:Continuous-variable-EPR-entangle}, we present the
simulation for continuous-variable EPR correlations. The simulation
of Bell nonlocality is explained in Sec. \ref{sec:CV-Bell-nonlocality}.A,
along with an analysis of the mechanism for Bell nonlocality based
on the evolution of the Q function in Sec.\ref{sec:CV-Bell-nonlocality}.B.
The different sorts of hidden loops that can occur for entangled states,
and explanation of the further hidden variable (HV) models $\mathcal{P}_{asym,A}$,
$\mathcal{P}_{B}$, $\mathcal{P}_{det,B}$ and $\mathcal{P}_{bell}$
that can be deduced from the simulation are given in Secs. \ref{sec:CV-Bell-nonlocality}.C
and D. A model for projection and collapse of the wave function is
presented in Sec. VIII, with causal relations for projection and Bell
nonlocality given in Sec. \ref{sec:Causal-structure-bell}. It is
shown through a series of Results in Secs. \ref{sec:CV-Bell-nonlocality},
\ref{sec:EPR-proj}, \ref{sec:CV-Bell-nonlocality} and \ref{sec:Causal-structure-bell}
that the simulations are consistent with three premises referred to
as \emph{weak macroscopic realism} (wMR). Based on the HV models,
we extend (in Secs. \ref{secQmodel-of}, \ref{sec:CV-Bell-nonlocality}
and \ref{sec:Causal-structure-bell}) to explain consistency with
\emph{weak local realism }(wLR). Proofs of the mutual consistency
of the projected states and of the causal consistency of the forward-backward
stochastic equations for the entangled system (as in Fig. 5) are given
in Sec. \ref{sec:Causal-structure-bell}.B, Sec. \ref{sec:Causal-structure-bell}.
C and Appendix H. The distinction between the hidden variables of
the Q-based model and those of Bell's local hidden variables is analyzed
in Sec. \ref{sec:Causal-structure-bell}, along with an explanation
of how Lorenz-invariance paradoxes may be resolved. Fine-tuning is
explained in Appendix J. In the Conclusion (Sec. \ref{sec:Conclusions}),
potential experiments are outlined.

\subsection{Definitions}

In Part I of paper, we consider measurement on a single-mode field,
and define in a rotating frame the quadrature phase amplitude $\hat{x}=\hat{a}+\hat{a}^{\dagger}$
where $\hat{a}^{\dagger}$ and $\hat{a}$ are the standard boson creation
and destruction operators \citep{yurke1986generating}. The complementary
observable is $\hat{p}=(\hat{a}-\hat{a}^{\dagger})/i$. Measurement
of $\hat{x}$ is treated as amplification of $\hat{x}$, by a factor
$G$, modeled by the interaction Hamiltonian $H_{amp}=\frac{i\hbar g}{2}\left[\hat{a}^{\dagger2}-\hat{a}^{2}\right]$
\citep{drummond2020retrocausal}.

In Part II of the paper, we treat entangled systems, $A$ and $B$,
described by a two-mode Q function. We consider quadrature phase amplitudes
$\hat{x}_{\theta A}=\hat{x}_{A}\cos\theta+\hat{p}_{A}\sin\theta$
and $\hat{x}_{\phi B}=\hat{x}_{B}\cos\phi+\hat{p}_{B}\sin\phi$ measured
at $A$ and $B$ respectively. Here, $\hat{x}_{K}$ and $\hat{p}_{K}$
(where $K=A,B$) are the $\hat{x}$ and $\hat{p}$ quadrature phase
amplitudes of system $K$.

\textbf{Definition (1): Q function:} The Q function of a single-mode
field in a quantum state $|\psi\rangle$ is defined as $Q(\alpha)=\frac{1}{\pi}|\langle\alpha|\psi\rangle|^{2}$
where $|\alpha\rangle$ is a coherent state \citep{Husimi1940}. The
Q function uniquely defines the quantum state. The phase-space coordinates
$\bm{\lambda}$ are the real coordinates $x$ and $p$, given by $\alpha=(x+ip)/2$.
In this paper, we denote the Q function of the quantum state at a
time $t$ as $Q(x,p,t)$.

The Q function of a two-mode field (modes $A$ and $B$) is defined
as $Q(\alpha,\beta)=\frac{1}{\pi^{2}}|\langle\beta|\langle\alpha|\psi\rangle|^{2}$
where $|\alpha\rangle$, $|\beta\rangle$ are coherent states for
$A$ and $B$ respectively. The phase-space coordinates $\bm{\lambda}$
are the real coordinates $x_{K}$ and $p_{K}$, given by $\alpha=(x_{A}+ip_{A})/2$
and $\beta=(x_{B}+ip_{B})/2$ . We denote the Q function of the quantum
state at a time $t$ as $Q(\bm{\lambda},t)$ where $\bm{\lambda}=(x_{A},p_{A},x_{B},p_{B})$.

\textbf{Definition (2): Forward-backward stochastic simulation:}\emph{
}Consider the equations $\frac{dp}{dt}=-gp+\xi_{p}$ and $\frac{dx}{dt_{-}}=-gx+\xi_{x}$
where $t_{-}=-t$, and $\xi_{x}(t)$, $\xi_{p}(t)$ are Gaussian noise
terms and $g$ is real \citep{drummond2020retrocausal,drummond2021time}.
The boundary conditions are determined by distributions given by
$Q$ functions which are sampled. The solutions $x(t)$ and $p(t)$
differ for each run. We refer to the $x(t)$ and $p(t)$ as ``\emph{trajectories}'',
and to the set of runs as a \emph{forward-backward stochastic equation
(FBSE) simulation}.

\textbf{Definition (3).} \textbf{Eigenstate of $\hat{x}$:} The eigenstate
of $\hat{x}$ is denoted $|x_{j}\rangle$ and is defined as the squeezed
state $|\frac{x_{j}}{2},r\rangle_{sq}=D(\frac{x_{j}}{2})S(r)|0\rangle$
in the limit where $r\rightarrow\infty$, where $x_{j}$ and $r$
are real. Here, $|0\rangle$ is the vacuum state of the field and
$D(\beta_{j})=e^{\beta_{j}\hat{a}^{\dagger}-\beta_{j}^{*}\hat{a}}$
and $S(z)=e^{\frac{1}{2}(z^{*}\hat{a}^{2}-z\hat{a}^{\dagger2})}$.

\textbf{Definition (4): Hidden Interference:} Consider the expansion
$|\psi_{sup}\rangle=\sum_{j}c_{j}|x_{j}\rangle$ of a quantum state
in terms of the measurement basis of $\hat{x}$ i.e. the eigenstates
of $\hat{x}$. Here, $c_{j}$ are probability amplitudes. The corresponding
Q function $Q_{sup}(x,p,t)$ is a sum of type $\sum_{j}|c_{j}|^{2}Q_{j}(x,p)+\mathcal{I}nt$
where $Q_{j}(x,p)$ is the Q function of $|x_{j}\rangle$. We will
refer to the extra terms $\mathcal{I}nt$ as ``hidden interference''
terms. These terms vanish in the marginal distribution defined as
$Q_{sup}(x,t)=\int dpQ_{sup}(x,p,t)$, in the limit of $r$ large
where $|x_{j}\rangle$ are eigenstates of $\hat{x}$, and are not
amplified by $H_{amp}$.

\textbf{Definition (5): Branches and the time $t_{m}$:} We will consider
the forward-backward stochastic simulation of the amplification $H_{amp}$
of $\hat{x}$ on the system prepared in the superposition $|\psi_{sup}\rangle$
defined in Definition (4). With sufficient time of evolution when
the system has become macroscopic, at a time we denote by $t_{m}$,
distinct groups of amplitudes $x(t_{m})$ emerge. Each group can be
identified with a distinct eigenvalue $x_{j}$. We refer to such a
group of amplitudes as the \emph{branch }$\mathcal{B}_{j}$ associated
with the eigenvalue $x_{j}$. The group of amplitudes $\mathcal{B}_{j}$
of a single branch will be associated with a single eigenvalue $x_{j}$.
In this paper, the branches are labelled by a variable $\widetilde{\lambda}_{x}$,
the value of which indicates the eigenvalue associated with the branch.
The value represents a measurable property of the system, at the time
$t_{m}$.

\textbf{Definition (6a): Weak Macroscopic Realism: First Premise (wMR(1)):}\emph{
}Consider a system $A$ that has $N$ (where $N\geq2$) macroscopically
distinct states available to it e.g. a system prepared at a time $t_{m}$
in a superposition of $N$ macroscopically distinct states. Suppose
a measurement $\hat{O}$ with $N$ discrete outcomes can distinguish
between the macroscopic states. For example, where $N=2$, we consider
two outcomes, $+1$ and $-1$, which identify and distinguish the
two states. The measurement $\hat{O}$ can be macroscopically coarse-grained,
as the requirement is only to distinguish between the states.

Macroscopic realism (MR) posits that the system, prior to the measurement
$\hat{O}$, is in \emph{one and only one }of the macroscopic states
\citep{leggett1985quantum}. To be more specific, MR posits that the
outcome of the measurement $\hat{O}$ is determined at the time $t_{m}$.
The system can be assigned a variable $\widetilde{\lambda}$, the
value of which indicates the outcome of $\hat{O}$ if the measurement
$\hat{O}$ takes place. Where $N=2$, the value of $\widetilde{\lambda}$
can be $+1$ or $-1$.

To define \emph{weak macroscopic realism} (wMR) \citep{thenabadu2022bipartite,hatharasinghe2023macroscopic,joseph2024macroscopic,fulton2022argument,fulton2024alternative,Fulton2024Weak},
we further qualify the definition of MR, to specify that at the time
$t_{m}$, any operations $U$ needed to fix measurement settings (e.g.
$\theta$ in the measurement where $\hat{O}\equiv\hat{x}_{\theta}=\hat{x}\cos\theta+\hat{p}\sin\theta$)
have been carried out. In this paper, we consider the system prepared
in a superposition $(|x_{1}\rangle+|-x_{1}\rangle)/\sqrt{2}$ of eigenstates
of $\hat{x}$, where $x_{1}$ is real and $|x_{1}|\rightarrow\infty$.
Here, $\hat{O}$ is defined as the sign of the outcome of $\hat{x}$.
To apply wMR, we assume that the operations $U$ have been implemented,
so that $\theta=0$. The amplification $H_{amp}$ of the system will
complete the measurement of $\hat{x}$. 

\emph{Comment: }The terminology ``weak'' is used to distinguish
alternative definitions based on stronger assumptions, such as in
classical mechanics where MR is defined for the system as it exists
\emph{prior} to the operations $U$. The stronger assumption of MR
is falsifiable \citep{thenabadu2020testing,thenabadu2022bipartite,jeong2009failure}.
The definition is also ``weaker'' than that of macro-realism, defined
by Leggett and Garg in Ref. \citep{leggett1985quantum}, which makes
the additional assumption of noninvasive measurability.

\textbf{Definition (6b): Weak Macroscopic Realism: Second Premise
(wMR(2)): }Consider two spacelike-separated systems $A$ and $B$
both stationary in a frame $F$. The systems may be entangled. Consider
the macroscopic system $A$ at the time $t_{m}$ described in Definition
(6a) of wMR(1), which has $N$ macroscopically distinct states available
to it. According to the Premise wMR(1), the outcome of the measurement
$\hat{O}$ is determined at the time $t_{m}$. We denote the value
of the outcome by the symbol $\widetilde{\lambda}^{A}$. We assume
all settings are fixed for system $A$. The value $\widetilde{\lambda}^{A}$
can be considered a property of the system $A$ at the time $t_{m}$,
and denoted $\widetilde{\lambda}^{A}(t_{m})$. The Premise wMR(2)
of weak macroscopic realism posits that the value of $\widetilde{\lambda}^{A}(t_{m})$
is not changed by any interactions or measurements that might take
place at system $B$ at a time \emph{after} $t_{m}$ \citep{fulton2024alternative}.
The property $\widetilde{\lambda}^{A}$ can only be changed if a local
interaction takes place at $A$, in which case the property is denoted
by $\widetilde{\lambda}^{A}(t)$, and is defined at a future time
$t>t_{m}$.

\textbf{Definition (6c): Weak Macroscopic realism: Third Premise (wMR(3)):
}Consider the system $A$ described in Definitions (6a) and (6b),
at a time $t_{mA}$. For example, in this paper, we will take $\hat{O}\equiv\hat{x}_{A}$
and assume that at a time $t_{mA}$ all operations that determine
the measurement setting ($\theta=0$) have been carried out. We may
consider that the system $A$ was initially prepared in the superposition
$\sum_{j}c_{j}|x_{j}^{A}\rangle$ given in Definition (4), and has
been amplified according to $H_{amp}$, so that the system at time
$t_{mA}$ is in the superposition $\sum_{j}c_{j}e^{-iH_{amp}t/\hbar}|x_{j}^{A}\rangle$
of macroscopically distinct states $e^{-iH_{amp}t/\hbar}|x_{j}^{A}\rangle$,
each with a definite outcome for $\hat{x}_{A}$.

According to wMR(1), the outcome for $\hat{x}_{A}$ is determined
at the time $t_{mA}$. The value of this outcome is denoted by a variable
$\widetilde{\lambda}_{x}^{A}$. The outcomes correspond to eigenvalues
of the measurement, in this case, the eigenvalues denoted $x_{j}^{A}$
(refer Definition (3)). Quantum mechanics predicts that the outcomes
of measurements at $B$ conditioned on an outcome $x_{j}^{A}$ at
$A$ are determined by a \emph{projected} quantum state that we denote
by $|\psi_{B}\rangle_{j}$.

The third premise wMR(3) of weak macroscopic realism introduced in
\citep{fulton2024alternative,Fulton2024Weak,fulton2022argument} posits
that the projected state $|\psi_{B}\rangle_{j}$ (as defined by its
predictions for $B$) is determined for the local system $B$ at (or
by) the time $t_{mA}$. For example, in the EPR paradox, the outcome
of $\hat{x}_{B}$ for system $B$ can be inferred from the outcome
of $\hat{x}_{A}$ at $A$. The Premise wMR(3) posits that the outcome
at $B$ is determined at (or by) the time $t_{mA}$ when the system
$A$ has (according to wMR(1)) a predetermined value for measurement
$\hat{x}_{A}$. Hence, there is an ``element of reality'' for the
measurement at $B$ defined at the time $t_{mA}$. There is no contradiction
with Bell's theorem, because the time $t_{mA}$ is \emph{after} the
settings at $A$ have been fixed \citep{fulton2024alternative}.

\textbf{Definition (7): Macroscopic causality:} As a consequence of
wMR (Definitions (6a) and (6b)), a variable $\widetilde{\lambda}$
is defined at a time $t_{m}$ for a macroscopic system $A$ in the
frame $F$. This variable signifies the predetermination of the outcome
of the measurement e.g. $\hat{O}$ that distinguishes the macroscopic
states of the system $A$. Macroscopic causality is the assertion
that such variables satisfy causality e.g. the value of $\widetilde{\lambda}$
is not changed retrocausally by any future event and is not changed
by any spacelike-separated event at system $B$ occurring after time
$t_{m}$.

\textbf{Definition (8a): Postselected state:} Consider the system
described by a Q function $Q(x,p,t)$ at time $t_{0}$. A correlation
exists between the values of $x$ and $p$, given by $Q(p|x)=Q(x,p,t_{0})/Q(x,t_{0})$
where $Q(x,t_{0})=\int dpQ(x,p,t_{0})$ is the marginal of $x$. Hence,
if we consider the solutions $x(t)$ that arise from the backward-propagating
trajectories of a given branch $\mathcal{B}_{j}$ (corresponding to
an outcome $x_{j}$, refer above), then these define a joint distribution
of variables $\{x(t),p(t)\}$ at the initial time $t_{0}$. The values
$x(t_{0})$ are consistent with the deterministic relation $x_{j}\rightarrow Gx_{j}$
associated with the amplification $H_{amp}$, in the evolution of
the Q function. The joint distribution is denoted $Q_{loop}(x,p,t_{0}|\mathcal{B}_{j})$.
We refer to this distribution as the postselected distribution, or
else as the postselected ``state''. The notations $Q_{loop}(x,p,t_{0}|x_{j})$
and $Q_{0}(x,p|x_{j})$ are also used to denote $Q_{loop}(x,p,t_{0}|\mathcal{B}_{j})$.
The notation $Q_{loop}(x,p,t_{0}|\widetilde{\lambda}_{x})$ is also
used, where the variable $\widetilde{\lambda}_{x}$ denotes a branch
associated with a particular outcome, $\widetilde{\lambda}_{x}\in\{x_{j}\}$.

\textbf{Definition (8b): Postselected State for system $B$ conditioned
on an outcome $x_{j}^{A}$ for $A$: Projected distribution:} Consider
the postselected state defined in Definition (8a). For a given branch
$\mathcal{B}_{j}^{A}$ that defines the outcome $x_{j}^{A}$ at $A$,
the postselected state as defined by the Q function of the \emph{bipartite}
system is $Q_{loop}(\bm{\lambda},t_{0}|\mathcal{B}_{j}^{A})$ where
we abbreviate $\bm{\lambda}=(x_{A},p_{A},x_{B},p_{A})$. To obtain
the distribution for system $B$ alone, we define 
\[
Q(x_{B},p_{B}|\mathcal{B}_{j})=\int dp_{A}dx_{A}Q_{loop}(\bm{\lambda},t_{0}|\mathcal{B}_{j})
\]
which corresponds to the \emph{projected} distribution (or state)
for $B$, given an outcome $x_{j}^{A}$ at $A$. The bipartite distribution
$Q_{loop}(\bm{\lambda},t_{0}|\mathcal{\mathcal{B}}_{j}^{A})$ is conditioned
on the outcome $x_{j}^{A}$ at $A$, and must be consistent with the
deterministic relations that account for amplification ($x_{j}^{A}\rightarrow Gx_{j}^{A}$,
$x_{k}^{B}\rightarrow Gx_{k}^{B}$), and also the deterministic relations
$\mathcal{D}_{1}$ that account for the correlations between the outcomes
$x_{k}^{B}$ at $B$ and $x_{j}^{A}$ at $A$.  Here, $\{x_{k}^{B}\}$
denotes the set of outcomes at $B$ for measurement of $\hat{x}_{B}$.
According to the causal model developed in this paper, the deterministic
relations exist for the system prepared at time $t_{0}$, since the
measurement settings are fixed, as consistent with the assumptions
of weak local realism defined below. Where the meaning is clear, we
denote $Q_{loop}(\bm{\lambda},t_{0}|\mathcal{\mathcal{B}}_{j}^{A})$
as $Q_{loop}(\bm{\lambda},t_{0}|x_{j}^{A})$.

\textbf{Definition (9a): Hidden loop:} Continuing from Definition
(8a), there is hence a connection between the backward-propagating
trajectories $x(t)$ defining a branch $\mathcal{B}_{j}$, and a \emph{correlated}
set of forward-propagating trajectories $p(t)$ that are determined
by the $Q_{loop}(x,p,t_{0}|\mathcal{B}_{j})$. This connection we
refer to as a ``loop''. The loop is ``hidden'' if the associated
distribution $Q_{loop}(x,p,t_{0}|\mathcal{B}_{j})$ cannot be equivalent
to a quantum state, or else if variables involved in the loop, such
as $p$, are not detectable in the model of measurement.

\textbf{Definition (9b): Local and nonlocal loops:} The local loop
is a loop (refer Definition (9a)) that connects backward and forward
amplitudes of the same system, $A$ or $B$. A nonlocal loop connects
backward and forward trajectories of different systems, $A$ and $B$.

\textbf{Definition (10a): Weak local realism: First Premise (wLR(1)):
}We consider a system $A$ at a time $t_{1}$, \emph{after} operations
$U$ needed to fix measurement settings have been carried out, so
that the system is prepared for a measurement $\hat{O}$ e.g. in this
paper $\hat{O}\equiv\hat{x}_{\theta A}=\hat{x}_{A}\cos\theta+\hat{p_{A}}\sin\theta$.
In a spin-Bell experiment, the time $t_{0}$ corresponds to the time
after a system (e.g. a photon or electron) has passed through an analyzer
(a polarizer or Stern-Gerlach apparatus), so that a final detection
will register the outcome of the polarization or spin along a direction
$\theta$, determined by the analyzer. In the examples of this paper,
the measurement is of $\hat{O}\equiv\hat{x}_{\theta A}$ and the ``analyzer''
corresponds to a phase-shift device that determines $\theta$. Unlike
the Definition of wMR, the system need\emph{ }not be macroscopic.

For such a system, two versions of weak local realism (wLR) can be
defined: \emph{Deterministic} wLR(1) and \emph{Probabilistic} wLR(1).
In the deterministic version, it is posited that the value for the
outcome of $\hat{O}$ is determined by the time $t_{0}$ \citep{fulton2024alternative,joseph2024macroscopic}.
We denote the value for the outcome by $\lambda^{A}$. In the probabilistic
version, it is posited that a \emph{set} of hidden variables $\{\lambda^{A}\}$
exist to describe the system, such that there is a definite probability
$P(x|\{\lambda^{A}\})$ for the outcome $x$ of $\hat{O}$, given
a set of values for $\{\lambda^{A}\}$.

\textbf{Definition (10b): Weak Local Realism: Second Premise (wLR(2)):
}Consider two spacelike-separated systems $A$ and $B$ both stationary
in a frame $F$. Consider the system $A$ at the time $t_{1A}$ after
the measurement settings at $A$ are fixed, as described in the Definition
(10a) of wLR(1).

According to the deterministic version of Premise wLR(1), the value
for the outcome $x_{\theta A}$ of the measurement $\hat{O}\equiv\hat{x}_{\theta A}$
on system $A$ is determined at time $t_{1A}$. We denote the value
by $\lambda_{x_{\theta}}^{A}$. Premise wLR(2) posits that this value
is not changed by any interactions or measurements that might take
place at system $B$ at a time after $t_{1A}$ \citep{fulton2024alternative}.
Similar to Premise wMR(2), the value can only be changed if there
is a local interaction at $A$.

According to the probabilistic version of wLR(2), one can define a
probability $P(x|\{\lambda^{A}\})$ for an outcome $x_{\theta A}$
of $\hat{O}$, given a set of values for the hidden variables of $\{\lambda^{A}\}$
that describe system $A$. The Premise wLR(2) posits that these probabilities
are unchanged by the interactions at $B$ that might take place at
a time $t>t_{1A}$.

\textbf{Definition (10c): Weak local Realism: Third Premise (wLR(3)):
}Consider the system $A$ described in Definitions (10a) and (10b)
of wLR(1) and (2). We consider measurement of $\hat{x}_{\theta A}$.
Unlike the Definition (5) for wMR, we do not need to assume the system
$A$ is macroscopic.

Consider the deterministic version of wLR, defined by Premise wLR(1)
(Definition (10a)). This Premise posits that the value for the outcome
of $\hat{x}_{\theta A}$ is determined at (or by) the time $t_{1A}$,
after the operations that fix the measurement settings for system
$A$ have been carried out. Consider the projected state $|\psi_{B}\rangle_{j}$
defined in the Definition (6c) wMR(3) above. Premise of wLR(3) posits
that this state (as defined by its predictions for $B$) is determined
at the time $t_{1A}$ \citep{fulton2024alternative}. In the EPR example
given for Definition (6c), this implies that the outcome for $\hat{x}_{B}$
is determined at (or by) the time $t_{1A}$, after the setting $\theta$
is fixed at $A$. In other words, an ``element of reality'' exists
for $B$ based on the prediction at $A$, at time $t_{1A}$.

Alternatively, we consider the probabilistic version of wLR, defined
by Premise wLR(1) in Definition (10a). This premise posits that probabilities
$P(x_{j}^{A}|\{\lambda^{A}\})$ for outcomes at $A$ are fixed at
the time $t_{1A}$. For any one of these outcomes $x_{j}^{A}$, a
projected state $|\psi_{B}\rangle_{j}$ for system $B$ is defined
by quantum mechanics, as in Definition (6c). Premise wLR(3) posits
that at the time $t_{1A}$, the system $B$ is described by the projected
state $|\psi_{B}\rangle_{j}$ (as defined by its predictions of outcomes
of future measurements at $B$) with probability $P(x_{j}^{A}|\{\lambda^{A}\})$.

\textbf{Definition (11): No-signaling:} No-signaling is defined as
the conditional independence of the outcome at $A$ and the setting
at $B$, given the setting at $A$ \citep{wood2015lesson}.

\section{\label{sec:Measurement-model}Measurement model}

In this paper, the system dynamics is described by a unified stochastic
model motivated by the Q function \citep{Husimi1940}. The examples
of this paper restrict to bosonic fields. Our approach requires the
measurement process to be included in the dynamics.

\subsection{Measurement model: parametric amplification}

First, restricting our analysis to a single mode of the field, we
define the complementary quadrature phase amplitude observables
\begin{eqnarray}
\hat{x} & = & \hat{a}+\hat{a}^{\dagger}\nonumber \\
\hat{p} & = & (\hat{a}-\hat{a}^{\dagger})/i\label{eq:quad}
\end{eqnarray}
in a rotating frame, where $\hat{a}$ is the boson destruction operator
\citep{yurke1986generating}. Hence $\hat{a}=(\hat{x}+i\hat{p})/2$.
This implies the uncertainty relation $\Delta\hat{x}\Delta\hat{p}\ge1$
for any state $|\psi\rangle$ of the single-mode field, where here
we denote $(\Delta\hat{O})^{2}=\langle\hat{O}^{2}\rangle-\langle\hat{O}\rangle^{2}$
as the variance of the observable $\hat{O}$.

We analyze the simplest measurement procedure that can take place
$-$ that of direct amplification to a macroscopic signal.  Measurement
is modeled by a measuring device (a meter) that produces macroscopic
outputs, namely the parametric amplifier \citep{drummond2020retrocausal}.
To measure $\hat{x}$, the system prepared in a state $|\psi\rangle$
is amplified according to the Hamiltonian \citep{Yuen1976}
\begin{equation}
H_{amp}=\frac{i\hbar g}{2}\left[\hat{a}^{\dagger2}-\hat{a}^{2}\right]\,\label{eq:ham-2-1}
\end{equation}
where $g>0$ is real. The system evolves under $H_{amp}$ for a time
$t_{f}$, to give the final measurement output. The measurement is
completed simply by the fields being amplified to become macroscopic,
and hence detectable to macroscopic devices or observers. The importance
of amplification is explained by Bohr, who regarded measurement as
amplification \citep{bohr1987essays}.

For $g>0$, the dynamics of $H_{amp}$ gives solutions that amplify
the ``position'' $\hat{x}$ but attenuate the orthogonal ``momentum''
quadrature $\hat{p}$. This is clear from the standard operator Heisenberg
equations which give solutions
\begin{eqnarray}
\hat{x}\left(t\right) & = & \hat{x}\left(0\right)e^{gt}\nonumber \\
\hat{p}\left(t\right) & = & \hat{p}\left(0\right)e^{-gt}\label{eq:amp}
\end{eqnarray}
$H_{amp}$ induces squeezing in $\hat{p}$ \citep{Yuen1976}. The
interaction $H_{amp}$ has been realized experimentally \citep{Wu1986generation,Yurke1989Observation}.
A related model of measurement based on amplification was presented
by Glauber, for a two-slit experiment \citep{glauber1986amplifiers,raymer1992information}.

\subsection{Equation of motion for the $Q$ function: Generalized Fokker-Planck
equation}

To solve for single-mode evolution, we use the Q function \citep{Husimi1940}
\begin{equation}
Q(\alpha)=\frac{1}{\pi}|\langle\alpha|\psi\rangle|^{2}\label{eq:QH-1}
\end{equation}
defined for a quantum state $|\psi\rangle$ with respect to the nonorthogonal
basis of coherent states $|\alpha\rangle$. The coherent state $|\alpha\rangle$
satisfies $\hat{a}|\alpha\rangle=\alpha|\alpha\rangle$ where $\alpha$
is a complex number. The phase-space coordinates $\bm{\lambda}$ are
the real coordinates $x$ and $p$, given by $\alpha=(x+ip)/2$. In
this paper, we denote the Q function of the quantum state at a time
$t$ as $Q(x,p,t)$. The single-mode Q function defines the quantum
state $|\psi\rangle$ uniquely as a positive probability distribution.
The moments evaluated from the Q function distribution are anti-normally
ordered operator moments \citep{drummond2014quantum}. The anti-normally
ordered variances $\sigma_{x}^{2}(t)$ and $\sigma_{p}^{2}(t)$ are
precisely the variances of $x$ and $p$ as defined by the Q function
$Q(x,p,t)$.

The dynamics for real variables $x$ and $p$ can be deduced by deriving
a generalized Fokker-Planck equation, which gives an equation of motion
of $Q$ for a given Hamiltonian $H$. In general terms, the Q function
probability density $Q\left(\bm{\lambda},t\right)$ for a phase-space
coordinate $\bm{\lambda}$ is defined with respect to a non-orthogonal
operator basis $\hat{\Lambda}\left(\bm{\lambda}\right)$ as 
\begin{equation}
Q\left(\bm{\lambda},t\right)\equiv Tr\left(\hat{\Lambda}\left(\bm{\lambda}\right)\hat{\rho}\left(t\right)\right)\label{eq:Q-1}
\end{equation}
where $\hat{\rho}$ is the density operator of the system. As $Q\left(\bm{\lambda},t\right)$
is normalized to unity, it is necessary to normalize the basis so
it integrates to unity, and the normalization condition is that $\int\hat{\Lambda}\left(\bm{\lambda}\right)d\bm{\lambda}=\hat{1}.$
The basis satisfies $\hat{\Lambda}^{2}\left(\bm{\lambda}\right)=\mathcal{N}\left(\bm{\lambda}\right)\hat{\Lambda}\left(\bm{\lambda}\right)$,
which is different to the condition for projectors that $\hat{P}^{2}=\hat{P}$,
because it is a continuous nonorthogonal basis.

From the Schrödinger equation, the dynamics of the probability distribution
is obtained from the usual equation $i\hbar d\rho/dt=\left[H,\rho\right]$.
As a result, one obtains an equation for the Q function time-evolution:
\begin{equation}
\frac{dQ\left(\bm{\lambda},t\right)}{dt}=\frac{i}{\hbar}Tr\left\{ \left[H,\hat{\Lambda}\left(\bm{\lambda}\right)\right]\rho\left(t\right)\right\} \label{eq:dynamics-1}
\end{equation}
This is equivalent to a zero-trace diffusion equation for the variables
$\bm{\lambda}$, of form $\dot{Q}\left(\bm{\lambda},t\right)=\mathcal{L}\left(\bm{\lambda}\right)Q\left(\bm{\lambda},t\right)$,
where $\mathcal{L}\left(\bm{\lambda}\right)$ is the differential
operator for Q function dynamics. For the single-mode, the phase-space
coordinates $\bm{\lambda}$ are the real coordinates $x$ and $p$.

For the system evolving according to the Hamiltonian $H_{amp}$ given
by (\ref{eq:ham-2-1}), a dynamical equation for the Q function can
be derived \citep{drummond2020retrocausal}. Applying the correspondence
rules to transform operators into differential operators, one obtains
a generalized Fokker-Planck type equation in terms of complex coherent
state variables $\alpha$: 
\begin{align}
\frac{dQ}{dt} & =-\left[g\frac{\partial}{\partial\alpha}\alpha^{\ast}+g\frac{\partial}{\partial\alpha^{*}}\alpha+\frac{g}{2}\frac{\partial^{2}}{\partial\alpha^{2}}+\frac{g}{2}\frac{\partial^{2}}{\partial\alpha^{*2}}\right]Q\,\label{eq:dq-1}
\end{align}
Here the differential operators act on all terms to the right \citep{Arnold1992-stochastic,Gardiner1997,Risken1996}.
Defining $\partial_{p}\equiv\partial/\partial p$ and $\partial_{x}\equiv\partial/\partial x$,
we obtain in real coordinates
\begin{equation}
\frac{dQ}{dt}=\left[\partial_{p}\left(gp\right)-\partial_{x}\left(gx\right)+g\left(\partial_{p}^{2}-\partial_{x}^{2}\right)\right]Q\,\label{eq:fp-q-1}
\end{equation}
The above equation demonstrates a diffusion matrix which is traceless
and equally divided into positive and negative definite parts, and
a drift matrix: 
\begin{equation}
\bm{A}=\left[\begin{array}{c}
\dot{x}\\
\dot{p}
\end{array}\right]=\left[\begin{array}{c}
gx\\
-gp
\end{array}\right]\label{eq:drift-1}
\end{equation}

\subsection{\label{sec:Stochastic-dynamics-1}Forward-backward stochastic equations}

Next, we derive from the equation of motion (\ref{eq:fp-q-1}) for
$Q(x,p)$ the forward-backward stochastic equations that determine
the dynamics $x(t)$ and $p(t)$ of the Q function amplitudes $\bm{\lambda}=(x,p)$.
 There is an equivalent time-symmetric stochastic action principle
for $Q\left(\bm{\lambda},t\right)$, leading to probabilistic path
integrals \citep{drummond2021time}. These have sample trajectories
$\bm{\lambda}$ that define the realistic path for all times.

In this paper, we derive two Theorems generalizing these results to
the case of an initial superposition state, where a conditional relation
implies a feedback between the forward-backward stochastic solutions.
The Theorems have a general applicability and are hence presented
as a distinct result in Appendix A. We prove the equivalence between
the averages calculated from the trajectories $\bm{\lambda}\left(t\right)$
($x$ and $p$) and those calculated from the Q function, $Q(x,p,t)$
(refer also Secs. V.B, IX.C and Appendix H). This result is independently
confirmed by statistical comparisons of the numerical simulations
with the Q function (Appendix B).

Using the equivalence theorems in Appendix A, we solve the measurement
dynamics for the dynamics of $H_{amp}$. Here, the $x$ and $p$ dynamics
decouple. However, to obtain a mathematically tractable equation for
the traceless noise matrix, we follow \citep{drummond2020retrocausal}
and the sign of $t$ is reversed in the amplified dynamics of $x$.
The corresponding integrated stochastic equations for $g>0$ are:
\begin{align}
p\left(t\right) & =p\left(t_{0}\left|x_{0}\right.\right)-\int_{t_{0}}^{t}gpdt'+\int_{t_{0}}^{t}dw_{p}\nonumber \\
x\left(t\right) & =x\left(t_{f}\right)-\int_{t}^{t_{f}}gxdt'+\int_{t}^{t_{f}}dw_{x}\label{eq:soln-dyn-1}
\end{align}
where the terms $dw_{p}$ and $dw_{x}$ are infinitesimal Gaussian
noise increments such that $\left\langle dw_{\mu}dw_{\nu}\right\rangle =2g\delta_{\mu\nu}dt$
\citep{Arnold1992-stochastic,Gardiner1997,Risken1996,ma1994solving}.
Both $x\left(t_{f}\right)$ and $p\left(t_{0}\left|x_{0}\right.\right)$
represent random boundary conditions with a specified marginal and
conditional probability, at the final time $t_{f}$ and the initial
time $t_{0}$ of the interaction, respectively. Here, $x_{0}\equiv x(t_{0})$.

Alternatively, in differential form, we obtain for $p$ the equation
\citep{drummond2020retrocausal,drummond2021objective}
\begin{equation}
\frac{dp}{dt}=-gp+\xi_{p}\label{eq:forwardSDE-2-1}
\end{equation}
propagating \emph{forward} in time, which has a boundary condition
determined by the Q function at the initial time $t_{0}=0$. The equation
for $x$ is
\begin{align}
\frac{dx}{dt_{-}} & =-gx+\xi_{x}\label{eq:backwardSDE-2-1}
\end{align}
where $t_{-}=-t$. The equation is solved \emph{backwards} in time,
and therefore has a boundary condition in the future, defined as the
time $t_{f}$ when the amplification is completed. Defining $\bm{\xi}=\left(\xi_{p},\xi_{x}\right)$,
the Gaussian random noises $\xi_{\mu}\left(t\right)$ satisfy: $\left\langle \xi_{\mu}\left(t\right)\xi_{\nu}\left(t'\right)\right\rangle =2g\delta_{\mu\nu}\delta\left(t-t'\right)$,
 where the noise terms are defined as $\xi_{\mu}\left(t\right)=\lim_{dt\rightarrow0}dw_{\mu}/dt$
\citep{Gardiner1997}.

Thus, there is a \emph{forward-backward stochastic differential equation}
(FBSDE), for individual trajectories. This describes two individual
stochastic trajectories, such that the average of the dynamical trajectories
equals the Q function averages. The trajectories are decoupled dynamically,
with decay and stochastic noise occurring in each of the time directions.
One propagates forward, and one backwards, in time. As there are random
noise inputs, and the boundary condition is a distribution given by
the $Q$ function which is sampled, a series of runs is required.
The solutions $x(t)$ and $p(t)$ differ for each run, and averages
are evaluated, to determine moments at a given time $t$. The $x(t)$
and $p(t)$ are ``\emph{trajectories}'' of the \emph{forward-backward
stochastic equation (FBSE) simulation }(Definition (1) in Sec. I.B).

In examples treated here, the trajectory in $x$ depends on a future
marginal, $Q(x,t_{f})$, while the trajectory in $p$ depends on a
conditional distribution, $Q(p|x)$ at the initial time $t_{0}$,
which depends on $x$ in the future (Figs. \ref{fig:summary-pics}
and \ref{fig:The-causal-structure-1}), giving a cyclic causal behavior.
It is possible to exchange these, by time-reversal symmetry. There
is an apparent Grandfather-type paradox arising from the present seeming
to depend on the future. This is resolved in Sec. \ref{sec:Causal-model-for}
by nature of the causal model.

Causal consistency, defined in the Introduction, follows from the
proofs and Results in Sec. V.B, Appendix A and Appendix H (refer Fig.
\ref{fig:causal-consistency-2-1} for a depiction and proposed experiment):
The quantum state $|\psi\rangle$ as determined uniquely by the Q
function $Q(x,p,t)$ depends on the time $t$ of evolution, starting
from an initial time $t=0$.  The consistency ensures that the trajectories
at large time $t_{f}\rightarrow\infty$ satisfy the conditional relations
connecting $x(t_{f})$ and $p(t_{f})$ for the amplified state, which
quantifies the end part of the ``loop'' (refer Figs. \ref{fig:The-causal-structure-1}
and \ref{fig:causal-consistency-2-1}). For the measurement interaction
$H_{amp}$, we find that at the final time $t_{f}\rightarrow\infty$,
the $p$ decouples completely from $x$, and there is no further feedback
in the simulation from $p$ to $x$: the ``loop'' is hence incomplete,
disconnecting at the end-time $t_{f}$.

\section{Forward-backward stochastic simulations\label{sec:Forward-backward-stochastic-simu}}

\subsection{Measurement on an eigenstate of $\hat{x}$}

We first consider the measurement of $\hat{x}$ on an eigenstate $|x_{j}\rangle$
of $\hat{x}$, where $x_{j}$ is the eigenvalue (Definition (3) in
Sec. I.B). We resolve problems with normalization, by  defining
the eigenstates in $\hat{x}$ as highly squeezed states. The squeezed
state is defined as \citep{Yuen1976}
\begin{equation}
|\beta_{j},z\rangle_{sq}=D(\beta_{j})S(z)|0\rangle\label{eq:sq}
\end{equation}
Here, $|0\rangle$ is the vacuum state satisfying $\hat{a}|0\rangle=0$,
and $D(\beta_{j})=e^{\beta_{j}\hat{a}^{\dagger}-\beta_{j}^{*}\hat{a}}$
and $S(z)=e^{\frac{1}{2}(z^{*}\hat{a}^{2}-z\hat{a}^{\dagger2})}$
are the displacement and squeezing operators, where $z$ and $\beta_{j}$
are complex numbers. For the state with squeezed fluctuations in $\hat{x}$,
we note that $z=r$ where $r$ is a real, positive number, referred
to as the \emph{squeeze parameter}.  Hence $(\Delta\hat{x})^{2}=e^{-2r}$
and $(\Delta\hat{p})^{2}=e^{2r}.$ The parameter $r$ determines the
variances and hence the amount of squeezing in $\hat{x}$. The mean
values are given by $\langle\hat{a}\rangle=\beta_{j}$ where $\beta_{j}=(x_{j}+ip_{j})/2$
and $x_{j}$ and $p_{j}$ are real. The eigenstate denoted $|x_{j}\rangle$
is thus defined as the limiting squeezed state
\begin{equation}
|x_{j}\rangle\equiv\lim_{r\rightarrow\infty}|\frac{x_{j}}{2},r\rangle_{sq}\label{eq:eigenstate-def}
\end{equation}
with $x_{j}$ real and $p_{j}=0$. 

\begin{figure}
\begin{centering}
\includegraphics[width=0.6\columnwidth]{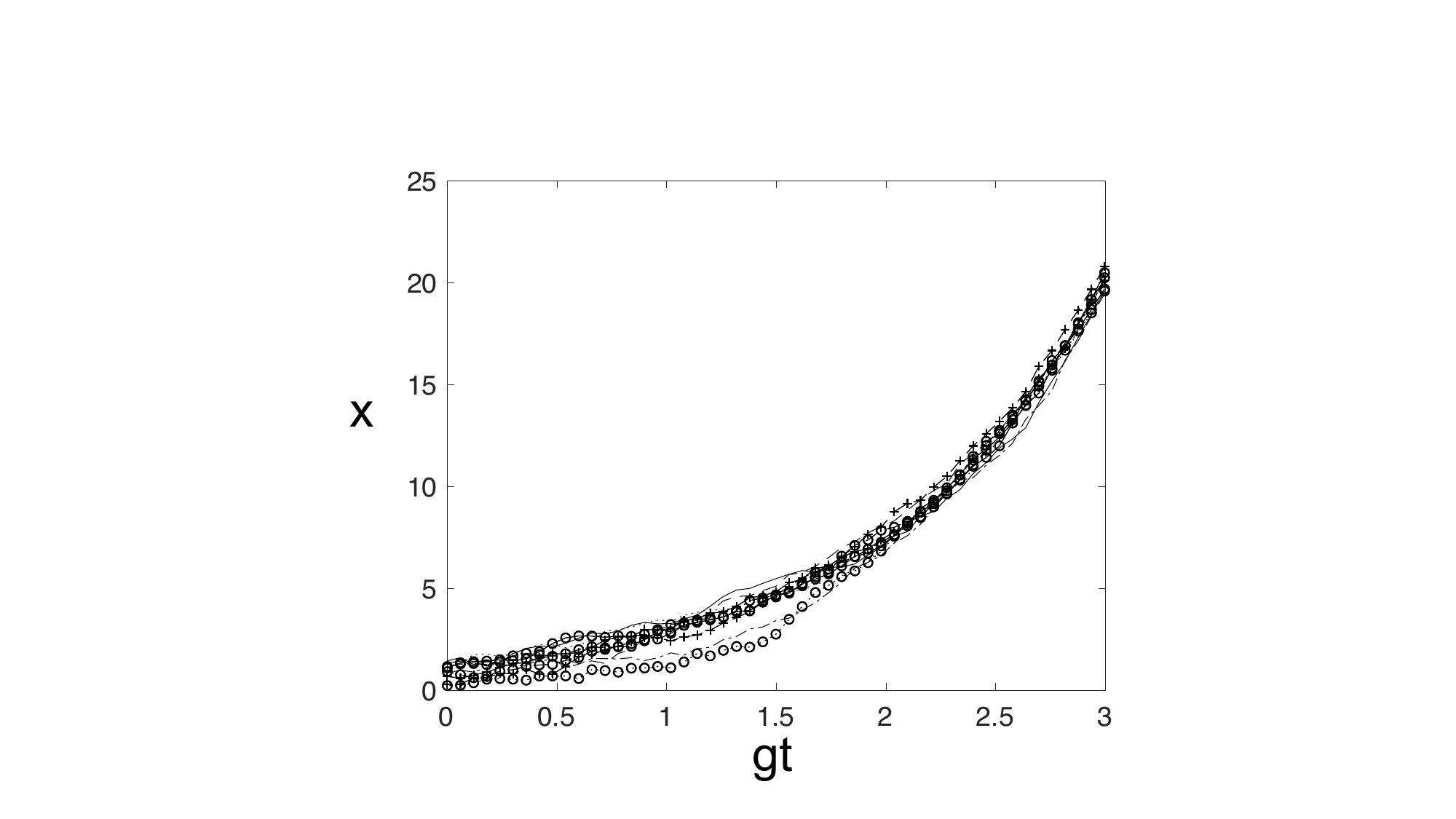}
\par\end{centering}
\caption{\textbf{\emph{Stochastic solutions $x(t)$ of the backward equation
(\ref{eq:backwardSDE-2-1}).}} Measurement of $\hat{x}$ on a system
prepared in an eigenstate $|x_{j}\rangle$ of $\hat{x}$. The mean
of $x(t)$ is amplified to $Gx_{j}$ but the variance $\sigma_{x}$
is unchanged. $r=2$, $x_{1}=1$ and $gt_{f}=3$.\label{fig:stochastic-eigenstate-x}\textcolor{blue}{}}
\end{figure}

\begin{figure}
\begin{centering}
\includegraphics[width=0.6\columnwidth]{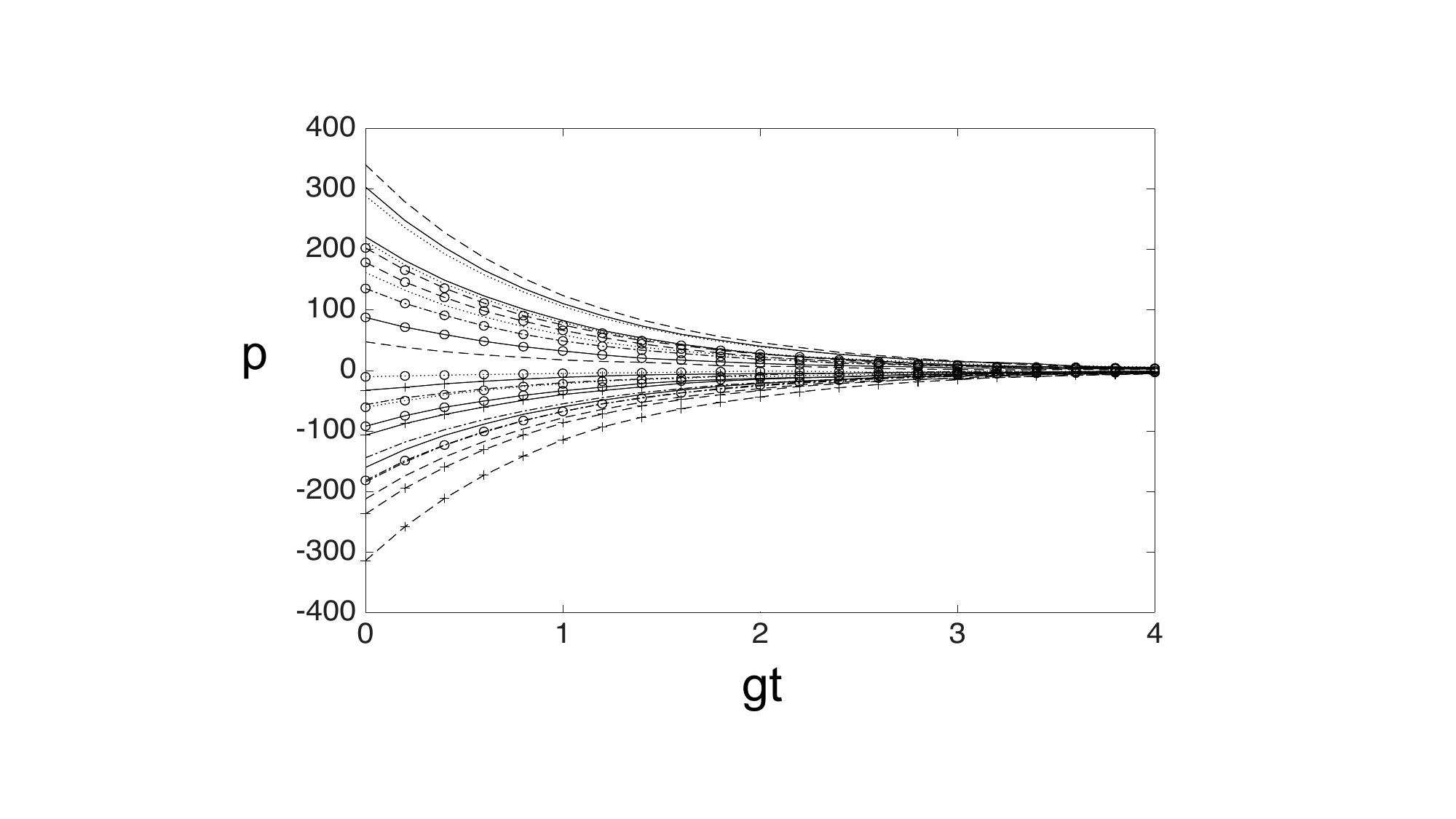}
\par\end{centering}
\caption{\textbf{\emph{Stochastic solutions $p(t)$ of the forward equation
(\ref{eq:forwardSDE-2-1}).}} Measurement of $\hat{x}$ as in Figure
\ref{fig:stochastic-eigenstate-x}. There is attenuation of the complementary
variable $p$ to the undetected (``hidden'') vacuum level, where
$\sigma_{p}^{2}=1$. \label{fig:stochastic-p-2}\textcolor{blue}{
}Similar results are obtained for all values of $x_{j}$.}
\end{figure}

The Q function of the eigenstate $|x_{j}\rangle$ is the bivariate
Gaussian (refer Appendix C)
\begin{equation}
Q_{x_{j}}(x,p)=\frac{e^{-p^{2}/2\sigma_{p}^{2}}}{2\pi\sigma_{x}\sigma_{p}}e^{-(x-x_{j})^{2}/2\sigma_{x}^{2}}\label{eq:q-sq-1}
\end{equation}
with variances $\sigma_{x}^{2}=1+e^{-2r}$ and $\sigma_{p}^{2}=1+e^{2r}$,
where  $r\rightarrow\infty$. In this limit, $\sigma_{x}^{2}=1$
and $\sigma_{p}^{2}$ is large.  The mean value of $x$ is the eigenvalue
$x_{j}$. The variance in $x$ is $\sigma_{x}^{2}=1$, which is the
level of the vacuum noise determined by the Heisenberg uncertainty
relation $\Delta\hat{x}\Delta\hat{p}\geq1$. The \emph{measured} variance
of $\hat{x}$ is zero (since for an eigenstate $\Delta\hat{x}=0$),
indicating that the fluctuations given by $\sigma_{x}$ are not detectable
in the measured output. We refer to these fluctuations, where $\sigma_{x}^{2}=1$,
as ``\emph{hidden}'' noise, symbolized by $\eta(t)$. 

In order to solve the forward-backward stochastic equations, we require
to evaluate the Q function for the system at the final time $t_{f}$,
after the amplification $H_{amp}$. The state of the amplified system
is given by
\begin{equation}
|\psi(t)\rangle=e^{-iH_{amp}t/\hbar}|\frac{x_{j}}{2},r\rangle_{sq}=|G(t)\frac{x_{j}}{2},r'\rangle_{sq}\label{eq:sq:amp}
\end{equation}
(refer Appendix C) where $r'=-gt+r$ and $G(t)=e^{gt}$ is the amplification
factor (Eq. (\ref{eq:amp})). The Q function for the amplified system
at a time $t$ is hence
\begin{equation}
Q(x,p,t)=\frac{e^{-p^{2}/2\sigma_{p}^{2}(t)}}{2\pi\sigma_{x}(t)\sigma_{p}(t)}e^{-(x-G(t)x_{j})^{2}/2\sigma_{x}^{2}(t)}\label{eq:Qamp}
\end{equation}
with variances
\begin{eqnarray}
\sigma_{x}^{2}(t) & = & 1+e^{-2r+2gt}\nonumber \\
\sigma_{p}^{2}(t) & = & 1+e^{2r-2gt}\label{eq:amp-var}
\end{eqnarray}
It is convenient to write $\sigma_{x}^{2}\left(t\right)=1+G(t)^{2}\left[\sigma_{x}^{2}\left(0\right)-1\right]$
and $\sigma_{p}^{2}(t)=1+[\sigma_{p}^{2}(0)-1]/G(t)^{2}$ where $G(t)=e^{gt}$.
We then see that as $G(t)\rightarrow\infty$, the ``hidden'' noise
$\eta(t)$ (with variance $\sigma_{x}^{2}(0)=1$) associated with
the eigenstate will \emph{not} be amplified (but any noise above this
level will be). The noise in $p$ decays to the hidden noise level
($\sigma_{p}^{2}(t)\rightarrow1$).

The forward-backward equations (\ref{eq:forwardSDE-2-1}) and (\ref{eq:backwardSDE-2-1})
can now be solved \citep{drummond2021time}. The boundary condition
for the forward equation is the marginal of $Q(x,p,t_{0})$ in $p$,
which is $Q(p,t_{0})=e^{-p^{2}/2\sigma_{p}^{2}}/(\sigma_{p}\sqrt{2\pi})$.
The boundary condition for the backward equation is the marginal $Q(x,t_{f}$)
of $Q(x,p,t_{f})$ in $x$, which is $Q(x,t_{f})=e^{-(x-Gx_{j})^{2}/2\sigma_{x}^{2}(t_{f})}/(\sigma_{x}(t_{f})\sqrt{2\pi})$.
Solutions are shown in Figures \ref{fig:stochastic-eigenstate-x}
and \ref{fig:stochastic-p-2}.

\subsection{Measurement on a superposition of eigenstates}

In order to address the measurement problem, we require to consider
measurement on a system prepared in a superposition of eigenstates.
Here, we consider the superposition
\begin{eqnarray}
|\psi_{sup}\rangle & = & c_{1}|x_{1}\rangle+c_{2}|x_{2}\rangle\nonumber \\
 & \rightarrow & Nc_{1}|\frac{x_{1}}{2},r\rangle_{sq}+c_{2}|\frac{x_{2}}{2},r\rangle_{sq}\label{eq:sup-sq}
\end{eqnarray}
of eigenstates of $\hat{x}$, approximated as a superposition of squeezed
states with $r\rightarrow\infty$. The $c_{j}$ are probability amplitudes.
Without loss of generality we take $c_{1}$ to be real and positive,
and write $c_{2}=|c_{2}|e^{i\varphi}$. The Q function is
\begin{eqnarray}
Q_{sup}(x,p,t_{0}) & = & N\frac{e^{-p^{2}/2\sigma_{p}^{2}}}{2\pi\sigma_{x}\sigma_{p}}\Bigr(|c_{1}|^{2}e^{-(x-x_{1})^{2}/2\sigma_{x}^{2}}\nonumber \\
 &  & +|c_{2}|^{2}e^{-(x-x_{2})^{2}/2\sigma_{x}^{2}}\Bigl)+\mathcal{I}nt\nonumber \\
\label{eq:Q-sup}
\end{eqnarray}
where $\mathcal{I}nt=N\frac{e^{-p^{2}/2\sigma_{p}^{2}}}{2\pi\sigma_{x}\sigma_{p}}2|c_{1}c_{2}|\mathcal{I}$.
The variances are $\sigma_{x}^{2}=1+e^{-2r}$ and $\sigma_{p}^{2}=1+e^{2r}$.
$N$ is the normalization factor.  We find
\begin{eqnarray}
\mathcal{I} & = & e^{-[(x-x_{1})^{2}+(x-x_{2})^{2}]/4\sigma_{x}^{2}}\mathcal{F}\label{eq:int-2}
\end{eqnarray}
where
\begin{eqnarray}
\mathcal{F} & = & (\cos\varphi)\cos[\frac{p}{2\sigma_{x}^{2}}(x_{1}-x_{2})]-(\sin\varphi)\sin[\frac{p}{2\sigma_{x}^{2}}(x_{1}-x_{2})]\nonumber \\
\label{eq:fringe-1}
\end{eqnarray}
and $N=\Bigl(1+2c_{1}|c_{2}|(\cos\varphi)e^{-\frac{(x_{1}-x_{2})^{2}}{8\sigma_{x}^{2}}\{1+\frac{\sigma_{p}^{2}}{\sigma_{x}^{2}}\}}\Bigl)^{-1}$.
NB: $N\rightarrow1$ as $\sigma_{p}^{2}\rightarrow\infty$, in the
limit of large $r$ when the squeezed states become eigenstates of
$\hat{x}$.

Taking the limit $r\rightarrow\infty$, we see that the Q function
has two Gaussian peaks with fixed variance $\sigma_{x}^{2}=1$, centered
at the eigenvalues $x_{1}$ and $x_{2}$. There is an additional interference
term which we symbolize by $\mathcal{I}nt$, corresponding to a central
Gaussian peak at $x=0$ modulated by a fringe pattern $\mathcal{F}$.
The Q function of the superposition differs from the Q function $Q_{mix}(x,p)$
of the mixture of eigenstates
\begin{equation}
\rho_{mix}=\sum_{j}|c_{j}|^{2}|x_{j}\rangle\langle x_{j}|\label{eq:mix-sup-1}
\end{equation}
only by the addition of the third term $\mathcal{I}nt$ \citep{Milburn1986Dissipative}.
\begin{figure}
\begin{centering}
\includegraphics[width=0.7\columnwidth]{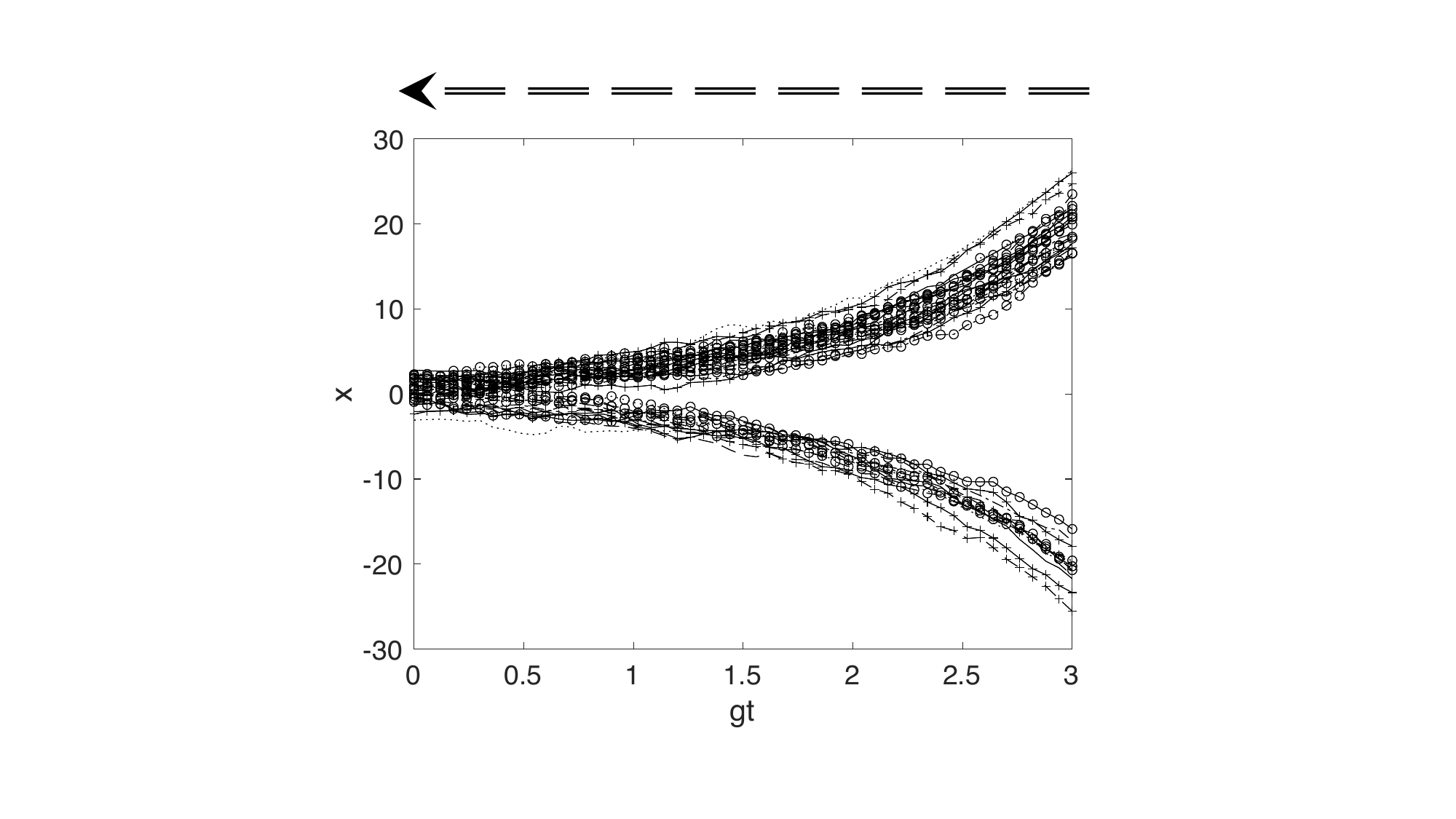}
\par\end{centering}
\begin{centering}
\negthickspace{}\includegraphics[width=0.7\columnwidth]{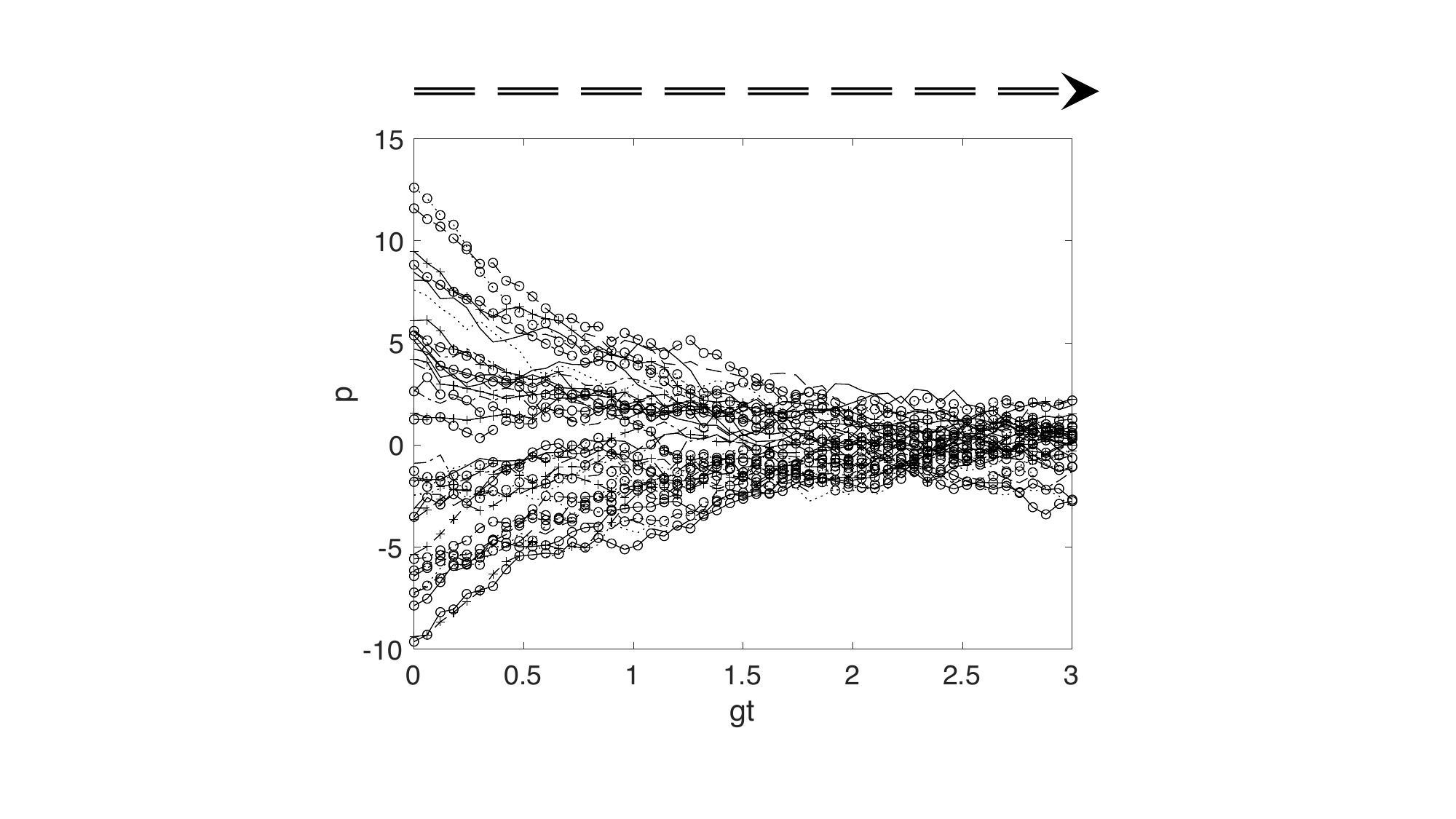}\negthickspace{}\negthickspace{}\negthickspace{}
\par\end{centering}
\caption{\textbf{\emph{Emergence of branches with amplification:}} Measurement
of $\hat{x}$ on a system prepared in a superposition given by $|\psi_{sup}\rangle$
of eigenstates of $\hat{x}$ (Eq. \ref{eq:sup-sq})). We take $c_{1}=-ic_{2}=1/\sqrt{2}$,
$x_{1}=-x_{2}=1$ and $gt_{f}=3$. The top figure gives individual
stochastic solutions (trajectories) of the backward equation (\ref{eq:backwardSDE-2-1}),
showing the emergence of branches associated with each eigenvalue
$x_{j}$. Here, the measurement is made on a microscopic superposition.
 We note the ``width'' of each branch is constant.  \label{fig:stochastic-sup-micro}\textcolor{blue}{}
The lower plot shows individual solutions of the forward equation
(\ref{eq:forwardSDE-2-1}) for the variable $p$. The values are attenuated
to an undetected (``hidden'') vacuum noise level, $\sigma_{p}^{2}=1$.
The superposition of eigenstates is approximated by (\ref{eq:sup-sq})
with $r=2$.}
\end{figure}

\begin{figure}
\begin{centering}
\includegraphics[width=0.7\columnwidth]{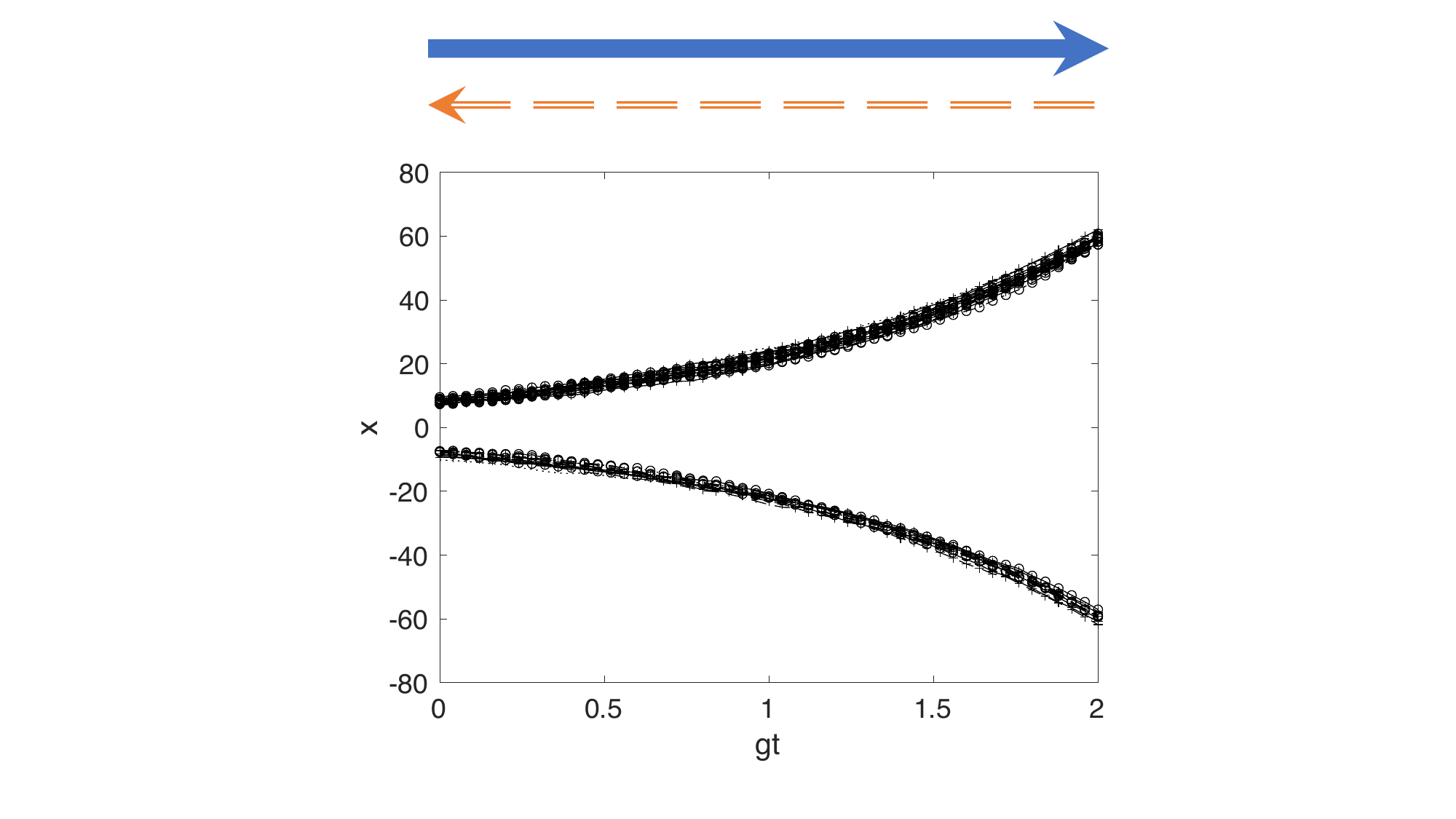}
\par\end{centering}
\begin{centering}
\includegraphics[width=0.7\columnwidth]{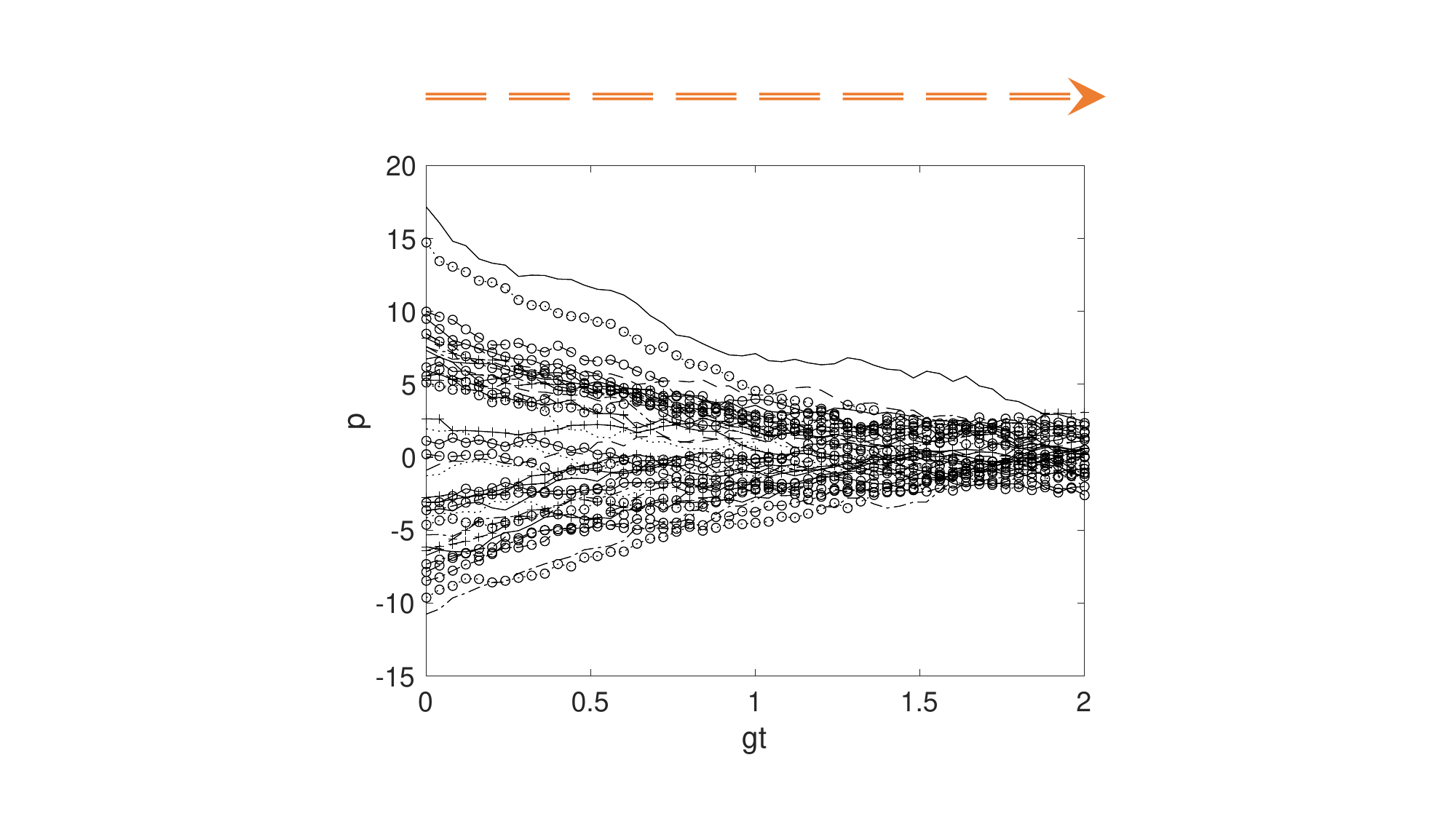}
\par\end{centering}
\caption{\textbf{\emph{Measurement on a macroscopic superposition:}} As in
Figure \ref{fig:stochastic-sup-micro}, but with $gt_{f}=2$ and $x_{1}=8$.
The top and lower figures show solutions of the backward and forward
equations (\ref{eq:backwardSDE-2-1}) and (\ref{eq:forwardSDE-2-1}),
respectively. The solid blue line denotes the causal relation $x_{j}\rightarrow Gx_{j}$
that amplifies the mean $x_{j}$ of the Gaussian terms, as evident
in the Q function (Eq. (\ref{eq:Q-sup-1})) of $|\psi_{sup}(t)\rangle$.
The individual trajectories propagate in the backward-time direction,
and are subject to input noises, $\delta x(t_{f})\equiv\eta(t_{f})$
and $\xi_{x}(t)$, as defined by Eqs. (\ref{eq:initialBC-1}) and
(\ref{eq:backwardSDE-2-1}). The noise $\delta x(t)$ gives the width
of each branch a time $t$, and is constant in average magnitude throughout
the amplification, as determined by the variance $\sigma_{x}$. The
orange dashed lines represent backward-propagating (right to left)
and forward-propagating (left to right) trajectories for $x$ and
$p$. In the $Q$ model of reality (Sec. \ref{secQmodel-of}), the
value given by $x(t_{f})$ is the detected outcome. The inferred
outcome for $\hat{x}$ is $x(t)/G$ giving either $x_{1}$ or $-x_{1}$
($G=e^{gt_{f}}$). The $p$ and $\delta x$ are not amplified and
are hence not measured. \label{fig:macro-sim-1}}
\end{figure}

\begin{figure}
\includegraphics[width=1\columnwidth]{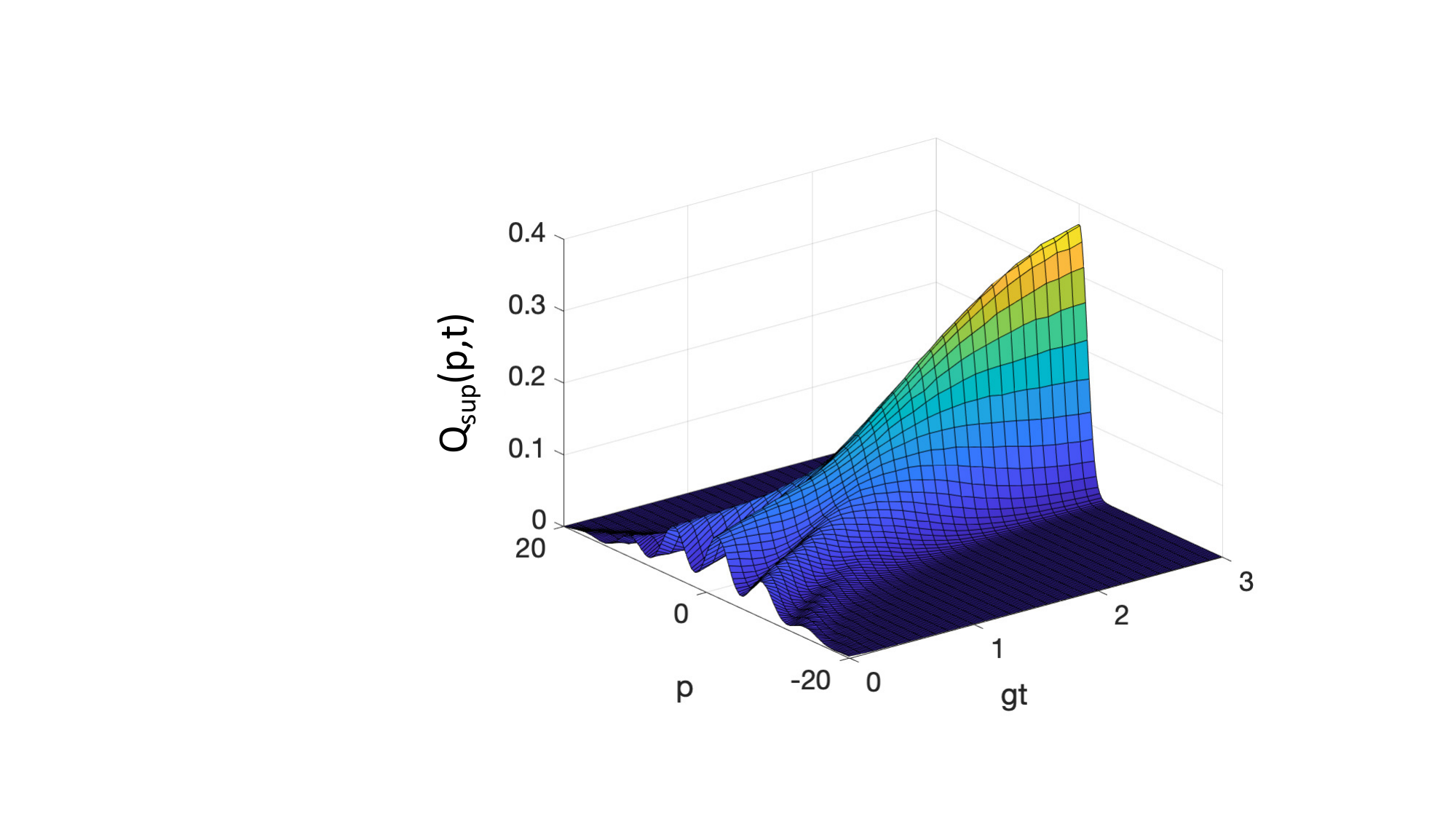}

\includegraphics[width=0.7\columnwidth]{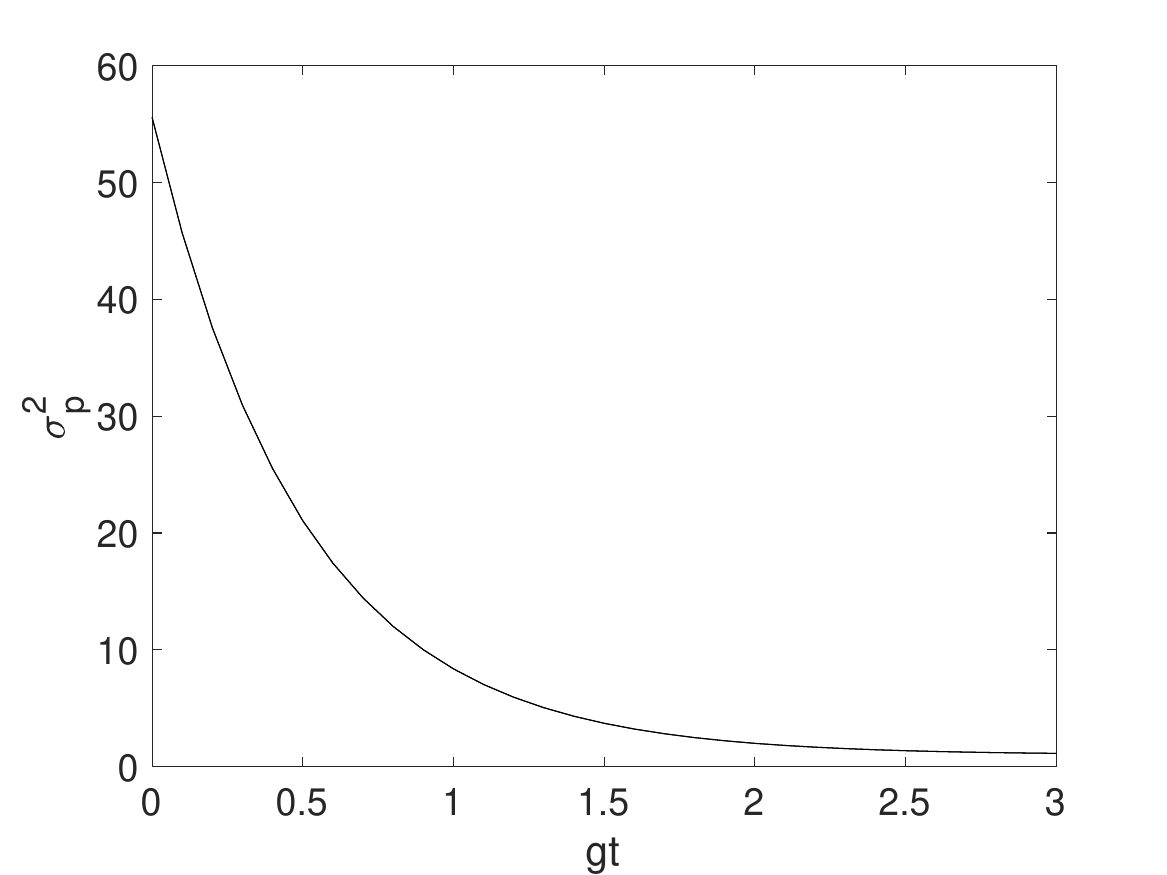}\caption{\textbf{\emph{Decay of the complementary variable: }}Solutions for
the complementary variable $p$, for measurement of $\hat{x}$ on
the system in the superposition $|\psi_{sup}\rangle$ of two eigenstates
of $\hat{x}$. Parameters are as for Figure \ref{fig:stochastic-sup-micro},
with $x_{1}=1$. \label{fig:p-trajectories-fringe} The top figure
shows the distribution $Q_{sup}(p,t)$ of the forward-propagating
trajectories of $p$, as evaluated from $10^{6}$ trajectories, a
sample of which is given in Figure \ref{fig:stochastic-sup-micro}.
Fringes are evident at $t=0$. The lower plot confirms the reduction
in the variance $\sigma_{p}^{2}(t)$ to the ``hidden'' unobservable
value of $1$, as $t_{f}\rightarrow\infty$.}
\end{figure}

\textbf{\emph{Forward solutions for $p$:}} The boundary condition
for the forward equation (\ref{eq:forwardSDE-2-1}) for $p$ is given
by the marginal $Q_{sup}(p,t_{0})=\int Q_{sup}(x,p,t_{0})dx$ for
$p$ at $t_{0}=0$. We find
\begin{eqnarray}
Q_{sup}(p,t_{0}) & = & N\frac{e^{-p^{2}/2\sigma_{p}^{2}}}{\sqrt{2\pi}\sigma_{p}}[1+2|c_{1}c_{2}|e^{-\frac{(x_{1}-x_{2})^{2}}{8\sigma_{x}^{2}}}\mathcal{F}]\nonumber \\
\label{eq:pmarg}
\end{eqnarray}
Taking $\cos\varphi=0$ ($c_{2}$ is imaginary) and $x_{1}=-x_{2}$,
this reduces to $\frac{e^{-p^{2}/2\sigma_{p}^{2}}}{\sqrt{2\pi}\sigma_{p}}\Bigl\{1-e^{-x_{1}^{2}/2\sigma_{x}^{2}}\sin(px_{1}/\sigma_{x}^{2})\Bigl\}$.
 Forward trajectories for the attenuated variable $p$ are shown
in Figures \ref{fig:stochastic-sup-micro} and \ref{fig:macro-sim-1}.
 From the solution (\ref{eq:pmarg}), we see that with an increasing
separation of the two eigenstates $|x_{1}\rangle$ and $|-x_{1}\rangle$
in the superposition $|\psi_{sup}\rangle$ (so that $|x_{1}|$ increases),
the fringes become less prominent. The evolution of $Q_{sup}(p,t)$
is shown in Figure \ref{fig:p-trajectories-fringe} (top panel). The
trajectories of $p$ decay to the hidden noise level $\sigma_{p}^{2}(t_{f})=1$
(Fig. \ref{fig:p-trajectories-fringe}, lower panel).

\textbf{\emph{Backward solutions for $x$:}} The boundary condition
for the backward equation (\ref{eq:backwardSDE-2-1}) determines the
dynamics of the amplified variable $x$, corresponding to the measured
quantity $\hat{x}$, which is of most interest. The boundary condition
is determined by the state $|\psi_{sup}(t)\rangle=e^{-iH_{amp}t/\hbar}|\psi_{sup}\rangle$
after amplification $H_{amp}$. We find
\begin{eqnarray}
|\psi_{sup}(t)\rangle & = & N(c_{1}|Gx_{1},r'\rangle_{sq}+c_{2}|Gx_{2},r'\rangle_{sq})\nonumber \\
\label{eq:amp-sup-3}
\end{eqnarray}
where $G=e^{gt}$ is the amplification factor, $r'=-gt+r$ and $t$
is the time of evolution (refer Appendix C). The $Q$ function of
the amplified system is
\begin{eqnarray}
Q_{sup}(x,p,t) & = & N_{t}\frac{e^{-p^{2}/2\sigma_{p}^{2}(t)}}{2\pi\sigma_{x}(t)\sigma_{p}(t)}\Bigr(|c_{1}|^{2}e^{-(x-G(t)x_{1})^{2}/2\sigma_{x}^{2}(t)}\nonumber \\
 &  & +|c_{2}|^{2}e^{-(x-G(t)x_{2})^{2}/2\sigma_{x}^{2}(t)}+2|c_{1}c_{2}|\mathcal{\mathcal{I}}(t)\Bigl)\nonumber \\
\label{eq:Q-sup-1}
\end{eqnarray}
where $G(t)=e^{gt}$, $N_{t}$ is a normalization constant and $\mathcal{I}(t)$
is given by $\mathcal{I}$ and $\mathcal{F}$ of Eq. (\ref{eq:int-2}),
but substituting $x_{j}\rightarrow G(t)x_{j}$, and $\sigma_{x}\rightarrow\sigma_{x}(t)$,
$\sigma_{p}\rightarrow\sigma_{p}(t)$. \textcolor{black}{The variances
$\sigma_{x}(t)$ and $\sigma_{p}(t)$ of the amplified state are given
by (}\ref{eq:amp-var}). Importantly, we note that the means $x_{1}$
and $x_{2}$ of the Gaussian functions amplify according to 
\begin{equation}
x_{j}\rightarrow X_{j}=Gx_{j}\label{eq:amp-g-1-1}
\end{equation}
where $G\equiv G(t)=e^{gt}$, which being a deterministic relation
can be reversed, so that $Gx_{j}\rightarrow x_{j}$. By contrast,
the variances in $x$ of the Gaussian distributions do not amplify.
The variances are 
\begin{equation}
\sigma_{x}^{2}(t)=1\label{eq:var-const}
\end{equation}
throughout the evolution, the fluctuations hence remaining constant
at the ``hidden level''. \textcolor{red}{}This is evident in
Figures \ref{fig:stochastic-sup-micro} and \ref{fig:macro-sim-1},
where the width of each band of trajectories, as associated with each
eigenvalue, is constant with $t$. By contrast, $\sigma_{p}^{2}$
is initially large as for an eigenstate $|x_{j}\rangle$, but reduces
to the hidden noise level $\sigma_{p}^{2}(t)=1$ as $t\rightarrow\infty$.
We see that the Gaussian functions become increasingly sharp as the
system amplifies. We must also examine how the interference term $\mathcal{I}(t)$
in the Q function (\ref{eq:Q-sup-1}) behaves on amplification.

The future boundary condition for the backward equation is given by
the marginal 
\begin{equation}
Q_{sup}(x,t)=\int Q_{sup}(x,p,t)dp\label{eq:fbc}
\end{equation}
at the final time $t=t_{f}$. This requires integration over the
variable $p$. In order to calculate the marginal, we evaluate $\mathcal{I}(x,t)=\int dp\mathcal{I}$,
where $\mathcal{I}$ is given by (\ref{eq:int-2}), which requires
the integration
\begin{eqnarray}
 &  & \int_{-\infty}^{\infty}dp\frac{e^{-p^{2}/2\sigma_{p}^{2}(t)}}{2\pi\sigma_{x}(t)\sigma_{p}(t)}\Bigl\{(\cos\varphi)\cos[\frac{p}{2\sigma_{x}^{2}(t)}G(t)(x_{1}-x_{2})]\nonumber \\
 &  & -(\sin\varphi)\sin[\frac{p}{2\sigma_{x}^{2}(t)}G(t)(x_{1}-x_{2})]\Bigr\}\label{eq:int}
\end{eqnarray}
The second integral proportional to $\sin\varphi$ is zero by symmetry.
The first term simplifies to
\begin{eqnarray}
 &  & \cos\varphi\int_{-\infty}^{\infty}dp\frac{e^{-p^{2}/2\sigma_{p}^{2}(t)}}{\sqrt{2\pi}\sigma_{p}}\cos[\frac{p}{2\sigma_{x}^{2}(t)}(x_{1}-x_{2})]\nonumber \\
 & = & (\cos\varphi)\thinspace e^{-G(t)^{2}(x_{1}-x_{2})^{2}\sigma_{p}^{2}(t)/8\sigma_{x}^{4}(t)}\label{eq:int-2-2}
\end{eqnarray}
This term gives the contribution of the interference to the marginal,
and becomes important for the mechanism of Bell violations in Secs.
\ref{sec:CV-Bell-nonlocality} and \ref{sec:Causal-structure-bell}.
It depends on the variance $\sigma_{p}$ of $p$, which tends to $\infty$
at the initial time $t_{0}$, since we have considered a superposition
$|\psi_{sup}\rangle=\sum_{j}c_{j}|x_{j}\rangle$ of eigenstates of
$\hat{x}$. The solution for the marginal is
\begin{eqnarray}
Q_{sup}(x,t) & = & N_{t}\frac{1}{\sqrt{2\pi}\sigma_{x}(t)}\Bigr(|c_{1}|^{2}e^{-(x-G(t)x_{1})^{2}/2\sigma_{x}^{2}(t)}\nonumber \\
 &  & +|c_{2}|^{2}e^{-(x-G(t)x_{2})^{2}/2\sigma_{x}^{2}(t)}+2|c_{1}c_{2}|\mathcal{I}(x,t)\Bigl)\nonumber \\
\label{eq:Qmarg-x-amp}
\end{eqnarray}
where
\begin{eqnarray}
\mathcal{I}(x,t) & = & \cos\varphi e^{-[(x-G(t)x_{1})^{2}+(x-G(t)x_{2})^{2}]/4\sigma_{x}^{2}(t)}\nonumber \\
 &  & e^{-G(t)^{2}(x_{1}-x_{2})^{2}\sigma_{p}^{2}(t)/8\sigma_{x}^{4}(t)}\label{eq:int-marg}
\end{eqnarray}
Here, $\sigma_{x}^{2}\left(t\right)=1+G(t)^{2}\left[\sigma_{x}^{2}\left(0\right)-1\right]$
and $\sigma_{p}^{2}(t)=1+[\sigma_{p}^{2}(0)-1]/G(t)^{2}$ (Eq. (\ref{eq:amp-var})).
We see that $\sigma_{p}(0)\rightarrow\infty$ and $\sigma_{x}^{2}\left(t\right)=1$.
Immediately, we see that the interference term $\mathcal{I}$ will
vanish at the initial time $t=0$ in the marginal for $x$. Hence
$\mathcal{I}(x,0)\rightarrow0$ in the limit $r\rightarrow\infty$,
which corresponds to the expansion of $|\psi_{sup}\rangle$ in terms
of the measurement basis, as in Eq. (\ref{eq:sup-sq}). 

The vanishing of the interference term $\mathcal{I}(x,0)$ in the
marginal is an important result, which is due mathematically to the
integral (\ref{eq:int-2-2}), which vanishes when $\sigma_{p}(t)$
is large. Regardless, on substituting $\sigma_{p}(t_{f})=\sigma_{x}(t_{f})=1$,
we see that  $\mathcal{I}(x,t_{f})$ rapidly de-amplifies with amplification
$G$, with $\mathcal{I}(x,t_{f})\rightarrow(\cos\varphi)e^{-x^{2}/2-G^{2}x_{1}^{2}}\rightarrow0$
for $x_{1}=-x_{2}$. In this way, we see that \emph{amplification
plays a similar role to decoherence in attenuating the interference
terms }that distinguish the superposition $|\psi_{sup}\rangle$ from
the mixture $\rho_{mix}$.

The generalization to expansions $|\psi_{sup}\rangle=\sum_{j}c_{j}|x_{j}\rangle$
is straightforward (refer Appendix C). This leads to the following
Definition (given as Definition (4) in Sec. I.B) and Result:

\textbf{\emph{Definition: Hidden interference terms:}} Consider measurement
of $\hat{x}$. The interference terms $\mathcal{I}$ appearing in
the Q function $Q(x,p,t)$ when the Q function is expanded in terms
of the Q functions of the eigenstates of $\hat{x}$ are referred to
as ``hidden'' interference terms. These terms are calculated by
writing $|\psi\rangle$ in terms of the eigenstates $|x_{j}\rangle$
(i.e. expanding $|\psi\rangle$ in the measurement basis) and calculating
the Q function directly from Eq. (\ref{eq:QH-1}).

\textbf{\emph{Result III.1: Interference terms and the Future Boundary
Condition: }}The hidden interference terms arising in the Q function
of a state $|\psi\rangle$ do not contribute to the future boundary
condition, given by Eq. (\ref{eq:fbc}) (proved above). Consider measurement
of $\hat{x}$ on the system $|\psi_{sup}\rangle=\sum_{j}c_{j}|x_{j}\rangle$
(Eq. (\ref{eq:sup-sq})) at time $t_{0}=0$.  From Eq. (\ref{eq:Qmarg-x-amp})
with $\mathcal{I}(x,t)\rightarrow0$ and $\sigma_{x}(t_{f})=1$, the
future boundary condition for the backward equation (\ref{eq:backwardSDE-2-1})
reduces to ($G=e^{gt_{f}})$\textbf{
\begin{eqnarray}
Q_{sup}(x,t_{f}) & = & \frac{1}{\sqrt{2\pi}\sigma_{x}(t_{f})}\sum_{j}|c_{j}|^{2}e^{-\left(x-Gx_{j}\right){}^{2}/2}\nonumber \\
\label{eq:fmarg-1}
\end{eqnarray}
}Solutions showing individual trajectories for the amplified variable
$x$ are given in Figures \ref{fig:stochastic-sup-micro} and \ref{fig:macro-sim-1}.

\emph{Comment: }The interference terms $\mathcal{I}(x,t)$ of the
\emph{marginal} $Q(x)$ vanish in the limit of large $r$, because
the integral (\ref{eq:int-2-2}) vanishes. However, the interference
$\mathcal{I}(t)$ does not vanish in the initial joint distribution
$Q(x,p,0)$. Moreover, the integral will not vanish where there is
a change of basis so that $\sigma_{p}(0)$ is no longer large, which
is relevant to Bell violations (refer Sec. VII). $\square$

\emph{Comment:} The future boundary condition (\ref{eq:fmarg-1})
is the same as that for the system prepared in the mixed state $\rho_{mix}$
(Eq. (\ref{eq:mix-sup-1})).

\textbf{\emph{Procedure of simulation:}} The Result III.1 defines
the procedure of the simulation. The interference terms in $Q(x,t)$
are justified to be ignored \emph{provided we write in the measurement
basis} as in Result III.1 where $r\rightarrow\infty$ (or else for
all $r$, if $Re\{c_{2}\}=0$ (refer (\ref{eq:sup-sq}))). Then, the
future boundary condition for simulation of (\ref{eq:backwardSDE-2-1})
is determined by the marginal (\ref{eq:fmarg-1}), \textbf{}which
comprises a set of Gaussian peaks (one for each outcome $x_{j}$)
with equal variance $\sigma_{x}^{2}(t_{f})=1$. In order to model
measurement, strong amplification is needed, and $gt_{f}$ is large.

We consider the superposition of two eigenstates $|x_{1}\rangle$
and $|x_{2}\rangle$ as in Eq. (\ref{eq:sup-sq}). The initial condition
for a given run of the simulation of Eq. (\ref{eq:backwardSDE-2-1})
is
\begin{equation}
x(t_{f})=Gx_{j}+\delta x(t_{f})\label{eq:initialBC-1}
\end{equation}
where the eigenvalue $x_{j}$ is selected to be $x_{1}$ or $x_{2}$
with probability $|c_{j}|^{2}$. The $\delta x(t_{f})$ is a noise
term, which we also label $\eta(t_{f})$, sampled from the Gaussian
distribution $\mathcal{G}(0,1)$ of mean $0$ and variance $\sigma^{2}=1$.
In this paper, we define the Gaussian distribution using the notation
$\mathcal{G}(\bar{x},\sigma^{2})=\frac{e^{-(x-\bar{x})^{2}/2\sigma^{2}}}{\sigma\sqrt{2\pi}}$,
where $\bar{x}$ is the mean and $\sigma^{2}$ is the variance of
$x$.

\begin{figure}
\begin{centering}
\includegraphics[width=0.7\columnwidth]{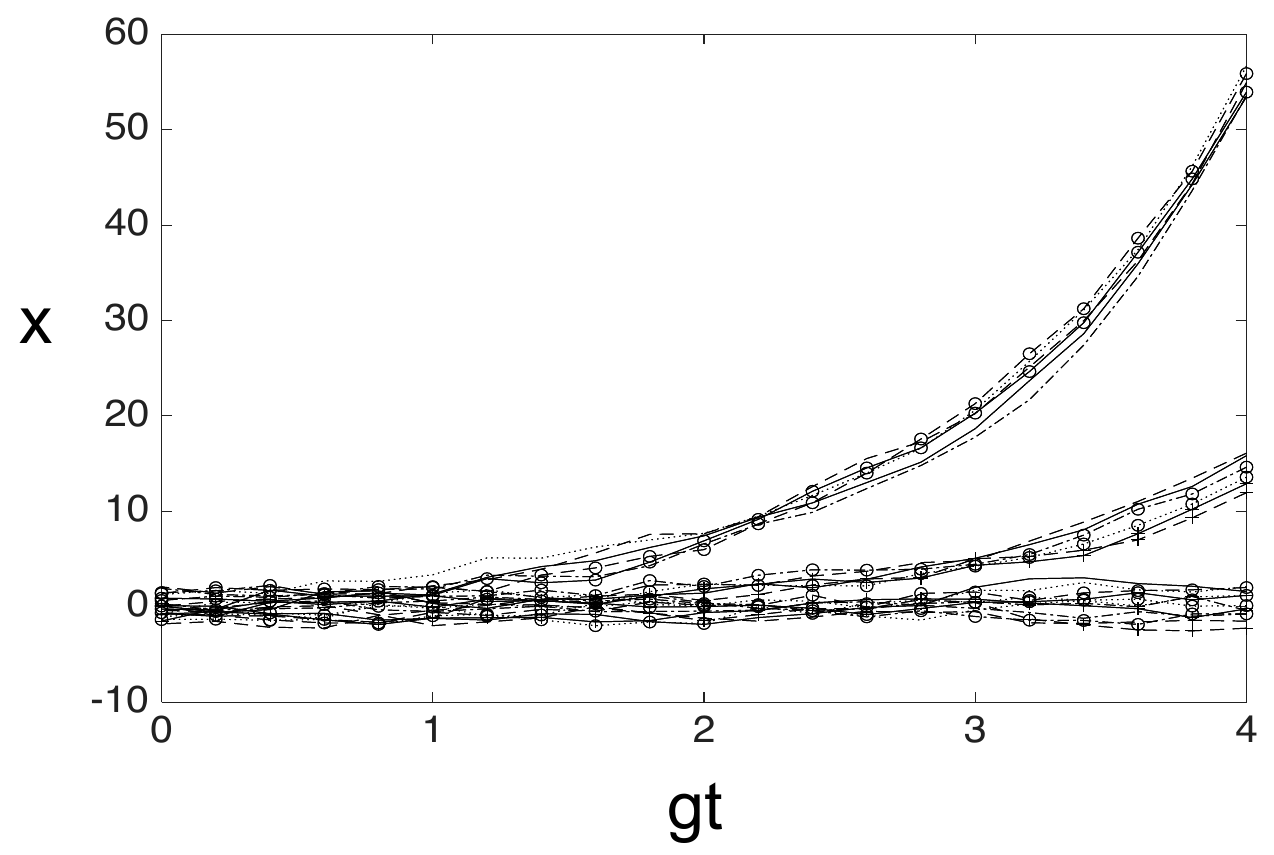}
\par\end{centering}
\caption{\textbf{\emph{Measurement on Eigenstates showing branches have constant
width independent of eigenvalue:}} Solutions showing the amplification
of $\hat{x}$ for the eigenstate $|x_{j}\rangle$, for different values
of $x_{j}$: The curves correspond to $x_{j}=1$ (top curve), $x_{j}=0.25$
(centre top), $x_{j}=0.01$ (centre lower) and $x_{j}=0$ (lowest
curve), with $gt_{f}=4$. The curves $x=0.01$ and $x_{j}=0$ cannot
be distinguished with the level of amplification given by $gt_{f}=4$.
The curve for $x_{j}=0$ is the solution $\delta x$ of Eq. (\ref{eq:delta-eq-1}),
validating the expansion $x(t)=x_{j}(t)+\delta x(t)$ of Eq. (\ref{eq:fullx})
in which $\delta x$ is a ``hidden'' noise term of constant average
magnitude independent of $x_{j}$. \label{fig:solutions-eigenstate-zero}\textcolor{blue}{}}
\end{figure}

\emph{Comment:} Importantly, we see that value $\delta x(t_{f})$
(denoted by $\eta(t)$ in Figure 2) is hence\emph{ independent} of
the choice of eigenvalue $x_{j}$ and independent of $t_{f}$ (Fig.
\ref{fig:solutions-eigenstate-zero}). This is relevant to understanding
how Grandfather paradoxes are avoided. The sampling is determined
by (\ref{eq:fmarg-1}), which is identical to that for the system
prepared in a mixture of eigenstates, $\rho_{mix}$, (Eq. (\ref{eq:mix-sup-1})).
We will see that the procedure of the simulation defines the causal
model depicted in Figure 2, and explained in Sec. \ref{sec:Causal-model-for}.

The simulations reveal \emph{branches}, as $gt_{f}\rightarrow\infty$
(Figs. \ref{fig:stochastic-sup-micro} and \ref{fig:macro-sim-1}).
This leads to the following Definition (given as Definition (5) in
Sec. I.B).

\textbf{\emph{Definition:}} Branches $\mathcal{B}_{j}$ are defined
as groups of amplitudes $x(t)$, each associated with a distinct eigenstate
$|x_{j}\rangle$ in the superposition $|\psi_{sup}\rangle$ (Figs.
\ref{fig:stochastic-sup-micro} and \ref{fig:macro-sim-1}).

\textbf{\emph{Result III.2: Branches emerge with amplification:}}

\emph{Proof:} The fluctuations with variance $\sigma_{x}$ about the
mean $x_{j}$, as associated with each Gaussian in the Q function
$Q_{sup}(x,p,t_{0}$) (Eq. (\ref{eq:Q-sup})) of the initial state,
are not amplified (Eq. (\ref{eq:Q-sup-1})) with the evolution. However,
the mean values $x_{j}$ are amplified according to $x_{j}\rightarrow X_{j}=Gx_{j}$,
where $G=e^{gt_{f}}$ (Eq. (\ref{eq:amp-g-1-1})). Hence,  as $gt\rightarrow\infty$,
we see distinct peaks centered at $Gx_{j}$ with relative probabilities
$|c_{j}|^{2}$. The eigenvalues $x_{j}$ are amplified, but the ``hidden''
noise $\delta x$ about these values is not. Hence, branches emerge
for $x(t)$ as $t$ increases. The hidden noise represents the ``width
$w\sim\delta x$ of the ``lines'' that form the branches, and is
constant, independent of $x_{j}$, as in Figure \ref{fig:stochastic-sup-micro}.
$\square$

\textbf{\emph{Result III.3: The hidden noise $\delta x$ (the ``width''
of the branch) is independent of the eigenvalue $x_{j}$:}}

\emph{Proof:} The width $w$ of each branch $\mathcal{B}_{j}$ becomes
indiscernible relative to the mean value $Gx_{j}$, as $gt\rightarrow\infty$.
The ``hidden'' vacuum fluctuation $\delta x(t)$ does not lead to
a macroscopic value. The amplitudes can be written
\begin{equation}
x(t)=x_{j}(t)+\delta x(t)\label{eq:fullx}
\end{equation}
where $x_{j}(t)=e^{gt}x_{j}$, the solutions for $\delta x(t)$ being
independent of $x_{j}$, satisfying the stochastic equation
\begin{equation}
\frac{d(\delta x(t))}{dt_{-}}=-g\delta x(t)+\xi_{x}(t)\label{eq:delta-eq-1}
\end{equation}
with input $\delta x(t_{f})\equiv\eta(t_{f})$ given by sampling the
Gaussian $\mathcal{G}(0,1)$. This is verified in Figure \ref{fig:solutions-eigenstate-zero},
where the $\delta x(t)$ can be taken as the solution of (\ref{eq:backwardSDE-2-1})
with $x_{j}=0$. The noise $\delta x(t)=\eta(t)$ adds equally to
the exponential lines given for the different eigenvalues $x_{j}$
in Figure \ref{fig:solutions-eigenstate-zero}. $\square$

\section{$Q$ model of reality\label{secQmodel-of}}

\subsection{Assumptions and basic results}

The solutions for $x(t_{f})$ in Figures \ref{fig:stochastic-sup-micro}
and \ref{fig:macro-sim-1} show two branches, associated with each
eigenstate of the superposition $|\psi_{sup}\rangle$. The stochastic
equations hence motivate a ``hidden-variable'' model for quantum
measurement, referred to as the $Q$-based reality model ($Q$-model
of reality) or Objective-field model \citep{drummond2020retrocausal,Friederich2021Introducing,drummond2021objective}.
We regard the amplitudes $\bm{\lambda}$ (i.e. $x$ and $p$) of the
Q function $Q(x,p)$ as representing a realization of the quantum
state, immediately prior to and during the measurement. The physical
assumptions made for the meter $H_{amp}$ that defines the measurement
of $\hat{x}$ are: 
\begin{itemize}
\item Measurement is identified as an amplification process (after measurement
settings are fixed).
\item The amplitudes $x$ and $p$ are phase-space variables, with distribution
$Q(x,p,t)$ at time $t$. We denote the values of $x$ and $p$ at
time $t$ as $x(t)$ and $p(t)$. Here, we denote $x(t_{f})$ as the
amplified variable.
\item The detected value is $x(t_{f})$, when macroscopic. The amplitudes
at the ``hidden'' vacuum level (i.e. where the variances $\sigma_{x}^{2}$
or $\sigma_{p}^{2}$ are $1$) are not observable. The inferred (or
``measured'') outcome is $x_{0}=x(t_{f})/G$ where $G$ is the amplification
factor.
\item The probability density function for the detection of $x(t_{f})$
is $Q(x,t_{f})=\int Q(x,p,t_{f})dp$ where $t_{f}\rightarrow\infty$
is the time of interaction with the amplifier.
\item Amplified variables satisfy future boundary conditions and attenuated
variables satisfy past boundary conditions. The amplifier does not
couple amplified and attenuated quadratures.
\end{itemize}
The last condition applies to the model of measurement given by $H_{amp}$
(Eq. \ref{eq:ham-2-1})) but may be adjusted when the measurement
model includes a coupling to a separate system that is a meter (refer
\citep{Reid2023Short}).  The condition is motivated by the stochastic
method (Appendices A and B) and by the causal consistency proved
in Secs. V.B, IX.C and Appendix H. In the model, the system evolves
under $H_{amp}$ for a time $t_{f}$, to give the final measurement
output. The measurement is completed simply by the fields being amplified
to become macroscopic, and hence detectable to macroscopic devices
or observers. No special ``collapse'' is required, only standard
physical processes. The analysis is hence compatible with Bell's
requirement \citep{bell1990against} that there should be no ``shifty
split'' between system and measurement apparatus.

We present a series of Results applying to the $Q$-based model of
reality.

\textbf{\emph{Result IV.1: The inferred outcomes are the eigenvalues:
}}Consider measurement of $\hat{x}$. After sufficient amplification,
the $x(t_{f})$ separate into branches each associated with a single
eigenstate $|x_{j}\rangle$ of $\hat{x}$ (Definition (5) of Sec.
I.B and Result III.2, Fig. \ref{fig:stochastic-sup-micro}). The detected
value according to the model of reality is $x(t_{f})$, where $t_{f}\rightarrow\infty$
so that $x(t_{f})$ is macroscopic, and the inferred outcome for $\hat{x}$
is $x(t_{f})/G$. The inferred outcome will always be one of the eigenvalues
$x_{j}$.

\emph{Proof:}   Since from Eq. (\ref{eq:fullx}) (based on the
solutions of (\ref{eq:backwardSDE-2-1})), we write $x(t)=e^{gt}x_{j}+\delta x(t)$,
we see that with sufficient amplification, the inferred outcome is
one of the eigenvalues $x_{j}$. The vacuum fluctuation $\delta x(t)$
does not lead to a macroscopic value and is undetectable, becoming
$\delta x(t)/G$.  As $t_{f}$ increases, each branch becomes identifiable
as a thin line, which represents a distinct value $x_{j}$ for $\hat{x}$.
$\square$

\textbf{\emph{Result IV.2: Incompleteness:}}\emph{ }The amplitudes
$x$ and $p$ of $Q(x,p,t)$ at a time $t$ (along with the forward-backward
stochastic equations, (\ref{eq:backwardSDE-2-1}) and (\ref{eq:forwardSDE-2-1}))
are not sufficient to completely define the state and future dynamics
of the system.

\emph{Proof:} Consider the measurement $\hat{x}$ on an eigenstate
$|x_{j}\rangle$ of $\hat{x}$ where $x_{j}\rightarrow0$. The measurement
always yields the outcome $x_{j}$, which is the eigenvalue, but which
is not identifiable from the amplitudes $x$ and $p$ alone (refer
Fig. \ref{fig:solutions-eigenstate-zero}). The value $x_{j}$ is
masked by the hidden Gaussian noise when $x_{j}$ is small (Fig. \ref{fig:stochastic-sup-micro}).
The future boundary condition for the backward equation is required
for the solutions $x(t)$. This requires knowledge of the Q function
$Q(x,p,t_{0}$), \emph{or} at least (in a hidden-variable model, refer
Sec. IV.G and Sec. V) \emph{knowledge of the Gaussian-mean value}
$x_{j}$. $\square$

\emph{Comment IV.2:} Our interpretation of the Q-model of reality
and associated causal model for measurement (Sec. V) includes that
the system at time $t_{0}$ is defined as prepared with respect to
a measurement basis, such as $|x_{j}\rangle$ of $\hat{x}$ \citep{chaves2018causal}.
The Q function uniquely describes the quantum state $|\psi\rangle$
which can be expanded as $|\psi\rangle=\sum_{j}c_{j}|x_{j}\rangle$,
from which the Q function will be of the form (\ref{eq:Q-sup}), a
sum of Gaussian functions in $x$ and $p$ with mean $x_{j}$, as
well as interference terms of type $Int$, as generalized for an arbitrary
number of states $|x_{j}\rangle$. The means $x_{j}$ and $c_{j}$
are hence in principle identifiable as part of the description of
the state at time $t_{0}$ (Appendix C). We show in Sec. V.A that
the dynamics of $x(t)$ as implied by $H_{amp}$ can be deduced from
the backward stochastic equation (\ref{eq:backwardSDE-2-1}), if the
$x_{j}$ and $c_{j}$ are known, based on the deterministic relation
$x_{j}\leftrightarrow Gx_{j}$ (which assumes the system at time $t_{0}$
once $H_{amp}$ is implemented to be associated probabilistically
with a particular $x_{j}$, refer Secs. IV.G and V.A).

\subsection{Derivation of Born's rule}

In the $Q$ model of reality, the density of the amplitudes $x(t_{f})$
as $gt_{f}\rightarrow\infty$ gives the probability for detection,
which leads to Born's rule. The rule follows  because the measurement
of $\hat{x}$ amplifies the variable $x$, but attenuates the complementary
variable $p$ and interference terms $\mathcal{I}$ (Result III.1).
 We give a derivation below.

\textbf{\emph{Result IV.3a: Born's rule:}} Consider a general superposition
$|\psi_{sup}\rangle=\sum_{j}c_{j}|x_{j}\rangle$ of eigenstates $|x_{j}\rangle$
of $\hat{x}$, where $c_{j}$ are probability amplitudes. Here, we
define the eigenstates as the squeezed states $|\frac{x_{j}}{2},r\rangle_{sq}$
(Eq. (\ref{eq:eigenstate-def})) in the limit where $r\rightarrow\infty$.
The final density $Q_{sup}(x,t_{f})$ given by the $Q$ function of
the amplified state $e^{-iH_{amp}t/\hbar}|\psi_{sup}\rangle$ leads
to Born's rule, that the probability of an outcome $x_{j}$ is $|c_{j}|^{2}$.

\emph{Proof:} In the limit of $gt\rightarrow\infty$, the marginal
distribution for the amplified state is\textbf{{} 
\begin{eqnarray*}
Q_{sup}(x,t) & = & \frac{1}{\sqrt{2\pi}\sigma_{x}(t)}\sum_{j}|c_{j}|^{2}e^{-\left(x-G(t)x_{j}\right){}^{2}/2\sigma_{x}^{2}(t)}
\end{eqnarray*}
}where $G=e^{gt}$ ($g>0$) and $\sigma_{x}^{2}(t)=1+e^{2gt-2r}$
(refer Eq. (\ref{eq:q-sq-1})). Considering the measured inferred
variable $x_{0}=x/G$ where $G=e^{gt}$, the rescaled distribution
becomes\textbf{{} 
\begin{eqnarray}
Q_{sc}(x_{0},t) & = & \frac{1}{\sqrt{2\pi}\sigma_{x_{0}}(t)}\sum_{j}|c_{j}|^{2}e^{-\left(x_{0}-x_{j}\right){}^{2}/2(e^{-2gt}+e^{-2r})}\nonumber \\
 & \rightarrow & \frac{1}{\sqrt{2\pi}\sigma_{x_{0}}(t)}\sum_{j}|c_{j}|^{2}e^{-\left(x_{0}-x_{j}\right){}^{2}/(2e^{-2r})}\label{eq:qsc}
\end{eqnarray}
}on taking $gt$ large.\textbf{ }The eigenstates of $\hat{x}$
correspond to $r\rightarrow\infty$ where the Gaussian peaks become
infinitely narrow. The separation between peaks is much larger than
the width of the Gaussian distributions. Hence, in the Q model, the
probability of observing $x_{j}$ is given by the weighting $|c_{j}|^{2}$,
in accordance with Born's rule. $\square$

\textbf{\emph{Result IV.3b: }}We demonstrate Born's rule for measurement
of the rotated quadrature
\begin{equation}
\hat{x}_{\theta}=\hat{x}\cos\theta+\hat{p}\sin\theta\equiv\hat{a}e^{-i\theta}+\hat{a}^{\dagger}e^{i\theta}\label{eq:xtheta}
\end{equation}
The complementary observable is $\hat{p}_{\theta}=-\hat{x}\sin\theta+\hat{p}\cos\theta$.

\emph{Proof: }We expand in eigenstates $|x_{\theta j}\rangle$ of
$\hat{x}_{\theta}$ as $|\psi_{sup,\theta}\rangle=\sum_{j}d_{j}|x_{\theta j}\rangle,$
where $d_{j}$ are probability amplitudes. The eigenstates are defined
similarly to $|x_{j}\rangle$ (Eq. (\ref{eq:eigenstate-def})) as
highly squeezed states in $\hat{x}_{\theta}$ (Appendix C). We define
real coordinates $x_{\theta}$ and $p_{\theta}$ as $\alpha e^{-i\theta}=(x_{\theta}+ip_{\theta})/2$.
Extending the derivation of the Q function of a squeezed state as
in Appendix C, the Q function $Q_{sup,\theta}(x_{\theta},p_{\theta})$
can be derived for $|\psi_{sup,\theta}\rangle$. Modeling the amplification
of $|\psi_{sup,\theta}\rangle$ by applying
\begin{equation}
H_{amp,\theta}=i\hbar g(\hat{a}^{\dagger2}e^{2i\theta}-\hat{a}^{2}e^{-2i\theta})/2\label{eq:hamtheta-1}
\end{equation}
(which gives $\hat{x}_{\theta}(t)=e^{gt}\hat{x}_{\theta}(0)$), the
Result is proved by evaluating the marginal $Q_{sup,\theta}(x_{\theta},t)$
of the Q function of the amplified state (following Eqs. (\ref{eq:Q-sup-1}-\ref{eq:Qmarg-x-amp})),
and applying the same analysis as for Result IV.3. $\square$

Summarizing, the probability density for an outcome $x_{j}$, given
$gt_{f}\rightarrow\infty$, is
\begin{equation}
P(x_{j})\equiv Q_{sc}(x_{j},t_{f})=|\langle x_{j}|\psi_{sup}\rangle|^{2}\label{eq:pxborn-1-1}
\end{equation}
in agreement with the quantum prediction (Born's rule). Here, $Q_{sc}(x_{j},t_{f})$
is the rescaled distribution, found by taking $Q_{sup}(x,t_{f})$,
rewriting with respect to the scaled variable $x_{0}=x/G$, and taking
the limit where $G=e^{gt}\rightarrow\infty$.

\textbf{\emph{Role of the Wigner function: }}The realization of Born's
rule for the \emph{continuous-valued} measurements $\hat{x}_{\theta}$
can be best demonstrated by the connection with the Wigner function,
$W(x,p)$. We find this elucidates the nature of the backward stochastic
equation (\ref{eq:backwardSDE-2-1}). It is known that the marginal
\begin{equation}
W(x)=\int W(x,p)dp\label{eq:marg-wx}
\end{equation}
of the Wigner function gives the distribution for observed outcomes
$x$ of $\hat{x}$.

To establish the connection with the Q distribution in the model of
reality, we write the $Q$ function $Q(\bm{\lambda},t_{0})$ for $|\psi\rangle$
at time $t_{0}$ \emph{with respect to the measurement basis.} We
consider measurement of $\hat{x}$ and expand $Q(\bm{\lambda},t_{0})$
in terms of the $Q$ functions of the $\hat{x}$ eigenstates (Definition
(3), Sec. I.B). We find the required form is obtained by writing
the $Q$ function in terms of the Wigner function. The $Q$ function
of a state $|\psi\rangle$ is the convolution of the Wigner function
$W(\beta)$: $Q(\alpha)=\frac{2}{\pi}\int W(\beta)e^{-2|\alpha-\beta|^{2}}d^{2}\beta$
\citep{Wigner1932,hillery1984distribution}. Using $\alpha=(x+ip)/2$,
we find
\begin{eqnarray}
Q(\bm{\lambda},t_{0}) & = & \frac{1}{2\pi}\int W(\bm{\lambda}_{0})e^{-(x-x_{0})^{2}/2}e^{-(p-p_{0})^{2}/2}d\bm{\lambda}_{0}\nonumber \\
\label{eq:QconvW}
\end{eqnarray}
where $\bm{\lambda}\equiv(x,p)$, $\bm{\lambda}_{0}\equiv(x_{0},p_{0})$,
$d\bm{\lambda}_{0}=dx_{0}dp_{0}$ and $W(\bm{\lambda}_{0})$ is the
Wigner function of the state $|\psi\rangle$ at time $t_{0}$. The
convolution ensures the positivity of $Q$, given $W$ can be negative.
The marginal $Q(x,t_{0})=\int dpQ(\bm{\lambda},t_{0})$ is
\begin{eqnarray}
Q(x,t_{0}) & = & \frac{1}{\sqrt{2\pi}}\int dx_{0}W(x_{0})e^{-(x-x_{0})^{2}/2}\label{eq:Q(x)-wig}
\end{eqnarray}
which we can rewrite as
\begin{eqnarray}
Q(x,t_{0}) & = & \int dx_{0}W(x_{0})Q_{x_{0}}(x)\label{eq:Q-Wmarg}
\end{eqnarray}
where we note that $Q_{x_{j}}(x)=\frac{e^{-(x-x_{j})^{2}/2}}{\sqrt{2\pi}}$
is the marginal in $x$ of the Q function $Q_{x_{j}}(x,p)$ of the
$x$-eigenstate $|x_{j}\rangle$ with eigenvalue $x_{j}$, given by
Eq. (\ref{eq:q-sq-1}) on putting $\sigma_{x}^{2}=1$ as valid for
eigenstates of $\hat{x}$ where $r\rightarrow\infty$. Comparing
$Q(x)$ with the marginal $Q_{sup}(x,0)$ of Eq. (\ref{eq:Qmarg-x-amp}),
and using Result III.1 where we show interference terms $\mathcal{I}$
do not contribute to final probabilities for $\hat{x}$, we identify
$W(x_{0})$ as the probabilities $|c_{j}|^{2}$ (taking the limit
of a continuum of states $|x_{j}\rangle$). Hence, $W(x_{0})$ is
the probability density for an outcome $x_{0}$ of $\hat{x}$. This
gives consistency with the earlier Result III.1 that ignores interference
in the future boundary. In summary, we arrive at the following two
Results.

\textbf{\emph{Result IV.4:}} \textbf{\emph{The future boundary condition
is determined by the marginal of the Wigner function of the state
at the time $t_{0}$:}} The Q function $Q(x,p,t_{0})$ represents
the quantum state $|\psi\rangle$ of the system at time $t_{0}$.
The marginal $Q(x,t_{f})=\int dpQ(x,p,t_{f})$ of the Q function $Q(x,p,t_{f})$
applies to the state of the amplified system after interaction with
$H_{amp}$, at a time $t_{f}$. Rewriting $Q(x,t_{f})$ in terms of
the scaled variable $x_{0}=x/G$ that represents the measured outcome,
we establish from Result IV.3 that the resulting function $Q_{sc}(x_{0},t_{f})$
gives the probability distribution for the outcomes $x_{0}$. Hence,
\begin{equation}
Q_{sc}(x_{0},t_{f})\rightarrow W(x_{0})\label{eq:Q-wig}
\end{equation}
as $G=e^{gt_{f}}\rightarrow\infty$, where $W(x_{0})$ is the marginal
of the Wigner function defined by (\ref{eq:marg-wx}). Hence,  the
marginal $W(x)$ of the Wigner function at the time $t_{0}$ determines
the future boundary condition for $x$.

\textbf{\emph{Result IV.5: Physical interpretation of the Wigner function:
}}The simulation gives a physical interpretation of the Wigner function.
The quantum-optics relation (\ref{eq:Q(x)-wig}) between the Wigner
and Q function can be derived from the simulation.

\emph{Proof:} The Q function at time $t_{0}$ is $Q(x,p,t_{0})$ represents
the quantum state at time $t_{0}$.  The marginal $W(x)$ of the
Wigner function corresponds to the marginal $Q_{sc}(x_{0},t_{f})$
of the Q function of the amplified state, when rescaled so that the
``hidden'' Gaussian noise is treated as indiscernible. The Wigner
function determines the future boundary condition. The simulation
of the backward equation (\ref{eq:backwardSDE-2-1}) proceeds by rewriting
$W(x_{0})$ in terms of $x'=Gx_{0}$, where $G=G(t_{f})=e^{gt_{f}}$,
to obtain the rescaled distribution $W_{sc}(x')$, and convoluting
with the ``hidden'' Gaussian noise of variance $\sigma_{x}=1$ that
describes the input $\delta x$ (refer (\ref{eq:initialBC-1})). This
is given by
\begin{eqnarray}
Q(x,t_{f}) & = & \frac{1}{\sqrt{2\pi}}\int dx'W_{sc}(x')e^{-(x-x')^{2}/2}\label{eq:QfBW}
\end{eqnarray}
We have seen from the simulation of the backward equation (Figs. 8
and \ref{fig:macro-sim-1}) that $Q(x,t_{0})$ retains the same distribution
as $Q(x,t_{f})$ over the Gaussian terms $e^{-(x-Gx_{j})^{2}/2}/\sqrt{2\pi}$,
except that the means decay ($x'=Gx_{0}\rightarrow x_{0}$) in the
negative time direction. Hence (since $G(t)=1$ at $t_{0}$ implying
$W_{sc}(x')=W(x_{0})$), we find from the simulation that
\begin{eqnarray}
Q(x,t_{0}) & = & \frac{1}{\sqrt{2\pi}}\int dx_{0}W(x_{0})e^{-(x-x_{0})^{2}/2}\label{eq:QinitialW}
\end{eqnarray}
This explains the fundamental relation (\ref{eq:QconvW}) used in
quantum optics, that gives the relationship between the Wigner and
the Q-function distributions. $\square$

\subsection{(Weak) macroscopic realism}

\begin{figure}
\begin{centering}
\includegraphics[width=0.7\columnwidth]{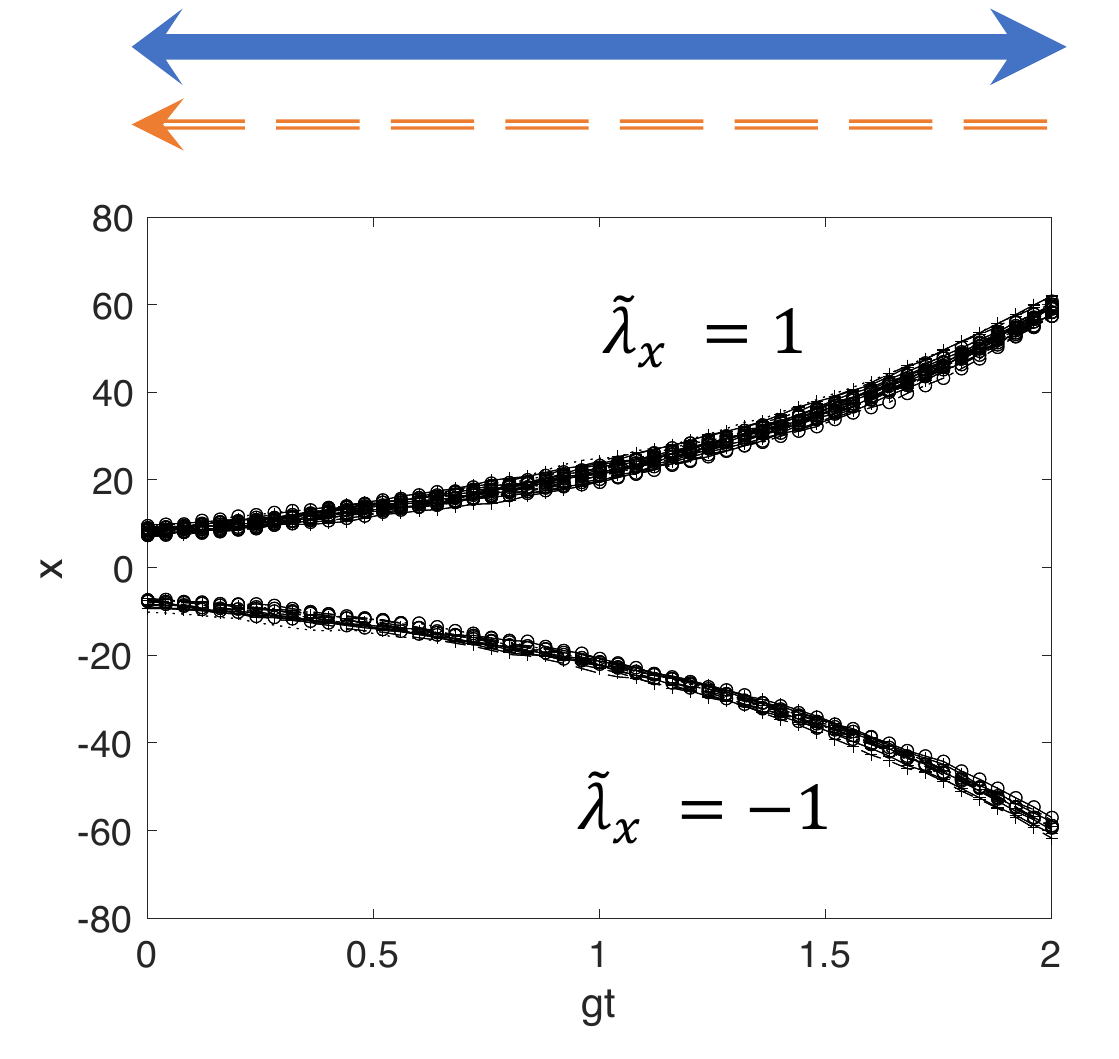}
\par\end{centering}
\caption{\textbf{\emph{Macroscopic realism:}} Trajectories modeling the measurement
of $\hat{x}$ on a macroscopic superposition $|\psi_{sup}\rangle$
of two eigenstates of $\hat{x}$, as in Figure \ref{fig:macro-sim-1}.
The solid blue arrow pointing forward in time denotes the causal relation
$x_{j}\rightarrow Gx_{j}$ that amplifies the mean of the Gaussian
peaks in the Q function of $|\psi_{sup}\rangle$. This relation is
deterministic, and hence can be reversed, giving a retrodiction $Gx_{j}\rightarrow x_{j}$,
indicated by the backward-pointing arrow. For the macroscopic superposition
depicted, the amplitudes $x(0)$ and $x(t_{f})$ of $x$ at the initial
and final times are on the same branch, $\widetilde{\lambda}_{x}$.
The value of $x(0)$ predetermines the outcome of $\hat{x}$ (Results
IV.6 and IV.7). \label{fig:macro-sim-1-1} The amplification, as
in $x_{j}\rightarrow Gx_{j}$ represents a macroscopic causal relation
(forward-pointing arrow).}
\end{figure}

Weak macroscopic realism is defined in Definition (6) of Sec. I.B.
We summarize below, for application to the system of Figures \ref{fig:macro-sim-1}
and \ref{fig:macro-sim-1-1}, where there is measurement of $\hat{x}$.

\textbf{\emph{Definition: Weak macroscopic realism (wMR):}}  Consider
a system in a superposition of two macroscopically distinct states
$\varphi_{1}$ and $\varphi_{2}$ that are distinguished by a binary
measurement $\hat{O}$. We assume that at a time $t_{0}$, any operations
$U$ needed to fix the measurement setting have been carried out.
Weak macroscopic realism posits that the outcome of $\hat{O}$ is
determined, at the time $t_{0}$ \citep{Fulton2024Weak,fulton2024alternative}.

In the simulation shown in Figure \ref{fig:macro-sim-1}, the macroscopically
distinct states are the eigenstates $|x_{1}\rangle$ and $|-x_{1}\rangle$,
where $|x_{1}|\gg1$. The binary-valued measurement $\hat{O}$ given
by the sign of the outcome of $\hat{x}$ distinguishes these states.
We refer to the outcomes of $\hat{O}$ as $+1$ and $-1$. The premise
of wMR posits that the system at time $t_{0}$ has a definite value
for the outcome (denoted by $\widetilde{\lambda}_{x}$) of the measurement
$\hat{O}$. That is, each realization of the system can be regarded
as \emph{either} in a state with $\widetilde{\lambda}_{x}=1$ \emph{or
}in a state with\emph{ }$\widetilde{\lambda}_{x}=-1$. 

\textbf{\emph{Result IV.6:}} \textbf{\emph{Weak macroscopic realism
holds in the Q model of reality: Premise wMR(1)}}

\emph{Proof:} We consider the system described above and depicted
in Figure \ref{fig:macro-sim-1-1}, which is prepared at time $t_{0}$
in the superposition $|\psi_{sup}\rangle$. There are two ways to
carry out the measurement of $\hat{x}$ (and hence of $\hat{O}$)
on the system at time $t_{0}$.

\textbf{(1) }The first is to directly detect the value $x(t_{0})$
of $x$ at time $t_{0}$, since this value is macroscopic by definition
of the macroscopic superposition state i.e. we take $t_{0}\equiv t_{f}$.
According to the assumptions of the $Q$ model of reality, the value
of $x(t_{f})$ determines the measurement outcome. For the macroscopic
superposition, the $x(t_{0})$ is associated with one or other branch,
which implies the outcome of $\hat{x}$ is determined by the value
$x(t_{0})$ at the time $t_{0}$, prior to detection. Hence, wMR holds
for the macroscopic superposition state at time $t_{0}$.

\textbf{(2)} The second way to measure $\hat{x}$ is to further amplify
the system, by applying $H_{amp}$, as in Figure \ref{fig:macro-sim-1-1},
and to then detect $x(t_{f})$ where $t_{f}>t_{0}$. According to
the Q model of reality, the macroscopic system is in one or other
branch at time $t_{0}$, as indicated by $x(t_{0})$. From the Figure,
we note that the trajectories propagating (backwards) from a given
branch $\widetilde{\lambda}_{x}$ at time $t_{f}$ are at time $t_{0}$
on the same branch. This follows from the deterministic relation $x_{j}\rightarrow Gx_{j}$
of the simulation (Eq. (\ref{eq:amp-g-1-1})), which, being deterministic,
can be reversed, so that $Gx_{j}\rightarrow x_{j}$, and is hence
indicated by the blue two-way arrow. Now, for wMR to fail, we would
require some instance where the value $x(t_{0})$ would not imply
the outcome for $\hat{x}$ as measured by the detection of $x(t_{f})$.
Since the branches are distinct at the time $t_{0}$, this can only
happen if the branch of $x(t_{0})$ is different to that $x(t_{f})$,
which appears to contradict the solutions of Figure \ref{fig:macro-sim-1-1}
which for macroscopic states are consistent with the relation $x_{j}\leftrightarrow Gx_{j}$.
We conclude consistency with wMR but the following Result IV.7 is
needed for clarity. $\square$

\textbf{\emph{Result IV.7: Assumption of branch continuity: }}Consider
the system described by the macroscopic superposition $|\psi_{sup}\rangle$
at a time $t_{m}$ (Fig. \ref{fig:macro-sim-1-1}). According to the
$Q$ model of reality, the system is, at time $t_{m}$, in one or
other of the branches $\widetilde{\lambda}_{x}$, which we denote
by $\widetilde{\lambda}_{x}(t_{m})$. Suppose a unitary operation
$U=e^{-iH_{amp}t/\hbar}$ then takes place, with Hamiltonian $H_{amp}$.
In quantum mechanics, the initial state for the operation $U$ is
$|\psi_{sup}\rangle$ (both branches). In the $Q$ model, we define
the \emph{assumption} of branch continuity in the macroscopic limit,
that the initial state for the operation $U$ is given by the set
of amplitudes of the same branch, $\widetilde{\lambda}_{x}(t_{m})$.
This assumption allows consistency with wMR in Result IV.6, part
(2).

\emph{Proof:} We prove that the assumption allows consistency with
wMR. Suppose for a system in a macroscopic superposition state $|\psi_{sup}\rangle$
at time $t_{0}\equiv t_{m1}$, as in Figure \ref{fig:macro-sim-1-1},
we consider successive $N-1$ operations $U$ that give further amplification,
to a time $t_{f}=Nt_{m}$. In the Q model of reality, the amplitude
$x(t_{0})$ would give the outcome of $\hat{x}$ if measured at time
$t_{0}$. For the macroscopic superposition we see from Figure \ref{fig:macro-sim-1-1}
that an individual system is on one or other branch. If we assume
branch continuity, then the individual macroscopic system will remain
on the same branch throughout the successive applications. If there
is a detection of $x(t_{f})$ indicating the branch $\widetilde{\lambda}_{x}(t_{m})$
corresponding to an eigenvalue $x_{j}$ at time $t_{f}$, then this
implies the system was on the same branch at the earlier times $(N-1)t_{m},...,t_{m}$,
which gives consistency with wMR. $\square$

\emph{Comment:} In the assumption of branch continuity, the predictions
for the probability of an outcome of $\hat{x}$ are the same as for
the quantum state $|\psi_{sup}\rangle$. Consider the state at time
$t_{m}$. We see from the Results IV.3 that in the Q model, the probability
for outcome $x_{j}$ is given by the probability density of amplitudes
$x(t_{f})$ at time $t_{f}$, which gives $|c_{j}|^{2}$ as correct
for $|\psi_{sup}\rangle$. The prediction is based on the marginal
$Q_{sup}(x,t_{f})$ (Result III.1) which is not changed over the ensemble
if we apply the assumption.

For an analysis of EPR correlations (Secs. \ref{sec:Continuous-variable-EPR-entangle}-\ref{sec:Causal-structure-bell}),
we extend the assumption Result IV.7.

\textbf{\emph{Result IV.8: Branch continuity for spacelike separated
systems $A$ and $B$ stationary in a frame $F$:}}\textbf{ }Suppose
system $A$ has been amplified so that branches $\widetilde{\lambda}_{x}^{A}$
of amplitudes $x_{A}(t_{mA})$ emerge at time $t_{mA}$. An operation
$U$ or interaction $H$ then takes place at $B$ at time $t_{m2}>t_{m1}$.
An individual system $A$ at time $t_{mA}$ is associated with one
or other branch, as determined by the value $x_{A}(t_{mA})$. The
Result IV.8 is presented as an assumption that states that the branch
for $A$ is unchanged by the operations or interactions at $B$.
We give a justification of this Result in Part II of the paper, based
on the properties of the bipartite Q function (refer to Results IX.2
and IX.3).

\subsection{Postselected state}

\begin{figure}
\begin{centering}
\includegraphics[width=1\columnwidth]{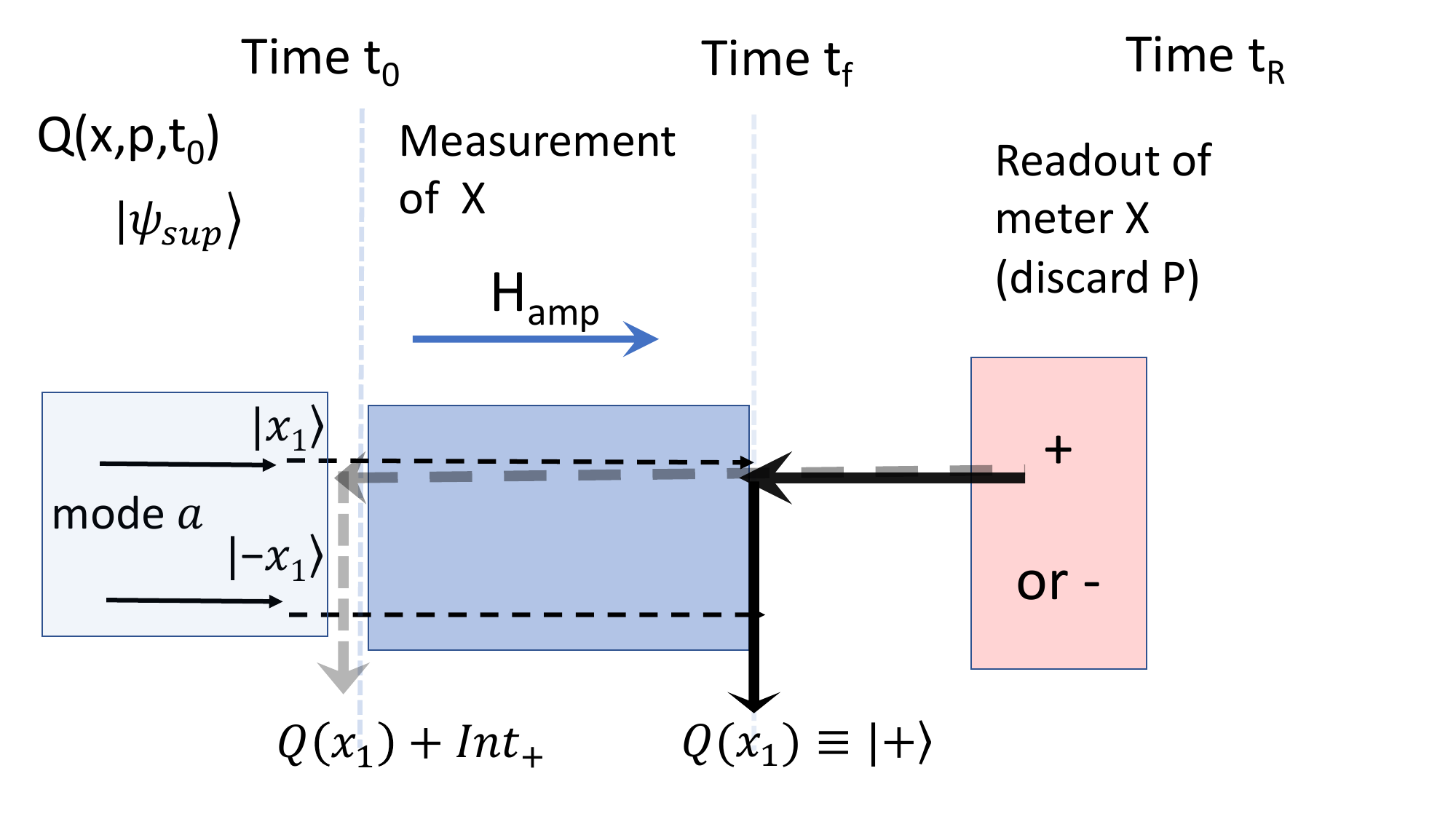}
\par\end{centering}
\caption{\textbf{\emph{Measurement by amplification:}} Diagram of the measurement
of $\hat{x}$ on a superposition of eigenstates $|x_{1}\rangle$ and
$|-x_{1}\rangle$ of $\hat{x}$. The measurement proceeds at a time
$t_{0}$ with the interaction $H_{amp}$, which amplifies $\hat{x}$
by the factor $G$.  The final stage of measurement is the irreversible
readout at time $t_{R}$, which gives the detected value $x(t_{f})$
as either $Gx_{1}$ or $-Gx_{1}$, discarding information about $\hat{p}$.
The measured outcome for $\hat{x}$ is $x(t_{f})/G$, implying outcomes
of either $x_{1}$ or $-x_{1}$. The solid black arrows show the conditioning
on the outcome $x_{1}$, to infer $|x_{1}\rangle$ for the macroscopic
superposition at time $t_{f}$. However, tracing the trajectories
back to the initial time $t_{0}$ (reverse dashed arrows) the postselected
state carries a contribution due to interference (denoted $Int_{+}$),
as evident by the fringes in Figure \ref{fig:post-select}. \label{fig:measurement-feedback}\textcolor{blue}{}}
\end{figure}

As part of analyzing the measurement problem, we seek to understand
the collapse of the wave function. With this motivation, we evaluate
the postselected state that is inferred at the initial time $t_{0}$,
\emph{given} a particular branch (with outcome $x_{j}$) at the later
time $t_{f}$ (Fig. \ref{fig:measurement-feedback} and Definition
(8a) of Sec. I.B). 

\textbf{\emph{Definition: The distribution postselected on a given
outcome $x_{j}$: }}The trajectories $x(t)$ for a given branch $\widetilde{\lambda}_{x}$
(associated with eigenvalue $x_{j}$) can be tracked from time $t_{f}$
to time $t_{0}$. This defines a set $\{x(t)\}$ of values of $x$
at time $t_{0}$, which we symbolize by $\mathcal{B}_{j}$. Associated
with this set for $x$ is a distribution for $p(t_{0})$, as determined
by the conditional distribution
\begin{equation}
Q_{0}(p|x)=Q(x,p,t_{0})/Q(x,t_{0})\label{eq:cond-1}
\end{equation}
 evaluated from the Q function $Q(x,p,t_{0})$ at $t_{0}=0$.\textbf{\emph{
}}Here, $Q(x,t_{0})=\int dpQ(x,p,t_{0})$. This defines a set of coordinates
$\{x(t_{0}),p(t_{0})\}$ and an associated distribution
\begin{eqnarray}
Q_{loop}(x,p,t_{0}|\mathcal{B}_{j}) & \equiv & Q\bigl(x(t_{0}),p(t_{0})|x(t_{0})\in\mathcal{B}_{j}\bigr)\nonumber \\
 & \equiv & Q_{loop}(x,p,t_{0}|x_{j})\label{eq:Qloop-postselect}
\end{eqnarray}
We define $Q_{loop}(x,p,t_{0}|\mathcal{B}_{j})$ as the \emph{postselected
state} or \emph{postselected distribution}, inferred on the outcome
$x_{j}$.

\emph{Notation:} Where it is clear that we restrict to the initial
time $t_{0}$, and refer to the branch associated with eigenvalue
$x_{j}$, we abbreviate to write $Q_{loop}(x,p,t_{0}|\mathcal{B}_{j})\equiv Q_{loop}(x,p|x_{j})$,
or else $Q_{0}(x,p|x_{j})$. Since the branch is symbolized by $\widetilde{\lambda}_{x}$,
with values $\widetilde{\lambda}_{x}\in\{x_{j}\}$, we sometimes write
$Q_{loop}(x,p|x_{j})\equiv Q_{loop}(x,p|\widetilde{\lambda}_{x})$.

\textbf{\emph{Result IV.9:}} \textbf{\emph{The postselected state
is not equivalent to the eigenstate}}: The postselected distribution
$Q_{loop}(x,p,t_{0}|x_{j})$  for a superposition $|\psi_{sup}\rangle=\sum_{k}c_{k}|x_{k}\rangle$
of eigenstates $|x_{k}\rangle$ of $\hat{x}$ (where $c_{j}\neq0$)
contains a sinusoidal term. Hence, the postselected state is not the
eigenstate $|x_{j}\rangle$.

\emph{Proof:} We prove by evaluating the postselected distribution
for $|\psi_{sup}\rangle=c_{1}|x_{1}\rangle+c_{2}|x_{2}\rangle$ of
Eq. (\ref{eq:sup-sq}), where we take $c_{1}$ to be real and positive,
and write $c_{2}=|c_{2}|e^{i\varphi}$. We trace the trajectories
of $x(t)$ from time $t_{f}$ to $t_{0}$, for each $x(t_{f})$ that
is on the branch with eigenvalue $x_{j}$. The superposition has just
two branches (Fig. \ref{fig:stochastic-sup-micro}). We take $x_{2}=-x_{1}$.
For sufficiently large $gt$, each $x(t_{f})$ is either positive
or negative, associated with the outcome $x_{1}$ or $-x_{1}$, which
we denote by $+$ or $-$.  We first consider just one branch,
$\mathcal{B}_{j}$ (of eigenvalue $x_{j}$). The simulation starts
at the future boundary. At time $t_{f}$, the distribution of the
values $x(t)$ of the branch $\mathcal{B}_{j}$ is given as $\frac{e^{-(x-e^{gt_{f}}x_{j})^{2}/2}}{\sqrt{2\pi}}$
($t_{0}<t<t_{f}$). The distribution of $\mathcal{B}_{j}$ at time
$t_{0}$ is $Q_{j}(x)=\frac{e^{-(x-x_{j})^{2}/2}}{\sqrt{2\pi}}$.
It is convenient to denote $\mathcal{B}_{1}$ and $\mathcal{B}_{2}$
by $\mathcal{B}_{+}$ and $\mathcal{B}_{-}$ respectively.

At time $t_{0}$, there is a connection of the values $\mathcal{B}_{+}$
with a set of values for $p$ determined by $Q_{0}(p|x)$ (Eq. (\ref{eq:cond-1})).
Denoting the distribution of the branch $\mathcal{B}_{\pm}$ at $t_{0}$
by $Q_{\pm}(x)=\frac{e^{-(x\mp x_{1})^{2}/2}}{\sqrt{2\pi}}$, we find
(using Eq. (\ref{eq:Qloop-postselect}))
\begin{eqnarray}
Q_{loop}(x,p,t_{0}|x\in\mathcal{B}_{\pm}) & = & Q_{\pm}(x)Q_{0}(p|x)\label{eq:branch-cond-1}
\end{eqnarray}
Calculation gives
\begin{eqnarray}
Q_{0}(p|x) & = & N\frac{e^{-p^{2}/2\sigma_{p}^{2}}}{\sqrt{2\pi}\sigma_{p}}\Bigr(1+2|c_{1}c_{2}|e^{-x^{2}/2}e^{-x_{1}^{2}/2}\frac{\mathcal{F}}{S_{G}}\Bigl)\nonumber \\
 & \rightarrow & N\frac{e^{-p^{2}/2\sigma_{p}^{2}}}{\sqrt{2\pi}\sigma_{p}}\Bigr(1+\frac{\mathcal{F}}{\cosh(xx_{1})}\Bigl)\label{eq:Q0limit-3-2-1}
\end{eqnarray}
where $Q(x,p,t_{0})$ is given by Eq. (\ref{eq:Q-sup}), $Q(x,0)$
is given by Eq. (\ref{eq:Qmarg-x-amp}), and 
\begin{equation}
\mathcal{F}=(\cos\varphi)\cos(x_{1}p)-(\sin\varphi)\sin(x_{1}p)\label{eq:F}
\end{equation}
is given by Eq. (\ref{eq:fringe-1}). The sinusoidal function $\mathcal{F}$
depends on $p$ but not $x$, and is independent of the branch. Also,
$S_{\mathcal{G}}=\sqrt{2\pi}(|c_{1}|^{2}Q_{+}(x)+|c_{2}|^{2}Q_{-}(x))$
and $N$ is a normalization factor (NB: $N\rightarrow1$ as $r\rightarrow\infty$
for (\ref{eq:Q-sup})). Fringes are evident, becoming finer for large
$x_{1}$ (when the superposition is macroscopic) and also increasingly
damped, provided $x\neq0$. The last line of (\ref{eq:Q0limit-3-2-1})
gives the special case where $|c_{1}|=|c_{2}|$. 

\begin{figure}
\begin{centering}
\includegraphics[width=0.8\columnwidth]{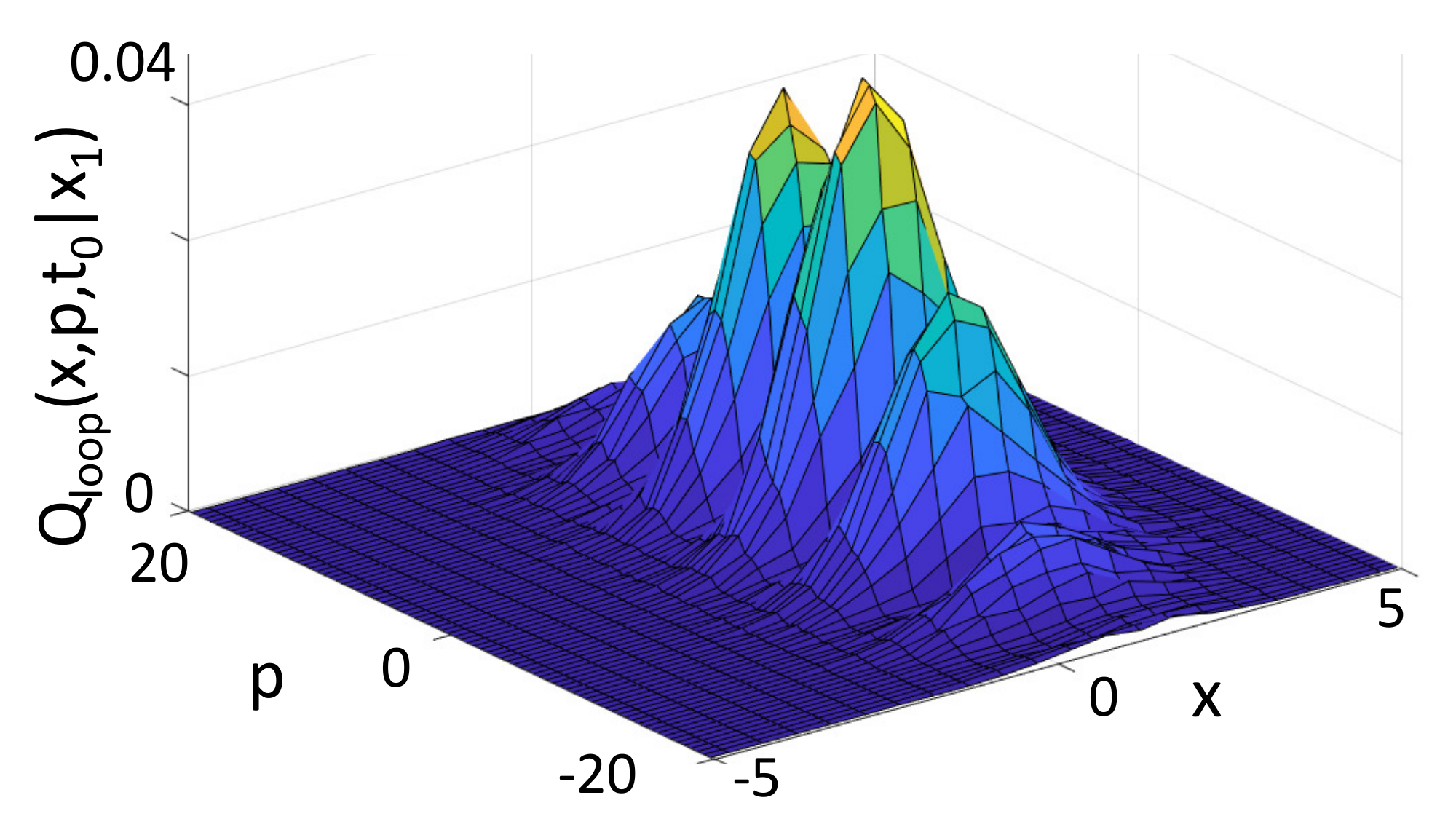}
\par\end{centering}
\caption{\textbf{\emph{Postselected state: }}The postselected distribution
$Q_{loop}(x,p,t_{0}|x_{1})$ at time $t_{0}=0$, given the branch
$x_{1}$ at time $t_{f}$, for the state $|\psi_{sup}\rangle$ of
Eq. (\ref{eq:sup-sq}) considered in Figure \ref{fig:stochastic-sup-micro},
where $x_{1}=1$. In practice, we trace back the trajectories $x(t)$
with a positive value for $x(t_{f})$. \label{fig:post-select}\textcolor{blue}{}}
\end{figure}

From (\ref{eq:branch-cond-1}), the postselected distribution is
\begin{equation}
Q_{loop}(x,p,t_{0}|\pm x_{1})=N(Q(\pm x_{1})+\mathcal{I}nt_{\pm})\label{eq:Qint+-1-1}
\end{equation}
where we take $|c_{1}|=|c_{2}|$ for simplicity. Here $Q(\pm x_{1})=\frac{e^{-p^{2}/2\sigma_{p}^{2}}}{2\pi\sigma_{x}\sigma_{p}}e^{-(x\mp x_{1})^{2}/2}$
is the Q function of the eigenstates $|\pm x_{1}\rangle$, and 
\begin{eqnarray}
\mathcal{I}nt_{\pm} & = & \frac{e^{-p^{2}/2\sigma_{p}^{2}}}{2\pi\sigma_{x}\sigma_{p}}2e^{-x^{2}/2\sigma_{x}^{2}}e^{-x_{1}^{2}/2\sigma_{x}^{2}}\frac{\mathcal{F}}{(1+Q_{\mp}(x)/Q_{\pm}(x))}\nonumber \\
 & = & \frac{e^{-p^{2}/2\sigma_{p}^{2}}}{2\pi\sigma_{x}\sigma_{p}}2e^{-x^{2}/2\sigma_{x}^{2}}e^{-x_{1}^{2}/2\sigma_{x}^{2}}\frac{\mathcal{F}}{(1+e^{\mp2x_{1}x})}\label{eq:int-3}
\end{eqnarray}
We confirm the Result IV.9: For example, the $Q_{loop}(x,p,t_{0}|x_{1})$
which is conditioned on the positive branch includes a contribution
from the bivariate Gaussian $Q(x_{1})$ with positive mean, as well
as a sinusoidal term $\mathcal{I}nt_{+}$ (Figures \ref{fig:measurement-feedback}
and \ref{fig:post-select}). $\square$

A plot of the distribution as evaluated numerically for the state
$|\psi_{sup}\rangle$ of Eq. (\ref{eq:sup-sq}) is presented in Figure
\ref{fig:post-select}. The fringe pattern arising from $\mathcal{F}$
is evident. The fringe distribution is independent of the outcome
(whether $x_{1}$ or $-x_{1}$), because the conditional $Q_{0}(p|x)$
is independent of sign of $x$. The marginal of the distribution hence
corresponds to that given by Figure \ref{fig:p-trajectories-fringe}
at time $t=0$.

\emph{Comment: }The fringes decay as the separation between the eigenstates
is larger, so that $|x_{1}|\gg1$. Hence, $Q_{loop}\bigl(x,p,t_{0}|x_{1}\bigr)$
becomes the Q function of the eigenstate $|x_{j}\rangle$ in the limit
of a macroscopic superposition.

\textbf{\emph{Result IV.10: The postselected state does not (generally)
correspond to a quantum state:}} The distribution $Q_{loop}\bigl(x,p|x_{j}\bigr)$
at the time $t_{0}$, associated with the branch $x_{j}$, is not
equivalent to the Q function of any quantum state. Hence, the postselected
``state'' is not a quantum state.

\begin{figure}
\begin{centering}
\par\end{centering}
\begin{centering}
\includegraphics[width=0.8\columnwidth]{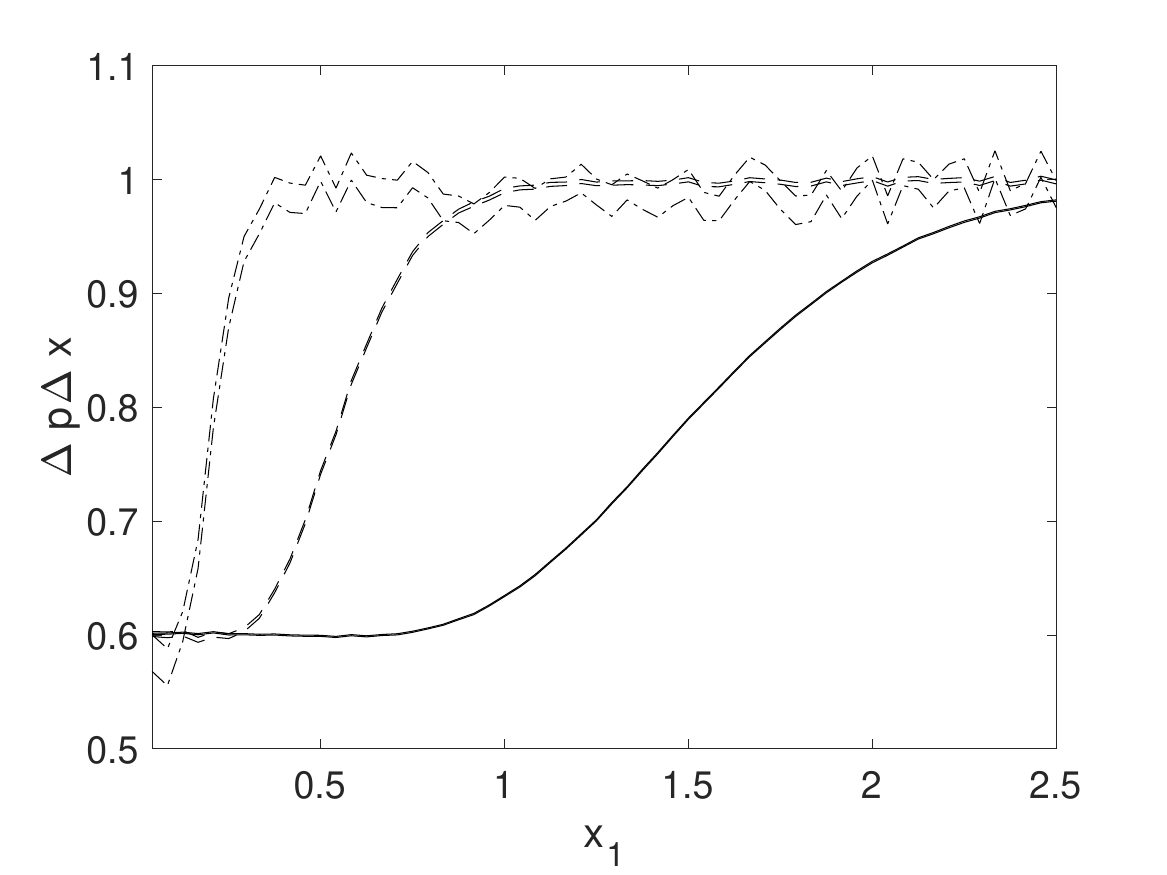}
\par\end{centering}
\caption{\textbf{\emph{Uncertainty product for the postselected distribution:
}}We plot the uncertainty product $\Delta\hat{x}\Delta\hat{p}$ associated
with the postselected distribution $Q_{loop}(x,p,t_{0}|x_{1})$ conditioned
on the outcome $x_{1}$ for $\hat{x}$, for the state $|\psi_{sup}\rangle$
of Figure \ref{fig:post-select}. Here, $t_{0}=0$. Here, $\Delta x$
and $\Delta p$ correspond to the observed uncertainties if $Q_{loop}(x,p,t_{0}|x_{1})$
were the Q function of a quantum state (we drop the operator hats
in $\Delta\hat{x}$ and $\Delta\hat{p}$ and write $\Delta\hat{x}\Delta\hat{p}=\Delta x\Delta p$
for convenience). The upper dashed-dotted line is for the superposition
$|\psi_{sup}\rangle$ of two eigenstates of $\hat{x}$, with $r=2$.
The dashed line is for $|\psi_{sup}(r)\rangle$ with $r=1$. The solid
line is for $r=0$. In each case, $gt=4$. The two parallel lines
indicate the upper and lower error bounds from sampling errors, with
$1.2\times10^{7}$ trajectories. \label{fig:cond-var-inferred}}
\end{figure}

\emph{Proof:} Let us assume $Q_{loop}\bigl(x,p,t_{0}|x_{1}\bigr)$
to be the $Q$ function of a quantum state $|\psi\rangle$. The measured
variances for $\hat{x}$ and $\hat{p}$ are $(\Delta\hat{x})^{2}=\sigma_{x,+}^{2}(0)-1$
and $(\Delta\hat{p})^{2}=\sigma_{p,+}^{2}(0)-1$, where $\sigma_{x,+}^{2}\equiv\sigma_{x,+}^{2}(0)$
and $\sigma_{p,+}^{2}\equiv\sigma_{p,+}^{2}(0)$ are the variances
of $x$ and $p$ as given by $Q_{loop}\bigl(x,p,t_{0}|x_{1}\bigr)$
(refer Sec. II). The variance product $\Delta\hat{x}\Delta\hat{p}$
is given in Figure \ref{fig:cond-var-inferred} versus $x_{1}$ (the
separation between the states of the superposition) for large $gt$.
From Figure \ref{fig:cond-var-inferred}, see that $\Delta\hat{x}\Delta\hat{p}<1$
for all $x_{1}$ (and $\alpha_{0}$) (although $\Delta\hat{x}\Delta\hat{p}\rightarrow1$
as $|x_{1}|\rightarrow\infty$). A contradiction is obtained with
the Heisenberg uncertainty relation $\Delta\hat{x}\Delta\hat{p}\geq1$.
Hence, the distribution $Q_{loop}\bigl(x,p,t_{0}|x_{1}\bigr)$ is
not a Q function and cannot represent a quantum state. $\square$

In summary, the post-selected distribution $Q_{loop}(x,p,t_{0}|x_{1})$
does not reflect a quantum state $|\psi\rangle$. The variables $x$
and $p$ are hence ``\emph{hidden variables}''. This is depicted
in Fig. \ref{fig:scat-1-1}, with reference to the Schrodinger-cat
paradox. The simulation shows how macroscopic realism can be consistent
with quantum mechanics for the system in a macroscopic superposition.
Yet, the postselected state for the macroscopic superposition is not
a quantum state, which highlights the argument for the incompleteness
of quantum mechanics, as put forward by Schrödinger \citep{schrodinger1935gegenwartige}.
However, the uncertainty product \emph{approaches} $1$ with increasing
separation $|x_{1}|$ of the eigenstates (Fig. \ref{fig:cond-var-inferred}),
where one has a true macroscopic superposition $-$ a ``cat state''.

\subsection{Hidden loop}

At the initial time $t_{0}$, there is a link between the backward
trajectories $x(t)$ of one branch and those going forward in time
for the complementary observable $p$. The link is determined by the
conditional distribution $Q_{0}(p|x)$ (Eq. (\ref{eq:cond-1})). This
enables definition of the joint distribution $Q_{loop}(x,p,t|x_{j})$
throughout the evolution, which is used to confirm causal consistency
(refers Secs. V.B and IX.C, Appendices B and H).

\begin{figure}[H]
\begin{centering}
\includegraphics[width=0.5\columnwidth]{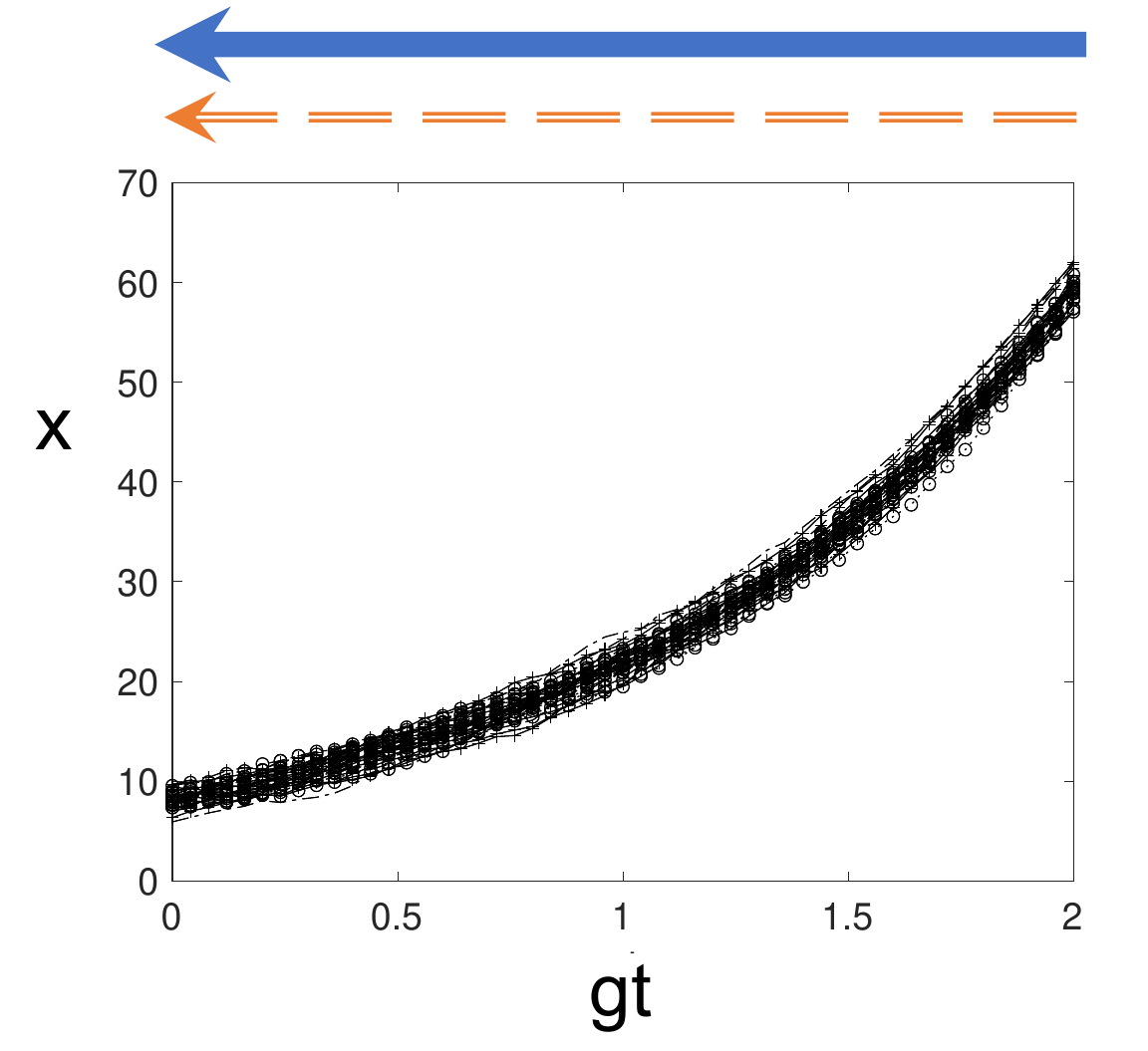}\includegraphics[width=0.5\columnwidth]{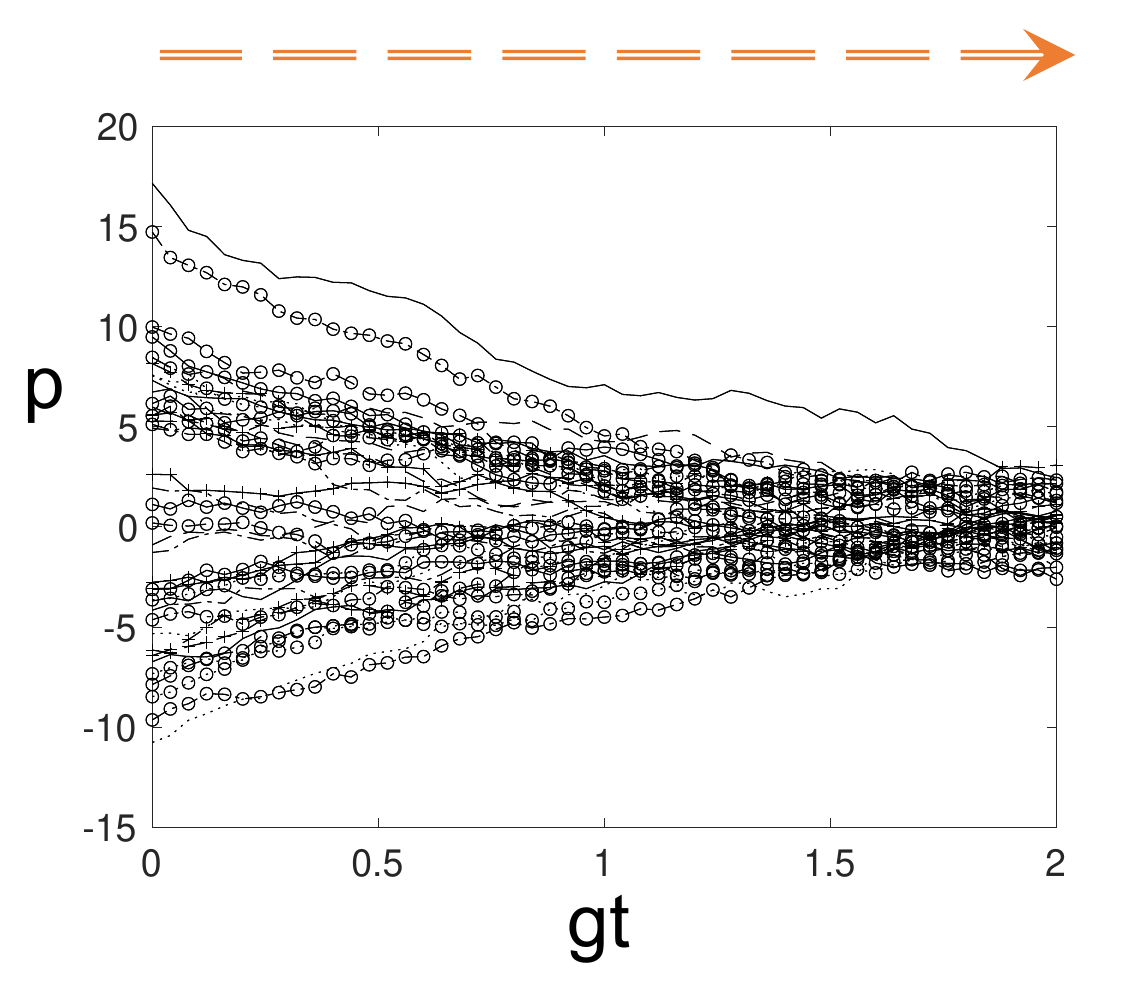}
\par\end{centering}
\centering{}\caption{\textbf{\emph{Connected trajectories forming a hidden loop:}} The
trajectories $x(t)$ of one branch for $\hat{x}$ can be traced backwards
in time to the initial time $t_{0}=0$ (left figure), where they couple
to certain forward-going trajectories for $p$ (right figure). Plots
are for the system prepared in a superposition $|\psi_{sup}\rangle$
of eigenstates $|x_{1}\rangle$ and $|-x_{1}\rangle$ of $\hat{x}$
as in Figure \ref{fig:macro-sim-1}, conditioned on an outcome $x_{1}$
for measurement of $\hat{x}$, with $x_{1}=8$ and $gt_{f}=2$. The
resulting distribution for $p$ is independent of the branch-value,
whether $x_{1}$ or $-x_{1}$.\label{fig:postx-trajectories-superposition}}
\end{figure}

\textbf{\emph{Definition: Hidden loop: }}For the set of values of
$x(t_{f})$ at the time $t_{f}$ on a given branch with outcome $x_{j}$,
we have defined the set $\{x(t_{0})\}$ at the initial time $t_{0}$,
found by tracing the backward trajectories $x(t)$. The set $\{x(t_{0})\}$is
denoted $\mathcal{B}_{j}$. The $\mathcal{B}_{j}$ defines a set of
\emph{connected forward-going $p$ trajectories}, found by propagating
each $p(t_{0})$ at time $t_{0}=0$ from the sample generated by $Q_{0}(p|x)$,
where $x\in\{x(t_{0})\}$, back to the future time $t_{f}$. We define
the sets of variables $\{x(t_{f}),x(t_{0}),p(t_{0}),p(t_{f})\}$ and
all intermediate values on the trajectories. We refer to such a set
of trajectories as a \emph{``loop}''. The distribution of the amplitudes
$\{x(t),p(t)\}$ at a given time $t$ ($t_{0}\leq t\leq t_{f}$) is
denoted 
\begin{equation}
Q_{loop}(x,p,t|x\in\mathcal{B}_{j})\label{eq:HL}
\end{equation}
Such a set of trajectories is plotted in Figure \ref{fig:postx-trajectories-superposition}.
The definition is summarized as Definition 9(a) in Sec. I.B.

\emph{Comment:} The term ``hidden'' is used, because the loop arises
from a distribution such as $Q_{loop}(x,p,t_{0}|x_{j})$ which cannot
correspond to a quantum state $|\psi\rangle$ (Result IV.10, Fig.
\ref{fig:scat-1-1}). The ``loop'' involves unobserved variables
$\delta x$ and the deamplified $p$. We note the term ``hidden causal
loop'' might also be used \citep{castagnoli2021unobservable}. However,
we avoid this term, because there is no backward information transfer
of the value $x_{j}$. The input at time $t_{f}$ is the stochastic
noise, while the observed value $e^{gt}x_{j}$ is governed by the
(causal) deterministic relation $x_{j}\rightarrow e^{gt}x_{j}$.

\textbf{\emph{Result IV.11:}} The coupling of the $p$ trajectories
with those for $x$ is determined by $Q(p|x)$, which is\emph{ }independent
of the sign of $x$ and is not sensitive to the outcome $x_{j}$.
This is evident in Figure \ref{fig:postx-trajectories-superposition}
which plots trajectories of $p$ conditioned on the outcome $x_{1}$.
The distribution of trajectories for $p$ is the same for both branches
$x_{1}$ and $-x_{1}$, and is hence given by Figure \ref{fig:macro-sim-1}
(lower panel).

\textbf{\emph{Result IV.12:}} \textbf{\emph{Hidden loops occur for
superpositions, not classical mixtures:}} The conditional distribution
$Q_{0}(p|x)$ for the mixture $\rho_{mix}$ (\ref{eq:mix-sup-1})
of eigenstates of $|x_{j}\rangle$ is given by
\begin{equation}
Q_{0}(p|x)=\frac{e^{-p^{2}/2\sigma_{p}^{2}}}{\sigma_{p}\sqrt{2\pi}}=Q(p)\label{eq:fact}
\end{equation}
which implies complete independence of the trajectories for $x$ and
$p$. There is no interference term $\mathcal{I}$, and the $Q_{0}(p|x)$
factorizes. Hence the hidden loop occurs only for the superposition
(not mixtures) of eigenstates $|x_{j}\rangle$ .

\subsection{Hidden variable models}

We have seen how the amplitudes $x$ and $p$ themselves do not give
a complete description (Result IV.2). Hidden variable (HV) models
can be deduced from the simulation however, which is useful in developing
a causal model for measurement (Sec. V) and for justifying weak local
realism (Definition (10a)). Consider the system in the superposition
$|\psi_{sup}\rangle$ of eigenstates $|x_{j}\rangle$ of $\hat{x}$,
as in Eq. (\ref{eq:sup-sq}). The system is prepared with respect
to the measurement basis $\hat{x}$ at the time $t_{0}$. 

\textbf{\emph{Result IV.13a:}} \textbf{\emph{Hidden variable model
$\mathcal{P}_{sup}$:}} A hidden variable model $\mathcal{P}_{sup}$
exists such that for any given amplitudes $x$ and $p$ at time $t_{0}$,
there is a definite probability $P(x_{j}|x,p)$ of an outcome $x_{j}$
of the measurement $\hat{x}$.

\emph{Proof:} The Q function of $|\psi_{sup}\rangle$ is $Q_{sup}(x,p,t_{0})$
of Eq. (\ref{eq:Q-sup}).  From postselection, we can determine
the distributions $Q_{loop}(x,p,t_{0}|\pm x_{1})$, since there is
a well defined set of $x$ and $p$ at the time $t_{0}$, given the
branch $x_{1}$ or $-x_{1}$. Using that for events $A$ and $B$,
the conditional probabilities satisfy
\begin{eqnarray}
P(A|B) & = & \frac{P(A\cup B)}{P(B)})=\frac{P(B|A)P(A)}{P(B)}\label{eq:cond-prob-2}
\end{eqnarray}
it follows that
\begin{equation}
P(\pm x_{1}|x,p)=\frac{Q_{loop}(x,p,t_{0}|\pm x_{1})P(\pm)}{Q_{sup}(x,p,t_{0})}\label{eq:Q_prob}
\end{equation}
which can be evaluated from the simulation, where $P(+)$ and $P(-)$
are the respective probabilities of the positive and negative outcomes.
Using (\ref{eq:branch-cond-1}) and that $Q(p|x)=Q_{sup}(x,p,t_{0})/Q_{sup}(x,t_{0})$
where from (\ref{eq:Qmarg-x-amp}) we see that $Q_{sup}(x,t_{0})=(Q_{+}(x)+Q_{-}(x))/2$,
we find
\begin{eqnarray}
P(\pm x_{1}|x,p) & = & \frac{Q_{\pm}(x)}{Q_{+}(x)+Q_{-}(x)}\label{eq:Q+1-1}
\end{eqnarray}
where $Q_{\pm}(x)=\frac{e^{-(x\mp x_{1})^{2}/2}}{\sqrt{2\pi}}$.
Hence, $P(x_{1}|x,p)+P(-x_{1}|x,p)=1$, as required. We note the distribution
is derived accounting for interference, but the final result is the
same as that obtained for the mixed state $\rho_{mix}$. 
$\square$

\textbf{\emph{Result IV.13b: (Deterministic) Hidden variable model
$\mathcal{P}_{det}$.}} There is another interpretation possible
from the postselected states (\ref{eq:Qint+-1-1}) as derived for
the superposition state $|\psi_{sup}\rangle=c_{1}|x_{1}\rangle+c_{2}|x_{2}\rangle$
of Eq. (\ref{eq:sup-sq}). Here, we take for simplicity $c_{1}$ to
be real and positive, $c_{2}=|c_{2}|e^{i\varphi}$, and$|c_{1}|=|c_{2}|$
and $x_{2}=-x_{1}$.

It can be posited that the system \emph{prior to measurement is in
a probabilistic mixture} of the postselected states i.e. is \emph{either}
in the state with distribution $Q_{loop}(x,p,t_{0}|+x_{1})$ with
probability $1/2$, \emph{or} in the state with distribution $Q_{loop}(x,p,t_{0}|-x_{1})$
with probability $1/2$. Defining $\mathcal{I}nt=\mathcal{I}nt_{+}+\mathcal{I}nt_{-}$,
we see that $\mathcal{I}nt=N\frac{e^{-p^{2}/2\sigma_{p}^{2}}}{2\pi\sigma_{x}\sigma_{p}}2e^{-x^{2}/2\sigma_{x}^{2}}e^{-x_{1}^{2}/2\sigma_{x}^{2}}\mathcal{F}$
where $\mathcal{F}$ is given by (\ref{eq:F}). Hence, since
\begin{equation}
Q_{sup}(x,p,t_{0})=\frac{Q(x_{1})}{2}+\frac{Q(-x_{1})}{2}+\frac{\mathcal{I}nt}{2}\label{eq:qsup-add}
\end{equation}
the probability distribution for the mixture is indeed given by the
Q function $Q_{sup}(x,p,t_{0})$, as required for the proposed model.

In the model, we note that for a system that could be described as
being in the state $Q_{loop}(x,p,t_{0}|+x_{1})$, the marginal $Q_{loop}(x,t_{0}|x_{1})=\int dpQ_{loop}(x,t_{0}|x_{1})$
in $x$ is 
\[
N\int dp(Q(x_{1})+\mathcal{I}nt_{+})
\]
Applying the techniques of Eqs. (\ref{eq:int})-(\ref{eq:int-2-2})
in Section III, we see that the contribution due to $\mathcal{I}nt_{+}$
will vanish. Hence, the marginal $Q_{loop}(x,t_{0}|x_{1})$ is as
for the eigenstate $|x_{1}\rangle$. Similarly, the marginal in $x$
for the state $Q_{loop}(x,p,t_{0}|-x_{1})$ is as for the eigenstate
$|-x_{1}\rangle$. In the proposed model, we make the assumption that
the backward stochastic equation (\ref{eq:backwardSDE-2-1}) applies
to describe the measurement of $\hat{x}$ on the system, which is
in the state given by \emph{either} $Q_{loop}(x,t_{0}|x_{1})$ \emph{or}
$Q_{loop}(x,t_{0}|-x_{1})$ at the time $t_{0}$. Then, the conclusion
is that the state $Q_{loop}(x,p,t_{0}|x_{1})$ will always give outcome
$x_{1}$; and the state $Q_{loop}(x,p,t_{0}|-x_{1})$ will always
give outcome $-x_{1}$. Hence, the system is at time $t_{0}$ in a
state with a predetermined outcome for $\hat{x}$, meaning either
definitely $x_{1}$ or definitely $-x_{1}$. For this reason, we refer
to the model as ``deterministic''. 

\subsection{Weak local realism (wLR)}

The concept of weak macroscopic realism (wMR) introduced in Result
IV.6 can be generalized to \emph{weak local realism}, as given by
Definition (10a) in Sec. I.B. Consider measurement of $\hat{x}$ on
a system at time $t_{0}$ (Fig. \ref{fig:measurement-feedback}).
The system is \emph{prepared with respect to the measurement basis}
$\hat{x}$ at $t_{0}$, meaning that the operations $U$ that fix
the measurement settings (e.g. the sign of $g$ in $H_{amp}$ (Eq.
(\ref{eq:ham-2-1}))) have been performed.

\textbf{\emph{Result IV.14:}} \textbf{\emph{Weak Local Realism, Premise
wLR(1): }}The Q-based model of reality is consistent with the assumption
of weak local realism (wLR). There are two versions of wLR, \emph{deterministic}
and \emph{probabilistic} (Refer Definition (10) in Sec. I.B). The
first assumes  a deterministic hidden variable (HV) model: there
is a predetermination of the outcome of $\hat{x}$ at the time $t_{0}$
(after settings are fixed). The second assumes a probabilistic HV
model: hidden variables exist for the system at the time $t_{0}$
(after settings are fixed) which imply definite probabilities for
outcomes of $\hat{x}$.

\emph{Justification:} In the Q-based simulation and model, the means
$x_{j}^{A}$ of the Gaussian functions in $Q(\bm{\lambda},t_{0})$
are amplified to $Gx_{j}^{A}$ (refer Eqs. (\ref{eq:Q-sup}) and (\ref{eq:Q-sup-1})).
The interference terms $\mathcal{I}$ in $Q(\lambda,t_{0})$ are not
amplified (Result III.1). Hence, the probability of an outcome $x_{j}$
of $\hat{x}$ for the system, when prepared with respect to the measurement
basis, is the same as for the mixture, $\rho_{mix}$, defined by Eq.
(\ref{eq:mix-sup-1}). For the mixture, the system \emph{can} be viewed
as being (with probability $|c_{j}|^{2}$) in a state with definite
outcome $x_{j}$.  Hence, a deterministic HV model for the observed
probabilities exists. Similarly, the HV model given by Result IV.13b
is a deterministic HV model deduced from the simulation, that accounts
for the interference terms, and is consistent with Premise wLR(1).

Alternatively, the probabilistic HV model $\mathcal{P}_{sup}$ can
be developed from the simulation (Result IV.13). At time $t_{0}$,
the system is in a HV state given by variables $\lambda\equiv(x,p)$,
with a probability density function $Q(\lambda,t_{0})$. For a given
set of values $\lambda$, there is defined a probability $P(x_{j}|x,p)$
for an outcome $x_{j}$, as given by Eqs. (\ref{eq:Q_prob}-\ref{eq:Q+1-1}).
This HV model accounts is more general than the deterministic HV model
of Result IV.13b and is accurately supported by the simulation.
$\square$

\section{Causal model for measurement, causal consistency, and the measurement
problem\label{sec:Causal-model-for}}

The simulations involve classical-like amplitudes $x$ and $p$ and
individual realizations, and hence we can identify ``cause and effect''
relations for quantum measurement, referred to as a ``causal model''.
A causal model is constructed based on the Procedure of the Simulation
(Sec. \ref{sec:Forward-backward-stochastic-simu}), as depicted in
Figure \ref{fig:causal-model}. This elucidates why retrocausality,
defined as the future affecting the past (as in a transfer of information
backward in time), is not present despite the use of future boundary
conditions. The causal model also explains how causal consistency
is achieved, and why there are no Grandfather paradoxes.

\subsection{Causal model}

A causal model is deduced to explain quantum measurement. We denote
the measurement by $\hat{X}$, where $\hat{X}$ is either $\hat{x}$
or $\hat{p}$ or $\hat{x}_{\theta}=\hat{x}\cos\theta+\hat{p}\sin\theta$.
The \emph{model parameters} are the measurement setting $\theta$,
and the variables that describe the state of the system, at the initial
time $t=0$ and at the final time $t=t_{f}$, as well as the noise
inputs $\xi_{1}(t)$ and $\xi_{2}(t)$ defined by (\ref{eq:backwardSDE-2-1})
and (\ref{eq:forwardSDE-2-1}), and the noise input $\delta x(t_{f})\equiv\eta(t)$
defined by the future boundary condition (\ref{eq:initialBC-1}) at
time $t_{f}$. Where we measure $\hat{x}$, the notation implies $\xi_{1}(t)\equiv\xi_{x}(t)$
and $\xi_{2}(t)\equiv\xi_{p}(t)$.

\begin{figure}
\begin{centering}
\includegraphics[width=1\columnwidth]{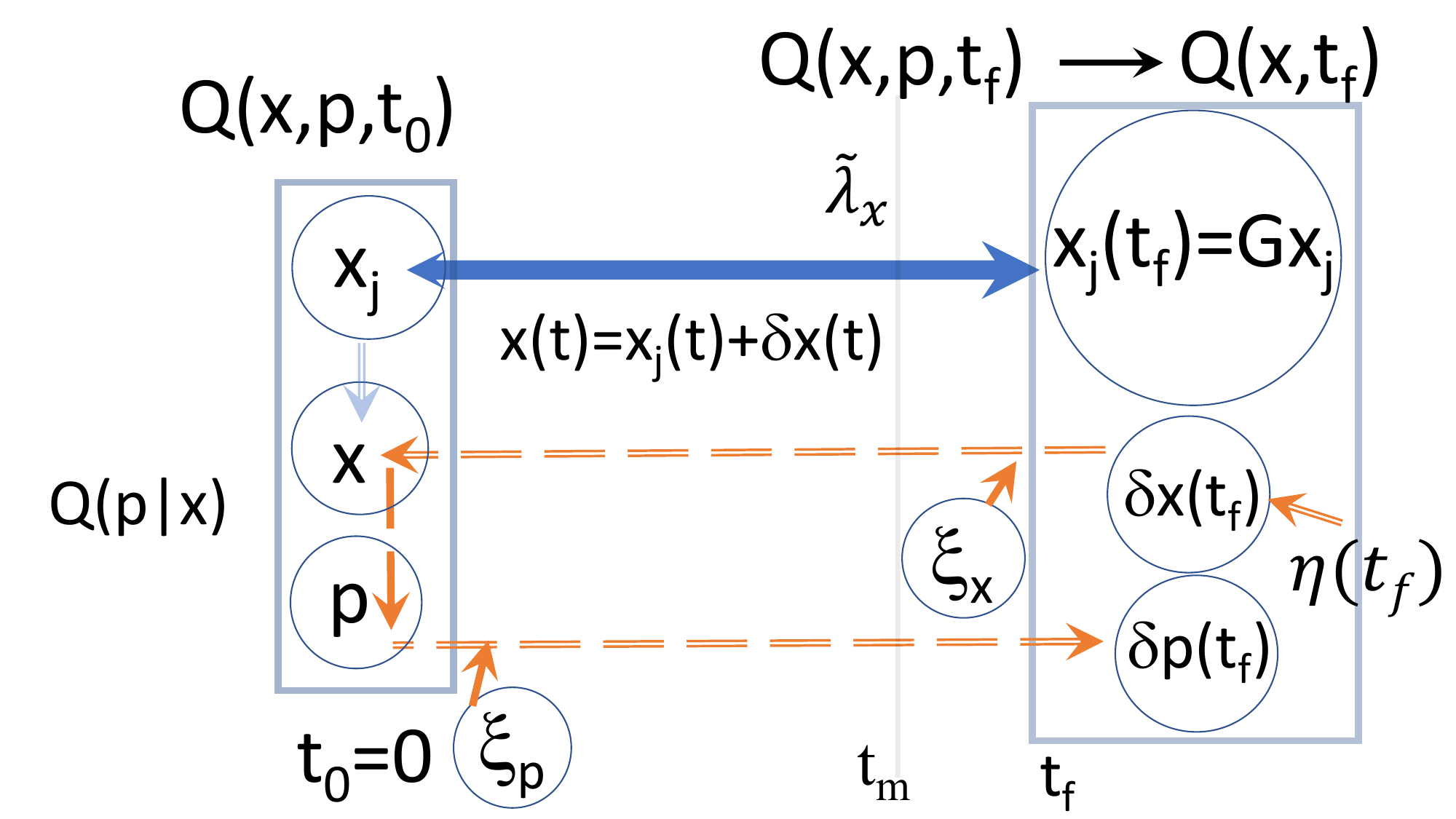}
\par\end{centering}
\caption{\textbf{\emph{Causal model for quantum measurement:}} Diagram of the
model inferred from the simulation of measurement $\hat{x}$ on a
superposition $|\psi_{sup}\rangle$ of eigenstates $|x_{j}\rangle$
of $\hat{x}$. The relation $x_{j}\rightarrow Gx_{j}$ (blue line)
is deterministic and causal. Branches emerge after sufficient amplification
when the system is macroscopic, at a time $t_{m}$, as depicted by
the value $\widetilde{\lambda}_{x}$ which determines the outcome
of the measurement $\hat{x}$ (Figs. \ref{fig:stochastic-sup-micro}
and \ref{fig:macro-sim-1}). The ``retrocausality'' present in the
model is embodied in the noise input $\delta x(t_{f})\equiv\eta(t_{f})$,
generated at the future boundary $t_{f}$. The properties of the input
$\delta x(t_{f})\equiv\eta(t_{f})$ at time $t_{f}$ are independent
of the value $t_{f}$ and of the eigenvalue $x_{j}$, and similarly
for $\xi_{x}$, which is determined by $H_{amp}$. Further, the properties
of the stochastic term $\delta x(t)$ remain unchanged with time $t$.
A hidden loop (orange dashed lines) is present for superpositions
$|\psi_{sup}\rangle$, but not mixtures, of $|x_{j}\rangle$. The
value $x_{j}$ can be inferred from $Gx_{j}$ (depicted by the two-way
blue arrow).\label{fig:causal-model}\textcolor{blue}{}}
\end{figure}

In the simulation, the setting $\theta$ is in principle adjustable,
as in the Hamiltonian (\ref{eq:hamtheta-1}). Experimentally, this
corresponds to a unitary operation $U_{\theta}=e^{-iH_{\theta}t/\hbar}$
where $H_{\theta}$ induces a phase shift. In the analysis below (Fig.
\ref{fig:causal-model}), we assume that at time $t_{0}$, the operation
$U_{\theta}$ has been carried out so that $g>0$, and the measurement
is to be of $\hat{x}$.

\textbf{\emph{Prepare with respect to the measurement basis:}} The
system variables at time $t_{0}=0$ are $x$ and $p$, with a distribution
given by the Q function $Q(x,p,t_{0}).$  The Q function for a state
of the system is unique but can be written as a superposition of the
eigenstates $|x_{\theta j}\rangle$ of $\hat{x}_{\theta}$, for any
choice of $\theta$. In an experiment, at the time $t_{0}$ after
the operation $U_{\theta}$ where we choose $\theta=0$, we assume
that the system is prepared for the amplification $H_{amp}$ (with
$g>0$) to finalize a measurement of $\hat{x}$. 

\emph{Assumption (1):} It is assumed as part of the Q model that the
physical state for the system at the time $t_{0}$, after the settings
are determined, is given by the Q function written \emph{with respect
to the coordinates of the measurement basis}, meaning in this case
that $Q$ is expanded in terms of the Q function of the eigenstates
$|x_{j}\rangle$ of $\hat{x}$ (as in Eq (\ref{eq:Q-sup})). Denoting
the Gaussian with mean $\bar{x}$ and variance $\sigma$ by $\mathcal{G}(\bar{x},\sigma)=\frac{e^{-(x-\bar{x})^{2}/2\sigma^{2}}}{\sqrt{2\pi}\sigma}$,
the Q function $Q(x,p,t_{0})$ for the superposition $|\psi_{sup}\rangle=\sum_{j}c_{j}|x_{j}\rangle$
is the sum of terms involving the Gaussians $\mathcal{G}(x_{j},\sigma_{x})$
associated with each eigenstate $|x_{j}\rangle$, where $\sigma_{x}=1$,
plus ``hidden'' interference terms $\mathcal{I}nt$.

\emph{Assumption (2):} The simplest causal model is based on the Procedure
of the Simulation (Sec. \ref{sec:Forward-backward-stochastic-simu})
and/ or the model $\mathcal{P}_{sup}$ (Result IV.13). The state of
the system at the time $t_{0}$ is identified by $x$ and $p$, along
with \emph{one} of the mean values, $x_{j}$. Model $\mathcal{P}_{sup}$
identifies a distribution $P(x_{j}|x,p)$ for each $x$ and $p$.

\textbf{\emph{Future boundary condition:}} The simulation begins
with the future boundary condition of the backward equation (\ref{eq:backwardSDE-2-1}).
The boundary condition is a probabilistic mixture of Gaussians $\mathcal{G}(Gx_{j},\sigma_{x})$
with $\sigma_{x}=1$ and an amplified mean $Gx_{j}$, with relative
probabilities $|c_{j}|^{2}$ (Result III.1) which is identical to
that of a mixed state
\begin{equation}
\rho_{mix}=|c_{1}|^{2}|x_{1}\rangle\langle x_{1}|+|c_{2}|^{2}|-x_{1}\rangle\langle-x_{1}|\label{eq:mix-3}
\end{equation}
Each individual trajectory (or run) is specified by a value of $x_{j}$,
the value being one of the set of eigenvalues $\{x_{j}\}$ of the
measurement $\hat{x}$. The relative weighting for $x_{j}$ in the
simulations is determined by the probability amplitudes, as in (\ref{eq:mix-3}).
The future boundary condition at time $t_{f}$ includes the input
fluctuation $\delta x(t_{f})$, that we depict by the stochastic symbol
$\eta(t_{f})$ in the Figures, and is not measured by amplification.
The value of $\delta x(t_{f})$ is sampled using a random Gaussian
function $\mathcal{\mathcal{G}}(0,1)$ with mean $0$ and variance
$\sigma_{x}^{2}=1$ at time $t_{f}$ (refer Procedure of Simulation,
Sec. \ref{sec:Forward-backward-stochastic-simu}). The backward equation
(\ref{eq:backwardSDE-2-1}) involves the noise inputs $\xi_{x}(t)$,
as well as the decay. Remarkably, there is no change to the properties
of the fluctuations $\delta x(t_{f})$ as the amplitudes evolve according
to (\ref{eq:backwardSDE-2-1})

\textbf{\emph{Causal deterministic relation:}} The future boundary
condition is hence determined by an initial condition at time $t_{0}$
which probabilistically specifies the value $x_{j}$ (according to
(\ref{eq:mix-3})). More generally, where the eigenvalues $x_{j}$
form a continuous set, we denote the set of outcomes by the continuous
variable $x_{0}$. The future boundary condition is then determined
by the marginal $W(x_{0})$ of the Wigner function $W(x_{0},p_{0})$
(Results IV. 4-5). The $W(x_{0})$ is the probability distribution
for possible outcomes forming a continuous set $\{x_{0}\}$, which
in the Figure \ref{fig:causal-model} are symbolized by the set $\{x_{j}\}$.
The amplification of the value $x_{j}$ is given as a deterministic
causal relation
\begin{equation}
x_{j}\rightarrow Gx_{j}\label{eq:G-amp}
\end{equation}
where here by causal, we mean forward in time. This causal deterministic
relation is represented by the blue forward-going solid arrow in Figure
\ref{fig:macro-sim-1}.  This is also justified, based on the model
$\mathcal{P}_{sup}$ (Result IV.13): for any $x,$$p$, a value $x_{j}$
can be selected for the initial state at time $t_{0}$, with probability
$P(x_{j}|x,p)$. We have noted that being deterministic, the relation
$x_{j}\rightarrow Gx_{j}$ gives a retrodictive, backward relation,
$Gx_{j}\rightarrow x_{j}$ . This is depicted by the solid blue two-way
arrows in Figure \ref{fig:causal-model}.

\textbf{\emph{Stochastic ``retrocausality'' and hidden loops:}}
We express the amplitude as $x(t)=x_{j}(t)+\delta x(t)$, where $x_{j}(t)$
satisfies the causal deterministic equation
\begin{equation}
\frac{dx_{j}(t)}{dt}=gx_{j}(t)\label{eq:delta}
\end{equation}
with boundary condition $x_{j}(0)=x_{j}$, and solution $x_{j}(t)=x_{j}(0)e^{gt}$
(solid blue line in Fig. \ref{fig:causal-model}). The exponential
amplification with respect to $t$ is evident in the trajectories
for $x$ plotted in Figures \ref{fig:stochastic-eigenstate-x}-\ref{fig:macro-sim-1}.

The ``retrocausality'' of the process is embodied stochastically,
in the noise term $\delta x(t)$ (Eq. (\ref{eq:fullx})). The noise
$\delta x$ satisfies the stochastic equation
\begin{equation}
\frac{d(\delta x(t))}{dt_{-}}=-g\delta x(t)+\xi_{x}(t)\label{eq:delta-eq}
\end{equation}
with input $\delta x(t_{f})$ given as a Gaussian $\mathcal{G}(0,1)$,
at the vacuum level. This is depicted by the orange reversed dashed
arrows in Figure \ref{fig:causal-model}. We have seen in Sec. \ref{sec:Forward-backward-stochastic-simu}
that $\delta x(t_{f})$ is added to the mean macroscopic value $x_{j}(t)$
in a way that is independent of the final time $t_{f}$, and independent
of the branch value, $x_{j}$. Solutions for $\delta x$ alone are
given in Figure \ref{fig:solutions-eigenstate-zero}. Similarly, the
properties of the noise $\xi_{x}$ do not depend on $t_{f}$ or $x_{j}$.

The stochastic behavior is also independent of the setting $\theta$,
i.e. whether $\hat{x}$ or $\hat{p}$ is measured. In considering
measurement of $\hat{p}$ where $\theta=-\pi/2$ (Eq. (\ref{eq:hamtheta-1})),
we would solve for trajectories $p(t)=p_{j}(t)+\delta p$, where $p_{j}$
satisfies
\begin{equation}
\frac{dp_{j}(t)}{dt}=gp_{j}(t)\label{eq:delta-1}
\end{equation}
 and the $\delta p(t)$ satisfies 
\begin{equation}
\frac{d(\delta p(t))}{dt_{-}}=-g\delta p(t)+\xi_{p}(t)\label{eq:delta-eq-2}
\end{equation}
with input $\delta p(t_{f})$ given as a Gaussian $\mathcal{G}(0,1)$,
at the vacuum level. The $\delta p(t)$ has the same initial condition
$\delta p(t_{f})$ and trajectory equations as $\delta x(t)$.

The hidden loop for measurement of $\hat{x}$ is depicted in Figure
\ref{fig:causal-model} and is based on the conditional distribution
$Q_{0}(p|x)$ (Eq. \ref{eq:cond-1})), which is determined from the
initial Q function $Q(x,p,t_{0})$.  The value of $x(t_{0})$ is
given as $x_{j}+\delta x(t_{0})$. The joint distribution for $\{x(t_{0}),p(t_{0})\}$
is given by $Q_{loop}(x,p,t_{0}|x_{j})$ (Eq.(\ref{eq:Qloop-postselect}))
as in Figure \ref{fig:The-causal-structure-1}. The values $p(t_{0})$
propagate forward in time according to (\ref{eq:forwardSDE-2-1}),
depicted by the orange dashed arrows in the forward-time direction.
Since they are at the hidden vacuum level at time $t_{f}$, both $p$
and $\delta x$ are unobservable.

In summary, although ``retrocausality'' is present in terms of the
equations (\ref{eq:backwardSDE-2-1}) for $x(t)$, the causal model
for the measurement shows that this reflects only in the way the backward
equation is solved, at a microscopic stochastic level. There is no
genuine retrocausality in the sense of a backward-in-time transfer
of information about the outcome $x_{j}$ or the setting $\theta$,
whether $\hat{x}$ or $\hat{p}$. The trajectory for $\delta x$ has
a stochastic solution involving the noise $\xi_{x}$, but the noise
inputs $\delta x(t_{f})$ and $\xi_{x}$ do not depend on $x_{j}(t)$
or the setting $\theta$.

\subsection{Causal consistency and causality at a macroscopic level}

The Q function, which represents the quantum state $|\psi(t)\rangle=e^{-iH_{amp}t/\hbar}$,
evolves causally, forward in time. The causal model must explain how
consistency is achieved between the causal evolution of the Q function
and the intrinsic retrocausality implicit in the forward-backward
stochastic dynamics.

\textbf{\emph{Definition: Causal consistency: }}We demonstrate the
causal consistency of the measurement model, by showing that the joint
distribution for $x(t)$ and $p(t)$ at a given time $t$ ($t_{0}<t<t_{f}$),
as provided by the hidden loops averaged over all branches $\mathcal{B}_{j}$,
is equivalent to the Q function $Q(x,p,t)$ at the time $t$. This
must be true for all choices of $t_{f}$ (the time of the future boundary).
The test of causal consistency is depicted in Figure \ref{fig:causal-consistency-2-1}.

\textbf{\emph{Result V.1: Causal consistency of the Q model:}}\textbf{
}The Q-based model of reality is causally consistent.

\emph{Proof:} Verification that the numerical solutions from the simulations
are causally consistent is given in the Appendix B. Here, we prove
this analytically.

The simulation for the state $\sum_{j}c_{j}|x_{j}\rangle$ begins
at the future boundary at time $t_{f}$, with a value $x(t)$ emanating
from a particular branch $\mathcal{B}_{j}$ (associated with eigenvalue
$x_{j}$), with probability $|c_{j}|^{2}$. For each branch, the distribution
of $x(t_{f})$ is given by the Gaussian $\frac{e^{-(x-Gx_{j})^{2}/2}}{\sqrt{2\pi}}$
where $G=e^{gt_{f}}$. We first consider just one branch, $x_{j}$.
At time $t$, the distribution is $\frac{e^{-(x-G(t)x_{j})^{2}/2}}{\sqrt{2\pi}}$
($t_{0}<t<t_{f}$) where $G(t)=e^{gt}$. The distribution at time
$t_{0}$ is $Q_{j}(x)=\frac{e^{-(x-x_{j})^{2}/2}}{\sqrt{2\pi}}$,
a set of values for $x$ that we denote by $\mathcal{B}_{j}$. At
time $t_{0}$, there is a connection with a set of values for $p$
determined by $Q_{0}(p|x)$ (Eq. (\ref{eq:cond-1})). For simplicity,
in this paper we consider the superposition of $|x_{1}\rangle$ and
$|x_{2}\rangle$ where there are just two branches, and take $|c_{1}|=|c_{2}|$,
$c_{2}=|c_{2}|e^{i\varphi}$, and $x_{2}=-x_{1}$, and denote the
two branches by $\mathcal{B}_{\pm}$. Following on from Eq. (\ref{eq:Qint+-1-1})
of the proof of Result IV.19, denoting the distribution of the branch
$\mathcal{B}_{\pm}$ at $t_{0}$ by $Q_{\pm}(x)=\frac{e^{-(x\mp x_{1})^{2}/2}}{\sqrt{2\pi}}$,
we consider the branch with eigenvalue $x_{j}=\pm x_{1}$, and find
\begin{eqnarray}
Q_{loop}(x,p,t_{0}|x\in\mathcal{B}_{\pm}) & = & Q_{\pm}(x)Q_{0}(p|x)\label{eq:branch-cond}
\end{eqnarray}
where $Q_{0}(p|x)$ is given by Eq. (\ref{eq:Q0limit-3-2-1}).

Continuing, we examine the \emph{forward evolution }of the distribution
of the $p$ amplitudes. The forward and backward equations are separable,
so that trajectories for $p$ propagate independently of those for
$x$. The marginal of $p$ is $Q_{sup}(p,t)=\int dxQ_{sup}(x,p,t)$
where $Q_{sup}(x,p,t)$ is given by Eq. (\ref{eq:Q-sup-1}). We find
\begin{eqnarray}
Q_{sup}(p,t) & = & N_{0}\frac{e^{-p^{2}/2\sigma_{p}^{2}(t)}}{\sqrt{2\pi}\sigma_{p}(t)}\Bigr(1+e^{-G(t)^{2}x_{1}^{2}}\mathcal{F}(t)\biggl)\nonumber \\
\label{eq:pmarg-2}
\end{eqnarray}
where
\[
\mathcal{F}(t)=(\cos\varphi)\cos(pG(t)x_{1})-(\sin\varphi)\sin(pG(t)x_{1})
\]
and $G(t)=e^{gt}$, $\sigma_{p}^{2}(t)=1+[\sigma_{p}^{2}(0)-1]/G(t)^{2}$
and $N_{0}$ is a normalization factor. The distribution is that evaluated
from the trajectories of $p$ as shown in Figures \ref{fig:stochastic-sup-micro}
and \ref{fig:macro-sim-1}, and, notably, is independent of the branch,
whether $\mathcal{B}_{+}$ or $\mathcal{B}_{-}$. 

The joint densities for $x(t)$ and $p(t)$ at a given time $t_{0}\leq t\leq t_{f}$
can now be evaluated for each branch, by examining the distribution
$Q_{loop}(x,p,t|x\in\mathcal{B}_{\pm})$, as in Eq. (\ref{eq:HL}),
which is the evolution of (\ref{eq:branch-cond}) (refer Fig. \ref{fig:causal-consistency-2-1}).
The value of $x_{1}$ is deterministically amplified to $G(t)x_{1}$
as the system evolves from the time $t_{0}$, so that
\begin{equation}
Q_{\pm}(x)\rightarrow Q_{G\pm}(x)=\frac{e^{-(x\mp G(t)x_{1})^{2}/2}}{\sqrt{2\pi}}\label{eq:qtrans}
\end{equation}
Hence, denoting $Q_{loop}(x,p,t|x\in\mathcal{B}_{\pm})$ by $Q_{loop\pm}(x,p,t)$
for simplicity of notation, we find $Q_{loop\pm}(x,p,t)=N_{t}Q_{G\pm}(x)Q_{t}(p|x)$
where $Q_{t}(p|x)=Q(x,p,t)/Q(x,t)$ is the conditional of the evolved
Q function at time $t$, and (noting Eq.(\ref{eq:Q-sup-1})) is obtained
from $Q_{0}(p|x)$ of Eq. (\ref{eq:Q0limit-3-2-1}) by replacing
$\mathcal{F}\rightarrow\mathcal{F}(t)$, $x_{1}\rightarrow Gx_{1}$,
$\sigma_{p}\rightarrow\sigma_{p}(t)$. Hence 
\begin{eqnarray}
Q_{loop\pm}(x,p,t) & = & N_{t}Q_{G\pm}(x)\frac{e^{-p^{2}/2\sigma_{p}^{2}(t)}}{\sqrt{2\pi}\sigma_{p}(t)}\nonumber \\
 &  & \times\Bigr(1+\frac{\mathcal{F}(t)}{\cosh(xG(t)x_{1})}\Bigl)\label{eq:loop-2}
\end{eqnarray}
where $N_{t}$ is a normalization factor. The Q function $Q_{sup}(x,p,t)$
for the state $|\psi_{sup}(t)\rangle=e^{-iH_{amp}t/\hbar}$ at time
$t$ corresponds to the joint density given as the average of that
for the branches. Hence
\begin{eqnarray}
Q_{sup}(x,p,t) & = & N_{t}\frac{e^{-p^{2}/2\sigma_{p}^{2}(t)}}{2\sqrt{2\pi}\sigma_{p}(t)}\Bigr(Q_{G+}(x)+Q_{G-}(x)\nonumber \\
 &  & +\{Q_{G+}(x)+Q_{G-}(x)\}\frac{\mathcal{F}(t)}{\cosh(xG(t)x_{1})}\Bigl)\nonumber \\
\label{eq:fullq}
\end{eqnarray}
which (on noting that $\sigma_{x}(t)=1$), reduces to the Q function
$Q_{sup}(x,p,t)$ given by Eq. (\ref{eq:Q-sup-1}), as required. We
note the proof follows because of the independence of the forward-distribution
for $p$ on the branch, and because of the simple deterministic relation
$x_{j}\rightarrow Gx_{j}$ for amplification. The proof for a superposition
with more branches follows along the same lines. $\square$

Several related Results for the Q model of reality apply.

\textbf{\emph{Result V.2a: Causality for macroscopic variables: No
retrocausality at a macroscopic level: }}As the system amplifies,
the superposition $\sum_{j}c_{j}|x_{j}\rangle$ becomes a macroscopic
superposition $\sum_{j}e^{-iH_{amp}t/\hbar}|x_{j}\rangle$ of amplified
states (Eq. (\ref{eq:amp-sup-3}) and Fig. \ref{fig:stochastic-sup-micro}).
The $x(t)$ becomes macroscopic and associated with one of the branches
$\widetilde{\lambda}_{x}$, for which the outcome of the measurement
$\hat{x}$ is specified, as one of the eigenvalues $x_{j}$ , consistent
with the premise of weak macroscopic realism (Result IV.6). The causal
relation is that the value $\widetilde{\lambda}_{x}$ determines the
measurement outcome \emph{subsequently}, if there is to be a final
detection, assuming there is no change of settings (Result IV.6).
The branch-value $\widetilde{\lambda}_{x}$ is not changed retrocausally
by the future boundary condition.

For such a macroscopic system, the deterministic relation $x_{j}\rightarrow Gx_{j}$
applies fully to macroscopic variables $x(t)$, since for the macroscopic
superposition, the $x_{j}$ are macroscopically distinct at the initial
time $t_{0}$. Hence, we obtain causal relations at the level of macroscopic
amplitudes $x_{j}$. This is illustrated in Figure \ref{fig:macro-sim-1-1},
where the system is in a macroscopic superposition at time $t_{0}$.
Each branch (symbolized by $\widetilde{\lambda}_{x}$) is distinct
throughout the evolution, and shows exponential amplification forward
in time (solid blue arrow in Fig. \ref{fig:macro-sim-1}). $\square$

We extend this Result in Secs. \ref{sec:CV-Bell-nonlocality} and
\ref{sec:Causal-structure-bell} to include that the model is consistent
with no-signaling.

\textbf{\emph{Result V.2b: No Grandfather paradoxes:}}\textbf{ }The
causal model explains why there is no Grandfather-type paradox arising
from the solutions. The Grandfather paradox is where the state of
a system at time $t$ is apparently affected by a future event. If
we define the ``state'' of the system as being determined by $Q(x,p,t)$,
then we have verified causal consistency (Result V.1), that $Q(x,p,t)$
is not changed by the future boundary condition. If we define the
``state'' of the system by the variable $\widetilde{\lambda}_{x}$
that denotes the branch for a physical observable $\hat{x}$, and
hence defines a measurable physical property for the system, then
we have verified (Result V.2a) that this value is not changed retrocausally.\textbf{}

\subsection{Measurement problem}

The solutions of this paper give a model for measurement contributing
an explanation of the measurement postulates. In the Q model, amplification
plays the key role in the quantum-to-classical transition. From Result
III.1 (Sec. III), we see that amplification  attenuates the interference
terms that distinguish the superposition from a mixed state.  The
macroscopic properties $\widetilde{\lambda}_{x}$ (given by the branches)
emerge as the system is amplified and the inferred outcome $x(t_{f})/G$
is one of the eigenvalues $x_{j}$. Born's rule is derived in Sec.
IV.B.

Amplification also explains the ``collapse'' of the wave-function.
As the system is amplified, the superposition $|\psi_{sup}\rangle$
evolves into a macroscopic superposition (Fig. \ref{fig:stochastic-sup-micro}).
Results from Sec. IV.D (Fig. \ref{fig:cond-var-inferred}) show that
for a macroscopic superposition, the postselected state conditioned
on the branch with outcome $x_{j}$ approaches the eigenstate $|x_{j}\rangle$.

However, the amplification $H_{amp}$ that creates the macroscopic
superposition state is a \emph{reversible} process. The complete
``collapse'' to the eigenstate must occur when the process is \emph{not
reversible}. We see that this corresponds to when information about
the complementary variable $p$ is removed at the final detection
(Fig. \ref{fig:measurement-feedback}). In the calculations (Sec.
IV.D), this corresponds to averaging the distribution $Q_{loop}(x,p,t_{0}|x_{j})$
for the inferred state (given an outcome $x_{j}$) over $p$. A
better model of the collapse allows for a coupling to a meter \citep{Reid2023Short}.
 A model for projection (and the collapse of the wave function) when
a measurement is made on a system $B$ by a meter $A$ is given in
Sec. \ref{sec:Causal-structure-bell}. 

\section{Continuous-variable EPR entanglement\label{sec:Continuous-variable-EPR-entangle}}

The stochastic equations can be applied to treat bipartite systems.
In this section, we examine EPR entanglement as detected using measurements
of $\hat{x}$ and $\hat{p}$.

Einstein, Podolsky and Rosen (EPR) presented an argument for the completion
of quantum mechanics in 1935 \citep{Einstein1935}. The argument can
be realized for two separated field modes $A$ and $B$ \citep{Reid:1989,Reid:2009_RMP81}.
Two modes prepared in a two-mode squeezed state
\begin{equation}
|\psi_{epr}\rangle=(1-\eta^{2})^{1/2}\sum_{n=0}^{\infty}\tanh^{n}r_{2}\thinspace|n\rangle_{A}|n\rangle_{B}\label{eq:tmss}
\end{equation}
at time $t_{0}=0$ possess EPR correlations. Here $\eta=\tanh r_{2}$
where $r_{2}$ is the two-mode squeeze parameter, and $|n\rangle_{A/B}$
are number states. Boson operators $\hat{a}$, $\hat{b}$ and quadrature
phase amplitudes $\hat{x}_{A}$, $\hat{p}_{A}$, $\hat{x}_{B}$ and
$\hat{p}_{B}$ are defined for each mode, where $\hat{x}_{A}=\hat{a}+\hat{a}^{\dagger}$,
$\hat{x}_{B}=\hat{b}+\hat{b}^{\dagger}$, $\hat{p}_{A}=(\hat{a}-\hat{a}^{\dagger})/i$,
$\hat{p}_{B}=(\hat{b}-\hat{b}^{\dagger})/i$.

The Q function for the two-mode quantum state $|\psi\rangle$ is defined
as $Q(\alpha,\beta)=\frac{1}{\pi^{2}}|\langle\alpha|\langle\beta|\psi\rangle|^{2}$
where $|\alpha\rangle$ and $|\beta\rangle$ are coherent states of
modes $A$ and $B$ respectively. Introducing real coordinates $x_{A}=\alpha+\alpha^{*}$,
$p_{A}=(\alpha-\alpha^{*})/i$, $x_{B}=\beta+\beta^{*}$, $p_{B}=(\beta-\beta^{*})/i$,
the $Q$ function of $|\psi_{epr}\rangle$ is
\begin{eqnarray}
Q_{epr}(\bm{\lambda},t_{0}) & = & \frac{(1-\eta^{2})}{16\pi^{2}}e^{-\frac{1}{8}(x_{A}-x_{B})^{2}(1+\eta)}e^{-\frac{1}{8}(p_{A}+p_{B})^{2}(1+\eta)}\nonumber \\
 &  & \times e^{-\frac{1}{8}(x_{A}+x_{B})^{2}(1-\eta)}e^{-\frac{1}{8}(p_{A}-p_{B})^{2}(1-\eta)}\label{eq:epr-q}
\end{eqnarray}
where $\bm{\lambda}=(x_{A},x_{B},p_{A},p_{B})$. As $r_{2}\rightarrow\infty$,
$|\psi_{epr}\rangle$ is an eigenstate of $\hat{x}_{-}=\hat{x}_{A}-\hat{x}_{B}$
and $\hat{p}_{+}=\hat{p}_{A}+\hat{p}_{B}$, giving a realization of
the EPR paradox \citep{Reid:2009_RMP81}.

Following Sec. \ref{sec:Measurement-model}, a measurement of $\hat{x}_{A}$
or $\hat{p}_{A}$ is modeled as a direct amplification, given by the
Hamiltonian
\begin{equation}
H_{amp}^{A}=i\hbar g(\hat{a}^{\dagger2}e^{2i\theta}-\hat{a}^{2}e^{-2i\theta})\label{eq:hampa}
\end{equation}
where $g$ is real and positive. The choice $\theta=0$ or $\pi/2$
allows measurement of $\hat{x}_{A}$ or $\hat{p}_{A}$ respectively.
Similarly, $H_{amp}^{B}=i\hbar g(\hat{b}^{\dagger2}e^{2i\phi}-\hat{b}^{2}e^{-2i\phi})/2$
with $\phi=0$ or $\pi/2$ allows measurement of $\hat{x}_{B}$ or
$\hat{p}_{B}$.

First we consider joint measurements of $\hat{x}_{A}$ and $\hat{x}_{B}$.
Following Sec. \ref{sec:Forward-backward-stochastic-simu}, the stochastic
equations for the measurements $\hat{x}_{A}$ and $\hat{x}_{B}$ are
($K\in\{A,B\}$)
\begin{equation}
\frac{dx_{K}}{dt_{-}}=-gx_{K}+\xi_{K1}\left(t\right)\label{eq:stoch1}
\end{equation}
with a boundary condition at the time $t_{f}$ and $\frac{dp_{K}}{dt}=-gp_{K}+\xi_{K2}\left(t\right)$
with a boundary condition at time $t_{0}$. We find $\left\langle \xi_{K\mu}\left(t\right)\xi_{K\nu}\left(t'\right)\right\rangle =2g\delta_{\mu\nu}\delta\left(t-t'\right)$.
The noise terms for $A$ and $B$ are independent.

\begin{figure}[H]
\begin{centering}
\includegraphics[width=0.6\columnwidth]{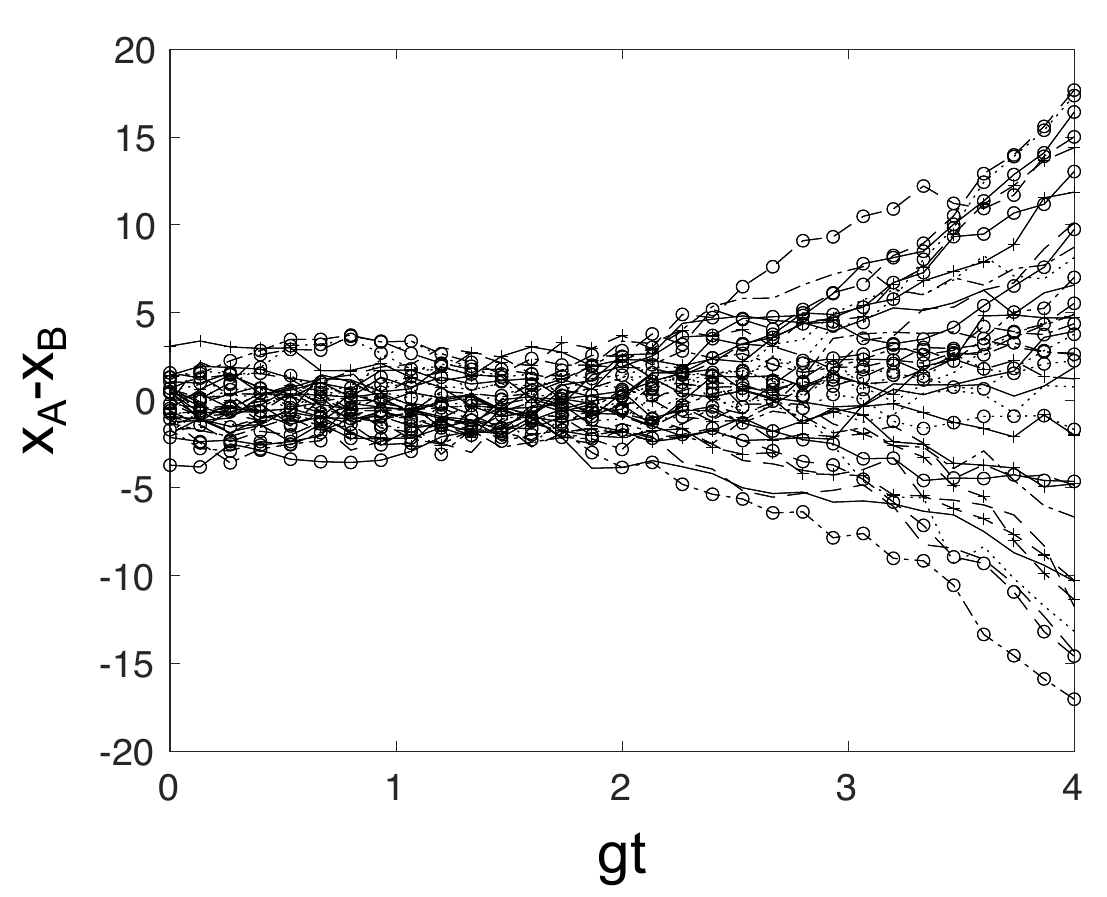}
\par\end{centering}
\centering{}\includegraphics[width=0.6\columnwidth]{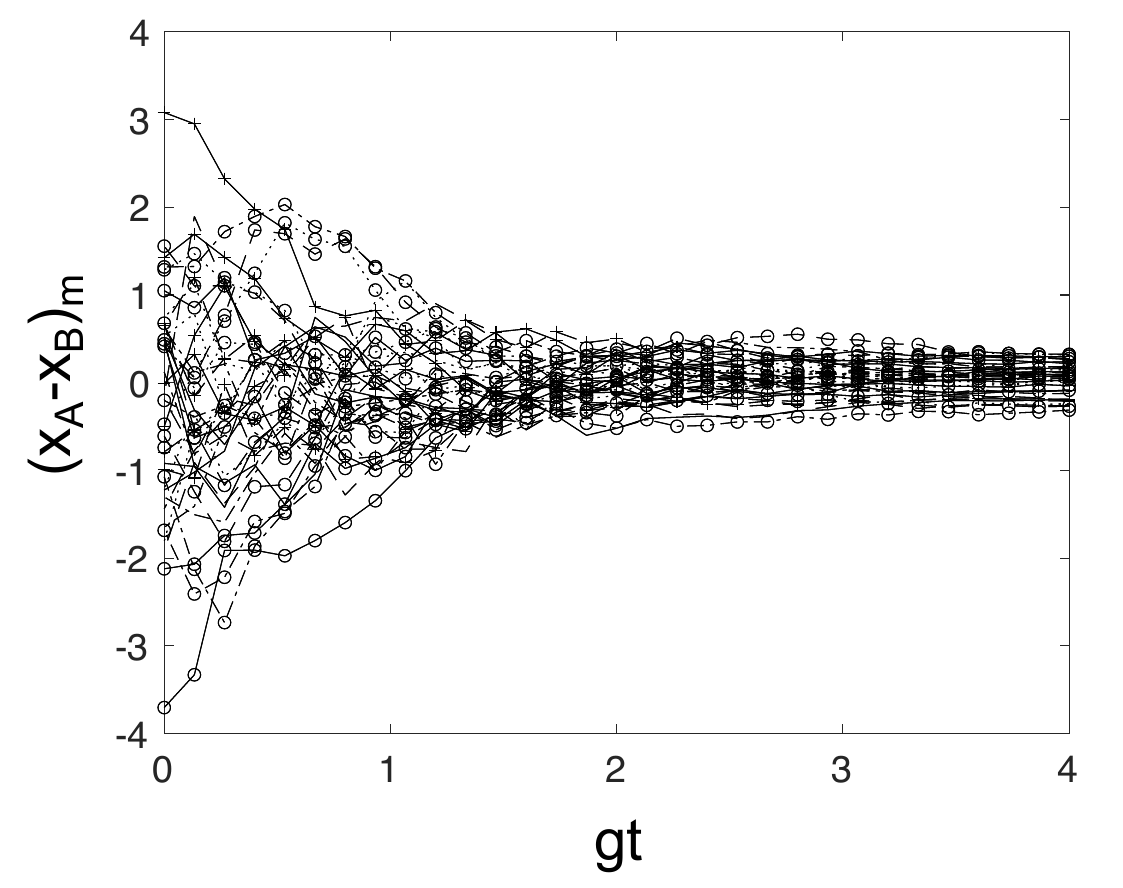}\caption{\textbf{\emph{EPR simulation:}} Simulations for the joint measurement
of $\hat{x}_{A}$ and $\hat{x}_{B}$ on the EPR state $|\psi_{epr}\rangle$.
The top figure shows trajectories of $x_{-}$. The lower figure shows
trajectories scaled according to $(x_{A}-x_{B})_{m}\equiv\widetilde{x}_{-}=x_{-}/e^{gt}$
which is the inferred (``measured'') outcome of the measurement
$\widetilde{x}_{-}=\hat{x}_{A}-\hat{x}_{B}$, as $gt\rightarrow\infty$.
 Here $r_{2}=2$. \textcolor{black}{\label{fig:correlation-xdiff-1}}}
\end{figure}

The equations can be solved by transforming to $x_{\pm}=x_{A}\pm x_{B}$,
$p_{\pm}=p_{A}\pm p_{B}$. The equations are 
\begin{align}
\frac{dx_{\pm}}{dt_{-}} & =-gx_{\pm}+\xi_{1\pm}\left(t\right)\label{eq:stochsetx}
\end{align}
with boundary conditions at the future time $t_{f}$. Those for $p_{+}$
and $p_{-}$ are $\frac{dp_{\pm}}{dt}=-gp_{\pm}+\xi_{2\pm}\left(t\right)$
with boundary conditions at the initial time $t_{0}=0$. Here $\left\langle \xi_{\mu+}\left(t\right)\xi_{\nu+}\left(t'\right)\right\rangle =4g\delta_{\mu\nu}\delta\left(t-t'\right)$
and $\left\langle \xi_{\mu-}\left(t\right)\xi_{\nu-}\left(t'\right)\right\rangle =4g\delta_{\mu\nu}\delta\left(t-t'\right)$
and cross-terms are zero. We examine the solutions for $x_{\pm}$,
since these are of most interest to us, representing the amplified,
or measured, quantities. As in previous examples, the dynamics for
the $x$ and $p$ separate, so that the amplification of $x$ is determined
by the marginals for $x_{A}$ and $x_{B}$. The original distribution
(\ref{eq:epr-q}) is a function of $x_{+}$, $x_{-}$, $p_{+}$ and
$p_{-}$. Transforming $Q_{epr}(\bm{\lambda},t_{0})$ to these variables
to obtain the function $Q_{epr,\pm}(x_{\pm},p_{\pm},t_{0})$, the
marginal for $x_{+}$ and $x_{-}$ at time $t$ (defined as $\int Q_{epr,\pm}(x_{\pm},p_{\pm},t)dp_{+}dp_{-}$
) is 
\begin{eqnarray}
Q_{epr,\pm}(x_{+},x_{-},t) & = & \frac{e^{-x_{-}^{2}/2\sigma_{-}^{2}(t)}e^{-x_{+}^{2}/2\sigma_{+}^{2}(t)}}{2\pi\sigma_{+}(t)\sigma_{-}(t)}\label{eq:bceprx}
\end{eqnarray}
Here the gain is $G\left(t\right)=e^{gt}$ and the variances are
$\sigma_{\pm}^{2}\left(t\right)=2+G^{2}\left(t\right)\left[\sigma_{\pm}^{2}\left(0\right)-2\right]$,
which we rewrite as
\begin{equation}
\sigma_{\pm}^{2}\left(t\right)=2(1+e^{2gt}e^{\pm2r_{2}})\label{eq:varpm}
\end{equation}
The function $Q_{epr,\pm}(x_{+},x_{-},t)$ is separable with respect
to $x_{-}$ and $x_{+}$. The value $\sigma_{\pm}^{2}(t)=2$ is the
hidden vacuum noise level associated with the Q function. The boundary
condition for the backward trajectories of $x_{\pm}$ is determined
by the marginal $Q_{epr,\pm}(x_{+},x_{-},t_{f})$ of the system at
time $t_{f}$.

\begin{figure}[H]
\begin{centering}
\includegraphics[width=0.6\columnwidth]{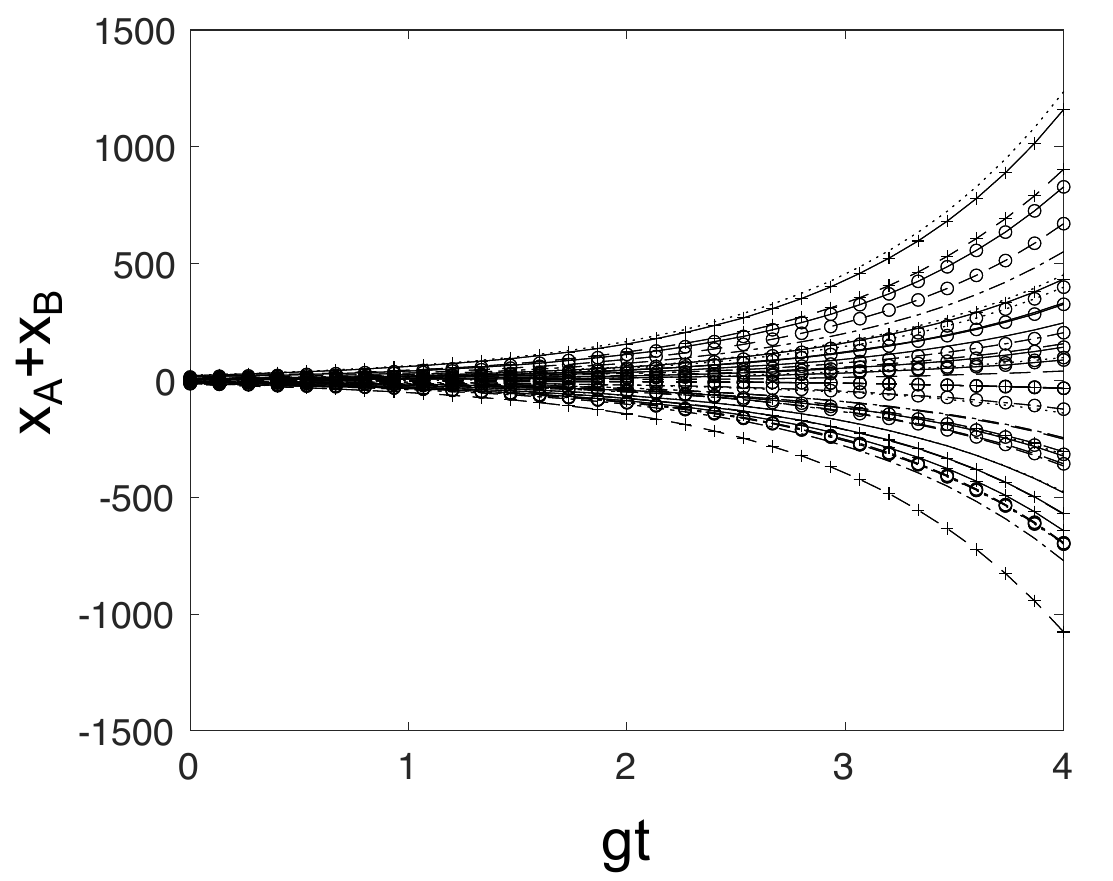}
\par\end{centering}
\begin{centering}
\includegraphics[width=0.6\columnwidth]{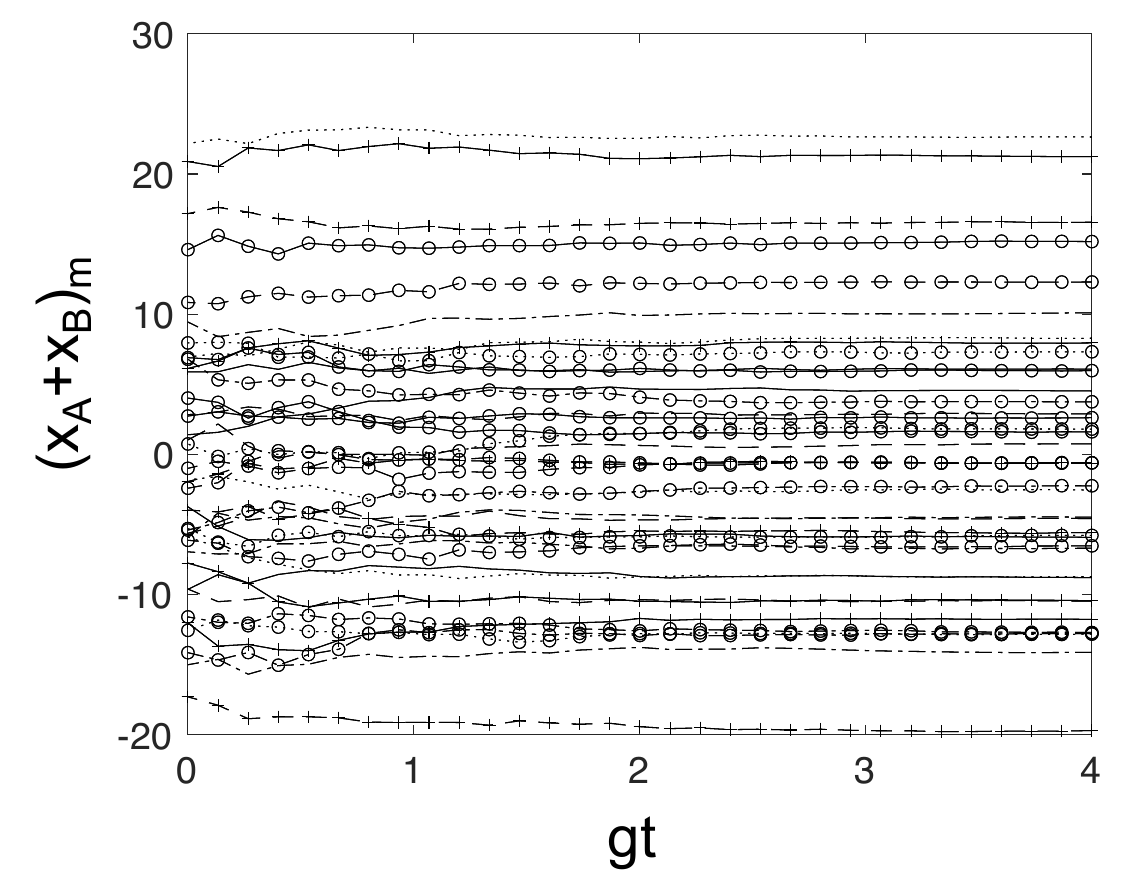}
\par\end{centering}
\centering{}\caption{\textbf{\emph{EPR simulation:}} Measurement of $\hat{x}_{A}$ and
$\hat{x}_{B}$ on the EPR state $|\psi_{epr}\rangle$. The top figure
shows trajectories of $x_{+}$. The lower figure plots trajectories
of the scaled value $(x_{A}+x_{B})_{m}\equiv\widetilde{x}_{+}=x_{+}/e^{gt}$,
which becomes the inferred (``measured'') outcome of the measurement
$\widetilde{x}_{+}=\hat{x}_{A}+\hat{x}_{B}$, as $gt\rightarrow\infty$.
Here, $r_{2}=2$.\label{fig:anticorrelation-xsum}}
\end{figure}

The results of the simulations for $x_{\pm}$ are shown in Figures
\ref{fig:correlation-xdiff-1} and \ref{fig:anticorrelation-xsum}.
The solutions for $x_{-}$ reveal the correlation for the EPR state
between $x_{A}$ and $x_{B}$. The variances $\sigma_{\pm}^{2}(t)$
increases with $gt$ due to the gain associated with the measurement.
However, the trajectories when scaled according to the amplification
factor $G=e^{gt}$ reveal a reduced variance $\sigma_{\pm}^{2}(t)/G^{2}=2(e^{-2gt}+e^{\pm2r_{2}})$.
We note that $\sigma_{-}(t)\rightarrow2(e^{-2gt}+e^{-2r_{2}})\rightarrow0$,
as consistent with the prediction of zero variance in the measured
quantity $\hat{x}_{-}=\hat{x}_{A}-\hat{x}_{B}$, when $r_{2}$ is
large. On the other hand, there is an enhanced variance in $x_{+}$
beyond that of the vacuum state (as the $x_{A}$ and $x_{B}$ are
uncorrelated), which becomes larger with the measurement, which amplifies
the field.

The agreement of the predictions from the simulations with quantum
mechanics can be demonstrated by introducing the scaled variables
$\widetilde{x}_{\pm}=x_{\pm}/e^{gt}$ and $\widetilde{x}_{A/B}=x_{A/B}/e^{gt}$.
These scaled variables correspond to the measured outcomes in the
Q model of reality (Result IV.1, Sec. IV), which we denoted by $x_{0}$
in the single-mode case. In the large amplification limit, $\sigma_{+}^{2}(t)\rightarrow2e^{2gt+2r_{2}}$.
The marginal (\ref{eq:bceprx}) at time $t_{f}$ becomes 
\begin{eqnarray}
Q_{sc}(\widetilde{x}_{+},\widetilde{x}_{-},t_{f}) & = & \frac{e^{-\widetilde{x}_{-}^{2}/(4(e^{-2gt}+e^{-2r_{2}})}e^{-\widetilde{x}_{+}^{2}/4e^{2r_{2}}}}{2\pi\widetilde{\sigma}_{+}(t_{f})\widetilde{\sigma}_{-}(t_{f})}\nonumber \\
 & \rightarrow & \frac{1}{4\pi}e^{-\widetilde{x}_{-}^{2}/(4e^{-2r_{2}})}e^{-\widetilde{x}_{+}^{2}/4e^{2r_{2}}}\label{eq:qlimit}
\end{eqnarray}
where we take $gt\rightarrow\infty$ and $\widetilde{\sigma}_{\pm}^{2}(t_{f})=2e^{\pm2r_{2}}$.
Once the amplification factor $G$ is accounted for in the measurement,
the final distributions for $x_{-}$ and $x_{+}$ are
\begin{equation}
Q_{sc}(\tilde{x}_{\mp},t_{f})=\frac{1}{2e^{\mp r_{2}}\sqrt{\pi}}e^{-\tilde{x}_{\mp}^{2}/(4e^{\mp2r_{2}})}\label{eq:q1}
\end{equation}
These distributions have variances of $2e^{\mp2r_{2}}$. The hidden
vacuum noise terms associated with the Q function, as given by $\sigma_{\pm}^{2}(t)=2$
in the expression (\ref{eq:varpm}) for $\sigma_{\pm}^{2}(t)$, have
become indiscernible. The limiting distributions give precisely the
distributions predicted from quantum mechanics. The two-mode squeezed
state $|\psi_{epr}\rangle$ predicts $[\Delta(\hat{x}_{\pm})]^{2}=2e^{\pm2r_{2}}$,
where $\hat{x}_{\pm}=\hat{x}_{A}\pm\hat{x}_{B}$, consistent with
the result given by Eq. (\ref{eq:q1}).
\begin{figure}
\begin{centering}
\includegraphics[width=0.7\columnwidth]{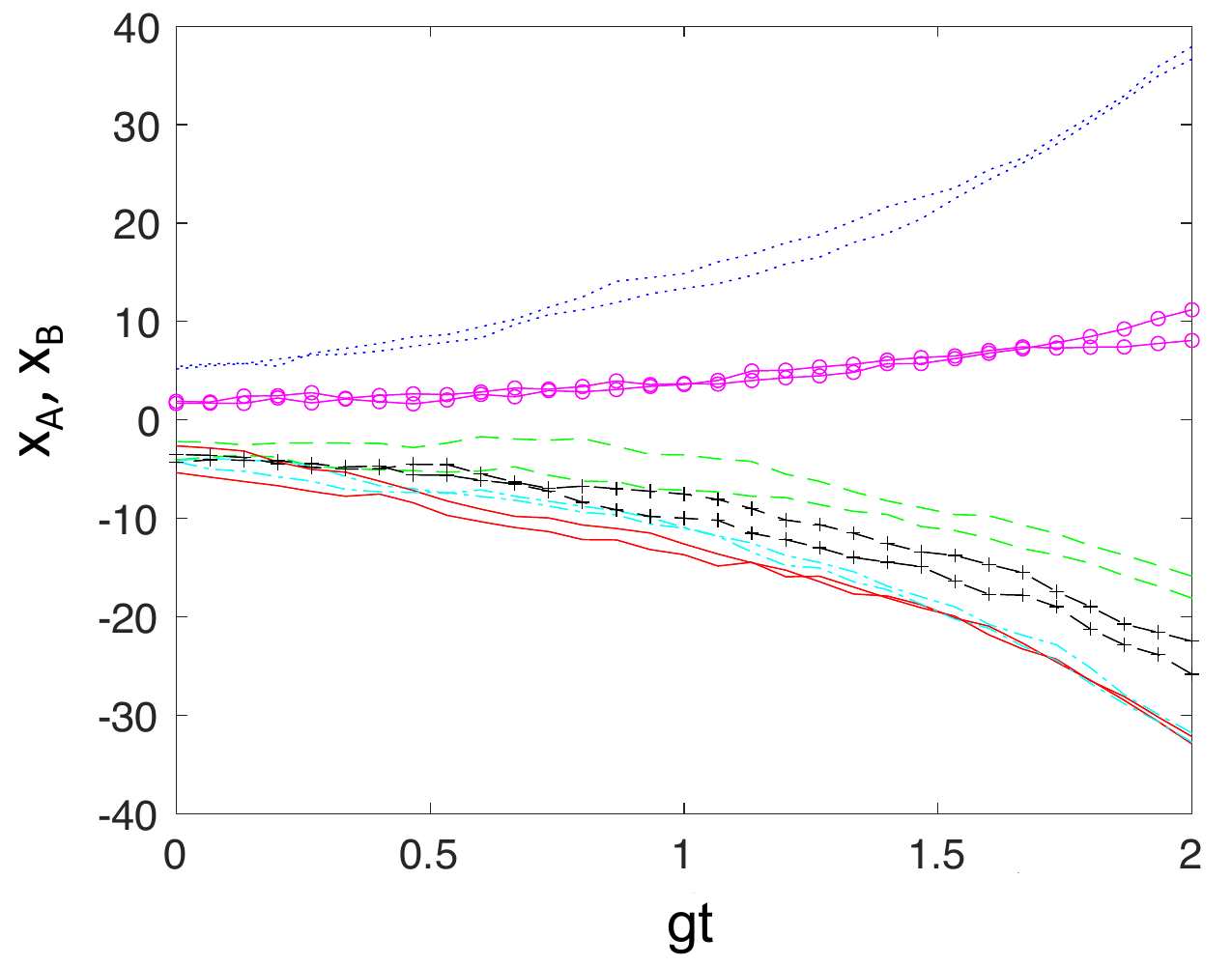}
\par\end{centering}
\caption{\textbf{\emph{Model of EPR correlations}} \textbf{\emph{and weak elements
of reality:}} We consider joint measurement of $\hat{x}_{A}$ and
$\hat{x}_{B}$ on the EPR state $|\psi_{epr}\rangle$, with $r_{2}=2$.
The figure shows trajectories for $x_{A}$ and $x_{B}$. Distinct
branches emerge as the system is amplified. The pair $(x_{A},x_{B})$
corresponding to the same run are shown with the same color and symbol.
The outcome of $\hat{x}_{A}$ is the value $x_{A}(t)/e^{gt}$. Similarly,
the outcome for $\hat{x}_{B}$ is $x_{B}(t)/e^{gt}$. For large $r$
and $G=e^{gt}$, the correlation between $x_{A}$ and $x_{B}$ is
perfect so that the outcome for $\hat{x}_{A}$ can be inferred from
$\hat{x}_{B}$.\label{fig:epr-3} The trajectories $p_{A}$ and $-p_{B}$
for the joint measurement of $\hat{p}_{A}$ and $-\hat{p}_{B}$ show
the same behavior as for $x_{A}$ and $x_{B}$, satisfying the same
equations.}
\end{figure}

The EPR argument is based not only on the perfect correlation between
$\hat{x}_{A}$ and $\hat{x}_{B}$, but also on the the perfect anti-correlation
between $\hat{p}_{A}$ and $\hat{p}_{B}$ \citep{Einstein1935}. The
stochastic equations for joint measurements of $\hat{p}_{A}$ and
$\hat{p}_{B}$ are
\begin{equation}
\frac{dp_{K}}{dt_{-}}=-gp_{K}+\xi_{K1}\left(t\right)\label{eq:stocbp}
\end{equation}
and $\frac{dx_{K}}{dt}=-gx_{K}+\xi_{K2}\left(t\right)$. The equations
of interest are those for $p_{K}$ since these give solutions for
the amplified variables that are measured. The marginal for $p_{\pm}$
for the initial state (\ref{eq:epr-q}) is
\begin{eqnarray}
Q_{epr,\pm}(p_{-},p_{+},0) & = & \frac{e^{-p_{-}^{2}/2\sigma_{+}^{2}(0)}e^{-p_{+}^{2}/2\sigma_{-}^{2}(0)}}{2\pi\sigma_{+}(0)\sigma_{-}(0)}\nonumber \\
\label{eq:q5}
\end{eqnarray}
where $\sigma_{\mp}^{2}(0)=2(1+e^{\mp2r_{2}})$. The fluctuations
are decreased in $p_{+}$. The marginal used for the backward equation
(\ref{eq:stocbp}) is
\begin{eqnarray}
Q_{epr,\pm}(p_{+},p_{-},t_{f}) & = & \frac{e^{-p_{-}^{2}/2\sigma_{+}^{2}(t_{f})}e^{-p_{+}^{2}/2\sigma_{-}^{2}(t_{f})}}{2\pi\sigma_{+}(t_{f})\sigma_{-}(t_{f})}\nonumber \\
\label{eq:q6}
\end{eqnarray}
where the gain is $G\left(t\right)=e^{gt}$ and the variances are
$\sigma_{\pm}^{2}\left(t\right)=2(1+e^{2gt}e^{\pm2r_{2}})$. The
$Q_{epr,\pm}(p_{-},p_{+},t_{f})$ is obtained from $Q_{epr,\pm}(x_{+},x_{-},t_{f})$
by replacing $x_{+}$ with $p_{-}$, and $x_{-}$ with $p_{+}$. The
stochastic solutions for $p_{\pm}$ are hence given by those of $x_{\pm}$,
but with $p_{\pm}$ replacing $x_{\mp}$. The results are in agreement
with those of quantum mechanics, showing an anticorrelation between
the outcomes of $\hat{p}_{A}$ and $\hat{p}_{B}$, as $r_{2}$ becomes
large. The simulation leads to the following conclusions.

\textbf{\emph{Result V1.1: Branches show EPR correlations:}} Branches
emerge as both systems are amplified, consistent with the prediction
of EPR correlations.

\emph{Proof:} First, we note that the interactions $H_{amp}^{A}$
and $H_{amp}^{B}$ (Eq. (\ref{eq:hampa})) that induce the choice
of settings and amplification at $A$ and $B$ are local, acting only
on the sets of variables $\{x_{A},p_{A}\}$ and $\{x_{B},p_{B}\}$
respectively. In the Q model, when the fields $A$ and $B$ are spatially
separated, we identify amplitudes $\{x_{A},p_{A}\}$ with field $A$,
and amplitudes $\{x_{B},p_{B}\}$ with field $B$.

The solutions for $x_{A}$ and $x_{B}$ are found as $x_{A}=(x_{+}+x_{-})/2$
and $x_{B}=(x_{+}-x_{-})/2$. For large amplification $gt$, the solutions
are dominated by those of $x_{+}$. In Figure \ref{fig:epr-3}, we
plot $x_{A}$ and $x_{B}$, using the same symbol (and color) for
the same run of the simulation. As the amplification due to measurement
increases, the ``branches'' (denoted $\widetilde{\lambda}_{x}^{A}$
and $\widetilde{\lambda}_{x}^{B}$) emerge, given by the ``lines''
in the Figure \ref{fig:epr-3} (Secs. III.B and IV.A). With amplification,
at a time $t_{m}$ when the values $x_{A}(t)$ and $x_{B}(t)$ are
macroscopic, the premise of weak macroscopic realism applies (Result
IV.6, Sec. IV.C) and the result of the measurements $\hat{x}_{A}$
and $\hat{x}_{B}$ are assumed determined. The value $x_{A}(t)/G$
or $x_{B}(t)/G$ given by the ``line'' as $gt\rightarrow\infty$
represents the outcome, if a detection is to be made. We see that
$x_{A}(t)-x_{B}(t)\rightarrow0$ as $r_{2}$ becomes large, implying
perfect correlation between the outcomes of $\hat{x}_{A}$ and $\hat{x}_{B}$,
in the Q model of reality. Similarly, for measurements of $\hat{p}_{A}$
and $\hat{p}_{B}$, we find branches (which we label $\widetilde{\lambda}_{p}^{A}$
and $\widetilde{\lambda}_{p}^{B}$) emerging for amplitudes $p_{A}(t)$
and $p_{B}(t)$ (Figure \ref{fig:epr-3}). The amplitudes satisfy
$p_{A}(t)+p_{B}(t)\rightarrow0$ as $r_{2}\rightarrow\infty$, implying
perfect anti-correlation between outcomes $\hat{p}_{A}$ and $\hat{p}_{B}$.
$\square$

\textbf{\emph{Result VI.2: EPR correlations are consistent with (weak)
macroscopic realism: Weak elements of reality:}}

\emph{Proof:} This follows from the proof of Result IV.1 and from
the definition of weak macroscopic realism (refer Definition 6(a)
of Sec. I.B and Result IV.6). If we consider measurement of $\hat{x}_{A}$
and $\hat{x}_{B}$, the trajectories separate and show distinct branches
as $gt\rightarrow\infty$, at a time we refer to as $t_{m}$, consistent
with the emergence of macroscopic superposition states as the system
is amplified (Result V2.a). Applying Result IV.6, weak macroscopic
realism holds: the outcomes are determined at the time $t_{m}$, denoted
by the value indicated by the ``lines'' in the Figure \ref{fig:epr-3}.
We refer to these values as ``\emph{weak elements of reality}''.
$\square$

The existence of the ``weak elements of reality'' does not conflict
with the known failure of the EPR ``elements of reality'' as shown
by Bell's theorem. This is because the weak elements of reality are
defined for the system after the local settings are fixed, at the
time $t_{m}$. Results VI.1 and VI.2 are hence consistent with proposals
that ``elements of reality'' exist for the EPR experiment, \emph{after}
the settings are fixed, in a way that does not conflict with Bell
nonlocality \citep{fulton2024alternative,Fulton2024Weak,mcguigan2024resolving}.

\section{CV Bell nonlocality\label{sec:CV-Bell-nonlocality}}

The demonstration of Bell nonlocality involves a choice of measurement
settings $\theta$ and $\phi$ for systems $A$ and $B$, respectively.
In this paper, we examine measurements of $\hat{x}$ and $\hat{p}$,
for which the outcomes are continuous variables (CV). It is known
that certain states will violate a Bell inequality for such measurements
\citep{Leonhardt:1995aa,gilchrist1998contradiction,gilchrist1999contradiction,Banaszek1999Testing}.
We define the rotated quadrature phase amplitudes
\begin{eqnarray}
\hat{x}_{\theta A} & = & \hat{x}_{A}\cos\theta+\hat{p}_{A}\sin\theta\nonumber \\
\hat{p}_{\theta A} & = & -\hat{x}_{A}\sin\theta+\hat{p}_{A}\cos\theta\label{eq:quad-1}
\end{eqnarray}
for system $A$ and
\begin{eqnarray}
\hat{x}_{\phi B} & = & \hat{x}_{B}\cos\phi+\hat{p}_{B}\sin\phi\nonumber \\
\hat{p}_{\phi B} & = & -\hat{x}_{B}\sin\phi+\hat{p}_{B}\cos\phi\label{eq:quad-2}
\end{eqnarray}
for system $B$.  In the proposed Bell test, the amplitudes $\hat{x}_{\theta A}$
and $\hat{x}_{\phi B}$ are measured, and the outcomes binned to be
$+1$ or $-1$, according to the sign $\mathcal{S}$ of the outcome.
Binary outcomes $\mathcal{S}=\pm1$ apply at each site, for given
settings $\theta$ and $\phi$. The joint probability $P_{++}^{AB}(\theta,\phi)$
for positive outcomes at $A$ and $B$, given $\theta$ and $\phi$,
can be determined experimentally. Bell's local hidden variables (LHV)
assumption is that this probability is given by
\begin{equation}
P_{++}^{AB}(\theta,\phi)=\int d\lambda\rho(\lambda)P_{A}(+|\lambda,\theta)P_{B}(+|\lambda,\phi)\label{eq:lhv}
\end{equation}
where $\rho(\lambda)$ is the distribution for a set of hidden variables
$\lambda$ that describe the hidden-variable state of the system,
and $P_{A}(+|\lambda,\theta)$ and $P_{B}(+|\lambda,\phi)$ are the
respective probabilities for an outcome $+1$ at $A$ and $B$, given
the values $\lambda$, $\theta$ and $\phi$ \citep{Bell1964,bell1966problem,clauser1978bell,brunner2014bell}.
The locality assumption is the independence of $P_{A}(+|\lambda,\theta)$
on $\phi$, and of $P_{B}(+|\lambda,\phi)$ on $\theta$. The condition
(\ref{eq:lhv}) allows a Bell inequality to be derived as for spins
\citep{Leonhardt:1995aa,gilchrist1998contradiction}. The Clauser-Horne
(CH) Bell inequality $S\leq1$ where
\begin{eqnarray}
S & = & \frac{P_{++}^{AB}(\theta,\phi)-P_{++}^{AB}(\theta,\phi')+P_{++}^{AB}(\theta',\phi)+P_{++}^{AB}(\theta',\phi')}{P_{+}^{A}(\theta')+P_{+}^{B}(\phi)}\nonumber \\
\label{eq:CH-bell}
\end{eqnarray}
follows, where $P_{+}^{A}(\theta)$ and $P_{+}^{B}(\phi)$ are the
marginal probabilities of obtaining $+1$ at site $A$ and $B$, with
the settings fixed at $\theta$ and $\phi$, respectively. The Clauser-Horne-Shimony-Holt
(CHSH) Bell inequality
\begin{equation}
|E(\theta,\phi)-E(\theta,\phi')+E(\theta',\phi)+E(\theta',\phi')|\leq2\label{eq:CHSH}
\end{equation}
 can also be derived, where $E(\theta,\phi)$ is the expected value
of the product of the outcome $\mathcal{S}$ at each site.

An example of a state that will violate the Bell inequalities is the
two-mode-squeezed cat state \citep{gilchrist1999contradiction}
\begin{equation}
|\psi_{Bell}\rangle=N_{0}^{2}e^{-iH_{I}t_{I}/\hbar}(|\alpha_{0}\rangle_{A}|\alpha_{0}\rangle_{B}+|-\alpha_{0}\rangle_{A}|-\alpha_{0}\rangle_{B})\label{eq:bell}
\end{equation}
where $H_{I}=i\kappa\hbar(\hat{a}^{\dagger}\hat{b}^{\dagger}-\hat{a}\hat{b})$,
$N_{0}^{2}=1/[2(1+e^{-4|\alpha_{0}|^{2}})]$ and $\alpha_{0}$ is
real. Violation of the CH-Bell inequality (\ref{eq:CH-bell}) with
$S=1.008$ for $\alpha_{0}\sim0.9$ and $\kappa t_{I}=0.6$ has been
predicted, for certain choices of $\theta$ and $\phi$. For example,
letting $\delta_{1}=$$\theta-\theta'$ and $\delta_{2}=\phi'-\phi$,
the authors of Ref. \citep{gilchrist1999contradiction} report violations
for $\theta=\phi=0.42\pi$ and $\delta_{1}=\delta_{2}=\delta=0.7\pi$.

\subsection{CV Bell simulation}

We next outline the simulation of the CV Bell violations. The system
is prepared in a bipartite state, such as $|\psi_{Bell}\rangle$,
at a time $t_{0}$. The Q function $Q_{Bell}(\alpha,\beta)=\frac{1}{\pi^{2}}|\langle\alpha|\langle\beta|\psi_{Bell}\rangle|^{2}$
for the state $|\psi_{Bell}\rangle$ is ($r_{0}=\kappa t_{I}$)

\begin{widetext} 
\begin{eqnarray}
Q_{Bell}(\alpha,\beta) & = & \frac{N_{0}^{2}e^{-p_{A}^{2}/4-p_{B}^{2}/4-\tanh r_{0}p_{A}p_{B}/2}}{16\pi^{2}\cosh^{2}r_{0}}\{e^{-(x_{A}-2\alpha_{0}e^{r_{0}})^{2}/4}e^{-(x_{B}-2\alpha_{0}e^{r_{0}})^{2}/4}\times e^{2(\tanh r_{0})[(x_{A}-2\alpha_{0}e^{r})(x_{B}-2\alpha_{0}e^{r_{0}})/4]}\nonumber \\
 &  & +e^{-(x_{A}+2\alpha_{0}e^{r_{0}})^{2}/4}e^{-(x_{B}+2\alpha_{0}e^{r_{0}})^{2}/4}\times e^{2(\tanh r_{0})[(x_{A}+2\alpha_{0}e^{r_{0}})(x_{B}+2\alpha_{0}e^{r_{0}})/4]}\nonumber \\
 &  & +2e^{-2(1-\tanh r_{0})\alpha_{0}^{2}e^{2r_{0}}}e^{-x_{A}^{2}/4}e^{-x_{B}^{2}/4}e^{\frac{1}{2}(\tanh r_{0})[x_{A}x_{B}]}\cos[(1-\tanh r_{0})\alpha_{0}e^{r_{0}}(p_{A}+p_{B})]\label{eq:qbell}
\end{eqnarray}

\end{widetext} This function has the form
\begin{eqnarray}
Q_{Bell}(\bm{\lambda},t_{0}) & \sim & G(p_{A},p_{B})\{G_{1}(x_{A},x_{B})+G_{2}(x_{A},x_{B})+I\}\nonumber \\
\label{eq:I}
\end{eqnarray}
where we denote the real variables $x_{A}=\alpha+\alpha^{*}$, $p_{A}=(\alpha-\alpha^{*})/i$,
$x_{B}=\beta+\beta^{*}$, $p_{B}=(\beta-\beta^{*})/i$ by $\bm{\lambda}=(x_{A},p_{A},x_{B},p_{B})$
as in Eq. (\ref{eq:epr-q}). Here, $G$ denotes a bivariate Gaussian
function and $I$ is a sinusoidal interference term arising due to
entanglement.

\textbf{\emph{Measurement settings:}} After preparation of the state
at time $t_{0}$, the field modes $A$ and $B$ are spatially separated.
Each field then interacts locally with a device that determines the
measurement setting. The amplitudes $\{x_{A},p_{A}\}$ represent
field $A$, and amplitudes $\{x_{B},p_{B}\}$ represent field $B$.
For measurements $\hat{x}_{\theta A}$ and $\hat{x}_{\phi B}$ given
by (\ref{eq:quad-1}) and (\ref{eq:quad-2}), the device that determines
the measurement settings $\theta$ and $\phi$ is a phase-shifter,
modeled by the interaction Hamiltonians 
\begin{equation}
H_{\theta}^{A}=\widetilde{g}\hat{n}_{A},\ H_{\phi}^{B}=\widetilde{g}\hat{n}_{B}\label{eq:phase-shiftham}
\end{equation}
for systems $A$ and $B$ respectively ($\widetilde{g}>0$). Here,
$\hat{n}_{A}$ and $\hat{n}_{B}$ are the number operators for modes
$A$ and $B$. Straightforward solutions of the operator equations
of motion yield for $H_{\theta}^{A}$ that $\dot{\hat{x}}_{A}=\widetilde{g}\hat{p}_{A}/\hbar$
and $\dot{\hat{p}}_{A}=-\widetilde{g}\hat{x}_{A}/\hbar$, and similarly
for $H_{\phi}^{B}$. The solutions are
\begin{eqnarray}
\hat{x}_{A}(t) & = & \hat{x}_{A}(0)\cos\theta+\hat{p}_{A}(0)\sin\theta\nonumber \\
\hat{x}_{B}(t) & = & \hat{x}_{B}(0)\cos\phi+\hat{p}_{B}(0)\sin\phi\label{eq:rot}
\end{eqnarray}
where $\theta=\widetilde{g}t_{A}/\hbar$ and $\phi=\widetilde{g}t_{B}/\hbar$,
the $t_{A}$ and $t_{B}$ being the times of interaction. This gives
the required transformations (\ref{eq:quad-1}) and (\ref{eq:quad-2}).
The interactions $H_{\theta}^{A}$ and $H_{\phi}^{B}$ correspond
to the unitary operations $U_{A}(\theta)=e^{-iH_{\theta}^{A}t_{A}/\hbar}=e^{-i\theta\hat{n}_{A}}$
and $U_{B}(\phi)=e^{-iH_{B}^{B}t_{B}/\hbar}=e^{-i\phi\hat{n}_{B}}$.
We may consider that the operations act simultaneously, or consecutively,
in which case one interaction only may be completed at a time $t_{1}$.
We denote the time at which both interactions are completed as the
time $t_{2}>t_{1}$.

\begin{figure}
\begin{centering}
\includegraphics[width=0.7\columnwidth]{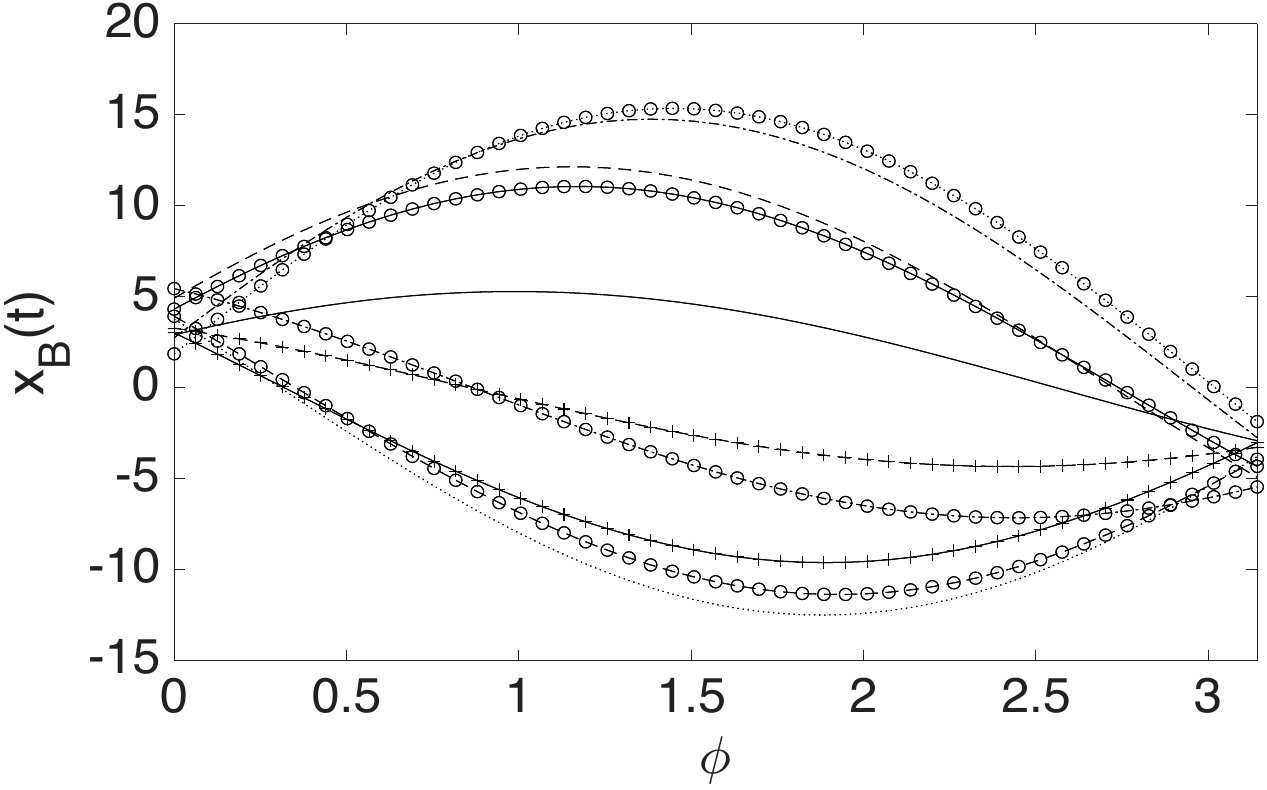}
\par\end{centering}
\begin{centering}
\includegraphics[width=0.7\columnwidth]{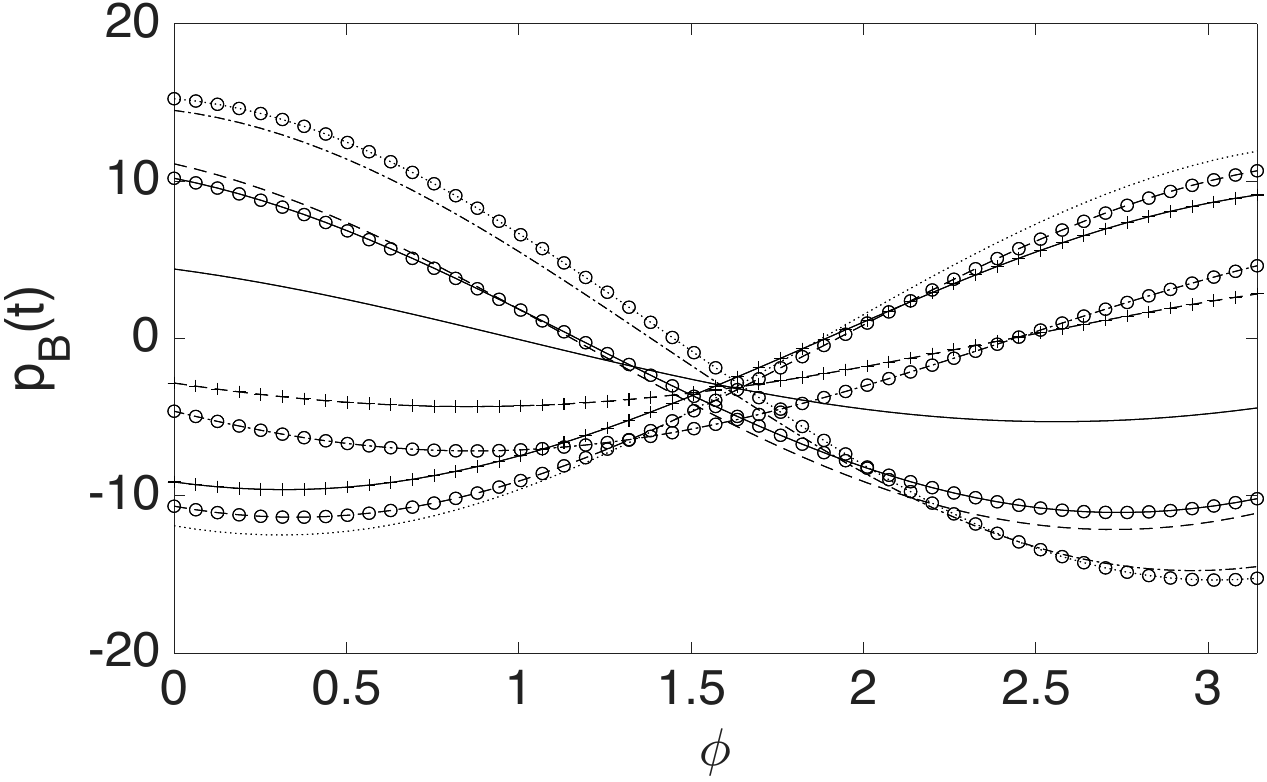}
\par\end{centering}
\caption{Evolution of amplitudes $x_{B}(t)$ and $p_{B}(t)$ as the system
$B$ interacts according to $H_{\phi}^{B}$ (Eq. \ref{eq:phase-shiftham})),
which induces a change of measurement setting $\phi=\widetilde{g}t/\hbar$
at $B$. The amplitudes evolve as (\ref{eq:rot-6-b-3}). Here, the
initial distribution of $x_{B}$ and $p_{B}$ is given by the Q function
(\ref{eq:q-sq-1}) of the squeezed state with $x_{k}=4$, $r=2$ (which
approximates an eigenstate $|x_{k}\rangle$ of $\hat{x}_{B}$). For
this approximate eigenstate, the variance of $p_{B}$ is large, that
of $x_{B}$ is small. After a time $t=\hbar\phi/\widetilde{g}$, the
system is prepared for measurement of $\hat{x}_{\phi B}$. Note that
at the intermediate times such as $\phi\sim0.7$, where $\phi\protect\neq n\pi/2$
($n$ is an integer), neither variance of $x_{B}$ or $p_{B}$ is
large. \label{fig:Evolution-settingB}}
\end{figure}

After interaction with both the phase-shifters, at a time $t_{2}$,
the state of the system initially in state $|\psi\rangle$ is
\begin{equation}
|\psi(t_{2})\rangle=e^{-iH_{\theta}^{A}t_{A}/\hbar}e^{-iH_{\phi}^{B}t_{B}/\hbar}|\psi\rangle\label{eq:state-rot}
\end{equation}
The corresponding Q function at time $t_{2}$ is
\begin{eqnarray}
Q(\bm{\lambda_{rot}},t_{2}) & = & \frac{1}{\pi^{2}}|\langle\alpha|\langle\beta|e^{-i\theta\hat{n}_{A}}e^{-i\phi\hat{n}_{B}}|\psi\rangle|^{2}\nonumber \\
 & = & \frac{1}{\pi^{2}}|\langle\alpha e^{i\theta}|\langle\beta e^{i\phi}|\psi\rangle|^{2}\label{eq:Q-rot}
\end{eqnarray}
where $\bm{\lambda}_{rot}=(x_{\theta A},p_{\theta A},x_{\phi B},p_{\phi B})$
represents the rotated coordinates: $x_{\theta A}=\alpha e^{i\theta}+\alpha^{*}e^{-i\theta}$,
$p_{\theta A}=(\alpha e^{i\theta}-\alpha^{*}e^{-i\theta})/i$, $x_{\phi B}=\beta e^{i\phi}+\beta^{*}e^{-i\phi}$,
and $p_{\phi B}=(\beta e^{i\phi}-\beta^{*}e^{-i\phi})/i$. The last
line follows since $e^{i\theta\hat{n}}|\alpha\rangle=|\alpha e^{i\theta}\rangle$.
From this result, it is shown in the Appendix D that the Q function
$Q(\lambda_{rot},t_{2})$ of the new state at time $t_{2}$ is that
found by substituting the coordinates $\bm{\lambda}=(x_{A},p_{A},x_{B},p_{B})$
in the original Q function $Q_{Bell}(\lambda,t_{0})$ for the new
coordinates $\bm{\lambda}_{rot}$, given by the rotation transformation
\begin{eqnarray}
x_{\theta A} & = & x_{A}\cos\theta+p_{A}\sin\theta\nonumber \\
p_{\theta A} & = & -x_{A}\sin\theta+p_{A}\cos\theta\label{eq:rot-6-a}
\end{eqnarray}
and
\begin{eqnarray}
x_{\phi B} & = & x_{B}\cos\phi+p_{B}\sin\phi\nonumber \\
p_{\phi B} & = & -x_{B}\sin\phi+p_{B}\cos\phi\label{eq:rot-6-b}
\end{eqnarray}
This is a general result, valid for arbitrary two-mode states. Alternatively,
we see that the Fokker-Planck equation of type (\ref{eq:fp-q-1})
corresponding to the interactions $H_{\theta}^{A}$ and $H_{\phi}^{B}$
contains drift terms only, without diffusion terms. The equivalent
equations for the amplitudes are hence deterministic (without stochastic
terms) \citep{Gardiner1997}. Consistent with the above analysis,
the evolution of the amplitudes $\lambda$ associated with the dynamics
of the Q function is hence given by 
\begin{eqnarray}
x_{A}(t_{A}) & = & x_{A}(0)\cos\theta+p_{A}(0)\sin\theta\nonumber \\
p_{A}(t_{A}) & = & -x_{A}(0)\sin\theta+p_{A}(0)\cos\theta\label{eq:rot-6-a-2}
\end{eqnarray}
and
\begin{eqnarray}
x_{B}(t_{B}) & = & x_{B}(0)\cos\phi+p_{B}(0)\sin\phi\nonumber \\
p_{B}(t_{B}) & = & -x_{B}(0)\sin\phi+p_{B}(0)\cos\phi\label{eq:rot-6-b-3}
\end{eqnarray}
where $\theta=\widetilde{g}t_{A}/\hbar$ and $\phi=\widetilde{g}t_{B}/\hbar$.
The final amplitudes are denoted $x_{\theta A}\equiv x_{A}(t_{A})$,
$p_{\theta A}\equiv p_{B}(t_{B})$, $x_{\phi B}\equiv x_{B}(t_{B})$
and $p_{\phi B}\equiv p_{B}(t_{B})$. We depict the dynamics  in
Figure \ref{fig:Evolution-settingB}. There is a change of setting
at $B$, for a time $t_{B}=t$. If the distribution of the amplitudes
at time $t_{0}=0$ is given by $Q(x_{A},p_{A},x_{B},p_{B},t_{0})$,
then the evolved amplitudes at time $t$ have a distribution $Q(x_{A},p_{A},x_{B},p_{B},t)$
obtained by substituting $x_{B}\rightarrow x_{B}\cos\phi-p_{B}\sin\phi$,
$p_{B}\rightarrow x_{B}\cos\phi+p_{B}\sin\phi$ in accordance with
(\ref{eq:rot-6-b-3}) (refer Appendix E).

\begin{figure}
\begin{centering}
\par\end{centering}
\begin{centering}
\includegraphics[width=1\columnwidth]{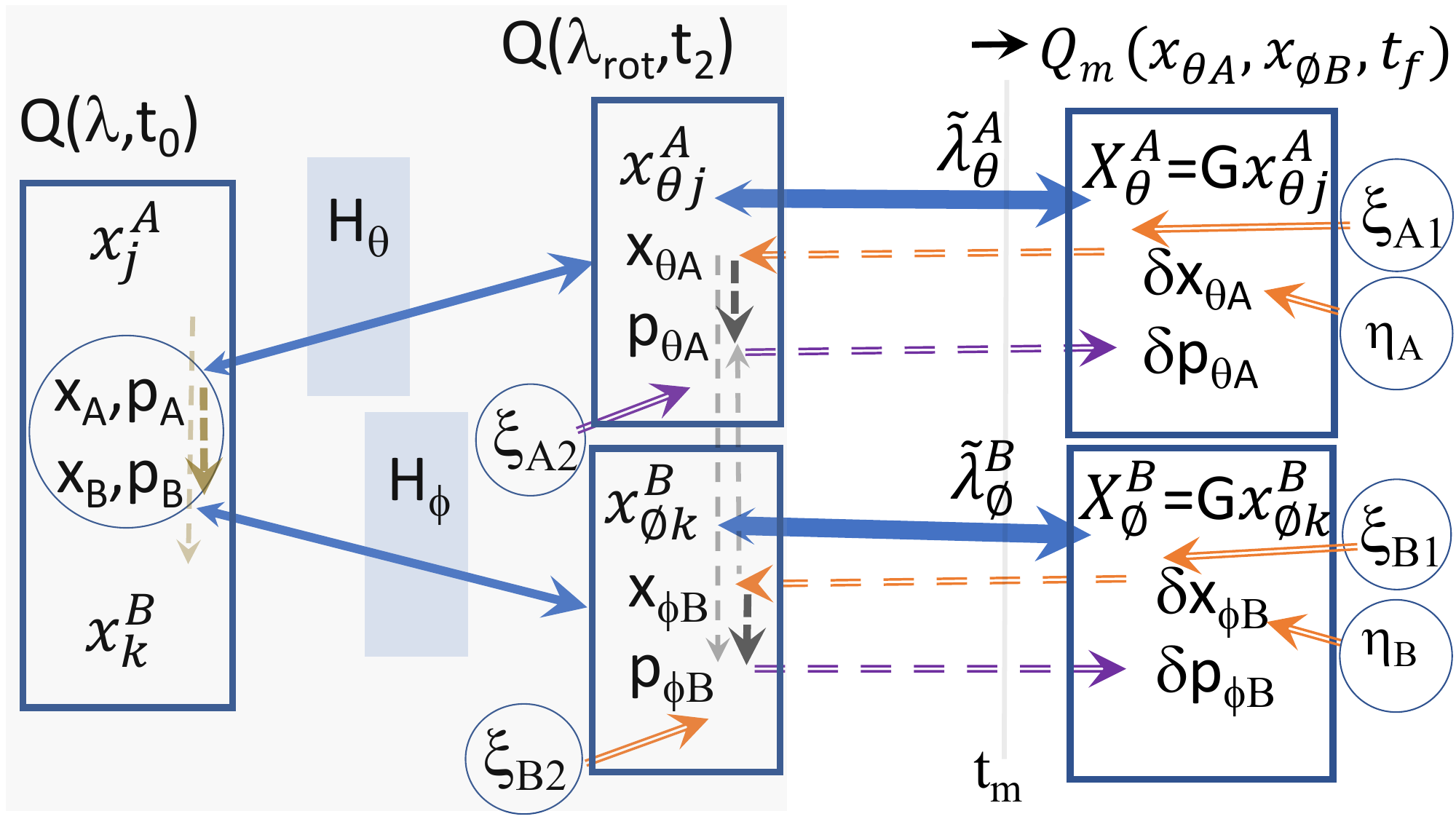}
\par\end{centering}
\caption{\textbf{\emph{Steps in the simulation of the measurement of CV EPR
or Bell nonlocality.}} The entangled state given by $Q(\lambda,t_{0})$
is generated at time $t_{0}$. The settings are changed according
to $H_{\theta}^{A}$ and $H_{\phi}^{B}$, which give causal local
reversible deterministic relations (thin blue two-way arrows). The
state at time $t_{2}$ is given by $Q(\lambda_{rot},t_{2}).$ Note
that while the amplitudes transform locally, the means $x_{j}^{A}\rightarrow x_{\theta j}^{A}$
and $x_{k}^{B}\rightarrow x_{\phi k}^{B}$ do not. After settings
are fixed, the measurements $\hat{x}_{\theta A}$ and $\hat{x}_{\phi B}$
are carried out by amplification, according to $H_{amp}^{A}$ and
$H_{amp}^{B}$, giving stochastic forward-backward dynamics, with
branches $\widetilde{\lambda}_{\theta}^{A}$ and $\widetilde{\lambda}_{\phi}^{B}$
emerging at time $t_{m}$. There is discontinuity at this stage, for
the forward-time dynamics. The solid thick blue two-way arrows denote
the deterministic relations ($x_{\theta j}^{A}\rightarrow Gx_{\theta j}^{A}$,
$x_{\phi k}^{B}\rightarrow Gx_{\phi k}^{B}$), in which the means
$x_{\theta j}^{A}$ and $x_{\phi k}^{B}$ of the Gaussians in $Q(\lambda_{rot},t_{2})$
are amplified to $Gx_{\theta j}^{A}$ and $Gx_{\phi k}^{B}$.  The
microscopic noise inputs $\eta_{A}(t_{f})$ and $\eta_{B}(t_{f})$
at the boundary at time $t_{f}$ give the reverse-time trajectories
$x_{\theta A}(t)=x_{\theta j}^{A}+\delta x_{\theta A}$ and $x_{\phi B}(t)=x_{\phi j}^{B}+\delta x_{\phi B}$
(orange dashed lines). \label{fig:sim} The $Q(\lambda_{rot},t_{2})$
and $Q(\lambda,t_{0})$ determine the correlations giving links between
trajectories (indicated by the faint brown lines at $t_{0}$). Local
loops between $x_{\theta A}$ and $p_{\theta A}$, and between $x_{\phi B}$
and $p_{\phi B}$, are depicted by dashed black vertical lines. \textcolor{red}{}}
\end{figure}

The transformations (\ref{eq:rot-6-a-2}) and (\ref{eq:rot-6-b-3})
corresponding to the setting-changes are \emph{local and deterministic}.
They give a \emph{causal relation}, from time $t_{0}$ to time $t_{2}>t_{0}$,
but are reversible. Being deterministic, \emph{there is a two-way
relation} (depicted by thin two-way arrows in Fig. \ref{fig:sim}),
since the values \textbf{$\bm{\lambda}_{rot}$ }of coordinates at
time $t_{2}$ allow a retrodictive inference of the values $\bm{\lambda}$
at $t_{0}$. Here, time $t_{2}$ is the time at which both settings
have been adjusted.

\textbf{\emph{Amplification:}} After both settings are fixed, at a
time $t_{2}$, the measurements of $\hat{x}_{\theta A}$ and $\hat{x}_{\phi B}$
proceed by amplifying $x_{\theta A}$ and $x_{\phi B}$ according
to $H_{amp}^{A}$ and $H_{amp}^{B}$ (Eq. (\ref{eq:hampa})) followed
by a final detection. The equations for the amplitudes $x_{\theta A}$
and $x_{\phi B}$ are \emph{local, stochastic and retrocausal}, given
by
\begin{align}
\frac{dx_{\theta A}}{dt_{-}} & =-gx_{\theta A}+\xi_{A1}\left(t\right)\nonumber \\
\frac{dx_{\phi B}}{dt_{-}} & =-gx_{\phi B}+\xi_{B1}\left(t\right)\label{eq:dyn-epr-x-1-1-2}
\end{align}
with a boundary condition at the future time $t_{f}$, where $t_{-}=-t$.
Here, we use the term retrocausal only to imply the propagation in
the negative time direction $t_{-}$. The equations for $p$ are $\frac{dp_{\theta A}}{dt}=-gp_{\theta A}+\xi_{A2}\left(t\right)$
and $\frac{dp_{\phi B}}{dt}=-gp_{\phi B}+\xi_{B2}\left(t\right)$,
which are causal with a boundary condition at the time $t_{2}$. The
Gaussian random noises satisfy $\left\langle \xi_{A\mu}\left(t\right)\xi_{A\nu}\left(t'\right)\right\rangle =2g\delta_{\mu\nu}\delta\left(t-t'\right)$
and $\left\langle \xi_{B\mu}\left(t\right)\xi_{B\nu}\left(t'\right)\right\rangle =2g\delta_{\mu\nu}\delta\left(t-t'\right)$
with cross terms zero. Here, we are imposing a new initial boundary
for Eq. (\ref{eq:dyn-epr-x-1-1-2}), at the time $t_{2}$, where a
\emph{discontinuity} is introduced. The backward-nature of the propagation
of the stochastic amplitudes clearly implies a discontinuity at the
time $t_{2}$ in the forward-time direction. The amplitudes do not
propagate continuously in the forward-time direction.

\textbf{\emph{Procedure of simulation: }}There is hence a two-step
process in the forward-time direction. At the time $t_{0}$, the system
is specified with respect to a particular measurement basis, in this
case $\hat{x}_{A}$ and $\hat{x}_{B}$ ($\theta=\phi=0$), as determined
by the preparation of the system. This implies that the Q function
$Q(\lambda,t_{0})$ is expanded with respect to this basis, in terms
of the Q function of $|x_{j}^{A}\rangle$ and $|x_{k}^{B}\rangle$,
the eigenstates of $\hat{x}_{A}$ and $\hat{x}_{B}$, as explained
in the causal model of Sec. \ref{sec:Causal-model-for}. The deterministic
transformations associated with $H_{\theta}^{A}$ and $H_{\phi}^{B}$
rotate the amplitudes, so that the Q function $Q(\lambda_{rot},t_{2})$
at time $t_{2}$ is expressed in the coordinates $\lambda_{rot}=(x_{\theta A},p_{\theta A},x_{\phi B},p_{\phi B}$)
of the new measurement basis, $\hat{x}_{\theta A}$ and $\hat{x}_{\phi B}$.

The $Q(\lambda_{rot},t_{2})$ is unique for the quantum state and
is equivalently expressible as an expansion in terms of the Q functions
of the eigenstates $|x_{\theta j}^{A}\rangle$ and $|x_{\phi k}^{B}\rangle$
of $\hat{x}_{\theta A}$ and $\hat{x}_{\phi B}$ (Appendix C). This
expansion $Q(\lambda_{rot},t_{2})$ involves bivariate Gaussian functions
$\mathcal{G}$ with means $x_{\theta j}^{A}$ and $x_{\phi k}^{B}$,
and interference terms $\mathcal{I}nt_{AB}$, similar to Eq. (\ref{eq:Q-sup}).
The future boundary condition for Eq. (\ref{eq:dyn-epr-x-1-1-2})
is determined by the marginal 
\begin{eqnarray}
Q_{m}(x_{\theta A},x_{\phi B},t) & = & \int dp_{\theta A}dp_{\phi B}Q(\lambda_{rot},t)\label{eq:marg-amp-bi}
\end{eqnarray}
of the Q function $Q(\lambda_{rot},t)$ of the amplified state where
$t\geq t_{2}$.  NB: This function is defined for the bipartite system
and hence differs from $Q(x,p,t)$ defined for a single system, in
Part I of the paper. We use the subscript $m$ to avoid confusion.

We confirm below in Result VII.3 that the interference terms of the
marginal (\ref{eq:marg-amp-bi}), as for (\ref{eq:Qmarg-x-amp}) in
Part I of the paper, vanish and as for Result III.1, the future boundary
condition for Eq. (\ref{eq:dyn-epr-x-1-1-2}) is the sum of Gaussian
terms. Extending the analysis of Results IV.3 and IV.4 of Sec. IV.B
to two modes, the final distribution $P(x_{\theta A},x_{\phi B})$
for the probability of outcomes of $\hat{x}_{\theta A}$ and $\hat{x}_{\phi B}$
is given by the rescaled marginal: 
\begin{equation}
P(x_{\theta A},x_{\phi B})\rightarrow Q_{sc}(x_{\theta A},x_{\phi B},t_{f})\label{eq:prob-q}
\end{equation}
as $t_{f}\rightarrow\infty$, which in turn is given by the marginal
$W(x_{\theta A},x_{\phi B})$ of the Wigner function of the state
$|\psi_{Bell}\rangle$. The Wigner function for $|\psi_{Bell}\rangle$
is derived in Ref. \citep{gilchrist1999contradiction}, revealing
a violation of the Bell inequality for certain choices of $\theta$
and $\phi$.

\emph{Paradox and projection:} Questions and paradoxes arise.  As
in Result III.1, the amplified quantities that appear in the scaled
distribution $Q_{sc}(x_{\theta A},x_{\phi B},t_{f})$ at a time $t_{f}$
are the \emph{means (}denoted\emph{ }$x_{\theta j}^{A}$ and $x_{\phi k}^{B}$
in Figure \ref{fig:sim}) of the bivariate Gaussian functions. These
means are amplified by each $H_{amp}$, and correspond to the final
detected outcomes, but are \emph{not} directly determinable from the
values of the amplitudes $x_{\theta A}(t)$ and $x_{\phi B}(t)$ at
time $t_{2}$ (Result IV.2).

Moreover, we show below that the transformation of the Gaussian means
$x_{j}^{A}\rightarrow x_{\theta j}^{A}$ and $x_{k}^{B}\rightarrow x_{\phi k}^{B}$
from time $t_{0}$ to time $t_{2}$ is nonlocal. In Part I, we show
consistency with a HV model where there is a probability for the $x_{\theta j}^{A}$
given values for the amplitudes $x_{A}$ and $p_{A}$, so that the
mean $x_{\theta j}^{A}$ can be considered part of the description
of the state at the time $t_{1}$, after the setting is fixed locally
at $A$. The interaction $H_{amp}^{A}$ is assumed to 'extract' and
amplify the mean. But the $H_{amp}^{A}$ is local, as is $H_{\theta}^{A}$.
How then can the change of mean, $x_{j}^{A}\rightarrow x_{\theta j}^{A}$,
be determined, if knowledge of $\phi$ is required? It seems that
there is no mechanism provided for the Bell nonlocality by the simulation,
other than that the violations are ``put in by hand'' as a future
boundary condition (FBC) (Sec. I).

In the following two Sections, VIII and IX, we will counter this argument
and elucidate the paradox by considering \emph{projection}. While
there is a discontinuity in the forward-time direction, \emph{the
amplitudes propagate backwards in a continuous way. This puts a constraint
on the possible states for one system (for future measurements), given
the setting of the other.} This also leads to a demonstration of the
consistency of the Bell nonlocality with the premises of weak macroscopic
realism and weak local realism, as defined by Definitions (6) and
(10) in Sec. I.B. These premises posit that real properties exist,
prior to the time $t_{f}$ at which the FBC is applied.

\subsection{Role of hidden interference: mechanism for Bell nonlocality}

 How does the simulation produce the Bell nonlocality? We examine
this in two parts. First, in this Section, we examine the dynamics
of the Q function, as it evolves through each step of the measurement
in the simulation, showing the role of the hidden interference. In
Sec. IX, we examine the trajectories of the forward-backward amplitudes,
emphasizing the role of projection.

The term $I$ that appears for the $|\psi_{Bell}\rangle$ (Eq. (\ref{eq:bell}))
includes a detectable type of interference which will contribute a
measurable central ``spot'' in the final detected joint probability
$P(x_{\theta A},x_{\phi B})$ for outcomes $x_{\theta A}$ and $x_{\phi B}$
of measurement $\hat{x}_{\theta A}$ and $\hat{x}_{\phi B}$. We hence
distinguish between the measurable interference $I$ present for a
superposition of coherent states, and the hidden interference $\mathcal{I}nt$
present in the superposition of eigenstates $|x_{j}\rangle$ of $\hat{x}$,
as in Eq. (\ref{eq:Q-sup}) with $r\rightarrow\infty$: The ``hidden''
interference is not detectable when the system undergoes amplification
$H_{amp}$ (Result III.1). The role of the hidden interference terms
is clarified by the simulation, from which we deduce a mechanism for
Bell violations.

\textbf{\emph{Definition: Hidden interference terms:}} We consider
measurements of $\hat{x}_{\theta A}$ and $\hat{x}_{\phi B}$. The
Q function is expressed in terms of the ``measurement basis'' i.e.
as a summation of the Q functions of the eigenfunctions $|x_{\theta A}\rangle$
and $|x_{\phi B}\rangle$ of $\hat{x}_{\theta A}$ and $\hat{x}_{\phi B}$).
These are defined as for the squeezed states (\ref{eq:q-sq-1}) but
with rotated axes $\theta$ and $\phi$ (Appendix C). The interference
terms appearing when the Q function for the state $|\psi\rangle$
is written with respect to the \emph{measurement} basis are referred
to as ``hidden'' interference terms. These are denoted by $\mathcal{I}nt_{AB}$
in the bipartite case.

\textbf{\emph{Result VII.1}}: \textbf{\emph{Mechanism for Bell violations
in the simulation}}: While the transformations $x_{A},p_{A}\rightarrow x_{\theta A},p_{\theta A}$
and $x_{B},p_{B}\rightarrow x_{\phi B},p_{\phi B}$ of the amplitudes
are local (refer (\ref{eq:rot-6-a}) and (\ref{eq:rot-6-b})), the
transformation of the means $x_{j}^{A}\rightarrow x_{\theta j}^{A}$
and $x_{k}^{B}\rightarrow x_{\phi k}^{B}$ of the Gaussian functions
is nonlocal (Fig. \ref{fig:sim}). This is explained by the dynamics
of the amplitudes in the simulation, as we show below.

\emph{Proof: }The transformations $x_{A},p_{A}\rightarrow x_{\theta A},p_{\theta A}$
and $x_{B},p_{B}\rightarrow x_{\phi B},p_{\phi B}$ given by (\ref{eq:rot-6-a})
and (\ref{eq:rot-6-b}) that arise due to the adjustment of measurement
settings are local and deterministic. Hence, if the amplitudes $x_{\theta A}$
and $x_{\phi B}$ were directly detectable in the final measurement,
a violation of the Bell inequality would not be possible. A Bell LHV
theory would hold, where $x_{A},p_{A},x_{B},p_{B}$ are the hidden
variables given by $\lambda$ in the expression (\ref{eq:lhv}). In
fact, the detected outcomes of the measurements (based on the amplifications
$H_{amp}^{A}$ and $H_{amp}^{B}$) are given by the means $x_{\theta j}^{A}$
and $x_{\phi k}^{B}$ of the Gaussians comprising the Q function $Q(\lambda,t_{2})$,
when this Q function is written with respect to the measurement basis.
Since the means $x_{\theta j}^{A}$ and $x_{\phi k}^{B}$ lead to
the outcomes via the deterministic relations $x_{\theta j}^{A}\rightarrow Gx_{\theta j}^{A}$
and $x_{\phi k}^{B}\rightarrow Gx_{\phi k}^{B}$, it is this stage
that leads to any violation of the Bell model (and which is hence
``nonlocal''). $\square$

\textbf{\emph{Result VII.2: Bipartite hidden interference and the
future boundary condition: }}The hidden interference terms $\mathcal{I}nt_{AB}$
do not contribute to the measured probabilities (as modeled by the
measurements $H_{amp}^{A}$ and $H_{amp}^{B}$). This is a bipartite
extension of Result III.1.

\emph{Proof:} To clarify the role of the hidden interference $\mathcal{I}nt_{AB}$,
we express the Q function in terms of the measurement basis. We take
the case where the system is prepared for measurements $\hat{x}_{\theta A}$
and $\hat{x}_{\phi B}$ (as at time $t_{2}$ in Fig. \ref{fig:sim}),
and expand the state for the system as
\begin{equation}
|\psi\rangle=\sum_{jk}c_{jk}|x_{\theta j}^{A}\rangle|x_{\phi k}^{B}\rangle\label{eq:state-gen}
\end{equation}
where $|x_{\theta j}^{A}\rangle$ and $|x_{\phi k}^{B}\rangle$ are
the eigenstates of $\hat{x}_{\theta A}$ and $\hat{x}_{\phi B}$ (refer
Appendix C for definitions). The Q function $Q(\alpha,\beta)=\frac{1}{\pi^{2}}\langle\alpha|\langle\beta|\rho|\beta\rangle|\alpha\rangle$
can be written
\begin{eqnarray}
Q(\alpha,\beta) & = & \frac{1}{\pi^{2}}\langle\alpha|\langle\beta|(\rho_{D}+\rho_{OD})|\beta\rangle|\alpha\rangle\label{eq:Q-D-OD}
\end{eqnarray}
where the density operator $\rho=|\psi\rangle\langle\psi|=\rho_{D}+\rho_{OD}$
is divided into a sum of diagonal and off-diagonal terms,
\begin{eqnarray}
\rho_{D} & = & \sum_{jk}|c_{jk}|^{2}|x_{\theta j}^{A}\rangle|x_{\phi k}^{B}\rangle\langle x_{\phi k}^{B}|\langle x_{\theta j}^{A}|\label{eq:mixab-1}
\end{eqnarray}
and
\begin{eqnarray}
\rho_{OD} & = & \sum_{j\neq j',k\neq k'}c_{jk}c_{j'k'}^{*}|x_{\theta j}^{A}\rangle|x_{\phi k}^{B}\rangle\langle x_{\phi k'}^{B}|\langle x_{\theta j'}^{A}|\nonumber \\
\end{eqnarray}
The Q function is hence 
\begin{eqnarray}
Q & = & \frac{1}{\pi^{2}}\sum_{jk}|c_{jk}|^{2}|\langle\alpha|\langle\beta|x_{\theta j}^{A}\rangle|x_{\phi k}^{B}\rangle|^{2}+\mathcal{I}nt_{AB}\nonumber \\
 & = & N\Bigl(\sum_{jk}|c_{jk}|^{2}\Bigl\{\mathcal{G}_{0,\sigma_{p_{\theta A}}}(p_{\theta A})\mathcal{G}_{x_{\theta j}^{A},1}(x_{\theta A})\nonumber \\
 &  & \times\mathcal{G}_{0,\sigma_{p_{\phi B}}}(p_{\phi B})\mathcal{G}_{x_{\phi k}^{B},1}(x_{\phi B})\Bigr\}+\mathcal{I}nt_{AB}\Bigr)\nonumber \\
\label{eq:Q-int}
\end{eqnarray}
Here we denote the Gaussian function of $x$ with mean $\bar{x}$
and variance $\sigma_{x}$ by
\begin{equation}
\mathcal{G}_{\bar{x},\sigma_{x}}(x)=\frac{1}{\sigma_{x}\sqrt{2\pi}}e^{-(x-\bar{x})^{2}/2\sigma_{x}}\label{eq:gauss}
\end{equation}
We have taken that the variance of $x_{\theta}$ for the eigenstate
$|x_{\theta}\rangle$ of $\hat{x}_{\theta}$ is $1$. Eq. (\ref{eq:Q-int})
is a two-mode generalization of Eq. (\ref{eq:Q-sup}). The summation
involving the Gaussian functions $\mathcal{G}$ arises from the diagonal
part of the density matrix $\rho_{D}$. The $\mathcal{I}nt_{AB}$
is the hidden interference term arising from $\rho_{OD}$.

In the simulation, the joint measurement of $\hat{x}_{\theta A}$
and $\hat{x}_{\phi B}$ is modeled by $H_{amp}^{A}$ and $H_{amp}^{B}$
acting independently on each system. As in the single mode case, the
means of the Gaussian functions of $x_{\theta A}$ and $x_{\phi B}$
are amplified, but their variances remain at $1$, while amplitudes
$p_{\theta A}$ and $p_{\phi B}$ deamplify. It is shown in Result
III.1 (Sec. III) that hidden interference terms are not detected in
the amplified field, a result that can be straightforwardly generalized
to apply to (\ref{eq:Q-int}). The probabilities $P(x_{\theta j}^{A},x_{\phi k}^{B})$
for joint outcomes $x_{\theta j}^{A}$ and $x_{\phi k}^{B}$ of $\hat{x}_{\theta A}$
and $\hat{x}_{\phi B}$ are known (in quantum mechanics) to be given
by the coefficients $|c_{jk}|^{2}$ in the diagonal expansion of $\rho$
(Eq. (\ref{eq:mixab-1}). This is in agreement with the simulation,
where the detected probabilities correspond to the coefficients of
the Gaussians in the expansion (\ref{eq:Q-int}), which are given
by $|c_{jk}|^{2}$ (Sec. IV.B). $\square$

To summarize, the observed probabilities are given by the mixed state
$\rho_{D}$ defined by the diagonal matrix (\ref{eq:mixab-1}). 
The hidden interference terms $Int_{AB}$ correspond to off-diagonal
terms that are not detected with amplification.

\textbf{\emph{Result VII.3: Changing the setting at one location.
No observable nonlocality:}} Changing the setting at one location,
say $B$ (leaving the other setting fixed) can change the hidden interference,
but not the observed probabilities, relative to a non-entangled state.
Hence, Bell nonlocality is not observed with a change of setting at
just one site. There are however changes to the hidden interference
terms.

\emph{Proof:} The analysis in Appendix F shows what happens to the
system prepared in $|\psi\rangle$ (Eq.(\ref{eq:state-gen})) when
we change the measurement setting at one location, say $B$. The transformed
state of $|\psi\rangle$ (i.e. of $\rho=|\psi\rangle\langle\psi|$)
is compared with the transformed state of the \emph{partially mixed}
state $\rho_{mix,A}$ given as
\begin{eqnarray}
\rho_{mix,A} & = & \sum_{l}\{\langle x_{\theta l}^{A}|\rho|x_{\theta l}^{A}\rangle\}|x_{\theta l}^{A}\rangle\langle x_{\theta l}^{A}|\nonumber \\
 & = & \sum_{l}|f_{l}|^{2}|\psi_{l}^{B}\rangle|x_{\theta l}^{A}\rangle\langle x_{\theta l}^{A}|\langle\psi_{l}^{B}|\label{eq:mix-asymA}
\end{eqnarray}
To understand the meaning of $\rho_{mix,A}$, we express the state
$|\psi\rangle$ of the system as $|\psi\rangle=\sum_{l}f_{l}|x_{\theta l}^{A}\rangle|\psi_{l}^{B}\rangle$,
where $\sum_{l}|f_{l}|^{2}=1$ and $|\psi_{l}^{B}\rangle$ are normalized
states for $B$. From the last line of (\ref{eq:mix-asymA}), we see
that $\rho_{mix,A}$ represents a system whereby a ``collapse''
to the eigenstates at $A$ has occurred. The $\rho_{mix,A}$ is hence
not entangled, although local superposition states remain possible
for $B$.

To model the change of setting at $B$ in the simulation, we transform
to the new basis of $B$. This changes the expansion of $\rho$ (and
of the Q function of the state $|\psi\rangle$), since in the Q model
we write with respect to the measurement basis (Secs. V.A and VII.A).
However, the changes to the diagonal part of $\rho$ ($\rho_{D}$
of Eq. (\ref{eq:Q-D-OD}))\emph{ }are the same for $|\psi\rangle$
and $\rho_{mix,A}$ (refer Appendix F). The new expansions of $\rho$
as given for $|\psi\rangle$ and $\rho_{mix,A}$ are different, but
the difference is only in the off-diagonal terms $\rho_{OD}$. This
difference will affect the hidden interference terms $\mathcal{I}nt_{AB}$
in the Q function, but these do not contribute to the observable probabilities
(Result VII.2). Hence, the observable joint probabilities for $|\psi\rangle$
are the same as given for the non-entangled state $\rho_{mix,A}$.
$\square$

\textbf{\emph{Result VII.4: Bell nonlocality: }}After a change of
setting at \emph{both }locations $A$ and $B$, the hidden interference
terms can contribute to measurable probabilities. This can lead to
Bell nonlocality.

\emph{Proof: } That changes of settings at both $A$ and $B$ can
lead to Bell nonlocality is illustrated by the spin Bell state (Appendix
G). We explicitly demonstrate how this occurs for the CV-Bell example
below (Result VII.5). $\square$

\textbf{\emph{Result VII.5: Dynamics of the mechanism for Bell nonlocality:}}\emph{
}We trace the dynamics of the amplitudes in the simulation depicted
in Figure \ref{fig:sim}, to demonstrate how Bell nonlocality arises
in accordance with Result VII.4.

\emph{Proof:} We consider first for simplicity that the state at time
$t_{0}$ is prepared, for measurements $\hat{x}_{A}$ and $\hat{x}_{B}$,
in the entangled state
\begin{equation}
|x_{j}^{A}\rangle|x_{k}^{B}\rangle+|x_{l}^{A}\rangle|x_{m}^{B}\rangle\label{eq:entbasis}
\end{equation}
Here, $|x_{j}^{A}\rangle$ and $|x_{k}^{B}\rangle$ are eigenstates
of $\hat{x}_{A}$ and $\hat{x}_{B}$ respectively, which we define
as highly squeezed states in $\hat{x}_{A}$ and $\hat{x}_{B}$ (Eq.
(\ref{eq:eigenstate-def})). The Q function is
\begin{eqnarray}
Q(\lambda,t_{0}) & = & N\Bigl(\mathcal{G}_{0,\sigma_{p_{A}}}(p_{A})\mathcal{G}_{0,\sigma_{p_{B}}}(p_{B})\Bigl\{\mathcal{G}_{x_{j}^{A},1}(x_{A})\mathcal{G}_{x_{k}^{B},1}(x_{B})\nonumber \\
 &  & +\mathcal{G}_{x_{l}^{A},1}(x_{A})\mathcal{G}_{x_{m}^{B},1}(x_{B})\Bigr\}+\mathcal{I}nt_{AB}\Bigr)\label{eq:qex}
\end{eqnarray}
where
\begin{eqnarray}
\mathcal{I}nt_{AB} & = & 2e^{-\frac{(\Delta_{A})^{2}}{2}}e^{-\frac{(\Delta_{B})^{2}}{2}}\mathcal{G}_{\frac{(x_{j}^{A}+x_{l}^{A})}{2},1}(x_{A})\mathcal{G}_{\frac{(x_{k}^{B}+x_{m}^{B})}{2},1}(x_{B})\nonumber \\
 &  & \times\mathcal{G}_{0,\sigma_{p_{A}}}(p_{A})\mathcal{G}_{0,\sigma_{p_{B}}}(p_{B})\times\mathcal{F_{AB}}\label{eq:int-1}
\end{eqnarray}
with the sinusoidal component 
\begin{eqnarray}
\mathcal{F_{AB}} & = & \cos[p_{A}\Delta_{A}/\sigma_{x_{A}}+p_{B}\Delta_{B}/\sigma_{x_{B}}]\label{eq:fringe1}
\end{eqnarray}
The $\mathcal{G}$ are the Gaussian functions defined by (\ref{eq:gauss}).
Here $\Delta_{A}=(x_{j}^{A}-x_{l}^{A}\bigl)/2\sigma_{x_{A}}$ and
$\Delta_{B}=(x_{k}^{B}-x_{m}^{B})/2\sigma_{x_{B}}$ are the separations
between the states of the superposition. The $\sigma_{x_{A}}^{2}$
and $\sigma_{p_{A}}^{2}$ are the variances of $x_{A}$ and $p_{A}$
in the Q functions of the squeezed states that approximate the eigenstates
$|x_{j}^{A}\rangle$ and $|x_{l}^{A}\rangle$. The $\sigma_{x_{B}}^{2}$
and $\sigma_{p_{B}}^{2}$ are defined similarly for eigenstates $|x_{k}^{B}\rangle$
and $|x_{m}^{B}\rangle$ of system $B$. Hence, $\sigma_{x_{A}}^{2}=\sigma_{x_{B}}^{2}=1$
and $\sigma_{p_{A}}^{2}\rightarrow\infty$, $\sigma_{p_{B}}^{2}\rightarrow\infty$.
The fringe term is modulated by Gaussian functions with peaks at the
average position of the means $x_{j}^{A}$ and $x_{l}^{A}$ (and $x_{k}^{B}$
and $x_{m}^{B}$), and is damped by the separation $\Delta_{A}$ (and
$\Delta_{B}$) between them.

From Results III.1 and VII.2, we know that the hidden interference
term $\mathcal{I}nt_{AB}$ does not contribute to the marginal defined
as the integral $Q_{m}(x_{A},x_{B},t_{0})=\int dp_{A}dp_{B}Q(\lambda,t_{0})$.
This is evident here as follows: When the cosine function in $\mathcal{F_{AB}}$
is expanded, the integral reduces to combination of integrals over
$p_{A}$ or $p_{B}$ individually. Those involving a sine function
are zero due to symmetry. Those involving the cosine function reduce
to integrals of type
\begin{equation}
\frac{1}{\sqrt{a\pi}}\int_{-\infty}^{\infty}e^{-p^{2}/a}\cos(kp)dp=e^{-ak^{2}/4}\label{eq:int-ex}
\end{equation}
as used in (\ref{eq:int-2-2}), which vanish since we take $a\equiv\sigma_{p_{A}}^{2}$
(or $\sigma_{p_{B}}^{2}$) to be large for the eigenstate. The vanishing
of the $\mathcal{I}nt_{AB}$ terms in the marginal $Q_{m}(x_{A},x_{B},t_{0})$
is consistent with Result VII.2, that the hidden terms $\mathcal{I}nt_{AB}$
do not contribute to the probabilities measured directly under $H_{amp}^{A}$
and $H_{amp}^{B}$.

If we change the setting at $B$ to $\phi$, then (following Sec.
VII.A) we substitute $x_{B}\rightarrow x_{\phi B}\cos\phi-p_{\phi B}\sin\phi$
and $p_{B}\rightarrow x_{\phi B}\sin\phi+p_{\phi B}\cos\phi$ in the
expression for $Q(\lambda,t_{0})$. The $Q(\lambda,t_{0})$ is
transformed to give a new function (refer Appendix E)
\begin{eqnarray}
Q(\lambda_{rot},t_{1}) & = & N_{1}\mathcal{G}_{0,\sigma_{p_{A}}}(p_{A})\Bigl\{\mathcal{G}_{x_{j}^{A},1}(x_{A})f_{x_{k}^{B}}(x_{\phi B},p_{\phi B})\nonumber \\
 &  & +\mathcal{G}_{x_{l}^{A},1}(x_{A})f_{x_{m}^{B}}(x_{\phi B},p_{\phi B})\Bigr\}+\mathcal{I}nt_{AB}(\phi)\nonumber \\
\label{eq:q1-1-2}
\end{eqnarray}
where $N_{1}$ is a normalization term. The bivariate distributions
$\mathcal{G}_{0,\sigma_{p_{B}}}(p_{B})\mathcal{G}_{x_{k}^{B},1}(x_{B})$
and $\mathcal{G}_{0,\sigma_{p_{B}}}(p_{B})\mathcal{G}_{x_{m}^{B},1}(x_{B})$
in (\ref{eq:qex}) are transformed to the new coordinates, as depicted
in Figure \ref{fig:Evolution-settingB}. These transformed distributions
are denoted by $f_{x_{k}^{B}}(x_{\phi B},p_{\phi B})$ and $f_{x_{m}^{B}}(x_{\phi B},p_{\phi B})$
respectively.  The transformed interference is  
\begin{eqnarray}
\mathcal{I}nt_{AB}(\phi) & = & 2e^{-\frac{(\Delta_{A})^{2}}{2}}\mathcal{G}_{\frac{(x_{j}^{A}+x_{l}^{A})}{2},1}(x_{A})\mathcal{G}_{0,\sigma_{p_{A}}}(p_{A})\nonumber \\
 &  & \times e^{-\frac{(\Delta_{B})^{2}}{2}}f_{\frac{(x_{k}^{B}+x_{m}^{B})}{2}}(x_{\phi B},p_{\phi B})\times\mathcal{F}_{AB}(\phi)\nonumber \\
\label{eq:int-bell-1-1-1-2}
\end{eqnarray}
where the fringe term is
\begin{eqnarray}
\mathcal{F}_{AB}(\phi) & = & \cos[\frac{p_{A}}{2}(x_{j}^{A}-x_{l}^{A}\bigl)+\frac{p_{B}(\phi)}{2}(x_{k}^{B}-x_{m}^{B}\bigl)]\nonumber \\
\end{eqnarray}
The fringe term $\mathcal{F}_{AB}(\phi)$ involves the rotated coordinate
$p_{B}(\phi)=x_{\phi B}\sin\phi+p_{\phi B}\cos\phi$ for $B$. However,
the $p_{A}$ is unchanged. We evaluate the new marginal defined as
$Q_{m}(x_{A},x_{\phi B},t_{1})=\int dp_{A}dp_{\phi B}Q(\lambda_{rotB},t_{1})$.
We can expand both terms in $\mathcal{F}_{AB}(\phi)$ as the sum of
two terms, one involving $\sin(p_{A}(x_{j}^{A}-x_{l}^{A})/2)$ and
the other $\cos(p_{A}(x_{j}^{A}-x_{l}^{A})/2)$. The sine integration
vanishes due to symmetry. The cosine integration vanishes in the limit
of $\sigma_{p_{A}}^{2}\rightarrow\infty$ (as for (\ref{eq:int-2-2})),
justified for the eigenstates $|x_{j}^{A}\rangle$ and $|x_{l}^{A}\rangle$.
Now, we see that the integration over $p_{A}$ ensures that $\mathcal{I}nt_{AB}$
vanishes from the marginal $Q_{m}(x_{A},x_{\phi B},t_{1})$: Hence,
the change of basis in $B$ changes the hidden interference term $\mathcal{I}nt_{AB}$,
but this change \emph{remains hidden}, since it does not contribute
to the marginal $Q_{m}(x_{A},x_{\phi B},t_{1})$ (consistent with
Result VII.3).

However, if there is a change of setting at \emph{both }sites $A$
and $B$, then \emph{both} $p_{A}$ and $p_{B}$ transform as in (\ref{eq:rot-6-a})
and (\ref{eq:rot-6-b}), meaning that both $p_{A}$ and $p_{B}$ are
rotated. The final transformed function is $Q(\lambda_{rot},t_{2})$
given by 
\begin{eqnarray}
Q(\lambda_{rot},t_{2}) & = & N_{2}\Bigl\{ f_{x_{j}^{A}}(x_{\theta A},p_{\theta A})f_{x_{k}^{B}}(x_{\phi B},p_{\phi B})\nonumber \\
 &  & +f_{x_{l}^{A}}(x_{\theta A},p_{\theta A})f_{x_{m}^{B}}(x_{\phi B},p_{\phi B})\nonumber \\
 &  & \thinspace\thinspace\thinspace\thinspace\thinspace\thinspace\thinspace\thinspace\thinspace+\mathcal{I}nt_{AB}(\theta,\phi)\Bigr\}\label{eq:q1-1-1}
\end{eqnarray}
where $N_{2}$ is the normalization constant. The interference term
is
\begin{eqnarray}
\mathcal{I}nt_{AB}(\theta,\phi) & = & f_{\frac{(x_{j}^{A}+x_{l}^{A})}{2}}(x_{\theta A},p_{\theta A})f_{\frac{(x_{k}^{B}+x_{m}^{B})}{2}}(x_{\phi B},p_{\phi B})\nonumber \\
 &  & \times e^{-\frac{(\Delta_{A})^{2}}{2}}e^{-\frac{(\Delta_{B})^{2}}{2}}\mathcal{F}_{AB}(\theta,\phi)\label{eq:bell-int-1-1-3}
\end{eqnarray}
where
\begin{eqnarray}
\mathcal{F}_{AB}(\theta,\phi) & = & \cos[\frac{p_{A}(\theta)}{2}(x_{j}^{A}-x_{l}^{A}\bigl)+\frac{p_{B}(\phi)}{2}(x_{k}^{B}-x_{m}^{B}\bigl)]\nonumber \\
\label{eq:fringe-2}
\end{eqnarray}
Here, $p_{A}(\theta)=x_{\theta A}\sin\theta+p_{\theta A}\cos\theta$
and $p_{B}(\phi)=x_{\phi B}\sin\phi+p_{\phi B}\cos\phi$. In order
to determine the marginal $Q_{m}(x_{\theta A},x_{\phi B},t_{2})=\int dp_{\theta A}dp_{\phi B}Q(\lambda_{rot},t_{2})$
at time $t_{2}$, it is necessary to integrate over $p_{\theta A}$
and $p_{\phi B}$. With both transformations, the integrals over $p_{\theta A}$
and $p_{\phi B}$ involving the cosine function \emph{no longer necessarily
vanish}, since with $\theta,\phi\neq n\pi$ (where $n$ is an integer)
the variances $\sigma_{p_{\theta A}}^{2}$ and $\sigma_{p\phi B}^{2}$
are \emph{no longer restricted to be very large}. This is evident
in Figure \ref{fig:Evolution-settingB}, and by Eq. (\ref{eq:int-ex})
where $a$ need not be large (refer Appendix E). \textcolor{blue}{}Hence,
we require $\theta,\phi\neq n\pi$ to observe nonlocality. This implies
that the rotated state after a change of setting is a superposition
of the original eigenstates. In conclusion, with such a selection
of settings, the interference term $\mathcal{I}nt_{AB}(\theta,\phi)$
is no longer ``hidden'', and can contribute nonzero terms to the
marginal $Q_{m}(x_{\theta A},x_{\phi B},t_{2})$. We conclude that
the interference terms may contribute to the measurable probability
distribution $P(x_{\theta A},x_{\phi B})$ for outcomes $x_{\theta A}$
and $x_{\phi B}$ of joint measurements $\hat{x}_{\theta A}$ and
$\hat{x}_{\phi B}$.

This result can be generalized to an arbitrary state prepared at time
$t_{0}$, by considering $|\psi\rangle=\sum_{jk}c_{jk}|x_{j}^{A}\rangle|x_{k}^{B}\rangle$
(refer Appendix C). $\square$

\emph{Comment: }It is the interference term $\mathcal{I}nt_{AB}$
present in the original Q function $Q(\lambda,t_{0})$ that leads
to the Bell nonlocality.

\emph{Proof:} The marginal $Q_{m}(x_{\theta A},x_{\phi B},t_{2})$,
defined by (\ref{eq:marg-amp-bi}) and used in the above proof, can
be expressed as the sum of three terms, $Q_{1}$, $Q_{2}$ and $Q_{3}$.
These are defined as the integration over $p_{\theta A}$ and $p_{\phi B}$
of the three terms appearing in $Q(\lambda_{rot},t_{2})$ in (\ref{eq:q1-1-1}).
The integration of the first two terms in $Q(\lambda_{rot},t_{2})$
gives $Q_{1}$ and $Q_{2}$, which lead to peaks in the distribution
$Q(x_{\theta A},x_{\phi B},t)$ corresponding to the states $|x_{j}^{A}\rangle|x_{k}^{B}\rangle$
and $|x_{l}^{A}\rangle|x_{m}^{B}\rangle$ respectively, as in a mixture
of the $|x_{j}^{A}\rangle|x_{k}^{B}\rangle$ and $|x_{l}^{A}\rangle|x_{m}^{B}\rangle$.
The expressions $Q_{1}$ and $Q_{2}$ are expressible as a product
of a function of $x_{\theta A}$ and a function of $x_{\phi B}$.
The $x_{\theta A}$ is a function only of the local variables $x_{A}$,
$p_{A}$, and $\theta$; $x_{\phi B}$ is a function only of local
variables $x_{B}$, $p_{B}$ and $\phi$. We see that any mixture
of the $|x_{j}^{A}\rangle|x_{k}^{B}\rangle$ and $|x_{l}^{A}\rangle|x_{m}^{B}\rangle$
would satisfy a Bell LHV model, where $x_{A}$, $p_{A}$, $x_{B}$
and $p_{B}$ play the role of hidden variables. The third term given
by $Q_{3}=\int dp_{\theta A}dp_{\phi B}\mathcal{I}nt_{AB}(\theta,\phi)$
does not factorize in this way, and hence leaves open the possibility
to violate a Bell inequality. $\square$

\textbf{\emph{Result VII.6: Nonlocality:}} Consistent with Results
VII.1 and VII.4, the effect due to the interference term $\mathcal{I}nt_{AB}(\theta,\phi)$
on the observed probabilities is \emph{``nonlocal'',} implying it
can lead to a violation of a Bell inequality.

\emph{Proof:} The proof follows from the above Comment. We note that
the impact of the change of setting $\theta$ at $A$ on any observed
probability that is due to $\mathcal{I}nt_{AB}$ is only present if
there is also a change of setting $\phi$ at $B$. This is evident
by the mathematical form of $\mathcal{I}nt_{AB}$ in Eq. (\ref{eq:bell-int-1-1-3}).
If $\phi=0$, there is no change to observed probabilities due to
any change of setting $\theta$ at $A$ that can be attributed to
$\mathcal{I}nt_{AB}$; if $\phi\neq0$, such a change becomes possible
(Result VII.5). Hence, the effect due to $\mathcal{I}nt_{AB}$ at
$A$ depends on $\phi$ at $B$. $\square$

\textbf{\emph{Result VII.7:}}\emph{ }\textbf{\emph{The CV Bell nonlocality
requires the settings to mix $x$ and $p$ locally}}: The constraint
$\theta,\phi\neq n\pi$ (where $n$ is an integer) is derived above
in Result VII.5 as a requirement for the Bell nonlocality.  This
implies the settings need to be changed (relative to the initial settings,
denoted $x_{A}$, $p_{A}$ and $x_{B}$, $p_{B}$) so that \emph{both}
the measured quadratures, $\hat{x}_{\theta}^{A}$ and $\hat{x}_{\phi}^{B}$,
are rotated to combine the local amplitudes $x_{A}$, $p_{A}$ and
$x_{B}$, $p_{B}$, respectively (so that the rotated states are superpositions
of the original eigenstates in $\hat{x}_{A}$ and $\hat{x}_{B}$).
This is consistent with quantum predictions.

\subsection{Local loops, nonlocal loops and correlation}

We see from Section IV.E that the ``hidden'' interference term $\mathcal{I}nt$
can contribute to a hidden loop, which distinguishes the superposition
from the mixed state (Figs. \ref{fig:postx-trajectories-superposition}
and \ref{fig:causal-model}). The correlations are more complex in
the two-mode case. However, we see that the term $\mathcal{I}nt_{AB}$
present in the Q function (\ref{eq:qex}) of the superposition $|x_{j}^{A}\rangle|x_{k}^{B}\rangle+|x_{l}^{A}\rangle|x_{m}^{B}\rangle$
implies that the conditional distribution $Q(p_{A}|x_{A})$ (unlike
that of a mixture of $|x_{j}^{A}\rangle|x_{k}^{B}\rangle$ and $|x_{l}^{A}\rangle|x_{m}^{B}\rangle$)
does not factorize to give $Q(p_{A}|x_{A})=Q(p_{A})$. Hence, a local
loop connecting variables $x_{A}$ and $p_{A}$ will be present.

\textbf{\emph{Definition: Local loop:}} We define a loop as local,
if it connects backward and forward-propagating variables for a single
system ($A$ or $B$).

\begin{figure}
\begin{centering}
\includegraphics[width=1\columnwidth]{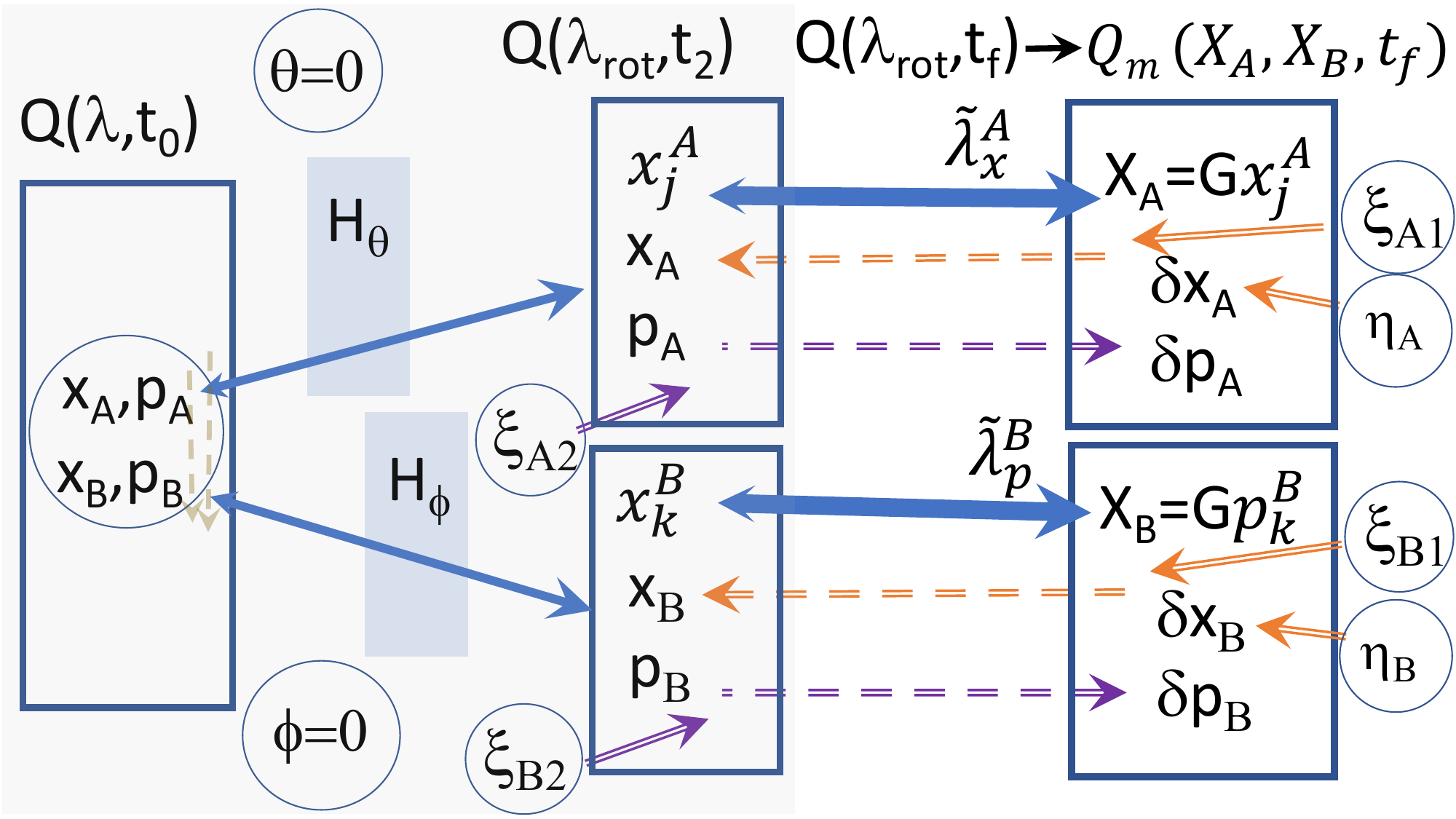}
\par\end{centering}
\medskip{}

\begin{centering}
\includegraphics[width=1\columnwidth]{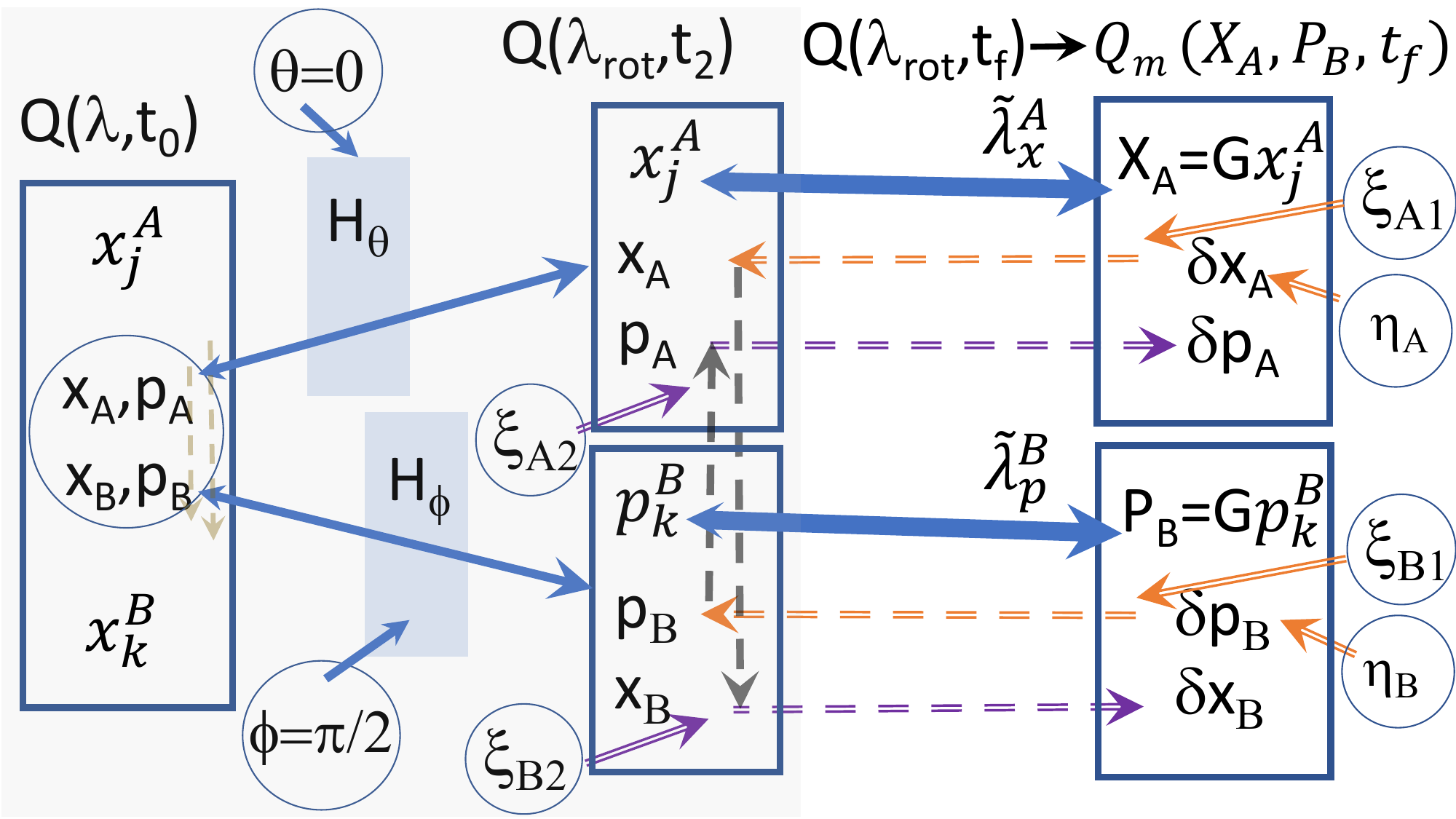}
\par\end{centering}
\caption{\textbf{\emph{Nonlocal loops and EPR states:}} Diagram as in Figure
\ref{fig:sim}, depicting the special case of the simulation of the
measurement of the CV EPR entanglement for $|\psi_{epr}\rangle$.
The system is prepared for measurements $\hat{x}_{A}$ and $\hat{x}_{B}$
at time $t_{0}$. The top figure depicts measurement of $\hat{x}_{A}$
and $\hat{x}_{B}$, where $x$ and $p$ trajectories are not connected.
The lower figure depicts measurement of $\hat{x}_{A}$ and $\hat{p}_{B}$,
showing nonlocal loops. \label{fig:epr-sim}  The local loops evident
in Figure \ref{fig:sim} for the Bell state are absent for the EPR
state $|\psi_{epr}\rangle$.}
\end{figure}

\textbf{\emph{Result VII.8: Local loops are associated with the Bell
violation: }}We see from Result VII.5 (Comment) that the Bell nonlocality
arises from hidden interference terms $\mathcal{I}nt_{AB}$. These
terms (as for $\mathcal{I}nt$ in Eq. (\ref{eq:Q-sup})) lead to a
correlation between the local coordinates. We see that $Q(p_{A}|x_{A})\neq Q(p_{A})$
and $Q(p_{B}|x_{B})\neq Q(p_{B})$. There are hence local loops for
each system $A$ and $B$. The local loops are depicted in Figure
\ref{fig:epr-sim}. Being a consequence of $\mathcal{I}nt_{AB}$,
they are present for the Bell nonlocal system.

\emph{Comment:} This is consistent with the observation that a local
loop does not exist for the CV EPR entangled state $|\psi_{epr}\rangle$
given by Eq. (\ref{eq:epr-q}) (Fig. \ref{fig:epr-sim}), for which
a local hidden variable theory (LHV) exists to explain measurements
of $\hat{x}_{\theta A}$ and $\hat{x}_{\phi B}$ \citep{bell2001john}.

For the two-mode case, it is possible to consider other correlations
e.g. we see that $Q(p_{B}|x_{A})\neq Q(p_{B})$ but also $Q(p_{A}|p_{B})\neq Q(p_{A})$
and $Q(x_{A}|x_{B})\neq Q(x_{A})$. The latter correlations, between
$x_{A}$ and $x_{B}$ (say), imply correlation between $\hat{x}_{A}$
and $\hat{x}_{B}$ that can clearly also have a classical nature.

\textbf{\emph{Definition: Nonlocal loop: }}We define a loop as nonlocal
if it connects backward and forward-propagating variables for systems
at separated sites.

\textbf{\emph{Correlation:}} There are constraints that exist between
the trajectories of different systems, due to the correlation between
the two systems, as defined by $Q(\lambda,t_{0})$. This occurs in
the EPR case $|\psi_{epr}\rangle$ (Eq. (\ref{eq:tmss}), when both
$\hat{x}_{A}$ and $\hat{x}_{B}$ are measured, since $Q(x_{A}|x_{B})\neq Q(x_{A})$
and $x_{A}$ and $x_{B}$ are correlated (Fig. \ref{fig:epr-sim},
top panel). These constraints ensure certain backward trajectories
are correlated between the two systems in a given run, as in Figure
\ref{fig:epr-3}, but can create nonlocal loops, for example in the
EPR case where $\hat{x}_{A}$ and $\hat{p}_{B}$ are measured. This
is depicted in Figure \ref{fig:epr-sim} (lower panel) for $|\psi_{epr}\rangle$,
where the existence of the perfect correlation between $\hat{x}_{A}$
and $\hat{x}_{B}$, and perfect anti-correlation between $\hat{p}_{A}$
and $\hat{p}_{B}$, allows for two nonlocal loops, and is also a signature
of entanglement \citep{duan2000inseparability}.

The above correlations and connections between the trajectories of
the simulation are determined by the $Q(\lambda_{rot},t_{2})$, which
are in turn determined by $Q(\lambda,t_{0})$, at the initial-time,
since the transformations (\ref{eq:rot-6-a}) and (\ref{eq:rot-6-b})
that change the measurement settings are deterministic (refer Appendix
H). This is depicted in Figures \ref{fig:sim} and \ref{fig:epr-sim}
by the vertical dashed lines at both times, $t_{1}$ and $t_{0}$.

\subsection{Hidden-variable (HV) models $\mathcal{P}_{B}$, $\mathcal{P}_{detB}$
and $\mathcal{P}_{Bell}$}

In Section IV.F, hidden-variable model $\mathcal{P}_{sup}$ was developed
from the simulation involving $|\psi_{sup}\rangle$. A similar approach
can be applied to the Bell simulation. In the proposed Bell test,
the amplitudes $\hat{x}_{\theta A}$ and $\hat{x}_{\phi B}$ are measured,
and the outcomes binned to be $+1$ or $-1$, according to the sign
$\mathcal{S}$ of the outcome. Binary outcomes $\mathcal{S}_{A}=\pm1$
and $\mathcal{S}_{B}=\pm1$ apply at each site, for given settings
$\theta$ and $\phi$.

\textbf{\emph{Result VII.9a:}}\textbf{ }\textbf{\emph{Hidden variable
model $\mathcal{P}_{B}$:}}\textbf{ }Unlike the forward trajectories,
the backward-propagating trajectories are continuous. Hence, for any
given joint ``spin'' outcomes $\mathcal{S_{A}}=\pm1$ and $\mathcal{S}_{B}=\pm1$
at $A$ and $B$ respectively, we can deduce a distribution $P(\mathbf{\lambda}_{rot}|\mathcal{S}_{A},\mathcal{S}_{B})$
for the amplitudes $\lambda_{rot}=(x_{\theta A},p_{\theta A},x_{\phi B},p_{\phi B})$
at time $t_{2}$. Here, $t_{2}$ is the time by which both the settings
$\theta$ and $\phi$ are fixed (Figure \ref{fig:sim}). The distribution
can be found by tracing the backward trajectories from $t_{f}$ to
the time $t_{2}$. Using that for events $A$ and $B$, the conditional
probabilities satisfy $P(A|B)=P(B|A)P(A)/P(B)$, it follows (refer
Result IV.13, Sec. IV.F) that for a given a set of values for the
variables $\lambda_{rot}=(x_{\theta A},p_{\theta A},x_{\phi B},p_{\phi B})$,
there exists a probability distribution 
\begin{equation}
P_{B}(\mathcal{S}_{A},\mathcal{S}_{B}|\lambda_{rot})\label{eq:model-b}
\end{equation}
for joint outcomes at $A$ and $B$. Hence, a probabilistic HV model
$\mathcal{P}_{B}$ exists, in which the system given by amplitudes
$\lambda_{rot}$ can be viewed as being in a state with a certain
probability for outcomes $\mathcal{S}_{A}$ and $\mathcal{S}_{B}$
at $A$ and $B$.

\emph{Proof:} Following Sec. IV.F, we identify $A$ with outcomes
$x_{\theta j}^{A},x_{\phi k}^{B}$, and $B$ with variables $\lambda_{rot}$.
Following Sec. IV.D, we define the distribution $Q_{loop}(\lambda_{rot},t_{2}|x_{\theta j}^{A},x_{\phi k}^{B})$
for the variables $\lambda_{rot}$ defined at time $t_{2}$, given
the amplitudes $x_{\theta A}(t)$, $x_{\phi B}(t)$ that are traced
back from the branches $\mathcal{B}_{Aj}$ and $\mathcal{B}_{Bk}$
associated with outcomes $x_{\theta j}^{A}$ and $x_{\phi k}^{B}$.
The probability distribution for outcomes $x_{\theta j}^{A}$, $x_{\phi k}^{B}$
given $\lambda_{rot}$ is 
\begin{eqnarray*}
P_{B}(x_{\theta j}^{A},x_{\phi k}^{B}|\lambda_{rot}) & = & \frac{Q_{loop}(\lambda_{rot},t_{2}|x_{\theta j}^{A},x_{\phi k}^{B})P(x_{\theta j}^{A},x_{\phi k}^{B})}{Q(\lambda_{rot},t_{2})}
\end{eqnarray*}
Extending Eq. (\ref{eq:branch-cond-1}) in Sec. IV.D to bipartite
case, we write 
\[
Q_{loop}(\lambda_{rot}|x_{\theta j}^{A},x_{\phi k}^{B})=Q_{jk}(x_{\theta A},x_{\phi B})Q(\lambda_{rot}|x_{\theta A},x_{\phi B})
\]
where $Q_{jk}(x_{\theta A},x_{\phi B})$ is the distribution at the
time $t_{2}$ associated with the branches $\mathcal{B}_{Aj}$ and
$\mathcal{B}_{Bk}$, and 
\[
Q(\lambda_{rot}|x_{\theta A},x_{\phi B})=Q(\lambda_{rot},t_{2})/Q_{m}(x_{\theta A},x_{\phi B},t_{2})
\]
is the conditional distribution (similar to $Q_{0}(p|x)$ in Eq. (\ref{eq:cond-1})),
defined from the initial state, $Q(\lambda_{rot},t_{2})$. Here, we
define the marginal $Q_{m}(x_{\theta A},x_{\phi B},t_{2})=\int dp_{\theta A}dp_{\phi B}Q(\lambda_{rot},t_{2})$,
which (extending Result III.1) is the summation of the $P(x_{\theta l}^{A},x_{\phi m}^{B})Q_{lm}(x_{\theta A},x_{\phi B})$
over all $l$ and $m$. Hence, 
\begin{eqnarray}
P_{B}(x_{\theta j}^{A},x_{\phi k}^{B}|\lambda_{rot}) & = & \frac{Q_{jk}(x_{\theta A},x_{\phi B})P(x_{\theta j}^{A},x_{\phi k}^{B})}{Q_{m}(x_{\theta A},x_{\phi B},t_{2})}\nonumber \\
\label{eq:probB-1}
\end{eqnarray}
Following Section IV.F, we find $Q_{jk}(x_{\theta A},x_{\phi B})=\frac{e^{-(x_{\theta A}-x_{\theta j}^{A})^{2}/2}}{\sqrt{2\pi}}\frac{e^{-(x_{\phi B}-x_{\phi k}^{B})^{2}/2}}{\sqrt{2\pi}}$.
The probability of outcomes $\mathcal{S}_{A}$ and $\mathcal{S}_{B}$
given the set of hidden variables $\lambda_{rot}$ can be found by
summing the $P_{B}(x_{\theta j}^{A},x_{\phi k}^{B}|\lambda_{rot})$
where outcomes $x_{\theta j}^{A}$, $x_{\phi k}^{B}$ have the appropriate
sign. $\square$

\emph{Comment:}\textbf{ }In Section IV.F (Result IV.13b), the deterministic
model $\mathcal{P}_{det}$ was also put forward, based on the postselected
state. A similar model $\mathcal{P}_{det,B}$ can be constructed in
the bipartite case, to describe the predetermination of the outcomes
$x_{\theta j}^{A}$ and $x_{\phi k}^{B}$ of $\hat{x}_{\theta A}$
and $\hat{x}_{\phi B}$ as given by the system defined at the time
$t_{2}$ in Figure \ref{fig:sim}, which is the time by which the
measurement settings have been fixed. The postselected states are
derived in Appendix H.

\textbf{\emph{Result VII.9b:}}\textbf{ }\textbf{\emph{Hidden variable
model $\mathcal{P}_{Bell}$:}}\textbf{ }A second model $\mathcal{P}_{Bell}$
traces back from branches at time $t_{f}$ to the time $t_{0}$, \emph{prior}
to the unitary operations that determine the setting choices $\theta$
and $\phi$. This is possible because the relations (\ref{eq:rot-6-a})
and (\ref{eq:rot-6-b}) are deterministic, allowing an algebraic transformation
of variables (Appendix H). Hence we define the HV model
\begin{equation}
P_{Bell}(\mathcal{S}_{A},\mathcal{S}_{B}|\lambda,\theta,\phi)\label{eq:model-bell}
\end{equation}
where $\lambda=(x_{A},p_{A},x_{B},p_{B})$ are the variables defined
for the state of the system at the time $t_{0}$, \emph{prior} to
the interactions that determine $\theta$ and $\phi$. Hence, a probabilistic
HV model $\mathcal{P}_{Bell}$ exists, in which the system given by
amplitudes $\lambda=(x_{A},p_{A},x_{B},p_{B})$ can be viewed as being
in a state with a certain probability for outcomes $\mathcal{S}_{A}$
and $\mathcal{S}_{B}$ at $A$ and $B$.

\emph{Proof:} We define
\begin{eqnarray}
P(x_{\theta j}^{A},x_{\phi k}^{B}|\lambda) & = & \frac{Q_{loop}(\lambda,t_{0}|x_{\theta j}^{A},x_{\phi k}^{B})P(x_{\theta j}^{A},x_{\phi k}^{B})}{Q(\lambda,t_{0})}\nonumber \\
\label{eq:probbell}
\end{eqnarray}
and evaluate the transformed probability density function from $Q_{loop}(\lambda_{rot},t_{2}|x_{\theta j}^{A},x_{\phi k}^{B})$.\textbf{
$\square$}

\textbf{\emph{Result VII.9c: Asymmetric local hidden variable model
$\mathcal{P}_{asym,A}$}}\emph{: }A third HV model is deduced from
Result VII.3. Suppose we fix the setting at system $A$ but allow
a change of setting $\phi$ at $B$. We consider measurements $\hat{x}_{A}$
and $\hat{x}_{\phi B}$. We see from Result VII.3 that a LHV model
exists based on $\rho_{mix,B}$ to correctly describe the outcomes
of $\hat{x}_{A}$ and $\hat{x}_{\phi B}$. Details are in Appendix
F.

\subsection{Weak macroscopic realism and weak local realism for bipartite systems}

We now extend Premises (1) of weak macroscopic realism (wMR) and weak
local realism (wLR), given by Definitions (6) and (10) in Sec. I.B,
and by Results IV.6 and IV.14, so that they apply to bipartite systems.
Consider the system of Figure \ref{fig:sim} prepared generally at
time $t_{2}$ for measurement of $\hat{x}_{\theta A}$ and $\hat{x}_{\phi B}$,
so that an independent amplification $H_{amp}$ for each system will
finalize the measurement. The possible outcomes at $A$ are denoted
by the set $\{x_{\theta j}^{A}\}$. The possible outcomes at $B$
are denoted by the set $\{x_{\phi k}^{B}\}$.

\textbf{\emph{Result VII.10: Weak Macroscopic Realism, Premise wMR(1):}}
The simulation (and Q model) of the bipartite system is consistent
with the Premise (1) of weak MR (Definition (6a) in Sec. I.B), which
posits that \emph{both} the values $x_{j}^{A}$ and $x_{k}^{B}$ for
the outcomes at $A$ and $B$ are determined at a time $t_{m}$, after
the settings are fixed, and after sufficient amplification, $H_{amp}^{A}$
and $H_{amp}^{B}$.

\emph{Proof:} Where there is a joint measurement as in the amplifications
$H_{amp}^{A}$ and $H_{amp}^{B}$, distinct branches emerge at a time
$t_{m}$ (as in Fig. \ref{fig:epr-3}, at time $t_{m}=2/g$ where
$G=e^{|g|t_{f}}\rightarrow\infty$). In the Q model, the measured
outcomes (if detection takes place) are given by $\widetilde{x}_{\theta A}=x_{\theta A}(t)/G$
and $\widetilde{x}_{\phi B}=x_{\phi B}(t)/G$ (Result IV.6 in Sec.
IV.C). We symbolize the predetermined values by variables $\widetilde{\lambda}_{\theta}^{A}$
and $\widetilde{\lambda}_{\phi}^{B}$. $\square$

\textbf{\emph{Result VII.11:}} \textbf{\emph{Weak Local Realism, Premise
wLR(1): }}At the time $t_{2}$, the system of Figure \ref{fig:sim}
has undergone the unitary transformations $U_{\theta}^{A}$ and $U_{\phi}^{B}$
that fix the settings to be $\theta$ and $\phi$. There are two versions
of wLR(1), \emph{deterministic} and \emph{probabilistic}, defined
in Definition (10a) (refer also Result IV.14, Sec. IV.G). The bipartite
simulation is consistent with the Premise (1) of both versions of
wLR, which posit that \emph{both} the outcomes at $A$ and $B$ are
determined (at least with a certain probability) at time $t_{2}$.

\emph{Proof:} The deterministic version of wLR posits consistency
with a hidden variable (HV) model in which the outcomes are determined
at the time $t_{2}$. In the Q model as applied to Figure \ref{fig:sim},
the means $x_{\theta j}^{A}$ and $x_{\phi k}^{B}$ of the Gaussian
functions in $Q(\bm{\lambda},t_{2})$ are amplified to $Gx_{\theta j}^{A}$
and $Gx_{\phi k}^{B}$, and the correlation between the outcomes is
hence determined by $Q(\bm{\lambda}_{rot},t_{2})$. The interference
terms $\mathcal{I}nt_{AB}$ in $Q(\bm{\lambda},t_{2})$ are not amplified
(Result VII.2, Sec. VII.B). Hence, the prediction for joint probabilities
of outcomes $\hat{x}_{A}$ and $\hat{x}_{B}$ is the same as for the
\emph{mixed} state, $\rho_{D}$ defined by Eq. (\ref{eq:mixab-1}),
for which the outcomes $x_{\theta j}^{A}$ and $x_{\phi k}^{B}$ can
be viewed as determined for an individual realization of the system.
 Hence a deterministic HV model consistent with wLR(1) exists. The
model $\mathcal{P}_{det,B}$ outlined above in Sec. VII.D provides
a more detailed deterministic HV model consistent with wLR(1), and
also accounts for the interference terms in the Q function.

Alternatively, for the probabilistic version of wLR, a hidden variable
(HV) model $\mathcal{P}_{B}$ (Eq. (\ref{eq:model-b})) can be developed
from the simulation. At time $t_{2}$, we consider the system to be
in a HV state $\lambda_{rot}=(x_{\theta A},p_{\theta A},x_{\phi B},p_{\phi B})$
with probability $Q(\bm{\lambda_{rot}},t_{2})$. It is possible to
construct a joint probability for the outcomes $x_{\theta j}^{A}$
and $x_{\phi k}^{B}$ (and hence of $\mathcal{S}_{A}$ and $\mathcal{S}_{B}$)
at the sites, for any given values of $\lambda_{rot}$ (Result VII.9a,
Eq. (\ref{eq:probB-1})).  Hence, a probabilistic HV model satisfying
the probabilistic version of wLR(1) exists. This model is more general
than $\mathcal{P}_{det,B}$, and is supported by the simulation. $\square$

\textbf{\emph{Result VII.12:}}\emph{ }\textbf{\emph{Deterministic
relations $\mathcal{D}$ for correlated outcomes:}} The Results specify
the outcomes $x_{\theta j}^{A}$ and $x_{\phi k}^{B}$ to be determined
at the time $t_{m}$, or else, for Result VII.11 (at least probabilistically)
at time $t_{2}$. Hence, the Results imply certain deterministic relations
$\mathcal{D}$ that specify the correlations between the outcomes
$x_{\theta j}^{A}$ and $x_{\phi k}^{B}$, \emph{at these times}.
For Result VII.1, the deterministic relations $\mathcal{D}$ will
apply to the system at time $t_{2}$,  because the settings at $A$
and $B$ are fixed at $\theta$ and $\phi$ at this time. The system
has been prepared, by the appropriate unitary operations, with respect
to the measurement basis $\hat{x}_{\theta A}$ and $\hat{x}_{\phi B}$.
 Specifically, examining state $|x_{j}^{A}\rangle|x_{k}^{B}\rangle+|x_{l}^{A}\rangle|x_{m}^{B}\rangle$
of Eq. (\ref{eq:entbasis}), the Q function (\ref{eq:qex}) shows
that the outcomes $x_{j}^{A}$ and $x_{k}^{B}$ are correlated, as
are those of $x_{l}^{A}$ and $x_{m}^{B}$.

\textbf{\emph{Result VII:13: Deterministic relations $\mathcal{D}$
for correlated outcomes after a single setting change: }}Suppose we
fix the setting at system $A$ but allow a change of setting $\phi$
at $B$. We consider measurements $\hat{x}_{A}$ and $\hat{x}_{\phi B}$.
We see from Result VII.3 that a local hidden variable (LHV) model
$\mathcal{P}_{asym,B}$ exists based on $\rho_{mix,B}$ to correctly
describe the outcomes of $\hat{x}_{A}$ and $\hat{x}_{\phi B}$ (Result
VII.9c and Appendix F). We hence deduce a\emph{ }set $\mathcal{D}$
of\emph{ }deterministic relations implied by $\mathcal{P}_{asym,B}$.
In the weak local realistic model (which is implemented by the Q-model
of reality), the values for the outcomes for $\hat{x}_{A}$ and $\hat{x}_{\phi B}$
are determined (at least probabilistically) as $x_{j}^{A}$ and $x_{\phi k}^{B}$,
at a time $t_{1}$ once the settings are fixed at $\theta=0$ and
$\phi$. The purpose of the relations $\mathcal{D}$ is to give the
correlation between the (hidden) variables, $x_{j}^{A}$ and $x_{\phi k}^{B}$.

\emph{Proof: }The $Q(\lambda,t_{0})$ denoting an initial state $|\psi(t_{0})\rangle$
prepared in the measurement basis with $\theta=\phi=0$ is transformed
to $Q(\lambda_{rotB},t_{1B})$ at the time $t_{1B}$, after the setting
change $\phi$ at $B$, where $\lambda_{rotB}=(x_{A},p_{A},x_{\phi B},p_{\phi B})$.
 The $Q(\lambda_{rotB},t_{1B})$ can be written in the new measurement
basis (Appendix E) using (\ref{eq:rot-6-b}). We can re-express in
the form (\ref{eq:Q-int}): a sum of Gaussian functions and interference
terms $\mathcal{I}nt_{AB}(\phi)$. This is done, by noting that the
initial quantum state $|\psi(t_{0})\rangle=\sum_{i,l}c_{il}|x_{i}^{A}\rangle|x_{l}^{B}\rangle$
is determined by the $c_{il}$. The $|\psi(t_{0})\rangle$ is defined
uniquely by $Q(\lambda,t_{0})$ and can in principle be determined
by $Q(\lambda,t_{0})$ (Appendix C). Hence, we can write in the
new basis, using $|x_{l}^{B}\rangle=\sum_{k}d_{lk}|x_{\phi k}^{B}\rangle$
where $d_{lk}$ are transformation coefficients given in Appendix
E, as functions of $\phi$. Hence $|\psi(t_{1B})\rangle=\sum_{j,k}e_{jk}|x_{j}^{A}\rangle|x_{\phi k}^{B}\rangle$,
where $e_{jk}=\sum_{l}c_{jl}d_{lk}$, can be determined solely from
the $c_{il}$ and $\phi$, which defines $Q(\lambda_{rotB},t_{1B})$
in the form (\ref{eq:q1-1-2}).

This also defines the joint probabilities of outcomes $x_{j}^{A}$
and $x_{\phi k}^{B}$, as $|e_{jk}|^{2}$. The deterministic relations
$\mathcal{D}_{1}$ that determine the correlations between outcomes
$x_{j}^{A}$ and $x_{\phi k}^{B}$ can hence be established from knowledge
of the initial state $Q(\lambda,t_{0})$ and $\phi$ at $B$. These
are $x_{j}^{A}\leftrightarrow x_{\phi k}^{B}$ with probability $|e_{jk}|^{2}$.
$\square$

\textbf{\emph{Result VII.13b:}}\textbf{ }\textbf{\emph{Compatibility
between wMR and wLR, and the deterministic relations}}\textbf{:} It
is remarked that weak macroscopic realism (wMR) is compatible with
weak local realism (wLR), and vice versa. In particular, the Q model
of reality supports both wMR, and the probabilistic version of wLR,
as evident from the simulations, which show consistency with wLR as
well as the emergence of branches. Hence, the causal model and the
above deterministic relations $\mathcal{D}$ are compatible with wMR.

\section{Measurement with a meter: state projection\label{sec:EPR-proj}}

\textcolor{blue}{}For the EPR system (\ref{eq:tmss}), it is well
known that the Wigner function provides a local hidden variable (LHV)
theory which explains the correlation between spacelike-separated
systems $A$ and $B$ \citep{bell2001john}.  However, the LHV model
cannot explain other correlated systems, such as $|\psi_{Bell}\rangle$
of Eq. (\ref{eq:bell}). 

In this section, we use the Q-based simulation to analyze measurement
by a meter, hence analyzing \emph{projection}, where a measurement
on one of two entangled systems ``collapses'' the state of the second
system. We show consistency with the Premises (3) of weak local and
macroscopic realism (Definitions (6c) and (10c) in Sec. I.B) which
leads to conclusions about the mechanism and the timing of the ``collapse
of the wave function''.

\subsection{Model for projection}

\begin{figure}
\begin{centering}
\par\end{centering}
\begin{centering}
\includegraphics[width=1\columnwidth]{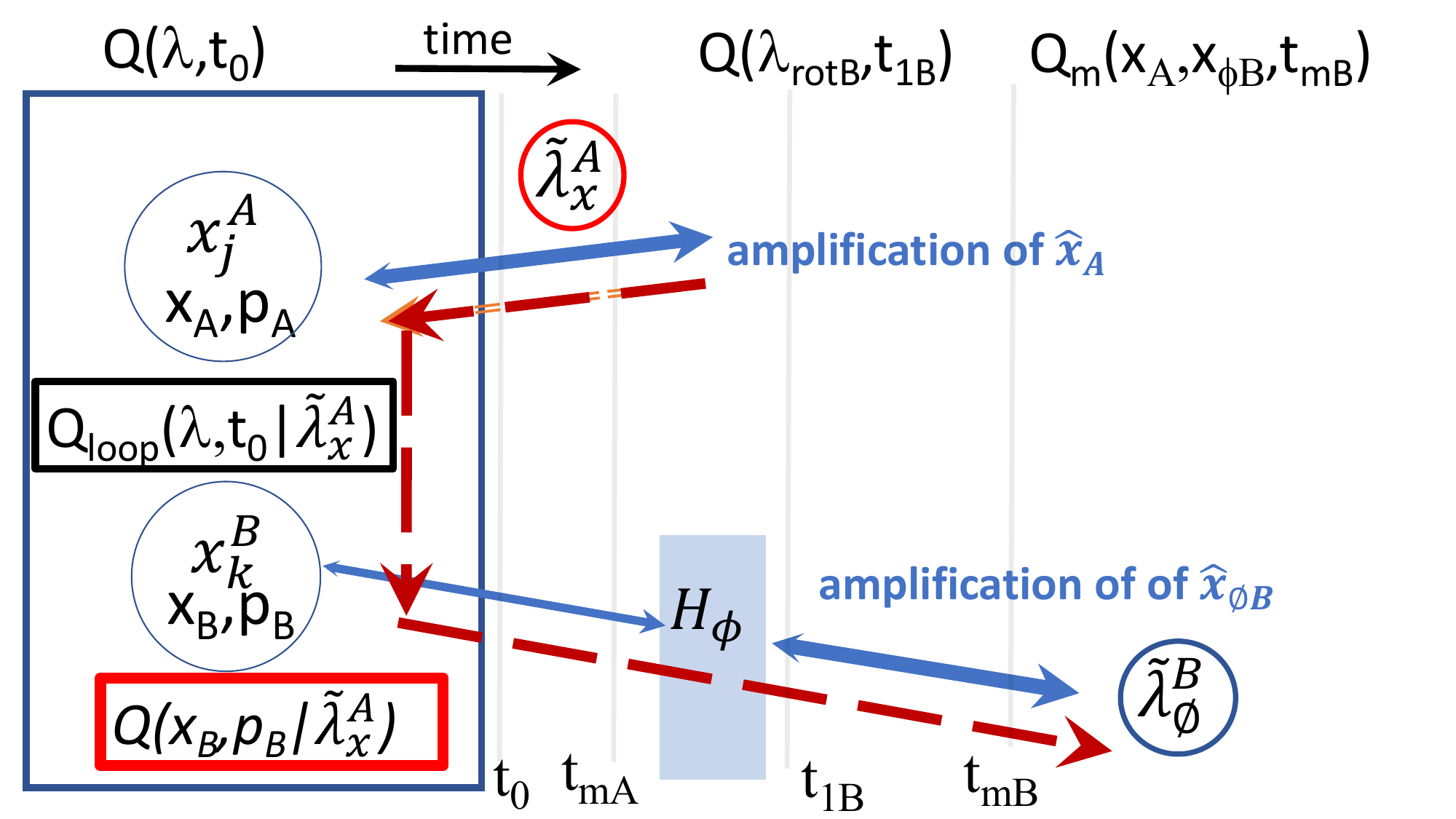}
\par\end{centering}
\caption{\textbf{\emph{Projection in the Q model:}} The system is prepared
at time $t_{0}$ with respect to the measurement basis of $\hat{x}_{A}$
and $\hat{x}_{B}$, so that $\theta=\phi=0$. The field $A$ is amplified
(at time $t_{mA}$) with the setting fixed at $\theta=0$.  The
trajectories $x_{A}(t)$ conditioned on the branch $\widetilde{\lambda}_{x}^{A}$
corresponding to an outcome $x_{j}^{A}$ trace back to time $t_{0}$
(orange dashed and solid blue line) where they correlate with certain
amplitudes, to determine the postselected distribution $Q_{loop}(\lambda,t_{0}|\widetilde{\lambda}_{x}^{A})\equiv Q_{loop}(\lambda,t_{0}|x_{j}^{A})$
(accounting for the deterministic relations $\mathcal{D}$ that constrain
correlations with outcomes $x_{k}^{B}$). The distribution $Q(x_{B},p_{B}|\widetilde{\lambda}_{x}^{A})\equiv Q(x_{B},p_{B}|x_{j}^{A})$
for system $B$ alone (Eq. (\ref{eq:cond-B-1})) corresponds to the
\emph{projected state} of $B$ given the outcome $x_{j}^{A}$ at $A$.
The projected state is determined at (or by) the time $t_{mA}$,
which can be prior to a setting change $H_{\phi}$ and final detection
at $B$. The projected state is consistent with the marginal $Q_{m}(x_{A},x_{\phi B},t_{mB})$
which gives the joint probability density of amplitudes $x_{A}(t_{mB})$
and $x_{\phi B}(t_{mB})$. \label{fig:projection-1}}
\end{figure}

 Figure \ref{fig:projection-1} shows a two-mode system prepared
at time $t_{0}$ for measurements $\hat{x}_{A}$ and $\hat{x}_{B}$
(where $\theta=\phi=0$). The system is described by a two-mode Q
function $Q(\lambda,t_{0})$, where $\lambda=(x_{A},p_{A},x_{B},p_{B}$).
The system might be in an EPR state such as given by Eq. $(\ref{eq:tmss}),$
or in a two-mode cat state similar to $|\psi_{Bell}\rangle$ of Eq.
(\ref{eq:bell}). We consider quadrature phase amplitude measurements
$\hat{x}_{\theta A}=\hat{x}_{A}\cos\theta+\hat{p}_{A}\sin\theta$
and $\hat{x}_{\phi B}=\hat{x}_{B}\cos\phi+\hat{p}_{B}\sin\phi$.

If there is a change of setting $\phi$ at $B$, then there are \emph{constraints
imposed} on system $B$ by the configuration at $A$. This gives a
model for \emph{projection}.

In Figure \ref{fig:projection-1}, the setting at $A$ has been fixed
at $\theta=0$, ready for a measurement of $\hat{x}_{A}$ (Fig. \ref{fig:projection-1}).
Let us suppose that system $A$ at the time $t_{mA}$ has been amplified
according to $H_{amp}^{A}$ (Eq. (\ref{eq:hampa})). The \emph{bipartite
postselected distribution} at the time $t_{0}$ conditioned on a branch
symbolized by $\widetilde{\lambda}_{x}^{A}$ with outcome $x_{j}^{A}$
for $\hat{x}_{A}$ at $A$ is denoted by (refer Eq. (\ref{eq:Qloop-postselect})
and Definition (8) in Sec. I.B)
\begin{equation}
Q_{loop}(\lambda,t_{0}|x_{j}^{A})\label{eq:qloop}
\end{equation}
The bipartite distribution $Q_{loop}(\lambda,t_{0}|x_{j}^{A})$ originates
from the backward-propagating trajectories that trace back to the
initial time $t_{0}$, from a given branch $x_{j}^{A}$ for $A$.
The initial state $Q(\lambda,t_{0})$ implies a restriction (red dashed
lines) on the state of system $B$, as depicted in Figure \ref{fig:projection-1}.

The $Q_{loop}$ is an explicit function of $\lambda=(x_{A},p_{A},x_{B},p_{B})$
but is derived within the framework of the Q model of reality, and
Results VII.10-12. Hence, its derivation must take into account the
possible outcomes $x_{k}^{B}$ at $B$, as given by the deterministic
relations $\mathcal{D}$ that specify the correlations between the
outcomes at $A$ and $B$ for the specified measurement basis. The
bipartite postselected distribution \emph{must be consistent with
any known deterministic relations $\mathcal{D}$ }that determine the
correlation between any of the possible outcomes $x_{k}^{B}$ at $B$,
and $x_{j}^{A}$ at $A$ (Results VII.10-12).

The state for system $B$ alone conditioned on the branch $x_{j}^{A}$
of $A$ is
\begin{equation}
Q(x_{B},p_{B}|x_{j}^{A})=\int dx_{A}dp_{A}Q_{loop}(\lambda,t_{0}|x_{j}^{A})\label{eq:cond-B-1}
\end{equation}
as calculated by integrating $Q_{loop}(\lambda,t_{0}|x_{j}^{A})$
over $x_{A}$ and $p_{A}$. The distribution (\ref{eq:cond-B-1})
constrains the outcomes for \textbf{\emph{$B$}}, including for any
future changes of settings at $B$. Figure \ref{fig:projection-1}
depicts a future change of setting at $B$ to $\phi$.

\textbf{\emph{Result VIII.1a: Retrodiction not retrocausality:}} The
distribution function $Q(x_{B},p_{B}|x_{j}^{A})$ for the projected
state for system $A$ is derived based on knowledge of the setting
and outcome $x_{j}^{A}$ at $A$ and of the initial distribution $Q(\lambda,t_{0})$.
No knowledge is required about the setting $\phi$ at $B$. Hence,
we claim the distribution is deduced by \emph{retrodiction} (\emph{not
retrocausality}), based on the relation $x_{j}^{A}\rightarrow Gx_{j}^{A}$
(Eq. \ref{eq:G-amp}), and the deterministic relations $\mathcal{D}$
that specify the correlations between outcomes $x_{j}^{A}$ and $x_{k}^{B}$
(assuming weak local or macroscopic realism, as in Results VII.1--12).
Here, we assume the setting is fixed at $A$.

\emph{Proof}: The proof follows since $Q(x_{B},p_{B}|x_{j}^{A})$
is derived directly from $Q_{loop}(\lambda,t_{0}|x_{j}^{A})$. The
method of derivation of $Q_{loop}(\lambda,t_{0}|x_{j}^{A})$ involves
tracking the trajectories $x_{A}(t)$ for the branch $x_{j}^{A}$
back to the original time $t_{0}$, but the nature of the backward
simulation means the distribution $Q_{loop}(\lambda,t_{0}|x_{j}^{A})$
can be deduced by retrodiction, by reversing the deterministic relation
$x_{j}^{A}\rightarrow Gx_{j}^{A}$ and similar deterministic relations
$\mathcal{D}$ between $x_{j}^{A}$ and $x_{k}^{B}$ (refer below).
This is because \emph{there is no change to the properties of the
noise} $\delta x_{A}\equiv\eta_{A}(t_{f})$, \emph{which is an input
at the future boundary}, as the system propagates in the backward-time
direction, from time $t_{f}$ (which determines the future boundary)
to time $t_{0}$ (refer Sec. III.B and V.A). $\square$

The projected distribution at $B$ however gives more information
than just the outcome for the measurement at $B$, given an outcome
at $A$. This is determined by the relations $\mathcal{D}$ as fixed
by the current setting at $B$ (Result VII.12). The projected distribution
gives information about the\emph{ state} for $B$, and hence information
about future measurements at $B$ after further setting-changes. This
is because the projected distribution is derived not only from $x_{j}^{A}$,
but from the Q function $Q(\lambda,t_{0})$, which includes noise
and interference terms. We clarify with the following Result. 

\textbf{\emph{Result VIII.1b: Future settings at $B$: }}The distribution
$Q(x_{B},p_{B}|x_{j}^{A})$ determines the probabilities of outcomes
at $B$, conditioned on the outcome $x_{j}^{A}$ of $\hat{x}_{A}$
at $A$, \emph{for any future choice of setting} $\phi$. The predictions
are consistent with the correlations between outcomes at $A$ and
$B$ as predicted by the Q function $Q(x_{A},x_{\phi B},t_{mB})$
of the system at time $t_{mB}$, after the setting-change and amplification
at $B$ (refer Figure \ref{fig:projection-1}).

\emph{Proof: }The proof is given in Appendix H (Result AH.3) and follows
because the change of setting at $B$ is a local transformation (\ref{eq:rot-6-b-1})
corresponding mathematically to changing coordinates in the distribution
$Q(x_{B},p_{B}|x_{j}^{A})$. $\square$

Hence, as expected, the distribution $Q(x_{B},p_{B}|x_{j}^{A})$ will
correspond to the projected state, given quantum mechanically. This
models the ``collapse to the projected state'' for bipartite systems.

\textbf{\emph{Result VIII.1c:}} \textbf{\emph{Projected state: }}The
probability distribution $Q(x_{B},p_{B}|x_{j}^{A})$ corresponds to
the Q function of the projected state for $B$ in quantum mechanics,
given the outcome $x_{j}^{A}$ for system $A$.

\emph{Proof:} Consider the general state $\sum_{jk}c_{jk}|x_{j}^{A}\rangle|x_{k}^{B}\rangle$
as in Eq. (\ref{eq:state-gen}). We consider that the systems $A$
and $B$ are prepared with respect to the measurement bases of $\hat{x}_{A}$
and $\hat{x}_{B}$. \textcolor{blue}{ }We first take the simplest
case of 
\begin{equation}
|x_{j}^{A}\rangle|x_{k}^{B}\rangle+|x_{l}^{A}\rangle|x_{m}^{B}\rangle\label{eq:state-3}
\end{equation}
as in (\ref{eq:entbasis}) where $j\neq l$, $k\neq m$, for which
according to quantum mechanics the projected state at $B$ given the
outcome $x_{j}^{A}$ at $A$ is $|x_{k}^{B}\rangle$. (This follows
because the eigenstates are orthogonal, proved in Appendix C). Applying
Result VII.11 and 12, the correlations between the outcomes at $A$
and $B$ define a set of deterministic relations $\mathcal{D}$ in
the Q model, in this case
\begin{equation}
x_{j}^{A}\leftrightarrow x_{k}^{B};x_{l}^{A}\leftrightarrow x_{m}^{B}\label{eq:Det}
\end{equation}
We consider outcome $x_{j}^{A}$. Following the procedure of Secs.
IV.D and Appendix H, we trace back the branch $\mathcal{B}_{j}^{A}$,
to obtain the distribution for $x_{A}$ at time $t_{0}$, as $Q_{x_{j}^{A}}(x_{A})=\frac{e^{-(x_{A}-x_{j}^{A})^{2}/2}}{\sqrt{2\pi}}$.
This gives for the inferred distribution at $t_{0}$,
\begin{eqnarray}
Q_{loop}(\lambda,t_{0}|x_{j}^{A}) & = & Q_{x_{j}^{A}}(x_{A})Q(p_{A},x_{B},p_{B}|x_{A})\nonumber \\
 & = & \frac{Q_{x_{j}^{A}}(x_{A})Q(\lambda,t_{0})}{Q_{m}(x_{A},t_{0})}\label{eq:qprojloop}
\end{eqnarray}
where $Q(\lambda,t_{0})$ is the Q function of the system at time
$t_{0}$ (Eq. (\ref{eq:qex})) and $Q_{m}(x_{A},t_{0})=\int dp_{A}dx_{B}dp_{B}Q(\lambda,t_{0})$
denotes the marginal distribution for $x_{A}$. However, a direct
evaluation does not fully define the projected state, due to the overlap
of the distributions of the eigenfunctions $|x_{j}^{A}\rangle$ and
$|x_{l}^{A}\rangle$ for small separations $|x_{j}^{A}-x_{l}^{A}|\lesssim1$.
The correct procedure also takes into account the deterministic relation
$\mathcal{D}$ that defines the correlation that $x_{j}^{A}$ implies
$x_{k}^{B}$ for $B$, which is embedded in the causal model, based
on the fixed settings at $t_{0}$. This relation rules out the bivariate
Gaussian in $Q(\lambda,t_{0})$ with means $x_{l}^{A}$ and $x_{m}^{B}$.
The interference term $\mathcal{I}nt_{AB}$ as given by Eq. (\ref{eq:int-1})
is shared between the two states of the superposition (\ref{eq:state-3}).
This term is a function of $p_{A}$ and cannot be inferred from the
outcome being $x_{j}^{A}$ or $x_{k}^{A}$: this term is not ruled
out. However, the projected state for $B$ is deduced by integrating
over $p_{A}$. We see from Result VII.5 that the integration over
$p_{A}$ leads to a vanishing of the term involving $\mathcal{I}nt_{AB}$.
The projected state in the Q model is hence
\begin{equation}
Q_{x_{j}^{A}}(x_{B},p_{B})=\mathcal{G}_{0,\sigma_{p_{B}}}(p_{B})\mathcal{G}_{x_{k}^{B},1}(x_{B})\label{eq:qsoln-proj}
\end{equation}
corresponding to the Q function of state $|x_{k}^{B}\rangle$ as
required. Here $\mathcal{G}$ are the Gaussian functions defined by
Eqs. (\ref{eq:qex}) and (\ref{eq:gauss}).

Generalizing, the projected state of $|\psi\rangle=\sum_{jk}c_{jk}|x_{j}^{A}\rangle|x_{k}^{B}\rangle$
given $x_{j}^{A}$ is found by rewriting $|\psi\rangle=\sum_{j}f_{j}|x_{j}^{A}\rangle|\psi_{j}^{B}\rangle$
where $|\psi_{j}^{B}\rangle=\sum_{k}d_{jk}|x_{k}^{B}\rangle$ is normalized
as in Eq. (\ref{eq:mix-asymA}). The projected state is $|\psi_{j}^{B}\rangle$.
Hence, we consider a superposition of type
\begin{equation}
|x_{j}^{A}\rangle(d_{j1}|x_{k1}^{B}\rangle+d_{j2}|x_{k2}^{B}\rangle+..)+|x_{l}^{A}\rangle|x_{m}^{B}\rangle\label{eq:state3}
\end{equation}
The Q function $Q(\lambda,t_{0})$ is of the form 
\begin{eqnarray}
Q(\lambda,t_{0}) & = & N\Bigl(\mathcal{G}_{0,\sigma_{p_{A}}}(p_{A})\mathcal{G}_{x_{j}^{A},1}(x_{A})Q_{Bj}(x_{B},p_{B})\nonumber \\
 &  & +\mathcal{G}_{0,\sigma_{p_{A}}}(p_{A})\mathcal{G}_{x_{l}^{A},1}(x_{A})\mathcal{G}_{0,\sigma_{p_{B}}}(p_{B})\mathcal{G}_{x_{m}^{B},1}(x_{B})\nonumber \\
 &  & +\mathcal{I}nt_{AB}\Bigr)\label{eq:qstatesec}
\end{eqnarray}
where $Q_{B_{j}}(x_{B},p_{B})$ is the Q function of the superposition
$d_{j1}|x_{k1}^{B}\rangle+d_{j2}|x_{k2}^{B}\rangle+..$, so that
\begin{eqnarray}
Q_{B_{j}}(x_{B},p_{B}) & = & \mathcal{G}_{0,\sigma_{p_{B}}}(p_{B})\{|d_{j1}|^{2}\mathcal{G}_{x_{k1}^{B},1}(x_{B},p_{B})\nonumber \\
 &  & +|d_{j2}|^{2}\mathcal{G}_{x_{k2}^{B},1}(x_{B},p_{B})+...+\mathcal{I}_{B}\}\nonumber \\
\label{eq:state7}
\end{eqnarray}
where here the interference term $\mathcal{I}_{B}$ is a sum of terms
such as $\mathcal{I}nt$ given for a single-mode two-state superposition,
in Eq. (\ref{eq:Q-sup}). The projected state conditioned on $x_{j}^{A}$
rules out the second Gaussian in (\ref{eq:qstatesec}), and the bipartite
interference $\mathcal{I}nt_{AB}$ vanishes over the integration with
respect to $p_{A}$. Integrating over the coordinates of $A$, the
projected distribution given the outcome $x_{j}^{A}$ at $A$ is hence
\begin{eqnarray}
Q_{x_{j}^{A}}(x_{B},p_{B}) & = & Q_{Bj}(x_{B},p_{B})\label{eq:prjgen}
\end{eqnarray}
in agreement with the Q function of the projected state $|\psi_{j}^{B}\rangle$.
The projected distribution $Q_{x_{j}^{A}}(x_{B},p_{B})$ describes
the projected state at time $t_{0}$. The predictions for the outcomes
if there is a change of setting is given by the transformation (\ref{eq:rot-6-b-1})
at $B$, which corresponds to the change of coordinates in $Q_{x_{j}^{A}}(x_{B},p_{B})$
(refer Appendix H). $\square$

The projection models the ``collapse to the projected state'' for
bipartite systems. What does this tell us about the process of ``collapse''
in the Q-model of reality?

\textbf{\emph{Result VIII.1d: Wavefunction collapse:}} The conditioned
state (\ref{eq:cond-B-1}) for $B$ alone has reduced information.
For example, fringe terms involving $p_{A}$ vanish in $Q(x_{B},p_{B}|x_{j}^{A})$
(refer above). Hence, in the Q-based simulation and model, the final
``collapse'' to the projected state occurs when an inference about
the state of the system $B$ is made, deduced from the detected outcome
$x_{j}^{A}$ at $A$. There is a loss of information about the complementary
variable $p_{A}$ of system $A$. The projected state does not fully
reflect the state of the system $B$ prior to the measurement at $A$
(Fig. \ref{fig:measurement-feedback}).

\begin{figure}
\begin{centering}
\includegraphics[width=1\columnwidth]{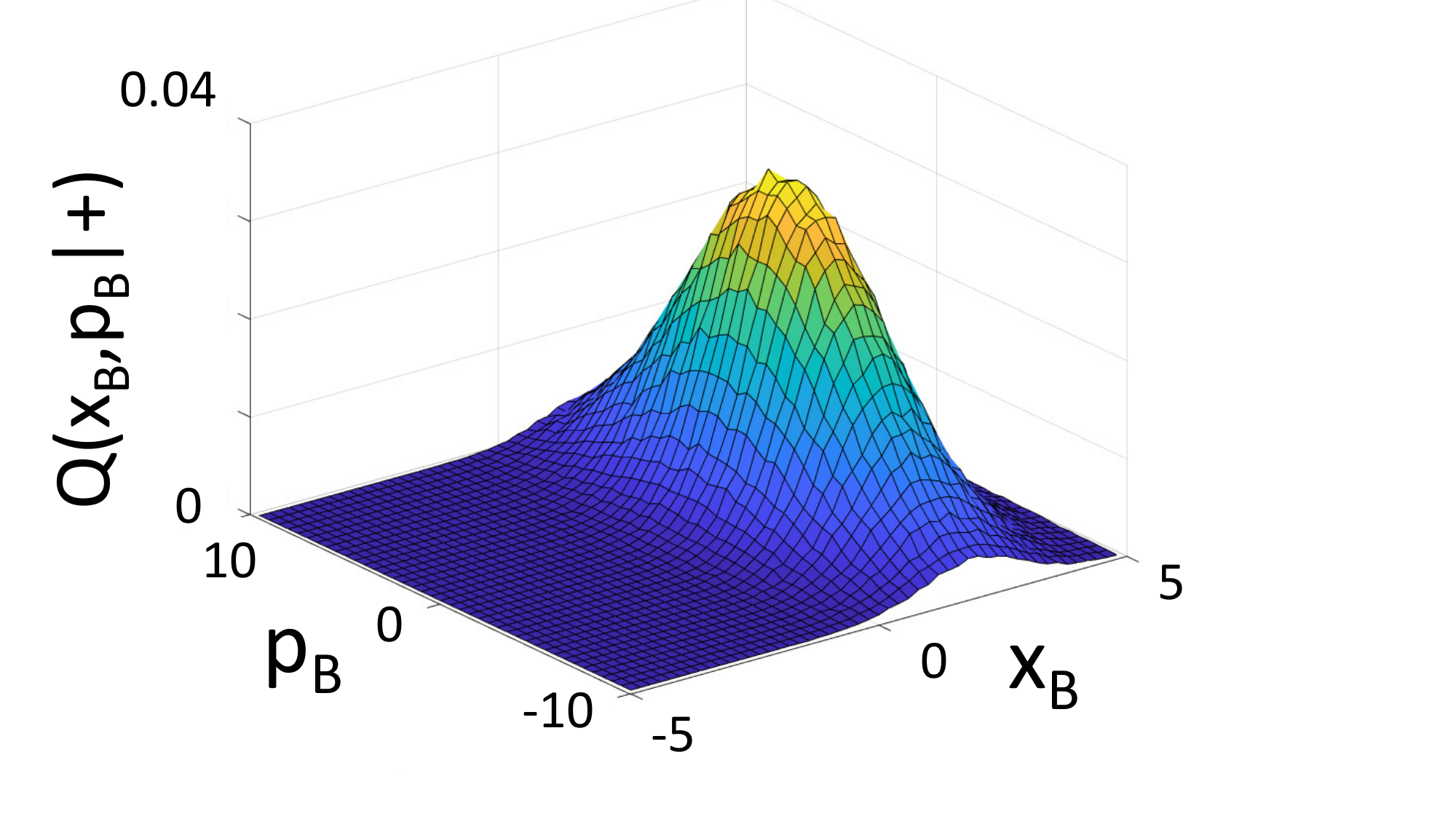}
\par\end{centering}
\caption{\textbf{\emph{Projected distribution and Collapse of the wave function:}}
The projected state for system $B$ entangled with a meter $A$, as
in Eq. (\ref{eq:ent-cat-1-1-1}). We plot $Q(x_{B},p_{B}|x_{1})\equiv Q(x_{B},p_{B}|+)$
for $B$ conditioned on the positive outcome $x_{1}$ of $\hat{x}_{A}$
on the meter $A$ i.e. we trace back trajectories with $x_{A}(t)>0$
at the time $t_{f}$, where $t_{f}$ is large (refer Fig. \ref{fig:projection-1}).
Here, we take $\alpha_{0}=2$, $x_{1}=2$ with $r=1.5$ and $1.2\times10^{6}$
trajectories. The $Q$ function $Q(x_{B},p_{B}|+)$ corresponds to
that of the squeezed state $|x_{1}/2,r\rangle$, which approximates
the eigenstate $|x_{1}\rangle$, modeling the collapse of system $B$
to the eigenstate $|x_{1}\rangle$ after measurement. \textcolor{blue}{\label{fig:ent-meter-infa-1-1}}}
\end{figure}

\subsection{Measurement by a meter}

We illustrate projection as applied to the measurement problem with
the example of the entangled Schrodinger-cat state
\begin{equation}
|\psi_{ent}\rangle_{M}=\frac{1}{\sqrt{2}}\{|\alpha_{0}\rangle|x_{1}\rangle+i|-\alpha_{0}\rangle|-x_{1}\rangle\}\label{eq:ent-meter-x-1-1}
\end{equation}
Here, $|\pm\alpha_{0}\rangle$ are macroscopic coherent states for
the field mode $A$ with $\alpha_{0}>0$ and large. The $|\pm x_{1}\rangle$
are eigenstates of $\hat{x}$ with eigenvalues $\pm x_{1}$. As in
Part 1 of the paper, we model the eigenstates $|x_{1}\rangle$ as
highly squeezed states in $\hat{x}$. The state (\ref{eq:ent-meter-x-1-1})
becomes 
\begin{equation}
|\psi_{ent}\rangle_{M}=\frac{1}{\sqrt{2}}\{|\alpha_{0}\rangle|\frac{x_{1}}{2},r\rangle_{sq}+i|-\alpha_{0}\rangle|-\frac{x_{1}}{2},r\rangle_{sq}\}\label{eq:ent-cat-1-1-1}
\end{equation}
where $|\frac{x_{1}}{2},r\rangle_{sq}$ is the squeezed state for
mode $B$ defined by Eqs. (\ref{eq:sq}) - (\ref{eq:q-sq-1}). For
$r$ large, the squeezed state $|\frac{x_{1}}{2},r\rangle_{sq}$ becomes
the eigenstate $|x_{1}\rangle$ of $\hat{x}$.

When $\alpha_{0}$ is large, the entangled state (\ref{eq:ent-cat-1-1-1})
models the state created when a \emph{macroscopic meter} (field $A$)
has coupled to a system $B$. The outcome of a measurement $\hat{x}_{A}$
on system $A$ gives (for $\alpha_{0}$ real and large) the outcome
of the sign of $\hat{x}$ for system $B$. According to the measurement
postulate, if the outcome for $\hat{x}_{A}$ at $A$ is positive ($+$)
then, for this superposition, the system $B$ ``collapses'' to the
eigenstate $|x_{1}\rangle$ (for $r$ large).

We evaluate the Q function for system $B$ conditioned on the branch
corresponding to the positive outcome $+$ for $\hat{x}_{A}$ for
$A$ i.e. $Q(x_{B},p_{B}|+)$. Here, because the system $A$ is macroscopic,
given by macroscopically distinct states $|\alpha_{0}\rangle$ and
$|-\alpha_{0}\rangle$ that become orthogonal as $\alpha_{0}\rightarrow\infty$,
the projected distribution $Q(x_{B},p_{B}|+)$ for $B$ given the
outcome $+$ at $A$ is deduced directly by integration of $Q_{loop}(x_{B},p_{B},p_{A}|+)$
over $p_{A}$, without the need for direct implementation of the deterministic
relations $\mathcal{D}$. The Q function $Q(x_{B},p_{B}|+)$ is plotted
in Figure \ref{fig:ent-meter-infa-1-1}, verifying equivalence to
the Q function of the squeezed state (modeling the eigenstate $|x_{1}\rangle$),
as expected from the meter-system state (\ref{eq:ent-meter-x-1-1})
and consistent with the measurement postulate. 

\subsection{Projection after a change of setting}

The projection procedure for one system, say $B$, conditioned on
an outcome at second system $A$, can be applied even when there has
been a change of setting at the system $A$ after the time $t_{0}$
at which the bipartite states are prepared. We examine this case,
in order to understand the violation of Bell inequalities in the next
Section.
\begin{figure}
\begin{centering}
\par\end{centering}
\begin{centering}
\includegraphics[width=1\columnwidth]{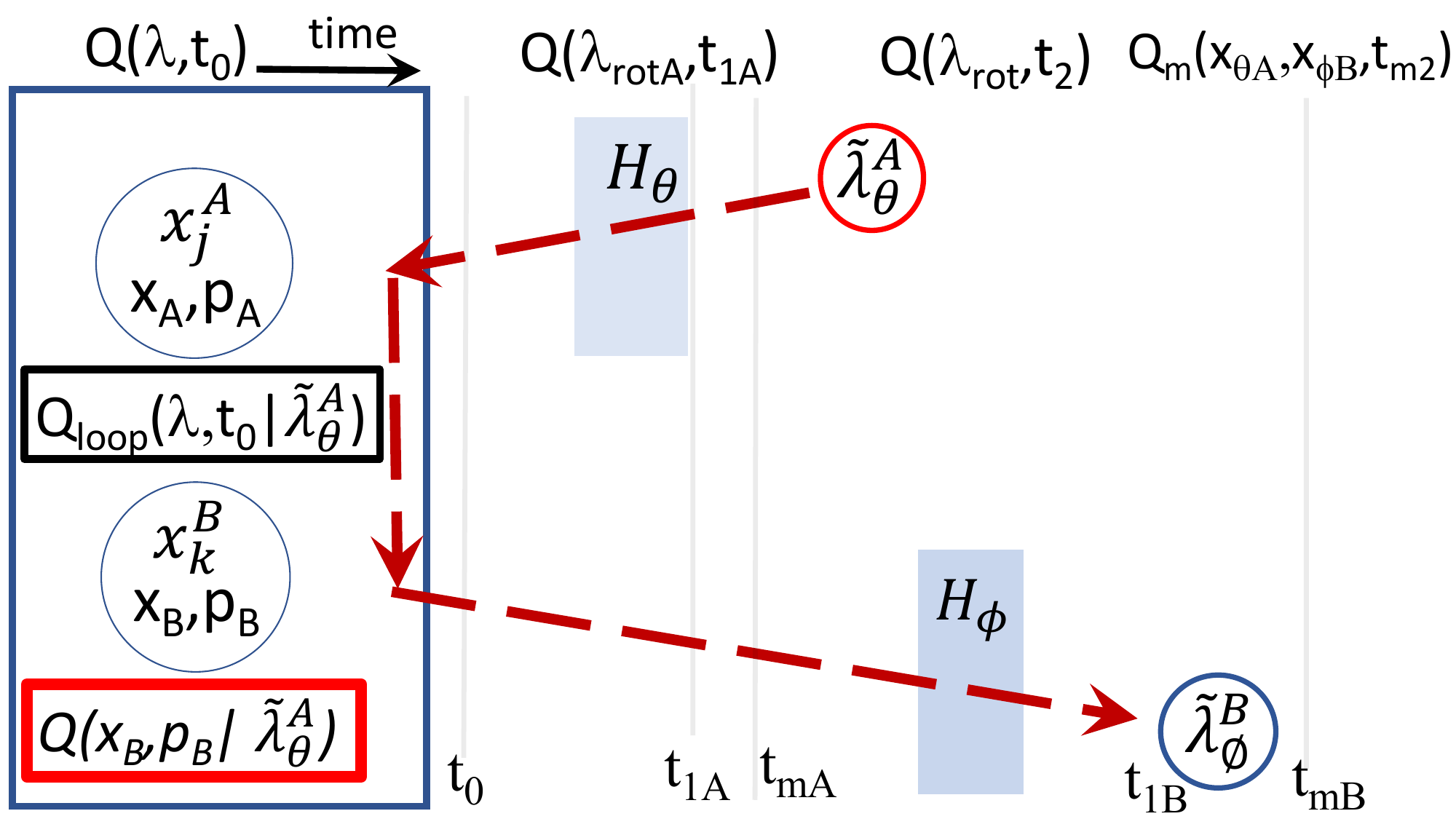}
\par\end{centering}
\caption{\textbf{\emph{Depiction of projection where there has been a setting
change $\theta$ at $A$}}\textbf{:} The projected state $Q(x_{B},p_{B}|\widetilde{\lambda}_{\theta}^{A}$)
is defined to describe the system at time $t_{0}$, based on the setting
change $\theta$ at $A$ and the Q function $Q(\lambda,t_{0})$. The
distribution $Q(x_{B},p_{B}|\widetilde{\lambda}_{\theta}^{A})$ defines
predictions for the system $B$, for further changes of setting $\phi$
that might take place, or have taken place, after the time $t_{0}$.
The derivation is explained in Result VIII.1e and depicted in Figure
\ref{fig:projection-setting-change}, involving the derivation of
$Q_{loop}(\lambda_{rotA},t_{1A}|x_{\theta j}^{A})$, followed by a
change of coordinates, equivalent to tracing trajectories back through
the interaction due to the setting-change $H_{\theta}$ (red dashed
line). The projected state $Q(x_{B},p_{B}|\widetilde{\lambda}_{\theta}^{A}$)
is consistent with the marginal distribution $Q_{m}(x_{\theta A},x_{\phi B},t_{mB})$
that defines the systems after both setting-changes. \label{fig:proj-setting-changeA}}
\end{figure}

\textbf{\emph{Result VIII.1e: Projected state at $B$ conditioned
on $A$ after a change of setting at $A$: }}A projected state and
distribution $Q(x_{B},p_{B}|x_{\theta j}^{A})$ can be defined for
the system $B$ at a time $t_{mA}$, given an outcome $x_{\theta j}^{A}$
for a measurement $\hat{x}_{\theta}^{A}$ (Figure \ref{fig:proj-setting-changeA}).
Here, we are allowing that there has been a change of setting to $\theta$
at $A$ after the initial preparation time $t_{0}$. The time $t_{mA}$
is the time after sufficient amplification at $A$, so that the outcome
for $\hat{x}_{\theta}^{A}$ is defined according to the assumptions
of weak macroscopic realism (wMR). 

\emph{Proof:} We allow for a setting change $\theta$ at $A$, but
consider system $B$ prior to any setting change at $B$. The $Q(\lambda,t_{0})$
is transformed to $Q(\lambda_{rotA},t_{1A})$ at the time $t_{1A}$,
after the setting change at $A$, where $\lambda_{rotA}=(x_{\theta A},p_{\theta A},x_{B},p_{B})$.
The $Q(\lambda_{rotA},t_{1A})$ can be written in the new measurement
basis as explained in Result VII.13 (exchanging the roles of $A$
and $B$). From Result VII.13, the deterministic relations $\mathcal{D}_{1}$
that determine the correlations between outcomes $x_{\theta j}^{A}$
and $x_{k}^{B}$ can hence be established from knowledge of the initial
state $Q(\lambda,t_{0})$ and $\theta$ at $A$. This is justified
because the local hidden variable model $\mathcal{P}_{asym,B}$ given
by the partially mixed state, in this case denoted $\rho_{mix,B}$,
as in Result VII.3 and Appendix F, applies to correctly describe the
outcomes of $\hat{x}_{\theta A}$ and $\hat{x}_{B}$.

Applying the technique given in proof of Result VIII.1c, the inferred
distribution $Q_{loop}(\lambda_{rotA},t_{1A}|x_{\theta j}^{A})$ conditioned
on the branch $\mathcal{B}_{\theta j}$ with outcome $x_{\theta j}^{A}$
will comprise the relevant Gaussian functions that are consistent
with deterministic relations $\mathcal{D}_{1}$, as well as relevant
interference terms $\mathcal{I}nt_{AB}$ (which will vanish over integration
of $p_{\theta A}$). This allows evaluation of the projected state
$Q(x_{B},p_{B}|x_{\theta j}^{A})$. The outcomes $x_{\theta j}^{A}$
and $x_{k}^{B}$ may not be maximally correlated, and the projected
state can be a superposition of eigenstates of $\hat{x}_{B}$, as
in Result VIII.1c.

\begin{figure}
\begin{centering}
\par\end{centering}
\begin{centering}
\includegraphics[width=1\columnwidth]{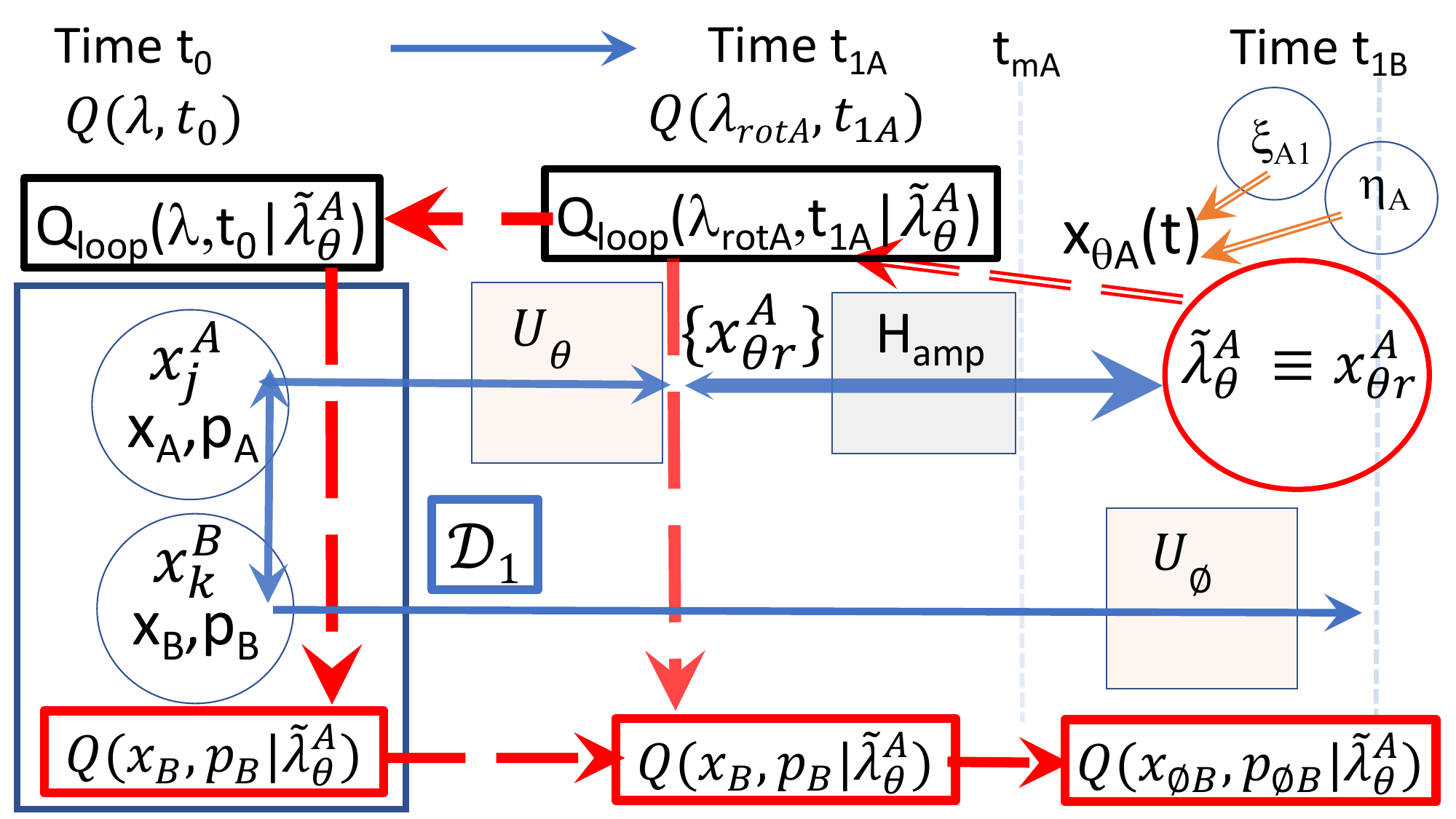}
\par\end{centering}
\caption{\textbf{\emph{Derivation of the projected state with a setting change
$\theta$ at $A$:}} The diagram depicts the inference of the projected
distribution $Q(x_{B},p_{B}|x_{\theta j}^{A})$ for system $B$, conditioned
on an outcome $x_{\theta r}^{A}$, symbolized by a branch $\widetilde{\lambda}_{\theta}^{A}$,
defined after the rotation $U_{\theta}$ and amplification $H_{amp}$.
The system is prepared at time $t_{0}$ for measurements $\hat{x}_{A}$
and $\hat{x}_{B}$ in the state $Q(\lambda,t_{0})$. The derivation
requires only knowledge of $\theta$ and the initial state, $Q(\lambda,t_{0})$
(given by the $c_{jk}$ defined in Result VII.13), and is based on
deterministic relations (blue arrows). The change of setting $\theta$
as given by the deterministic transformation (\ref{eq:rot-6-a-1})
at $A$ is consistent with deterministic relations $\mathcal{D}_{1}$
that account for the correlations between $x_{k}^{B}$ and $x_{\theta r}^{A}$
in $Q(\lambda_{rotA},t_{1A})$. Result VIII.1e and associated Comment
(1) explain the calculation of $Q(x_{B},p_{B}|x_{\theta j}^{A})$,
based on tracking trajectories $x_{\theta A}(t)$ associated with
the branch $\widetilde{\lambda}_{\theta}^{A}\equiv x_{\theta j}^{A}$
to deduce $Q_{loop}(\lambda_{rotA},t_{1A}|x_{\theta j}^{A})$, where
deterministic relations are accounted for. The projected distribution
$Q(x_{B},p_{B}|x_{\theta j}^{A})$ is evaluated by integrating over
coordinates of $A$, from either $Q_{loop}(\lambda_{rotA},t_{1A}|x_{\theta j}^{A})$,
or by converting to the distribution $Q_{loop}(\lambda,t_{0}|x_{\theta j}^{A})$
at time $t_{0}$. The probabilities of outcomes $x_{\phi k}^{B}$
conditioned on the outcome $\widetilde{\lambda}_{\theta}^{A}=x_{\theta r}^{A}$
at $A$, for a setting-change $\phi$ at $B$ are given by $Q(x_{\phi B},p_{\phi B}|x_{\theta j}^{A})$,
consistent with the coordinate transformation (\ref{eq:rot-6-b-1})
for $B$. \label{fig:projection-setting-change}\textcolor{red}{}}
\end{figure}

There is consistency with the projected state that can be evaluated
by tracking the trajectories of the branch $\mathcal{B}_{\theta j}$
back to the initial time $t_{0}$ (Fig. \ref{fig:projection-setting-change}).
To obtain the inferred distribution $Q_{loop}(\lambda,t_{0}|x_{\theta j}^{A})$
at time $t_{0}$, the amplitudes $x_{\theta A}(t_{1})$, $p_{\theta A}(t_{1})$
of $\mathcal{B}_{\theta j}$ are transformed according to (\ref{eq:rot-6-a-2})
to $x_{A}(t)$ and $p_{A}(t)$, corresponding to a change of setting,
but which is equivalent to changing coordinates in $Q_{loop}(\lambda_{rotA},t_{1A}|x_{\theta j}^{A})$.
 Mathematically, the distribution for $B$ conditioned on $x_{\theta j}^{A}$
does not change under a coordinate transformation, and hence the projected
distribution is $Q(x_{B},p_{B}|x_{\theta j}^{A})$ (refer Appendix
H). $\square$

We illustrate the proof below by calculating the projected state at
$B$ where the initial state of the system at time $t_{0}$ is the
entangled state $|x_{i}^{A}\rangle|x_{k}^{B}\rangle+|x_{l}^{A}\rangle|x_{m}^{B}\rangle$
given by Eq. (\ref{eq:state-3}). This leads to the following Comment.

\emph{Comment (1): Entangled states give projected states that are
superpositions:} Where there is entanglement, the projected state
at $B$ given an outcome at $A$ is a superposition, at least for
one of the outcomes and for particular choices of setting $\theta$.
For an entangled state, this is also true that the projected state
at $A$ given an outcome at $B$. The result would not hold for a
non-entangled state e.g. consider $|x_{i}^{A}\rangle(|x_{k}^{B}\rangle+|x_{m}^{B}\rangle)$
for which the projected state at $B$ given outcome $x_{\theta j}^{A}$
is $|x_{k}^{B}\rangle+|x_{m}^{B}\rangle$ but the projected state
for $A$ given an outcome $x_{\phi q}^{B}$ after a setting change
$\phi$ at $B$ is $|x_{i}^{A}\rangle$.

We demonstrate by considering the initial state $|\psi\rangle$ at
time $t_{0}$ to be $|x_{i}^{A}\rangle|x_{k}^{B}\rangle+|x_{l}^{A}\rangle|x_{m}^{B}\rangle$.
Changing the basis at $A$ to $\theta$ gives $|x_{i}^{A}\rangle=\sum_{r}d_{ir}|x_{\theta r}^{A}\rangle$,
$|x_{l}^{A}\rangle=\sum_{r}d_{lr}|x_{\theta r}^{A}\rangle$ so that
the expression for the state becomes 
\begin{eqnarray}
|\psi\rangle & \rightarrow & \sum_{r}|x_{\theta r}^{A}\rangle(d_{ir}|x_{k}^{B}\rangle+d_{lr}|x_{m}^{B}\rangle)\label{eq:qproj-1}
\end{eqnarray}
where the $d_{kr}$ are defined in Appendix E. This state is given
by the Q function $Q(\lambda_{rotA},t_{1A})$ in Figures \ref{fig:proj-setting-changeA}
and \ref{fig:projection-setting-change}, which is derived from the
initial state $Q(\lambda,t_{0})$ using knowledge of the coefficients
$c_{i}$ and $\theta$, as explained in the proof above. The deterministic
relations $\mathcal{D}_{1}$ are (Result VII.13)
\begin{equation}
x_{k}^{A}\leftrightarrow x_{\theta r}^{A};x_{m}^{B}\leftrightarrow x_{\theta r}^{A}\label{eq:D1}
\end{equation}
where $x_{k}^{B}\leftrightarrow x_{\theta r}^{A}$ occurs with probability
$|d_{ir}|^{2}$, and $x_{m}^{B}\leftrightarrow x_{\theta r}^{A}$
occurs with probability $|d_{lr}|^{2}$. From (\ref{eq:qproj-1}),
the projected state given an outcome $x_{\theta j}^{A}$ at $A$ for
$\hat{x}_{\theta}^{A}$ is the superposition 
\[
|\psi_{B}\rangle=d_{ij}|x_{k}^{B}\rangle+d_{lj}|x_{m}^{B}\rangle
\]
provided $d_{ij},d_{lj}\neq0$. This corresponds to the projected
distribution $Q(x_{B},p_{B}|x_{\theta j}^{A})$ derived from the distribution
$Q_{loop}(\lambda_{rotA},t_{1A}|x_{\theta j}^{A})$, which is deduced
from the branch $\widetilde{\lambda}_{\theta}^{A}\equiv x_{\theta j}^{A}$
associated with outcome $x_{\theta j}^{A}$ (red double dashed lines
in Fig. \ref{fig:projection-setting-change}), as explained in the
techniques of Result VIII.1c. The projected distribution $Q(x_{B},p_{B}|x_{\theta j}^{A})$
is derived on integrating $Q_{loop}(\lambda_{rotA},t_{1A}|x_{\theta j}^{A})$
over coordinates $x_{\theta A}$ and $p_{\theta A}$ (vertical dashed
lines in Fig. \ref{fig:projection-setting-change}). The same projected
distribution $Q(x_{B},p_{B}|x_{\theta j}^{A})$ is derived, if we
transform the coordinates of $A$ to $x_{A}$ and $p_{A}$ (top horizontal
red dashed lines in Fig. \ref{fig:projection-setting-change}). This
corresponds to the tracing of the trajectories of $A$ through the
reversible transformation (\ref{eq:rot-6-a-1}) given by the measurement-setting
adjustment $H_{\theta}$.

Considering the projected state of $A$ given an outcome $x_{s}^{B}$
at $B$, we find the projected state at $A$ is the superposition
of type $|\psi_{A}\rangle=d_{ks}|x_{i}^{A}\rangle+d_{ms}|x_{l}^{A}\rangle$.

\emph{Comment (2):} The inference of the projected state at $A$ requires
knowledge of the setting $\theta$ at $A$, and the deterministic
relations. The inference is based on retrodiction. The postselected
state $Q_{loop}(\lambda_{rotA},t_{1A}|x_{\theta j}^{A})$ is derived
solely from knowledge of $x_{\theta j}^{A}$ and the relation $x_{\theta j}^{A}\leftrightarrow Gx_{\theta j}^{A}$,
where $G$ is the amplification factor from $H_{amp}$, as well as
the deterministic relations $\mathcal{D}_{1}$, and the setting transformation
(\ref{eq:rot-6-a-1}). The specific value of the noise inputs $\eta$
or $\xi$ is not required. This is because there is no change to the
noise properties as the system propagates in the backward-time direction.

\emph{Comment (3): }The appearance of the superposition $|\psi_{B}\rangle$
in Comment (1) above is consistent with Result VII.5. A change of
setting at $A$ changes the interference term $\mathcal{I}nt_{AB}$
in the Q function of $|x_{j}^{A}\rangle|x_{k}^{B}\rangle+|x_{l}^{A}\rangle|x_{m}^{B}\rangle$
at time $t_{0}$, so that the integration over $x_{\theta A}$ and
$p_{\theta A}$ leads to a contribution that need not be zero. This
means that the interference term can contribute to the projected state
calculated from $Q_{loop}(\lambda_{rotA},t_{1A}|x_{\theta j}^{A})$,
and the resulting Q function $Q(x_{B},p_{B}|x_{\theta j}^{A})$ can
contain an interference term $\mathcal{I}_{B}$, thus allowing for
the superposition $|\psi_{B}\rangle$. Similarly, in reverse, going
from time $t_{1A}$ to $t_{0}$, the projected distribution $Q(x_{B},p_{B}|x_{\theta j}^{A})$
calculated by integrating $Q_{loop}(\lambda,t_{0}|x_{\theta j}^{A})$
over $x_{A}$ and $p_{A}$ can contain interference terms $\mathcal{I}_{B}$
in $B$.

\emph{Comment (4): Analogy for spin systems:} The procedure can be
illustrated by an analogy with basis changes for the system prepared
at time $t_{0}$ in the spin Bell state
\begin{equation}
|\psi_{bell},t_{0}\rangle=\frac{1}{\sqrt{2}}(|+\rangle_{zA}|-\rangle_{zB}-|-\rangle_{zA}|+\rangle_{zB})\label{eq:bell-1}
\end{equation}
which corresponds in the analogy as $Q(\lambda,t_{0})$. Here $|\pm\rangle_{zK}$
are the eigenstates of $\hat{\sigma_{z}}$ for system $K\in\{A,B\}$.
We take $\hat{\sigma}_{\theta}^{K}=\hat{\sigma}_{z}^{K}\cos\theta+\hat{\sigma}_{x}^{K}\sin\theta$.
At $A$, a change of setting from spin $z$ to spin $\hat{\sigma}_{\theta}^{A}$
changes the basis: the new basis states are $|\pm\rangle_{\theta A}=\cos\frac{\theta}{2}|\pm\rangle_{zA}\pm\sin\frac{\theta}{2}|\mp\rangle_{zA}$.
The state at time $t_{1A}$ is 
\begin{eqnarray}
|\psi_{bell},t_{1A}\rangle & = & \frac{1}{\sqrt{2}}\{|+\rangle_{\theta A}(\cos\frac{\theta}{2}|-\rangle_{zB}-\sin\frac{\theta}{2}|+\rangle_{zB})\nonumber \\
 &  & \ \ -|-\rangle_{\theta A}(\cos\frac{\theta}{2}|+\rangle_{zB}+\sin\frac{\theta}{2}|-\rangle_{zB})\}\nonumber \\
\label{eq:state-bell-2}
\end{eqnarray}
corresponding in the analogy to $Q(\lambda_{rotA},t_{1A})$. This
gives the expression in terms of the measurement basis for spin $\hat{\sigma}_{\theta}^{A}$
at $A$ and spin $z$ at $B$. The state conditioned on the $+$ outcome
at $A$ is $|+\rangle_{\theta A}|\psi_{Bsup}\rangle$ where $|\psi_{Bsup}\rangle=\cos\frac{\theta}{2}|-\rangle_{zB}-\sin\frac{\theta}{2}|+\rangle_{zB}$.
The projected state at $B$ given a $+$ outcome at $A$ is the superposition
$|\psi_{Bsup}\rangle$, in analogy with the Q function $Q(x_{B},p_{B}|x_{\theta j}^{A})$.
There is no direct analogy with $Q_{loop}(\lambda_{rotA},t_{1A}|x_{\theta j}^{A})$.
The change of basis back to spin $\sigma_{z}^{A}$ at $A$ gives 
\begin{equation}
(\cos\frac{\theta}{2}|+\rangle_{zA}+\sin\frac{\theta}{2}|-\rangle_{zA})|\psi_{Bsup}\rangle\label{eq:change-basis-bell}
\end{equation}
which corresponds to $Q_{loop}(\lambda,t_{0}|x_{\theta j}^{A})$.
The projected state for $B$ is unchanged, given by $|\psi_{Bsup}\rangle$
and hence $Q(x_{B},p_{B}|x_{\theta j}^{A})$. The projected state
describes the predictions at $B$, for future setting changes. e.g.
if there is then a change of setting to $\phi$ at $B$, then the
state at $B$ is
\begin{eqnarray*}
|\psi_{Bsup,\phi}\rangle & = & \cos(\frac{\theta-\phi}{2})|-\rangle_{\phi B}+\sin(\frac{\theta-\phi}{2})|+\rangle_{\phi B}
\end{eqnarray*}
where $|\pm\rangle_{\phi B}=\cos\frac{\phi}{2}|\pm\rangle_{zB}\pm\sin\frac{\phi}{2}|\mp\rangle_{zB}$.
The expected value of the spin product given $+$ at $A$ is hence
$E_{+}(\theta,\phi)=-\cos(\theta-\phi)$. The expected value conditioned
on a negative spin $\hat{\sigma}_{\theta}^{A}$ at $A$ can be similarly
calculated, to give agreement with the overall expected value, $E(\theta,\phi)=-\cos(\theta-\phi)$.

We consider the case where $A$ and $B$ are reversed, so that a setting
change $\phi$ takes place first at $B$. The projected state for
$A$ given a positive spin outcome for $\hat{\sigma}_{\phi}^{B}$
is the superposition $|\psi_{Asup}\rangle=(\cos\frac{\phi}{2}|-\rangle_{zA}-\sin\frac{\phi}{2}|+\rangle_{zA})$.
There is hence a \emph{mutual consistency} of the projected states
with respect to the initial state $|\psi_{bell}\rangle$ (refer Result
IX.3).

We compare with the mixed state $\rho_{mix}^{AB}$ defined as the
50/50 mixture of $|+\rangle_{zA}|-\rangle_{zB}$ and $|-\rangle_{zA}|+\rangle_{zB}$.
In the analogy, the initial state $Q(\lambda,t_{0})$ has no interference
term. At $A$, a change of setting from spin $z$ to spin $\hat{\sigma}_{\theta}^{A}$
changes the basis to $|\pm\rangle_{\theta A}=\cos\frac{\theta}{2}|\pm\rangle_{zA}\pm\sin\frac{\theta}{2}|\mp\rangle_{zA}$,
but the projected state for $B$ given a positive spin outcome at
$A$ is a mixture of up and down spin states at $B$.

\subsection{Timing of the Projection}

In the Introduction, we raised the question as to the timing of the
collapse of the wave function. From the Results, we see there is consistency
with Premises (3) of \emph{weak macroscopic realism} (wMR) and \emph{weak
local realism} (wLR), defined in Definitions (6c) and (10c) in Sec.
I.B, which imply a constraint on the time of projection.

\textbf{\emph{Result VIII.2:}} \textbf{\emph{Weak Macroscopic Realism,
Premise wMR(3): }}Consider the systems $A$ and $B$ prepared at time
$t_{0}$ with respect to a particular measurement basis which, without
loss of generality, we assume to be $\hat{x}_{A}$ and $\hat{x}_{B}$
as in (Figure \ref{fig:projection-1}). Suppose the measurement setting
at $A$ has been fixed and the system $A$ amplified, so that distinct
branches symbolized by $\widetilde{\lambda}_{x}^{A}$ have emerged
at a time $t_{mA}$ (Figure \ref{fig:projection-1}). The Premise
(1) of weak macroscopic realism asserts that the outcome $x_{j}^{A}$
for $\hat{x}_{A}$ at $A$ is specified at time $t_{mA}$ (Result
IV.6).  The Q model of reality and simulations are consistent with
the third premise of wMR, Premise wMR(3), defined below.

The Premise wMR(3) of \emph{weak macroscopic realism} posits that
the constraints given by the projected distributions $Q(x_{B},p_{B}|\widetilde{\lambda}_{x}^{A})\equiv Q(x_{B},p_{B}|x_{j}^{A})$
on the outcomes for $B$ describe the system $B$ at (or by) the time
$t_{mA}$, after sufficient amplification at $A$ has taken place
so that weak macroscopic realism (Premise wMR(1)) can be applied.
The $Q(x_{B},p_{B}|x_{j}^{A})$ implies predictions for $B$, for
any future setting change $\phi$ at $B$ that might take place. 

\emph{Proof:} This follows from Result VIII.1a. The trajectories for
$x_{A}(t_{mA})$ associated with the branch $\widetilde{\lambda}_{x}^{A}$
corresponding to outcome $x_{j}^{A}$ are defined at the time $t_{mA}$.
NB: The noise properties $\delta x(t)$ are unchanged with the reverse
propagation and the branch has fixed width due to the noise $\delta x(t)$
(Results III.2-3). The Q function $Q(\bm{\lambda},t_{0})$ for the
entangled system is defined at the earlier time $t_{0}$. The distribution
$Q_{loop}(\lambda|x_{j}^{A})$ can be deduced from $Q(\lambda,t_{0})$,
and hence so can $Q(x_{B},p_{B}|x_{j}^{A})$. Hence the $Q(x_{B},p_{B}|x_{j}^{A})$
is determined at time $t_{mA}$. The second statement of Premise wMR(3)
follows from Result VIII.1b.

If a setting change $\theta$ has taken place at $A$ at a time $t_{2}$
between the times $t_{0}$ and $t_{mA}$, then the distribution $Q(x_{B},p_{B}|x_{\theta j}^{A})$
can be deduced from the $Q_{loop}(\lambda_{rotA},t_{2}|x_{\theta j}^{A})$
where $\lambda_{rotA}=(x_{\theta A},p_{\theta A},x_{B},p_{B})$ or
else equivalently by tracing the amplitudes $x_{A}(t)$ back to the
time $t_{0}$, to deduce $Q_{loop}(\lambda,t_{0}|x_{\theta j}^{A})$,
as explained in Result VIII.1e. $\square$

\emph{Comment:} The projected state $Q(x_{B},p_{B}|x_{j}^{A})$ is
defined at the time $t_{mA}$, and it has been assumed that no setting
change has taken place at $B$, as in Figure \ref{fig:projection-1}.
Hence, the last statement refers to predictions for future changes
at $B$. In fact the projected distribution $Q(x_{B},p_{B}|x_{j}^{A})$
is defined to describe the system $B$ \emph{at the time $t_{0}$.
}We see from Result VIII.1b (and Appendix H) that this distribution
$Q(x_{B},p_{B}|x_{j}^{A})$ then defines the predictions at $B$ conditioned
on outcome $x_{j}^{A}$ at $A$ for any time $t>t_{0}$, which may
be prior to $t_{mA}$.

We generalize the Result, to apply to \emph{weak local realism }(wLR),
as introduced in Definition 10 in Sec. I.B and in Sec. IV.G (Result
IV.14).

\textbf{\emph{Result VIII.3:}} \textbf{\emph{Weak Local Realism, Premise
wLR(3): }}Suppose the measurement setting at $A$ has been fixed at
a time $t_{1A}$, meaning that the unitary operations $U_{\theta}^{A}$
that fix the setting as $\theta$ have been carried out. In Figure
\ref{fig:projection-1}, the time $t_{1A}$ corresponds to time $t_{0}$.
There are two versions of weak local realism, deterministic and probabilistic,
as explained in Result IV.14 (Sec. IV.G), for which the Premises (3)
are defined below. The Q model of reality and simulations are consistent
with the Premises wLR(3).

(i) The Premise wLR(3) of the \emph{deterministic }version of weak
local realism is based on Premise wLR(1), that the value $x_{\theta j}^{A}$
for the outcome of $\hat{x}_{\theta A}$ is determined for system
$A$ at time $t_{1A}$. The Premise wLR(3) asserts that the system
$B$ is accordingly constrained according to $Q(x_{B},p_{B}|x_{\theta j}^{A})$
at (or by) the time $t_{1A}$, when the setting at $A$ is specified
(provided the setting at $A$ remains fixed).

(ii) The Premise wLR(3) of the \emph{probabilistic} version of weak
local realism assumes that hidden variables $\{\lambda_{\theta}^{A}\}$
exist to describe the system $A$ at time $t_{1A}$, after settings
at $A$ are fixed. There exists a projected distribution $Q(x_{B},p_{B}|\{\lambda_{\theta}^{A}\})$
defined for system $B$ given that system $A$ is specified by $\{\lambda_{\theta}^{A}\}$.
The probabilistic version of Premise wLR(1) (Result IV.14, Definition
10a) asserts that there exists a probability $P(x_{\theta j}^{A}|\{\lambda_{\theta}^{A}\})$
of an outcome $x_{\theta j}^{A}$ for $\hat{x}_{\theta A}$ at $A$,
given the system $A$ is described by $\{\lambda_{\theta}^{A}\}$.
The projected distribution $Q(x_{B},p_{B}|\{\lambda_{\theta}^{A}\})$
is the mixture of projected distributions $Q(x_{B},p_{B}|x_{\theta j}^{A})$
weighted by $P(x_{\theta j}^{A}|\{\lambda_{\theta}^{A}\})$.

\emph{Proof:} (i) First, we consider the deterministic version of
wLR(3), where the outcome $x_{\theta j}^{A}$ at $A$ is assumed determined
at the time $t_{1A}$ (Result IV.14 for Premise wLR(1)), after the
setting at $A$ is fixed. Provided the setting at $A$ is fixed, we
know from Result VII.3 and Appendix F that the quantum predictions
(and hence those of the Q model and simulation) are consistent with
a LHV model $\mathcal{P}_{asym,A}$, based on the partially mixed
state $\rho_{mix,A}$ given by (\ref{eq:mix-asymA}). In the model
$\mathcal{P}_{asym,A}$, each realization of the system $A$ is in
a state with a definite predetermined outcome $x_{\theta j}^{A}$
for $\hat{x}_{\theta A}$ at $A$ at the time $t_{1A}$. The state
of system $B$ is the projected state for $B$, as correlated with
a given outcome $x_{\theta j}^{A}$: Hence, in the model $\mathcal{P}_{asym,A}$,
the predictions for $B$ are defined at the time $t_{1A}$, consistent
with Premise wLR(3). The Result follows, since the predictions of
the Q model (and of quantum mechanics) do not contradict those of
the model $\mathcal{P}_{asym,A}$ when the setting at $A$ is fixed.

(ii) Second, we assume the probabilistic version of wLR. This version
is directly verified by the Q model. The probabilities $P(x_{\theta j}^{A}|\{\lambda_{\theta}^{A}\})$
for system $A$ at time $t_{1A}$, where the variables $\lambda_{\theta}^{A}$
correspond to $x_{\theta A}$, $p_{\theta A}$, can be deduced (by
integration over variables for $B$) from Result VII.9a, Eq. (\ref{eq:probB-1}),
where a HV distribution $P(x_{\theta j}^{A},x_{\phi k}^{B}|\lambda_{rot})$
is derived consistently with the Q model an the predictions of quantum
mechanics. Here, $\lambda_{rot}=(x_{\theta A},p_{\theta A},x_{B},p_{B})$,
as we take $\phi=0$.  Hence, considering the system $A$ at time
$t_{1A}$ as defined by the distribution $Q(\lambda_{rotA},t_{1A})$,
the probability of an outcome $x_{j}^{A}$ at $A$ is known from the
values of the local hidden variable, $x_{\theta A}$ and $p_{\theta A}$.
For each $x_{j}^{A}$, the projected distribution $Q(x_{B},p_{B}|x_{j}^{A})$
can be deduced at $A$ based on knowledge of the initial state $Q(\lambda,t_{0})$
and any local setting change $\theta$ at $A$ (Result VIII1a). Hence,
the overall projected distribution for system $B$ can be deduced
by Alice at $A$ given $\{\lambda_{\theta}^{A}\}$ at a time $t_{1A}$
(within the framework of wLR) as the weighted mixture given by $\sum_{x_{j}^{A}}P(x_{j}^{A}|\{\lambda_{\theta}^{A}\})Q(x_{B},p_{B}|x_{j}^{A})$.
The total probability distribution $P(x_{\theta j}^{A})$ for an outcome
$x_{j}^{A}$ at $A$ is $\int dx_{\theta A}dp_{\theta A}P(\{\lambda_{\theta}^{A}\})P(x_{\theta j}^{A}|\{\lambda_{\theta}^{A}\})$
where $P(\{\lambda_{\theta}^{A}\})$ is the probability of variables
$\{\lambda_{\theta}^{A}\}=\{x_{\theta A},p_{\theta A}\}$ as given
by the marginal $Q_{m}(x_{\theta A},p_{\theta A})=\int dx_{B}dp_{B}Q(\lambda_{rot},t_{1A})$.
This correctly describes the projected state at $B$ given outcome
$x_{j}^{A}$ at $A$, according to quantum mechanics. Hence, the
Premise wLR(3) applies. $\square$

\textbf{\emph{Result VIII.4: Distinction between cause and correlation
in projection}}\textbf{: }We have considered correlated systems, where
the state of system $B$ can be determined by a measurement $\hat{x}_{A}$
at $A$. The measurement of $\hat{x}_{A}$ at $A$ involves, after
the setting is fixed at a time $t_{0}$, an amplification given by
$H_{amp}^{A}$, followed by, at a time after $t_{mA}$ (as depicted
in Fig. \ref{fig:projection-1}), a final detection. If we assume
weak local or macroscopic realism (wLR), as in the Q model, it is
not the case that either the amplification $H_{amp}$ or final detection
at $A$ \emph{cause} the outcome, or the projected state $Q(x_{B},p_{B}|\{\lambda_{\theta}^{A}\})$,
at $B$.

\emph{Proof: }From Results VII.12 and VII.13, we see that the deterministic
relations $\mathcal{D}$ that give the correlations between the outcomes
at $A$ and $B$ are determined at, or by, the time $t_{0}$, which
is before $H_{amp}^{A}$ and the final detection at $A$. According
to wLR (Result VIII.3), the projected state is specified for system
$B$, at (or by) the time $t_{0}$, when the settings at $A$ are
fixed. $\square$

\emph{Comment: No nonlocality of the meter:} A meter is defined as
a system such as $A$ in the Schrödinger cat example (Eq. (\ref{eq:ent-meter-x-1-1})),
which is correlated with $B$ so that the outcome at $A$ gives information
about the outcome of a measurement at $B$. The meter may be spatially
separated from $B$. The meter is configured so that \emph{its settings
are fixed}. Prior interaction with the system $B$ has generated a
state such as (\ref{eq:ent-meter-x-1-1}), where the measurement basis
for $A$ is fixed at $\hat{x}_{A}$. The meter $A$ is also macroscopic.
Hence, the systems $A$ and $B$ are analogous to those described
in Figure \ref{fig:projection-1} at the time $t_{mA}$. Result VIII.4
implies that the final part of the measurement to detect the outcome
at $A$ does not cause the outcome at $B$.

\section{Causal model for Bell nonlocality\label{sec:Causal-structure-bell}}

In the Q model of reality, the $x_{A},x_{B},p_{A},p_{B}$ are hidden
variables. The simulation can be examined to identify a causal (cause-and-effect)
model consistent with Bell nonlocality.  We have demonstrated in
the previous section how the projected state is deduced by retrodiction
based on deterministic relations, and knowledge of the initial state
$Q(\lambda,t_{0})$, only (Results VIII.1a and e). Yet, as we show
in this section, the projected state is sufficient to explain Bell
violations in the Q-based model, where the measurement is carried
out by amplification $H_{amp}$.

The test of a Bell inequality involves preparation of the Bell state.
 The fields are then spatially separated and the settings locally
adjusted as in Figure \ref{fig:sim}, in four sets of runs, to measure
$E(\theta,\phi)$, $E(\theta',\phi)$, $E(\theta,\phi')$ and $E(\theta',\phi')$
as given in (\ref{eq:CHSH}). To gain insight into the dynamics that
leads to the Bell nonlocality, we first analyze the simulation where
one adjusts the settings at the two locations, $A$ and $B$, \emph{sequentially}.
This corresponds to an alternative Scenario II for a Bell test, where
the system is prepared with respect to the measurement basis $\theta$
and $\phi$ at a time $t_{0}$. Then there are three types of runs:
(1) where settings remain fixed at $\theta$ and $\phi$; (2) runs
with a change of setting at one site only, to measure $E(\theta,\phi')$
and $E(\theta',\phi)$; (3) runs with a change of setting at both
sites, to measure $E(\theta',\phi')$. \textcolor{red}{}

\begin{figure}
\begin{centering}
\includegraphics[width=1\columnwidth]{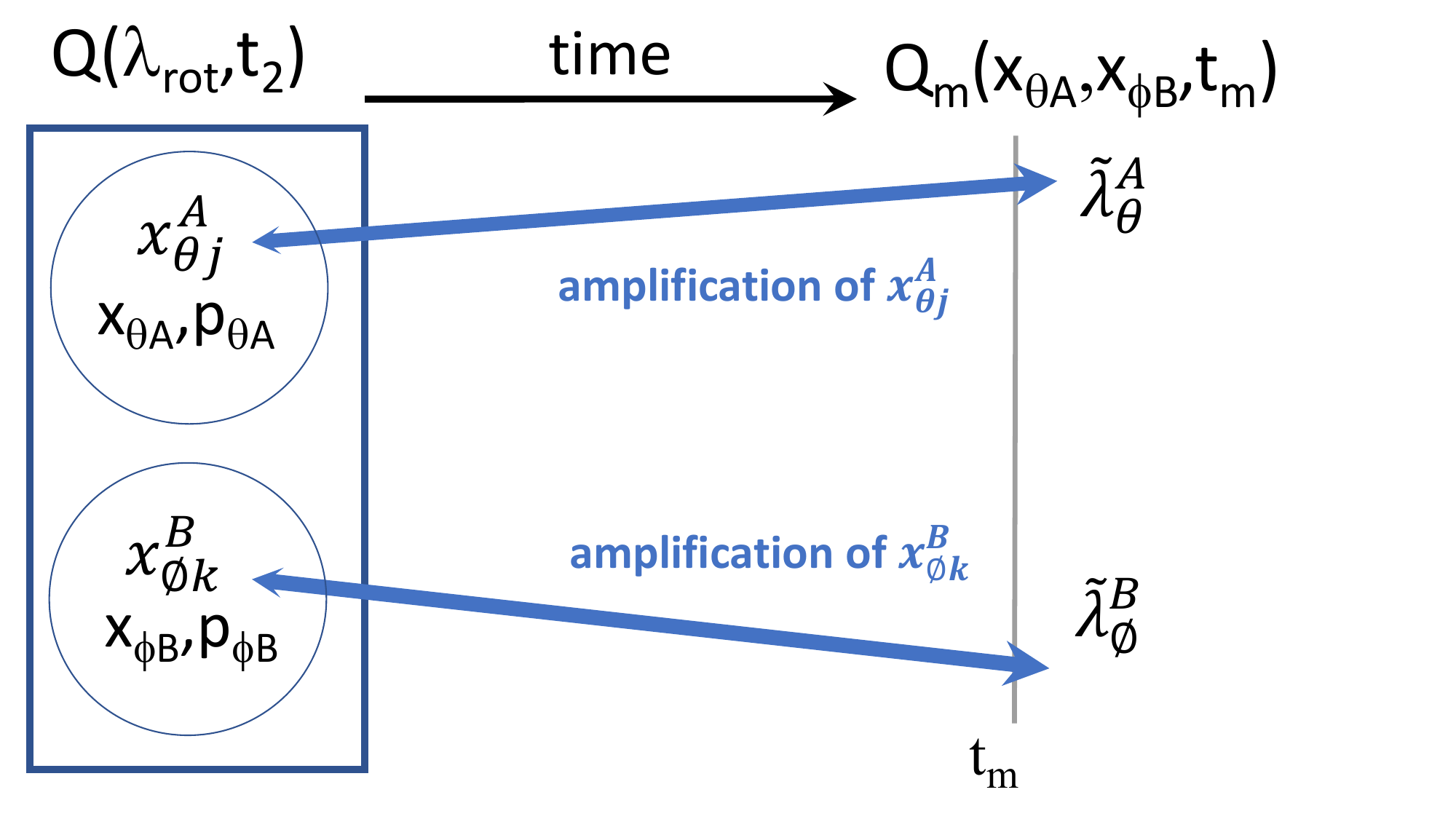}
\par\end{centering}
\caption{\textbf{\emph{Weak macroscopic realism (wMR) and weak local realism
(wLR):}} At the time $t_{2}$, the system has been prepared for measurement
of $\hat{x}_{\theta A}$ and $\hat{x}_{\phi B}$. The distribution
$Q_{m}(x_{\theta A},x_{\phi B},t_{m})$ gives the probability density
of amplitudes $x_{\theta A}(t)$ and $x_{\phi B}(t)$ at time $t_{m}$
after amplification $H_{amp}$ at each site. The blue two-way arrows
denote the causal deterministic relations in which the means $x_{\theta j}^{A}$
and $x_{\phi k}^{B}$ of the Gaussians in $Q(\lambda_{rot},t_{2})$
are amplified to $Gx_{\theta j}^{A}$ and $Gx_{\phi k}^{B}$. The
interference terms $\mathcal{I}nt_{AB}$ in $Q(\lambda_{rot},t_{2})$
do not amplify. Weak macroscopic realism posits that after sufficient
amplification, at time $t_{m}$, the outcomes of $\hat{x}_{\theta A}$
and $\hat{x}_{\phi B}$ are determined. We symbolize these values
by $\widetilde{\lambda}_{\theta}^{A}$ and $\widetilde{\lambda}_{\phi}^{B}$.
According to wLR, the values exist probabilistically as in hidden-variable
model $\mathcal{P}_{B}$ of Eq. (\ref{eq:model-b}), at the earlier
time $t_{2}$, after settings are fixed.\label{fig:epr-sim-xx-1}}
\end{figure}

\subsection{Preparation: measurement-basis of $\hat{x}_{A}$ and $\hat{x}_{B}$}

In the Bell test, we consider first that system is prepared at time
$t_{0}$, with respect to the measurement basis $\hat{x}_{A}$ and
$\hat{x}_{B}$. We ask what the Q model indicates if there is measurement
of $\hat{x}_{A}$ and $\hat{x}_{B}$, without a change of settings.
This corresponds to Figure \ref{fig:epr-sim-xx-1}, with $\theta=\phi=0$
(and $t_{0}=t_{2}$). 

In Sec. VII.E, Results VII.10 and VII.11, we have extended Premises
(1) of weak macroscopic realism (wMR) and weak local realism (wLR),
given by Definitions (6) and (10) in Sec. I.B, and by Results IV.6
and IV.14, so that they apply to the bipartite system. This implies
that once the measurement settings are fixed, as in Figure \ref{fig:epr-sim-xx-1}
at the time $t_{2}$, and an amplification $H_{amp}$ has taken place
at each site, then the values for the outcomes of $\hat{x}_{\theta A}$
and $\hat{x}_{\phi B}$ (given by $x_{\theta j}^{A}$ and $x_{\phi k}^{B}$)
are determined, as in the branches of Figure \ref{fig:epr-3}. Assuming
wLR , the system is viewed as being in a HV state where there is a
certain probability for outcomes, at the time $t_{2}$ when settings
are fixed.

\subsection{Changing the setting at one location: projection and no-signaling}

 Figure \ref{fig:projection-1} shows the system prepared at time
$t_{2}$ for measurements $\hat{x}_{A}$ and $\hat{x}_{B}$, where
$\theta=\phi=0$. Suppose the measurement setting is then changed
at one location, $B$, to $\phi\neq0$, with the setting at $A$ fixed.
We ask: does the change of setting at $B$ influence the system, or
outcome, at $A$ i.e. is there a nonlocal effect at $A$? We show
that the simulations modeling the Bell violations are consistent with
no-signaling.

\subsubsection{Mutual consistency of projection}

\begin{figure}
\begin{centering}
\par\end{centering}
\begin{centering}
\includegraphics[width=1\columnwidth]{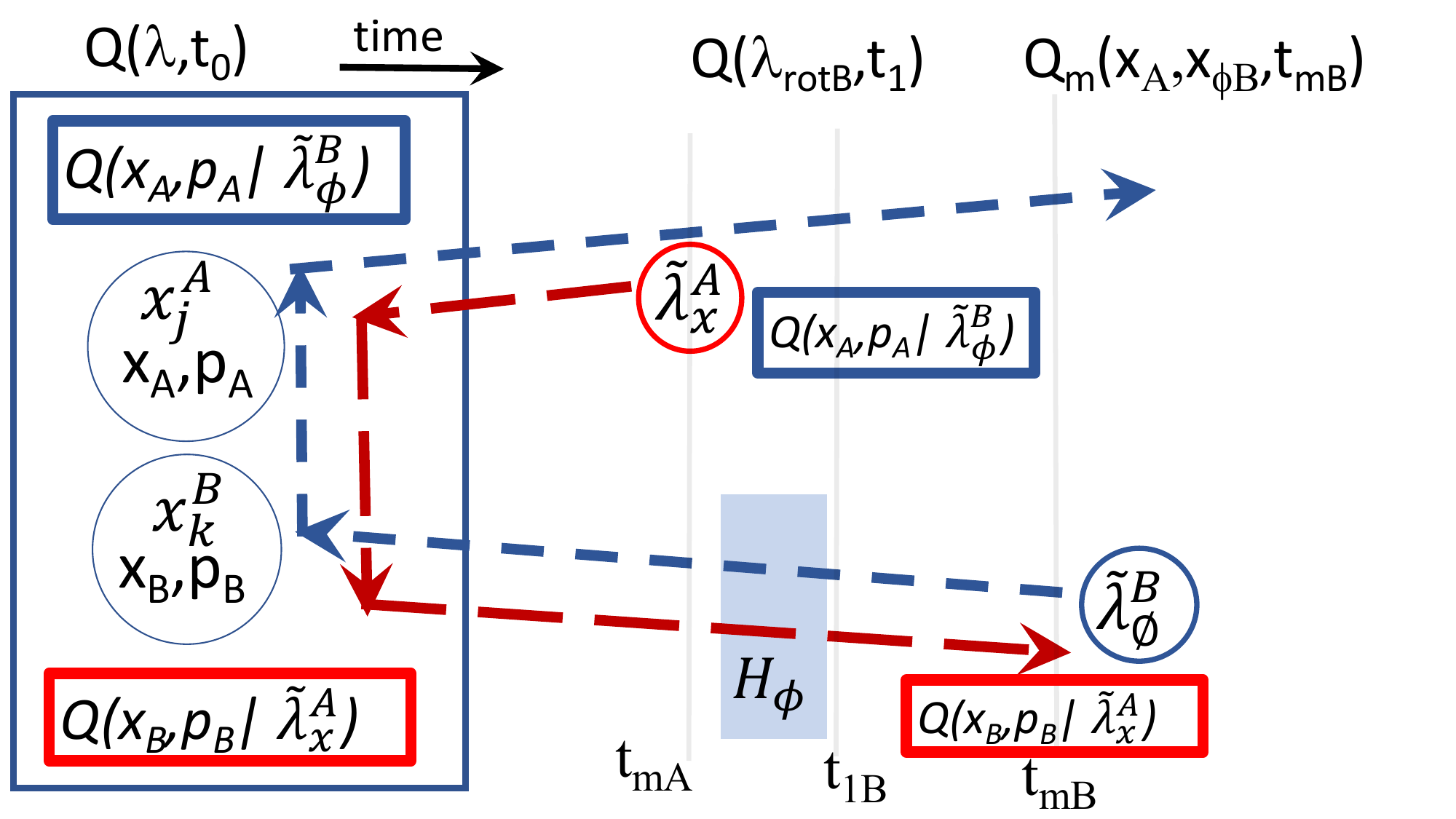}
\par\end{centering}
\caption{\textbf{\emph{Mutual projection and no-signaling: }}Depiction of the
constraints on the system $A$ of Figure \ref{fig:projection-1} due
to the change of setting at $B$ to $\phi$ at time $t_{1B}$, while
the setting at $A$ is fixed at $x$. At the time $t_{mB}$, after
amplification at $B$, the value $\widetilde{\lambda}_{\phi}^{B}$
determines the outcome for $\hat{x}_{\phi B}$. The amplitudes at
$B$ conditioned on branch $\widetilde{\lambda}_{\phi}^{B}$ correlate
with certain amplitudes for $A$ at $t_{0}$, as determined by $Q(\lambda,t_{0})$.
This ensures (blue dashed arrows) the projected state at $A$ is given
as $Q(x_{A},p_{A}/\widetilde{\lambda}_{\phi}^{B})$. There is also
projection from $A$ to $B$ (red dashed arrows) as in Figure \ref{fig:projection-1}.
The mutual projection is consistent with the density of amplitudes
at $t_{mB}$, as given by the marginal $Q_{m}(x_{A},x_{\phi B},t_{mB})$
of the Q function of the amplified state. The change of setting at
$B$ induces a change (blue dashed arrows) to the hidden interference
terms $\mathcal{I}nt_{AB}$ of the Q distribution $Q(\lambda_{rotB},t_{1B})$,
but does not change the observed probabilities (given by $Q(x_{A},p_{A}/\widetilde{\lambda}_{\phi}^{B})$)
over what can be explained by a local transformation at $B$.\label{fig:feedback-settingB}
Hence, in the Q model, the value $\widetilde{\lambda}_{x}^{A}$ giving
the outcome of $\hat{x}_{A}$ is unchanged, which implies \emph{no-signaling}.}
\end{figure}

\textcolor{red}{}First, we note that there are mutual constraints
on the two systems, due to projection, depicted by the red and blue
dashed lines in Fig. \ref{fig:feedback-settingB}). If there is a
change of setting to $\phi$ at $B$, then there are constraints imposed
on system $B$ by the fixed setting at $A$ (Result VIII.1, Sec.
VIII.A) as depicted in Figure \ref{fig:projection-1}, and by the
red dashed line in Figure \ref{fig:feedback-settingB}. The state
for system $B$ conditioned on the branch $\widetilde{\lambda}_{x}^{A}$
of $A$ is the projected state, distribution $Q(x_{B},p_{B}|\widetilde{\lambda}_{x}^{A})$,
given by Eq. (\ref{eq:cond-1}).

However, there is also a constraint to $A$ from $B$ (blue dashed
line in Fig. \ref{fig:feedback-settingB}). After amplification at
$B$, at time $t_{mB}$, there is a value, symbolized by $\widetilde{\lambda}_{\phi}^{B}$,
for the outcome of measurement $\hat{x}_{\phi B}$. The state at $A$
conditioned on the outcome associated with $\widetilde{\lambda}_{\phi}^{B}$
at $B$ is determined by the initial Q function and any relevant deterministic
relations, as explained in Sec. VIII.C. This defines the projected
state 
\begin{equation}
Q(x_{A},p_{A}|\widetilde{\lambda}_{\phi}^{B})\label{eq:cond-A}
\end{equation}
at $A$, which is the Q function for the reduced state of $A$ given
the outcome $\widetilde{\lambda}_{\phi}^{B}$ at $B$ (Result VIII.1c).

The $Q(x_{A},p_{A}|\widetilde{\lambda}_{\phi}^{B})$ and $Q(x_{B},p_{B}|\widetilde{\lambda}_{\theta}^{A})$
are mutually consistent with $Q(\bm{\lambda},t_{0})$ and the Q function
at time $t_{mB}$. Figure \ref{fig:feedback-settingB} depicts the
mutual consistency. 

\textbf{\emph{Definition: Mutual consistency of projection: }}The
mutual consistency of the projected probability distributions $Q(x_{A},p_{A}|\widetilde{\lambda}_{\phi}^{B})$
and $Q(x_{B},p_{B}|\widetilde{\lambda}_{x}^{A})$, as depicted in
Figure \ref{fig:feedback-settingB}, means that they co-exist for
the same system, independently of the time-order of the setting changes
$\theta$ and $\phi$ at $A$ and $B$. The distributions are defined
consistently with the Q function describing the bipartite system at
all times of the dynamics, including at time $t_{0}$.

\begin{figure}
\begin{centering}
\par\end{centering}
\begin{centering}
\includegraphics[width=1\columnwidth]{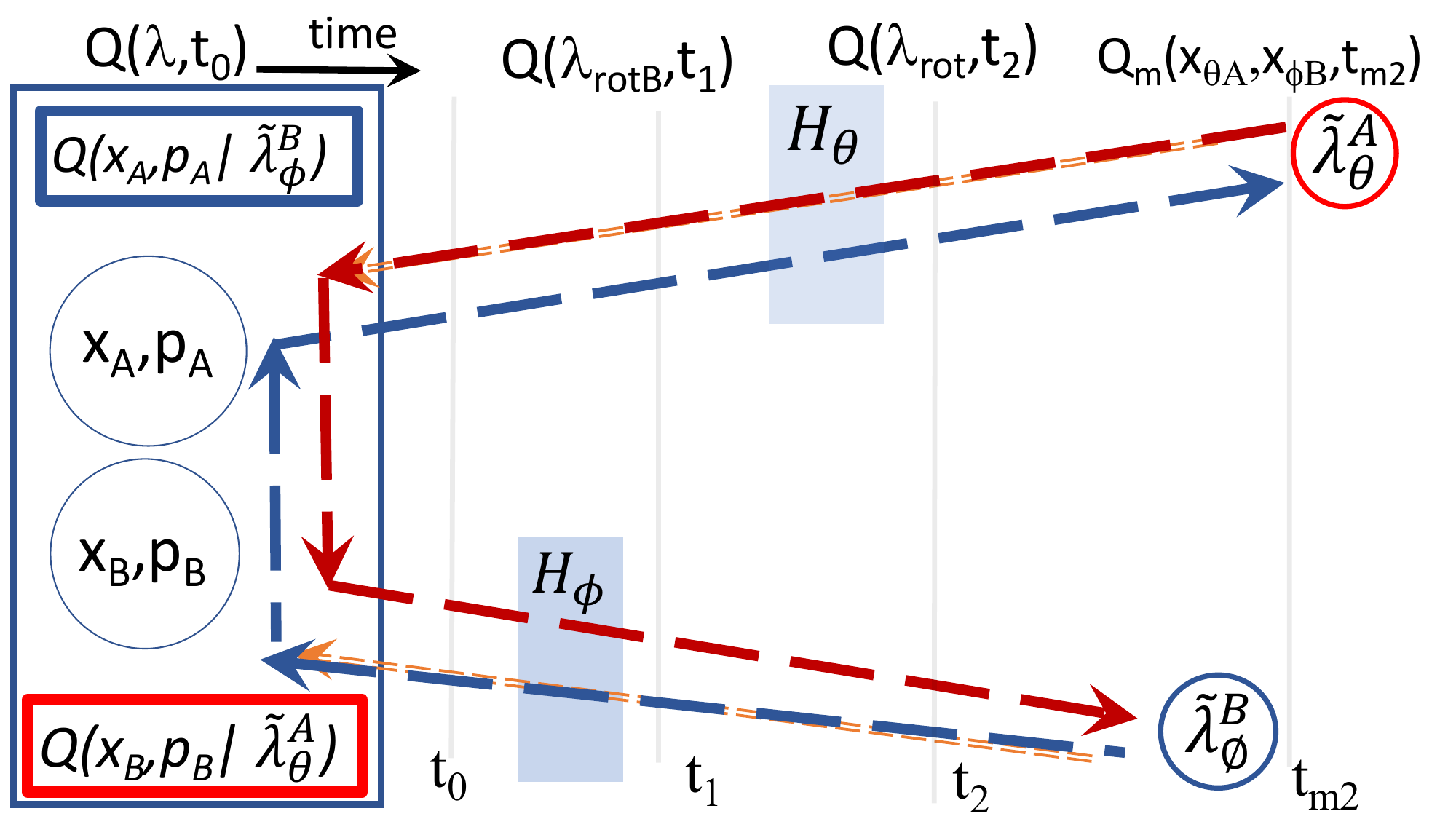}
\par\end{centering}
\caption{\textbf{\emph{Depiction of the mutual consistency of projection}}.
There is a change of setting $\phi$ at $B$ and a change of setting
$\theta$ at $A$. The Q function changes after each change of setting,
as described in Eqs. (\ref{eq:q1-1-2}) and (\ref{eq:q1-1-1}). The
projected states $Q(x_{A},p_{A}|\widetilde{\lambda}_{\phi}^{B})$
and $Q(x_{B},p_{B}|\widetilde{\lambda}_{\theta}^{A}$) are mutually
consistent with the Q function $Q(\lambda,t_{0})$. Each is derived
based only on the knowledge of the local setting (and outcome) and
the initial state $Q(\lambda,t_{0})$. The distributions $Q(x_{A},p_{A}|\widetilde{\lambda}_{\phi}^{B})$
and $Q(x_{B},p_{B}|\widetilde{\lambda}_{\theta}^{A}$) are hence independent
of the time order of the setting-changes, $H_{\phi}$ and $H_{\theta}$.
Any Bell-nonlocal effect is demonstrated after changes of setting
occur at both locations. This avoids time-order paradoxes, which arise
when in one frame of reference the setting is changed at $B$ first,
and in another frame of reference the setting is changed at $A$ first
(refer Sec. IX.C). \label{fig:mutual-Bell-nonlocality}}
\end{figure}

\textbf{\emph{Result IX.1a: Mutual Consistency of projection:}} The
projected probability distributions $Q(x_{A},p_{A}|\widetilde{\lambda}_{\phi}^{B})$
and $Q(x_{B},p_{B}|\widetilde{\lambda}_{x}^{A})$ depicted in Figure
\ref{fig:feedback-settingB} are mutually consistent. We can generalize
the consistency to the case depicted in Figures 5 and \ref{fig:mutual-Bell-nonlocality},
where a setting change $\theta$ has occurred at $A$ after the time
$t_{0}$ and prior to the time $t_{mA}$ depicted in Figure \ref{fig:feedback-settingB},
so that the projected distribution $Q(x_{B},p_{B}|\widetilde{\lambda}_{\theta}^{A})$
is defined (Result VIII.1e).

\emph{Proof:} From Results VIII.1a and VIII.1e given in Sec. VIII,
we see that the evaluation of $Q(x_{B},p_{B}|\widetilde{\lambda}_{\theta}^{A})\equiv Q(x_{B},p_{B}|x_{\theta j}^{A})$
requires the knowledge of the setting $\theta$ at $A$ and of the
initial Q function $Q(\lambda,t_{0})$. The evaluation does not require
knowledge of $\phi$, the setting at $B$. It is hence established
locally at $A$, independent of whether a setting change $\phi$ has
already occurred at $B$. The distribution $Q(x_{B},p_{B}|\widetilde{\lambda}_{\theta}^{A})$
gives predictions of probabilities at $B$ conditioned on the outcome
$x_{\theta j}^{A}$, for all choices of setting $\phi$ at $B$ that
may occur after time $t_{mA}$, or that have occurred after time $t_{0}$.
This is because the probability distribution for the measurement $\hat{x}_{\phi B}$
at $B$ is given by transforming from coordinates $x_{B}$, $p_{B}$
to $x_{\phi B}$ and $p_{\phi B}$, as explained in Results VIII.1b
and AH.2. A similar result applies to $Q(x_{A},p_{A}|\widetilde{\lambda}_{\phi}^{B})\equiv Q(x_{A},p_{A}|x_{\phi k}^{B})$
where $x_{\phi k}^{B}$ is the outcome at $B$.

The proof of causal consistency (Results IX.9 and AH.1) demonstrates
how the projected distributions are consistent with $Q(\lambda,t_{0})$.
Details are given by Results AH in Appendix H. $\square$

\emph{Comment:} From the above, we see how the Q-based model allows
a mutual consistency of projected states, where the initial state
at time $t_{0}$ is given as a probabilistic distribution over variables
$\lambda=(x_{A},p_{A},x_{B},p_{B})$.

\textbf{\emph{Result IX.1b: Paradox of the mutual consistency of projection:}}\emph{
}A paradox seemingly arises. Consider Figure \ref{fig:mutual-Bell-nonlocality}
where there are two changes of settings, $\theta$ at $A$ and $\phi$
at $B$. We see from Result VIII.1e and the related Comments that
the projected distribution $Q(x_{B},p_{B}|\widetilde{\lambda}_{\theta}^{A})\equiv Q(x_{B},p_{B}|x_{\theta j}^{A})$
conditioned on an outcome $x_{\theta j}^{A}$ at $A$ is based on
a\emph{ }set $\mathcal{D}_{1}$ of\emph{ deterministic} \emph{relations}
implied by the asymmetric local hidden variable (LHV) model $\mathcal{P}_{asym,B}$
given by $\rho_{mix,B}$ as in Result VII.3, which correctly describes
the outcomes of $\hat{x}_{\theta A}$ and $\hat{x}_{B}$ (refer Result
VII.13). These deterministic relations determine the correlations
between outcomes of $\hat{x}_{\theta A}$ and $\hat{x}_{B}$, but
assume one setting change only, in this case at $A$. The deterministic
relations $\mathcal{D}_{1}$ are reversible, allowing inference of
the $Q(x_{B},p_{B}|x_{\theta j}^{A})$ at the initial time.

The paradox is how can the second projected distribution $Q(x_{A},p_{A}|x_{\phi k}^{B})$
be consistent with $Q(x_{B},p_{B}|x_{\theta j}^{A})$, since the derivation
of $Q(x_{A},p_{A}|x_{\phi k}^{B})$ assumes a LHV model $\mathcal{P}_{asym.A}$
based on $\rho_{mix,A}$ defined by Result VII.3, which also assumes
a change of setting at \emph{one} location, in this case $B$, only.
A second set $\mathcal{D}_{2}$ of deterministic relations is used
for derivation of $Q(x_{A},p_{A}|x_{\phi k}^{B})$, but both sets
of deterministic relations are consistent with LHV models. How is
the existence of the mutually compatible projected states consistent
with the violation of LHV models, as implied by Bell's theorem?

\begin{figure}
\begin{centering}
\par\end{centering}
\begin{centering}
\includegraphics[width=1\columnwidth]{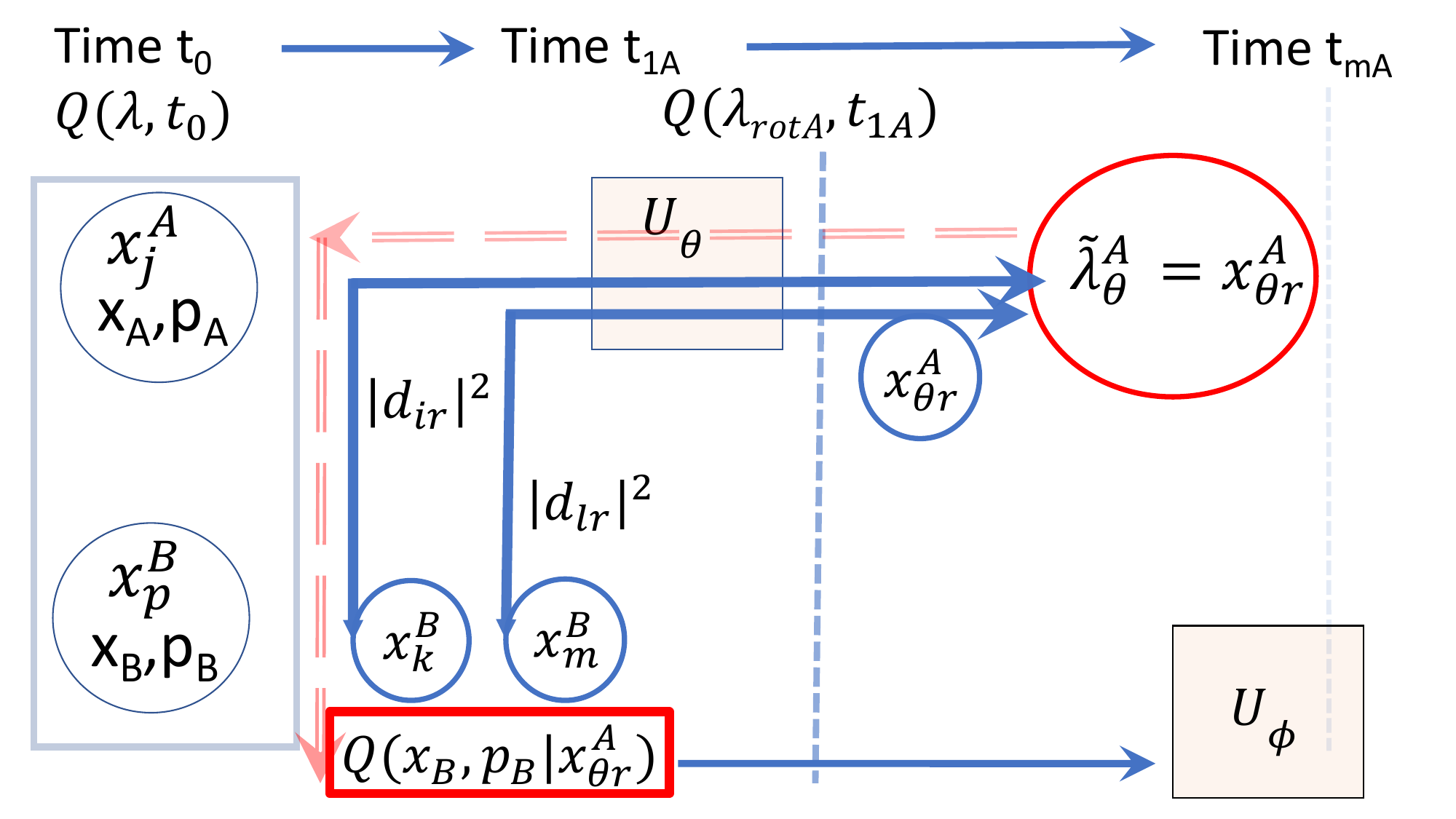}
\par\end{centering}
\caption{\textbf{\emph{Deterministic relations and the mutual consistency of
projection:}} The diagram depicts the evaluation of the projected
distribution $Q(x_{B},p_{B}|x_{\theta r}^{A})$ for system $B$, conditioned
on an outcome $x_{\theta j}^{A}$, as in Figure \ref{fig:projection-setting-change},
and how this is consistent with evaluation of $Q(x_{A},p_{A}|x_{\phi s}^{B})$
at $B$. We consider that $Q(\lambda,t_{0})$ represents the entangled
state $|x_{i}^{A}\rangle|x_{k}^{B}\rangle+|x_{l}^{A}\rangle|x_{m}^{B}\rangle$.
 For certain $\theta$, outcomes $x_{k}^{B}$ and $x_{m}^{B}$ can
both be consistent with a specific outcome $x_{\theta r}^{A}$ at
$A$, according to deterministic relations $\mathcal{D}_{1}$ depicted
by the solid thick blue arrows (Result IX.1). The relations $\mathcal{D}_{1}$
are $x_{k}^{B}\leftrightarrow x_{\theta r}^{A}$ with probability
$|d_{ir}|^{2}$; $x_{m}^{B}\leftrightarrow x_{\theta r}^{A}$ with
probability $|d_{lr}|^{2}$. Hence, retrodiction implies the outcome
at $B$ given outcome $x_{\theta r}^{A}$ at $A$ would be either
$x_{k}^{B}$ or $x_{m}^{B}$. Knowledge of $Q(\lambda,t_{0})$ and
$Q(\lambda_{rotA},t_{1A})$ allows inference that the projected state
$Q(x_{B},p_{B}|x_{\theta j}^{A})$ is a superposition  $|\psi_{B}\rangle=d_{ij}|x_{k}^{B}\rangle+d_{lj}|x_{m}^{B}\rangle$.
Similarly, deterministic relations $\mathcal{D}_{2}$ imply  that
for certain choices of $\phi$, the projected state $Q(x_{A},p_{A}|x_{\phi s}^{B})$
for system $A$ given an outcome $x_{\phi s}^{B}$ at $B$ is a superposition
$|\psi_{A}\rangle=d_{ks}|x_{i}^{A}\rangle+d_{ms}|x_{l}^{A}\rangle$.
\label{fig:projection-smutual-setting}Each projected state is deduced
from deterministic relations, and knowledge of the initial state $Q(\lambda,t_{0})$
as well as the setting at the other site, and is sufficient to explain
Bell violations.}
\end{figure}

\emph{Resolution of the paradox:} The resolution is that the projected
state for $B$ given an outcome at $A$ is inferred for the system
$B$ as it is \emph{defined at the initial time} $t_{0}$, prior to
any change of setting at $B$ (Fig. \ref{fig:projection-smutual-setting}).
Similarly the projected state of $A$ given an outcome at $B$ is
inferred to describe the system $A$ at the time $t_{0}$. The deterministic
relations $\mathcal{D}_{1}$ and $\mathcal{D}_{2}$ (Results VII.13)
provide a set of relations for outcomes in the Q model, but the 
relations \emph{connect only the initial outcomes ($x_{j}^{A}$ or
$x_{k}^{B}$) of one system to the final outcomes ($x_{\phi}^{B}$
or $x_{\theta}^{A}$) of the other,} and are hence defined consistently
over both setting-changes. They provide relations that are weaker
than what could be deduced from the full LHV models $\mathcal{P}_{aym,A}$
and $\mathcal{P}_{aym,B}$, because they do not distinguish between
mixtures and superposition states. 

We illustrate by considering that $Q(\lambda,t_{0})$ represents the
entangled state $|\psi(t_{0})\rangle=c_{1}|x_{i}^{A}\rangle|x_{k}^{B}\rangle+c_{2}|x_{l}^{A}\rangle|x_{m}^{B}\rangle$
($c_{1},c_{2}\neq0$) defined in Comment (1) of Result VIII.1e. Changing
the basis at $A$ gives 
\begin{eqnarray*}
|\psi\rangle & \rightarrow & \sum_{r}|x_{\theta r}^{A}\rangle(c_{1}d_{ir}|x_{k}^{B}\rangle+c_{2}d_{lr}|x_{m}^{B}\rangle)
\end{eqnarray*}
where the $d_{ir}$ are defined in Appendix E, implying a projected
state at $B$ of $(c_{1}d_{ir}|x_{k}^{B}\rangle+c_{2}d_{lr}|x_{m}^{B}\rangle)$.
Taking $c_{1}=c_{2},$ the relations $\mathcal{D}_{1}$ are $x_{k}^{B}\leftrightarrow x_{\theta r}^{A}$
with probability $|d_{ir}|^{2}$; $x_{m}^{B}\leftrightarrow x_{\theta r}^{A}$
with probability $|d_{lr}|^{2}$. The relations are based on the model
$\mathcal{P}_{asym,B}$ (Result VII.3) where system $B$ is in a mixed
state and cannot account for superpositions states at $B$. For certain
choices of $\theta$, the outcomes $x_{k}^{B}$ and $x_{m}^{B}$ can
both be consistent with a specific outcome $x_{\theta r}^{A}$ at
$A$. Hence, simple retrodiction implies the outcome at $B$ could
be either $x_{k}^{B}$ or $x_{m}^{B}$.

The relations $\mathcal{D}_{1}$ can also appear for a state $\rho_{mix,AB}$
which is a mixture of $|x_{i}^{A}\rangle|x_{k}^{B}\rangle$ and $|x_{l}^{A}\rangle|x_{m}^{B}\rangle$.
Knowledge of $Q(\lambda,t_{0})$ and $Q(\lambda_{rotA},t_{1A})$ allows
inference that the projected state $Q(x_{B},p_{B}|x_{\theta r}^{A})$
is a \emph{superposition}  $|\psi_{B}\rangle=c_{1}d_{ir}|x_{k}^{B}\rangle+c_{2}d_{lr}|x_{m}^{B}\rangle$,
rather than a mixture of $|x_{k}^{B}\rangle$ and $|x_{m}^{B}\rangle$,
where here we assume $d_{ir},d_{lr}$ are nonzero for the choice of
$\theta$. The deterministic relations $\mathcal{D}_{1}$ themselves
provide a weak framework only, and do not provide the distinction
between the superposition and mixture.

If we consider a change of setting to $\phi$ at $B$, then we would
write 
\begin{eqnarray*}
|\psi\rangle & \rightarrow & \sum_{s}(c_{1}d_{ks}|x_{i}^{A}\rangle+c_{2}d_{ms}|x_{l}^{A}\rangle)|x_{\phi s}^{B}\rangle
\end{eqnarray*}
Given an outcome $x_{\phi s}^{B}$ at $B$, the possible correlated
values $x_{i}^{A}$ and $x_{l}^{A}$ at $A$, at the initial time
$t_{0}$. The deterministic relations $\mathcal{D}_{2}$ are $x_{i}^{A}\leftrightarrow x_{\phi s}^{B}$
with probability $|d_{ks}|^{2}$; $x_{l}^{A}\leftrightarrow x_{\phi s}^{B}$
with probability $|d_{ms}|^{2}$. For certain choices of $\phi$,
such that $d_{ks},d_{ms}\neq0$, the projected state $Q(x_{A},p_{A}|x_{\phi s}^{B})$
for system $A$ given an outcome $x_{\phi s}^{B}$ at $B$ is a superposition
$|\psi_{A}\rangle=d_{ks}|x_{i}^{A}\rangle+d_{ms}|x_{l}^{A}\rangle$.
 The projected states are mutually consistent, and consistent
with $Q(\lambda,t_{0})$: This is evident since the $|\psi\rangle$
can be written simultaneously with respect to both bases (refer Comments
(1) and (3) of VIII.1e), and the deterministic relations do not specify
the full nature of the projected state. The superposition nature of
the projected states that arise for an entangled state $Q(\lambda,t_{0})$
ensures the correlations for future choices of setting can differ
from those predicted, for example, by a non-entangled mixed state
$\rho_{mix,AB}$ or $\rho_{mix}^{AB}$ defined in Comment (3), allowing
a violation of Bell inequalities.

\emph{Comment:} The choice of setting-change that enables the distinction
between the entangled and non-entangled state requires $d_{ij},d_{lj}\neq0$,
$d_{kq},d_{mq}\neq0$. This is such to create a superposition of the
eigenstates in the original measurement basis, consistent with Results
VII.5,7, 8.

\subsubsection{No-signaling}

The following Result complements the earlier Result V.2a (``no macroscopic
retrocausality'') of Sec. V. It shows consistency with the second
Premise of wMR (Definition (6b) in Sec. I.B), completing the earlier
Results IV.6 and VIII.2 for wMR. We assume spacelike-separated systems
$A$ and $B$.

\textbf{\emph{Result IX.2: Weak Macroscopic Realism: Premise wMR(2):
No-signaling:}} Without loss of generality, we assume that the system
is prepared for measurements $\hat{x}_{A}$ and $\hat{x}_{B}$ at
time $t_{0}$, as in Figure \ref{fig:feedback-settingB}. The system
$A$ is amplified, so that at time $t_{mA}$ the value $\widetilde{\lambda}_{x}^{A}$
gives the outcome for $A$ if $\hat{x}_{A}$ is to be measured. The
second Premise of wMR posits that the value $\widetilde{\lambda}_{x}^{A}$
for the outcome at $A$ is not changed by interactions or dynamics
that may then occur (after $t_{mA}$) at system $B$. Hence, a change
of setting $\phi$ at $B$ does not change the measurable outcome
$\widetilde{\lambda}_{x}^{A}$ for $\hat{x}_{\theta A}$. (NB: We
can generalize to consider the system prepared for measurements $\hat{x}_{\theta A}$
and $\hat{x}_{\phi B}$ at time $t_{0}$.)

\emph{Proof:} We assume the frame $F$ in which both systems are stationary.
The setting at $A$ is fixed, so the simulation is consistent with
the outcome $\widetilde{\lambda}_{x}^{A}$ of $A$ being unchanged
if we apply the assumption of Result IV.8. It has been shown in the
literature (e.g. refer Ref. \citep{fulton2024alternative}) that this
assumption, which applies after measurement settings are fixed, is
consistent with Bell nonlocality. The Result can also be proved directly,
as in the Corollary of Result IX.3 below. $\square$

\emph{Comment:} The Result IX.2 explains no-signaling:\emph{ No-signaling}
is defined (Sec. I.B) as the conditional independence of the outcome
at $A$ and the setting at $B$, given the setting at $A$ \citep{wood2015lesson}.
We consider the set-up in Figure \ref{fig:feedback-settingB}. In
the simulation and Q model, weak macroscopic realism holds, and the
value of $\widetilde{\lambda}_{x}^{A}$ gives the outcome for measurement
$\hat{x}_{A}$. This value is not changed by the change of setting
at $B$. Hence, there is no signaling.

\emph{Comment:} While the branch \emph{value} $\widetilde{\lambda}_{x}^{A}$
does not change, there can be changes in hidden interference terms.
The detailed mechanism for nonlocality in the Q model is explained
in Result IX.4 below.

Result IX.2 can be generalized to hold at the earlier time $t_{1A}$
(which corresponds to time $t_{0}$ in Figure \ref{fig:feedback-settingB}),
just after the unitary operation determining the setting at $A$ \citep{fulton2024alternative,joseph2024macroscopic}.
We hence complete Results IV.14 and VIII.3 for weak local realism,
by showing consistency with the second Premise of wLR (Definition
(10b) of Sec. I.B).

\textbf{\emph{Result IX.3: Weak Local Realism: Premise wLR(2):}} We
consider the system at the time $t_{1A}$, when the unitary operation
$U_{\theta}^{A}$ that determines the setting to be $\theta$ at $A$
has been carried out. In Figure \ref{fig:feedback-settingB}, this
corresponds to the time $t_{1A}\equiv t_{0}$, and it is taken that
$\theta=0$, so the system $A$ is prepared for the measurement $\hat{x}_{\theta A}\equiv\hat{x}_{A}$.
We assume there is subsequently a change of setting $\phi$ at $B$.
There are two versions of weak local realism (wLR): deterministic
and probabilistic. The Q-model simulation of the Bell nonlocality
is consistent with both versions of the second Premise wLR(2) as given
by Definition (10b) in Sec. I.B.

(i) For the first version, the Premise wLR(2) posits that at time
$t_{1A}$ the value $x_{\theta j}^{A}$ for the outcome of the measurement
$\hat{x}_{\theta A}$ is determined, and this value is unchanged by
any subsequent interactions such as $H_{\phi}$ occurring at $B$.

(ii) The second version of Premise wLR(2) posits that the system $A$
is described by hidden variables $\{\lambda_{\theta}^{A}\}$ at the
time $t_{1A}$. There is a probability $P(x_{\theta j}^{A}|\{\lambda_{\theta}^{A}\})$
for an outcome $x_{\theta j}^{A}$ of the measurement at $A$, given
the hidden-variable state $\{\lambda_{\theta}^{A}\}$. For the second
version, the Premise wLR(2) posits that these probabilities are unchanged
by any subsequent interactions at $B$ occurring after the time $t_{1A}$.

\emph{Proof:} Consider the first version of wLR: Result VII.3 (Sec.
VII) shows that the probabilities for the joint outcomes of $\hat{x}_{A}$
and $\hat{x}_{\phi B}$ are indistinguishable from those of a non-entangled
state $\rho_{mix,A}$ (Eq. (\ref{eq:mix-asymA})), for which there
is a fixed predetermined value $\lambda_{A}$ for the outcome of system
$A$ which is unchanged by a local change of setting $\phi$ occurring
at $B$. The quantum predictions (and those of the simulation) arising
from a change of basis at the single location $B$ only can hence
be modeled by a LHV theory, denoted model $\mathcal{P}_{asym,A}$,
the details of which are given in Appendix F. Hence there is no inconsistency
with the deterministic version of Premise wLR(2), as given by the
prediction for the joint probabilities.

Consider the second version of wLR: The consistency with the probabilistic
version of Premise wLR(2) is illustrated for an entangled state in
Result VII.5. Without loss of generality, we consider $\theta=\phi=0$.
The hidden variables $\{\lambda\}$ for both systems are the $x_{A}$,
$p_{A}$, $x_{B}$, $p_{B}$ of $Q(\lambda,t_{0})$. We define $\{\lambda_{\theta}^{A}\}$
as $x_{A},p_{A}$, since here $\theta=0$ (refer Results VII.11 and
VIII.3). For the given $\{\lambda_{\theta}^{A}\}$ at time $t_{0}$,
the probabilities of outcomes at $A$ are defined by the Q function
for system $A$, which is the marginal $Q_{A}(x_{A},p_{A},t_{0})=\int Q(\lambda,t_{0})dx_{B}dp_{B}$
(refer Appendix H). The HV model $\mathcal{P}_{sup}$ can be developed
as in Sec. IV.F, to define a probability $P(x_{j}^{A}|\{\lambda_{\theta}^{A}\})$
of an outcome $x_{j}^{A}$ given $\{\lambda_{\theta}^{A}\}$. The
change in setting to $\phi$ at $B$ induces a new $Q(\bm{\lambda}_{rotB},t_{1})$
(e.g. as in Eq. (\ref{eq:q1-1-2})) with a change in $\mathcal{I}nt_{AB}$
(Eq. (\ref{eq:int-bell-1-1-1-2})) but which does not transform the
variables $x_{A}$, $p_{A}$. The change of setting at $B$ is equivalent
to a change of coordinates from $x_{B}$, $p_{B}$ to $x_{\phi B}$,
$p_{\phi B}$, but this does not change $Q(x_{A},p_{A})$, and does
not change the probabilities for $A$, if the setting at $A$ is fixed.
 Hence there is consistency with Premise wLR(2). $\square$

\emph{Corollary: Proof of consistency with Premise wMR(2):} We prove
consistency of the Q model with Premise wMR(2). The above proof for
probabilistic wLR is general and applies to any system, including
where wMR Premise (1) can be applied. We can hence consider the time
$t_{m}$ at which the system $A$ has been amplified, so that the
 branches corresponding to the amplified eigenvalues $Gx_{j}^{A}$
are distinct, as in Figures \ref{fig:The-causal-structure-1} and
\ref{fig:macro-sim-1}, so that any given amplitude $x_{A}(t_{m})$
is associated with just one of the branches (note Results IV.6 and
7). In the proof above, we define the variables $\{\lambda_{\theta}^{A}\}\equiv\{x_{A}(t_{m}),p_{A}(t_{m})\}$,
where we assume $\theta=0$ without loss of generality, and note that
for each pair, the value of $x_{A}(t_{m})$ defines a particular branch
$\mathcal{B}_{x_{A}(t_{m})}$, and hence a particular outcome, which
we denote by $x_{j}^{A}(x_{A}(t_{m}))$. Hence, the $P(x_{j}^{A}|\{\lambda_{\theta}^{A}\})$
is either $1$ or $0$, depending on whether $x_{j}^{A}=x_{j}^{A}(x_{A}(t_{m}))$
or not. This is a special case of the above proof, which implies the
probability $P(x_{j}^{A}|\{\lambda_{\theta}^{A}\})$ does not change
when a change of setting $\phi$ is implemented. Hence, in the Q model,
the branch and outcome does not change at $A$. $\square$

\subsubsection{Projection as a mechanism for Bell nonlocality}

\textbf{\emph{Result IX.4: Microscopic mechanism for Bell nonlocality,
in the presence of no-signaling: }}In the Q-model, the change of
setting to $\phi$ at $B$ (while keeping the setting at $A$ fixed)
leads to changes in the unobservable ``hidden'' interference terms
of the reduced Q function for $A$. There is no immediate observable
nonlocal effect. The simulation which accounts for the forward-backward
trajectories is consistent with the above Results and those in Sec.
VII.B. The hidden interference terms for the state at $A$ can lead
to an observable Bell nonlocality, if there is a further change of
setting at $A$.

\begin{figure}
\begin{centering}
\par\end{centering}
\begin{centering}
\includegraphics[width=1\columnwidth]{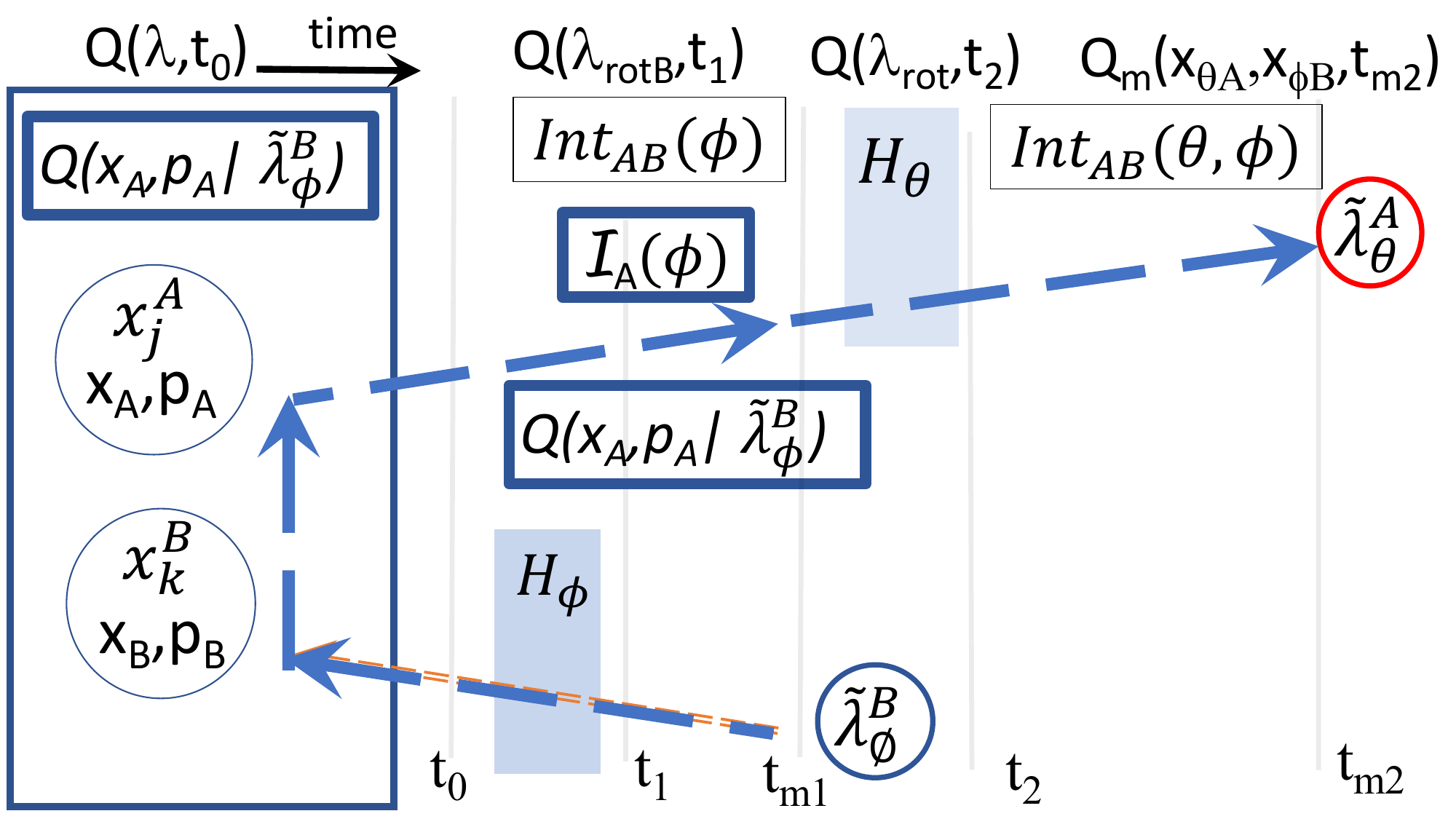}
\par\end{centering}
\caption{\textbf{\emph{Mechanism for Bell nonlocality:}} Depiction of Bell
nonlocality, when there is a change of setting at $B$ to $\phi$,
and then a change of setting $\theta$ at $A$. The change of setting
$\phi$ at $B$ induces a change to the state, given by $Q(\lambda_{rotB},t_{1})$.
However, the observable correlations if made at the time $t_{1}$
can be explained as a local transformation at $B$ (as induced for
the non-entangled state $\rho_{mix,A}$ of (\ref{eq:mix-asymA})).
However, the Q function $Q(\lambda,t_{0})$ for the entangled state
is different to that of $\rho_{mix,A}$. \label{fig:settings-AB}
This difference can appear as unobservable interference terms $\mathcal{I}nt_{AB}(\phi)$
in $Q(\lambda_{rotB},t_{1})$ (refer Eq. (\ref{eq:int-bell-1-1-1-2})),
and also in the projected distribution $Q(x_{A},p_{A}|\widetilde{\lambda}_{\phi}^{B})$
which describes the state for $A$ conditioned on the outcome $\widetilde{\lambda}_{\phi}^{B}$
after the setting change at $B$ (blue reversed dashed lines): For
certain $\phi$, the $Q(x_{A},p_{A}|\widetilde{\lambda}_{\phi}^{B})$
is a superposition state, with hidden interference terms $\mathcal{I}_{A}(\phi)$
not present for the system prepared in $\rho_{mix,A}$. The difference
can manifest with a \emph{further} change of setting $\theta$ at
$A$, for certain choices of $\theta$, when Bell nonlocality can
arise, detectable in the joint correlations $\langle\widetilde{\lambda}_{\theta}^{A}\widetilde{\lambda}_{\phi}^{B}\rangle$
at time $t_{m2}$. }
\end{figure}

\emph{Proof:} The mechanism is depicted in Figures \ref{fig:settings-AB}
and \ref{fig:mutual-Bell-nonlocality}. The state at time $t_{0}$
is given by the Q function $Q(\bm{\lambda},t_{0})$ written with respect
to the measurement basis, for $\hat{x}_{A}$ and $\hat{x}_{B}$. There
is then a change of setting at $B$ to $\phi$.  Assuming the system
$B$ is then amplified, since the $\hat{x}_{\phi B}$ is to be measured,
the trajectories $x_{\phi B}(t_{m1})$ conditioned on a branch symbolized
by $\widetilde{\lambda}_{\phi}^{B}$ can be traced to the time $t_{0}$
(blue dashed arrows in Fig. \ref{fig:settings-AB}). This requires
first tracing to time $t_{1}$, to infer distribution $Q_{loop}(\lambda_{rotB},t_{1}|\widetilde{\lambda}_{\phi}^{B})$
(refer Result VIII.1e and Fig. \ref{fig:projection-setting-change}).
A  deterministic transformation then combines the $x_{\phi B}(t_{1})$
with the $p_{\phi B}(t_{1})$ according to the inverse of (\ref{eq:rot-6-b})
to infer the postselected distribution $Q_{loop}(\lambda,t_{0}|\widetilde{\lambda}_{\phi}^{B})$,
where $\lambda=(x_{A},p_{A},x_{B},p_{B})$ at the initial time $t_{0}$.
The projected distribution $Q(x_{A},p_{A}|\widetilde{\lambda}_{\phi}^{B})$
for system $A$ can be evaluated by integrating over $x_{A}$ and
$p_{B}$. This corresponds to the projected state for $A$ given the
outcome $\widetilde{\lambda}_{\phi}^{B}$ and gives the predictions
for future measurements at $A$, with the setting fixed at $\phi$,
at $B$ (Results VIII.1b and VIII.1c). Where there is entanglement,
we find that this distribution $Q(x_{A},p_{A}|\widetilde{\lambda}_{\phi}^{B})$
for the state $A$ can contain hidden interference terms that can
lead to Bell nonlocality.

First, to explain the no-signaling, we know from Sec. VII (Results
VII.3 and VII.5) that the change of setting at $B$ changes the hidden
interference term $\mathcal{I}nt_{AB}$ (Eq. (\ref{eq:int-bell-1-1-1-2})).
This change however does not manifest in the \emph{measured} probabilities,
if the setting at $A$ is fixed. The new state $Q(\bm{\lambda}_{rotB},t_{1})$
has predictions for joint probabilities $P(x_{A},x_{\phi B})$ identical
to those of a non-entangled state $\rho_{mix,A}$ (Eq. (\ref{eq:mix-asymA}))
for which a LHV model $\mathcal{P}_{asym,A}$ (Appendix F) is possible.
Consistent with this, measurements of $\hat{x}_{A}$ on the system
$A$ given by the reduced state $Q(x_{A},p_{A}|\widetilde{\lambda}_{\phi}^{B})$
will give results indistinguishable from the state $\rho_{mix,A}$.
This explains consistency with no-signaling: There is no observable
Bell nonlocality at this stage.

However, different to $\rho_{mix,A}$, the projected distribution
$Q(x_{A},p_{A}|\widetilde{\lambda}_{\phi}^{B})$ for the Bell state
has extra interference terms that reflect the system $A$ is in a
superposition, rather than a mixture, of eigenstates $|x_{j}\rangle$
of $\hat{x}_{A}$. We see from the analysis of Eq. (\ref{eq:int-bell-1-1-1-2})
that the integration over $p_{B}$ can lead to nonzero integrals if
the rotation of coordinates due to the change of setting $\phi$ combines
both $x_{B}$ and $p_{B}$. Hence, different to $\rho_{mix,A}$, interference
terms arising from $\mathcal{I}nt_{AB}$ can be present in $Q(x_{A},p_{A}|\widetilde{\lambda}_{\phi}^{B})$.
This occurs for certain choices of $\phi$ only.

In short, the system $A$ as given by projected distribution $Q(x_{A},p_{A}|\widetilde{\lambda}_{\phi}^{B})$
is a\emph{ }superposition rather than a mixture.\emph{ However, in
order to distinguish the superposition from the mixture, a further
change of setting at $A$ is required. }This is because the Q function
for the superposition differs from that of the mixture only by the
presence of the hidden interference terms (refer Result III.1).\emph{
}Hence, the\emph{ }superposition effects ultimately give different
predictions to those of the LHV model $\mathcal{P}_{asym,A}$, leading
to Bell nonlocality, but only if a further change of setting at $A$
occurs. $\square$

\emph{Comment: }The projected distribution $Q(x_{A},p_{A}|\widetilde{\lambda}_{\phi}^{B})$
is sufficient to determine the Bell nonlocality, for future measurements
at $A$. From Result VIII.1a, we see that the projected distribution
$Q(x_{A},p_{A}|\widetilde{\lambda}_{\phi}^{B})$ is evaluated based
on knowledge of the initial Q function $Q(\lambda,t_{0})$, using
\emph{continuous individual backward} trajectories, applying first
the deterministic relation $x_{\phi k}^{B}\leftrightarrow Gx_{\phi k}^{B}$
due to $H_{amp}^{B}$ (refer Appendix H), ensuring consistency with
deterministic relations $\mathcal{D}$ due to correlations (Result
VII.13), and then the deterministic transformation (\ref{eq:rot-6-b})
due to the setting-change at $B$. This continuity cannot apply in
the forward direction however, where there is a discontinuity at the
time $t_{1}$.

\emph{Comment:} As depicted in Figure \ref{fig:feedback-settingB},
the trajectories and projected distribution $Q(x_{A},p_{A}|\widetilde{\lambda}_{\phi}^{B})$
are consistent with the Q function $Q(\bm{\lambda}_{rotB},t_{1})$
for the joint system at time $t_{1}$, after the change of setting
at $B$. This leads to the following Results.

\textbf{\emph{Result IX.5a: Forward-backward stochastic simulation
of Bell nonlocality:}} The Bell nonlocality can be described by the
forward-backward stochastic trajectories consistently with the dynamics
of the Q function summarized by the Results in Sec. VII. The proof
of causal consistency for the bipartite system is given in Result
IX.7 below.\textcolor{red}{}

\textbf{\emph{Result IX.5b:}} Consistent with Result VII.4 based on
the dynamics of the Q function, the stochastic solutions show that
the Bell nonlocality emerges when there is a change of measurement
setting $\phi$ and $\theta$, at both locations, $A$ and $B$ respectively.

\subsection{Bell nonlocality, causal consistency, and resolution of paradoxes}

 In summary, the Bell nonlocality arises because of changes to the
interference terms $\mathcal{I}nt_{AB}$ when there is a change of
setting at both locations. A careful study of the final joint probabilities
as in Sec. VII.B after changes of settings at $A$ and $B$ shows
there is a conversion of ``hidden'' interference terms $\mathcal{I}nt_{AB}$
into terms $\mathcal{I}nt_{AB}(\theta,\phi)$ which contribute to
the \emph{observable} probabilities on measurement (e.g. Eqs. (\ref{eq:q1-1-1})
and (\ref{eq:bell-int-1-1-3})). We summarize the following Results
about nonlocality.

\begin{figure}
\begin{centering}
\includegraphics[width=1\columnwidth]{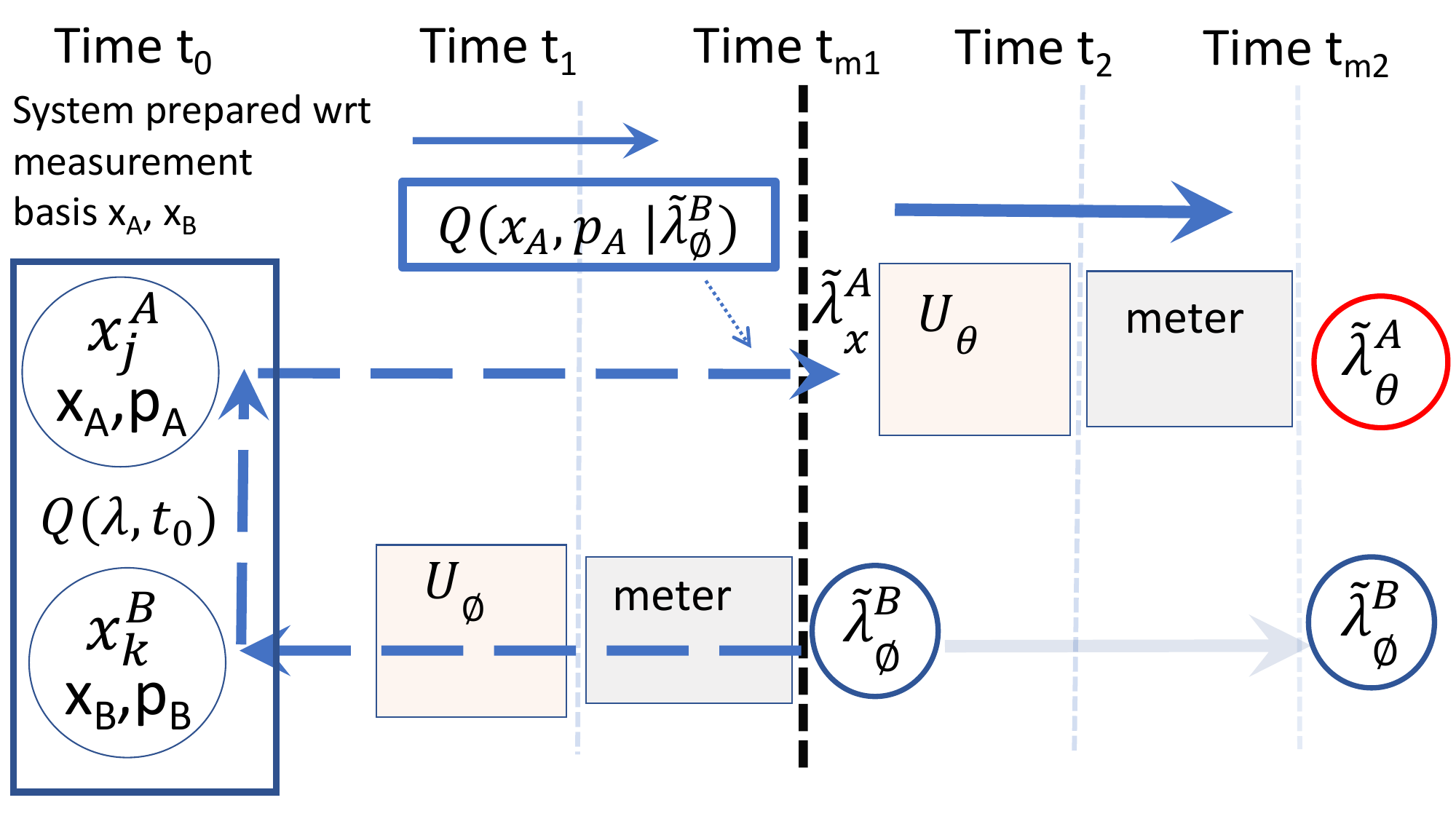}
\par\end{centering}
\caption{\textbf{\emph{Bell nonlocality is consistent with the three Premises
of weak macroscopic realism.}} The diagram sketches the relations
based on the trajectories of the simulation of measurements $\hat{x}_{\theta A}$
and $\hat{x}_{\phi B}$ on spacelike-separated systems $A$ and $B$,
with all systems stationary in a frame $F$. Here, $U_{\theta}$
and $U_{\phi}$ represent the unitary operations that determine the
settings $\theta$ and $\phi$. The meter corresponds to the amplification
$H_{amp}$. The first premise (Premise wMR(1)) asserts that the outcome
of the measurement $\hat{x}_{\phi B}$ is determined at time $t_{m1}$,
(vertical black dashed line), after the amplification due to the meter
at $B$. The value for the outcome is denoted $\widetilde{\lambda}_{\phi}^{B}$.
The second premise (Premise wMR(2)) asserts that this value is not
changed by any later interaction at $A$ (e.g. due to the meter at
$A$) at a time $t>t_{m1}$ (indicated by faint blue horizontal arrow).
 The third premise (Premise wMR(3)) asserts that at time $t_{m1}$,
a projected distribution $Q(x_{A},p_{A}|\widetilde{\lambda}_{\phi}^{B})$
is defined for system $A$, which determines the outcomes of $\hat{x}_{\theta A}$
of system $A$ (for any future change of setting $\theta$ at $A$)
conditioned on the outcome $\widetilde{\lambda}_{\phi}^{B}$ at $B$
(indicated by dashed blue arrows). \label{fig:epr-bohm-premises}}
\end{figure}

\begin{figure}
\begin{centering}
\includegraphics[width=1\columnwidth]{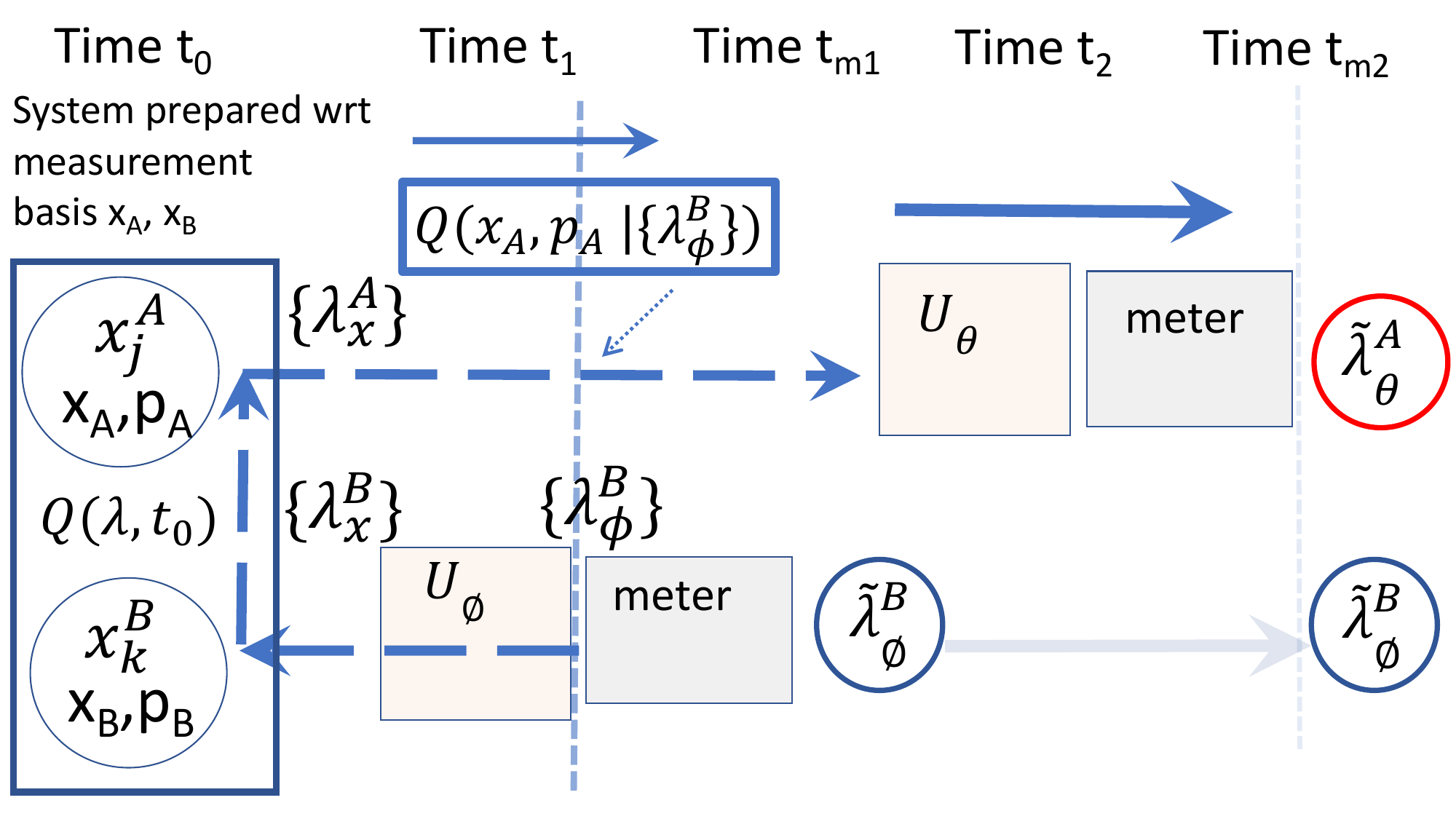}
\par\end{centering}
\caption{\label{fig:epr-bohm-premises-1-1} \textbf{\emph{Bell nonlocality
is consistent with the three Premises of weak local realism.}} The
first premise (Premise wLR(1)) asserts that the outcome (or, at least,
the probabilities for outcomes) of the measurement $\hat{x}_{\phi}^{B}$
at $B$ is determined at time $t_{1}$, after the operation $U_{\phi}$
(vertical light blue dashed line). In the simulation, based on the
hidden variable model $\mathcal{P}_{B}$ (Eq. \ref{eq:model-b})),
the probabilities of outcomes for $\hat{x}_{\phi}^{B}$ are determined
by variables $\{x_{\phi B},p_{\phi B}\}$ which we symbolize by $\{\lambda_{\phi}^{B}\}$.
The second premise (Premise wLR(2)) asserts that the outcome (or
else, the probabilities of outcomes) are not changed by any later
interaction (e.g. $U_{\theta}$) at $A$.  The third premise (Premise
wLR(3)) posits that at time $t_{1}$, a projected Q function $Q(x_{A},p_{A}|\{\lambda_{\phi}^{B}\})$
is defined, which determines the state of system $A$ (for any future
change of setting $\theta$ at $A$) conditioned on the variables
$\{\lambda_{\phi}^{B}\}$ at $B$ (blue dashed arrows).}
\end{figure}
\textbf{\emph{Weak local reality:}} The Bell simulation is consistent
with the assumptions of weak local realism and weak macroscopic realism,
presented as the three Premises in Definitions (6) and (10) in Sec.
I.B. The consistency has been demonstrated in the earlier Results
(IV.6, IV.14, VII.10, VII.11, VIII.2, VIII.3, IX.2, IX.3) and is summarized
in Figures \ref{fig:epr-bohm-premises} and \ref{fig:epr-bohm-premises-1-1}.

\begin{figure}
\begin{centering}
\includegraphics[width=1\columnwidth]{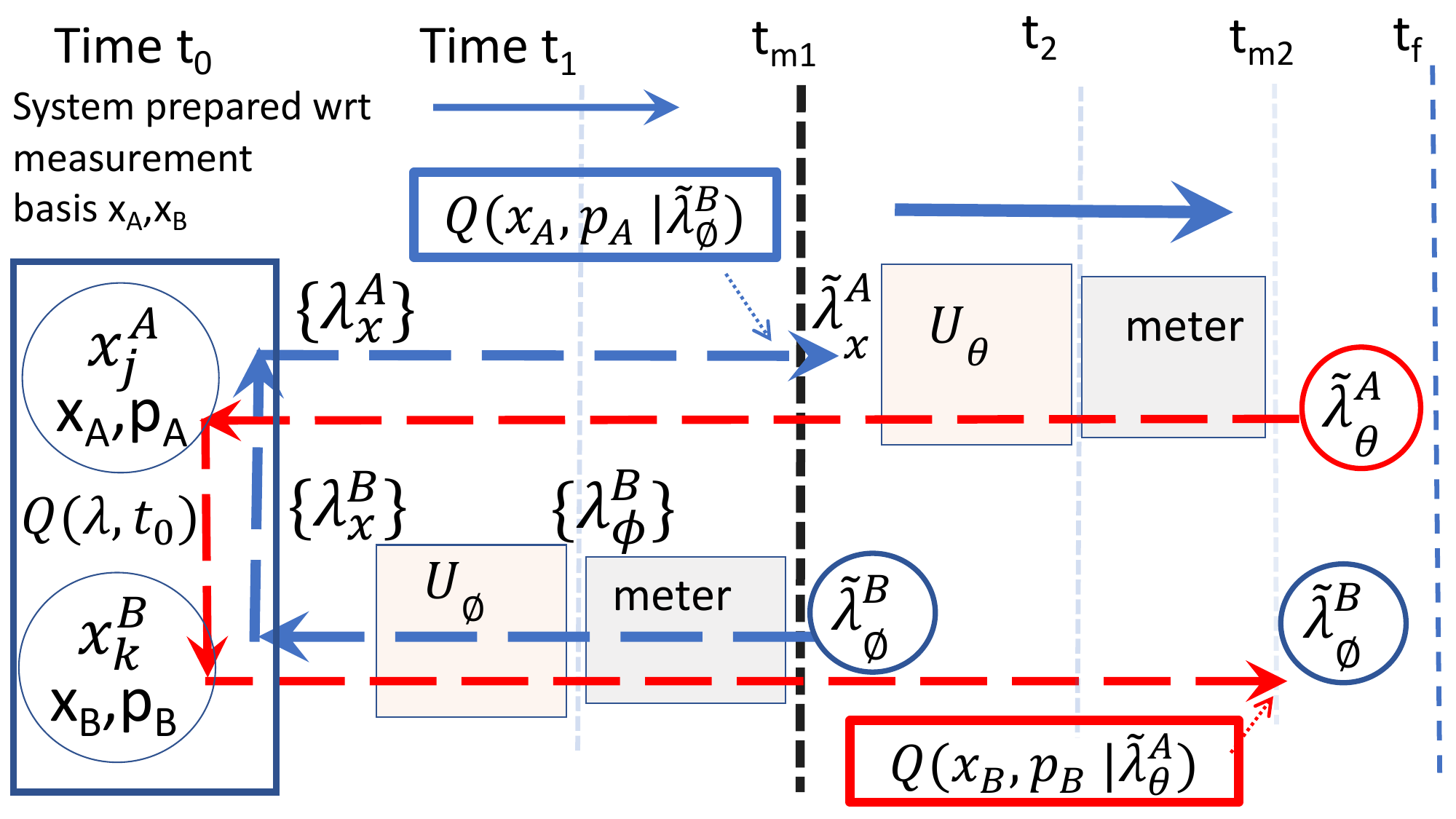}
\par\end{centering}
\caption{\textbf{\emph{Resolving the paradox of the future boundary condition:
}}\emph{The timing of the emergence of Bell nonlocality:}\textbf{\emph{
}}The system is as described for Figure \ref{fig:epr-bohm-premises}.
A ``hidden'' nonlocality emerges at (or by) the time $t_{m1}$
(vertical black dashed line), prior to the specification of the future
boundary at time $t_{f}>t_{m2}$.   According to Premise wMR(3),
the $Q$ function $Q(x_{A},p_{A}|\widetilde{\lambda}_{\phi}^{B})$
which determines the state of system $A$ (for any future change of
setting $\theta$ induced by $U_{\theta}$ at $A$) conditioned on
the outcome $\widetilde{\lambda}_{\phi}^{B}$ is determined at (or
by) time $t_{m1}$ (blue dashed lines). \label{fig:epr-bohm-premises-1}Hence,
the nonlocality is determined at time $t_{m1}$, when the conditional
state $Q(x_{A},p_{A}|\widetilde{\lambda}_{\phi}^{B})$ is specified,
but \emph{only over both operations} $U_{\phi}$ and $U_{\theta}$
at time $t_{m2}$ does an observable Bell nonlocality emerge. The
\emph{values} for the outcomes that demonstrate the nonlocality are
hence determined at (or by) the time $t_{m2}<t_{f}$.}
\end{figure}

\textbf{\emph{Result IX.6: Causal order in the Bell experiment: There
is no retrocausality, where the future affects the past:}}\emph{ }This
is an extension of Results V.2a and V.2b. Suppose the change of setting
$\phi$ at $B$ occurs at time $t_{1}$, illustrated in Figures \ref{fig:epr-bohm-premises}
and \ref{fig:epr-bohm-premises-1-1}). The change of setting at $B$
does not change the value of the macroscopic variables $\widetilde{\lambda}_{x}^{A}$
or $\widetilde{\lambda}_{x}^{B}$ that are defined \emph{prior to
}or \emph{at the time }$t_{1}$. There is hence no genuine retrocausality,
where a future change of setting affects past macroscopic properties.
We have seen from the dynamics outlined in Result IX.4 that the \emph{impact
on $A$ of the change of setting at $B$ occurs only for future events,
after further setting changes at }\textbf{\emph{$A$.}} A change of
setting at $A$ is required for any change to macroscopic values (e.g.
$\widetilde{\lambda}_{x}^{A}$) at $A$. This is illustrated in Figures
\ref{fig:epr-bohm-premises} and \ref{fig:epr-bohm-premises-1}, where
we see that the macroscopic values $\widetilde{\lambda}_{x}^{A}$,
$\widetilde{\lambda}_{x}^{B}$, $\widetilde{\lambda}_{\theta}^{A}$
and $\widetilde{\lambda}_{\phi}^{B}$ only change when there is a
local interaction. 

\textbf{\emph{Result IX.7: Causal consistency:}} This is the bipartite
extension of Result V.1 (Sec. V.B). At any given time $t$, the Q
function $Q(\lambda,t)$ is defined as a probability distribution,
which uniquely represents the quantum state and which hence evolves
causally in the forward-time direction. The distribution $Q(\lambda,t)$
is consistent with the probability densities of the forward-backward
trajectories. The causal consistency is depicted in Figure \ref{fig:causal-consistency-2-1}.
We extend the proof presented for Result V.1 to bipartite systems
in Appendix H.

\textbf{\emph{Result IX.8: Resolving the paradox of the future boundary
condition: Timing of emergence of Bell nonlocality: }}\emph{The Bell
violations are not ``put in by hand'', as part of the future boundary
condition: }In the causal model (proposed in Sec. V), the correlations
that lead to detection of Bell nonlocality do not arise from the future
boundary condition at the time $t_{f}$, as indicated in Figure \ref{fig:epr-bohm-premises-1}.
They are determined at a time $t_{Bell}$ prior to $t_{f}$. 

\emph{So, when do the Bell correlations emerge?} We see from Result
VII.12 that the correlation between the outcomes $\hat{x}_{\theta A}$
and $\hat{x}_{\phi B}$ are determined at the time $t_{Bell}=t_{2}$,
after the interactions $H_{\theta}^{A}$ and $H_{\phi}^{B}$ (refer
Fig. \ref{fig:sim}). This follows from Results III.1 and VII.2, and
the proposed causal and HV model, $\mathcal{P}_{B}$ (Sec. VII.D),
which apply after both settings $\theta$ and $\phi$ are fixed. The
correlation between the outcomes $\hat{x}_{\theta A}$ and $\hat{x}_{\phi B}$
is directly determined by the correlation between the means, $x_{\theta j}^{A}$
and $x_{\phi k}^{B}$, of the Gaussians in the Q function $Q(\lambda_{rot},t_{2}$),
defined at time $t_{2}$. We refer to this time as the ``time of
emergence of Bell nonlocality''.

However, we also see that, if the setting-changes are sequential,
a ``hidden'' Bell nonlocality emerges at an earlier time, after
the change of just one setting (Result IX.4). This is evident by Eqs.
(\ref{eq:Q-int})--(\ref{eq:q1-1-1}) (Sec. VII.B) and is depicted
in Figure \ref{fig:epr-bohm-premises-1}. First, we assume wMR. According
to Premise wMR(3), the distribution $Q(x_{A},p_{A}|\widetilde{\lambda}_{\phi}^{B})$
for outcomes at $A$ given an outcome $\widetilde{\lambda}_{\phi}^{B}$
at $B$ is determined at (or by) the time $t_{m1}$, when system $B$
has been amplified (and hence a branch $\widetilde{\lambda}_{\phi}^{B}$
and outcome specified). Since $Q(x_{A},p_{A}|\widetilde{\lambda}_{\phi}^{B})$
\emph{fully} determines the outcomes for $A$, which will exhibit
the Bell nonlocality after a further setting change $\theta$ is made,
it can be argued that the nonlocality has emerged at (or by) this
time $t_{Bell}=t_{m1}$. On the other hand, if we assume wLR, the
distribution for outcomes at $A$ can be defined earlier, at the time
$t_{1}$ when the setting at $B$ is fixed, so that $t_{Bell}=t_{1}$.

This leaves us with two final Results.

\textbf{\emph{Result IX.9: Paradox of time-order: Lorentz invariance:
}}At first glance, the concepts of weak macroscopic realism and weak
local realism appear paradoxical. In Figures \ref{fig:epr-bohm-premises}
and \ref{fig:epr-bohm-premises-1-1}, we have considered that the
setting of $A$ is changed to $\theta$, \emph{after} that of $B$,
but for spacelike-separated systems such as $A$ and $B$, there is
no specified time order of events, which can change in different frames.
According to wMR, the value of outcome at $B$ is fixed and determined
(as $\widetilde{\lambda}_{\phi}^{B}$) at time $t_{m1}$, and this
conditions the outcome at $A$ at the later time $t_{m2}$ after the
setting change $\theta$ at $A$. But in the different frame, the
outcome at $A$ is fixed and determined first, prior to that of $B$.
Hence, there is a paradox.

The paradox may be explained by the symmetry and mutual consistency
allowed by the projected states of the Q function (Fig. \ref{fig:mutual-Bell-nonlocality}).
The outcome at $A$ (say) is determined not only by the nonlocal setting
change at $B$, but by the local setting change at $A$ (and vice
versa). The reduced states given by $Q(x_{A},p_{A}|\widetilde{\lambda}_{\phi}^{B})$
and $Q(x_{A},p_{A}|\widetilde{\lambda}_{\phi}^{B})$ (blue and red
dashed lines) in Figure \ref{fig:mutual-Bell-nonlocality} are independent
of the time order of the setting interactions $H_{\theta}$ and $H_{\phi}$.
They are consistent with each other and the initial state $Q(\lambda,t_{0})$,
and with the final joint probabilities determined by the marginal
$Q(x_{\theta A},x_{\phi B},t_{mA})$ of the final Q function of the
amplified state, which is also independent of the time order of the
settings. 

\textbf{\emph{Result IX.10: Breakdown of Bell's assumptions: Failure
of statistical independence or local causality?}} In the Introduction
we ask what assumptions of Bell's local hidden variable (LHV) model
break down, according to the Q-based model? Bell's local hidden variable
condition is
\begin{equation}
P_{++}^{AB}(\theta,\phi)=\int d\lambda\rho(\lambda)P_{A}(+|\lambda,\theta)P_{B}(+|\lambda,\phi)\label{eq:Bell-expression}
\end{equation}
as defined by Eq. (\ref{eq:lhv}), where $P_{++}^{AB}(\theta,\phi)$
is the probability of obtaining $+1$ at both sites with settings
$\theta$ and $\phi$; $\rho(\lambda)$ is the distribution for hidden
variables $\lambda$, and $P_{A}(+|\lambda,\theta)$ ($P_{B}(+|\lambda,\phi)$)
is the probability for $+1$ at $A$ ($B$), given $\lambda$ and
$\theta$ ($\phi)$. Bell nonlocality is the breakdown of the condition
(\ref{eq:Bell-expression}).

Bell's assumption may break down due to a failure of the existence
of a\emph{ }distribution\emph{ }$\rho(\lambda)$ that defines hidden
variables $\lambda$ for the system, as it exists at a time $t_{0}$
prior to the implementation of the setting transformations. Such a
distribution is \emph{statistically independent} of any future choice
of measurement settings. Another possibility is a breakdown of the
assumption of the factorization in the integrand (\emph{local causality}).

In the simulations, the hidden variables $\lambda$ are the coordinates
$x_{A}$, $p_{A}$, $x_{B}$ and $p_{B}$ of the Q function. The interactions
$H_{\theta}$ and $H_{\phi}$ that determine the measurement settings
at $A$ and $B$ respectively are local and deterministic.  The amplitudes
$x_{\theta}^{A}$, $x_{\phi}^{B}$ of the simulation if directly measurable
would lead to moments satisfying Bell's condition, with $\rho(\lambda)\equiv Q(\lambda,t_{0})$
where $\lambda\equiv(x_{A},p_{A},x_{B},p_{B})$. However, this is
not the case, because of the hidden noise, which is not measured,
associated with noise inputs at the future boundary, or else with
hidden interference (Sec. III.B).

The Bell violations require changes of settings $\theta$ and $\phi$
at both sites, so that the Bell HV state $\{\lambda\}$ is considered
with respect to rotations to two different bases, one for each system
$A$ and $B$. After both unitary rotations, we have seen from Results
VII.5 and IX.4 that interference terms that are ``hidden'' at the
original time $t_{0}$ \emph{}can contribute to probabilities for
observable outcomes. For certain conditions, the contribution can
be nonlocal (Result VII.6), implying
\begin{equation}
P_{++}^{AB}(\theta,\phi)\neq\int d\lambda Q(\lambda,t_{0})P_{A}(+|\lambda,\theta)P_{B}(+|\lambda,\phi)\label{eq:bell-cond}
\end{equation}

It is possible to deduce the reason for the breakdown of the condition
(\ref{eq:bell-cond}) in the Q model, by evaluating from the simulation
a probability $P_{++}^{AB}(\lambda,\theta,\phi)$ for outcomes given
the settings $\theta$ and $\phi$, and given a set of values of the
hidden variables $\lambda$, where $\lambda\equiv(x_{A},p_{A},x_{B},p_{B})$
are defined by the Q function. We can write
\begin{equation}
P_{++}^{AB}(\theta,\phi)=\int d\lambda Q(\lambda,t_{0})P_{++}^{AB}(\lambda,\theta,\phi)\label{eq:bell-cond-1}
\end{equation}
The evaluation is explained in Sec. VII.D, where a probability $P_{Bell}(\mathcal{S}_{A},\mathcal{S}_{B}|\lambda,\theta,\phi)$
is deduced for spin outcomes $\mathcal{S}_{A}$ and $\mathcal{S}_{B}$
at $A$ and $B$, given the hidden variables $\lambda\equiv(x_{A},p_{A},x_{B},p_{B})$
and settings $\theta$ and $\phi$. We obtain $P_{++}^{AB}(\lambda,\theta,\phi)=P_{Bell}(1,1|\lambda,\theta,\phi)$.
This provides a model $\mathcal{P}_{Bell}$ to explain the violations,
with a statistically independent\emph{ }distribution $\rho(\lambda)=Q(\lambda,t_{0})$
that describes the system at time $t_{0}$, prior to the dynamics
associated with the choice of settings. The causal consistency of
the model is depicted in Figure \ref{fig:causal-consistency-2-1},
and proved in Appendix H. At any time $t$, the Q function is defined
as a fixed distribution, consistent with the forward-backward trajectories.
The locality assumption implies that it is possible to write $P_{++}^{AB}(\lambda,\theta,\phi)=P_{A}(+|\lambda,\theta)P_{B}(+|\lambda,\phi)$.
Hence, the Q-based model implies a breakdown of the locality assumption.

\section{Conclusions\label{sec:Conclusions}}

A summary of the main results of this paper has been given in the
Introduction. The simulations we present are motivated by the Q function
in quantum optics and contribute toward an explanation of how a hidden
variable theory may complete quantum mechanics, while remaining consistent
with the predictions of quantum mechanics, in particular, allowing
violation of Bell inequalities. In the model, the measurement proceeds
as amplification, which we show plays a role similar to decoherence:
The interference terms that distinguish a superposition from a mixed
state decay with amplification, and hence do not contribute to directly
measured probabilities (Results III.1 and VII.2).

In Secs. II-IV, we prove and demonstrate an equivalence of the evolution
of the Q function $Q(\mathbf{x,p},t)$ (which represents the quantum
state) to a forward-backward stochastic dynamics of amplitudes, \textbf{$\mathbf{x}$
}and $\mathbf{p}$. The Q-based model is presented in Sec. IV, with
a derivation of Born's rule and the role of the Wigner function. We
treat entangled systems, modeling EPR and Bell correlations, in Secs.
VI-IX. A causal (``cause-and-effect'') model for the dynamics is
given in Secs. V and IX, along with a proof of causal consistency
and explanation of why there is no retrocausality at a macroscopic
level (Results V.1,2, IX.6-9 and Appendix H).

The ``collapse of the wave function'' is analyzed by examining a
postselected state conditioned on an outcome $\lambda$ (Results IV.9,
10). We include in Sec. VIII a treatment of projection, analyzing
a Schrodinger-cat state where a system is entangled with a meter.
We conclude that the system is inferred to be in an eigenstate, because
the observer conditions on the value of an amplified variable $x$,
losing information about the complementary variable $p$ (Result VIII.1).
We examine the forward-backward dynamics of the variables associated
with a given branch and outcome $\lambda$, thus deducing ``hidden
loops'' (Result IV.12).

The simulations are consistent with a Q-based model of reality in
which three weak local realistic premises (\emph{weak local realism},
wLR) hold (Results IV.14, VII.11, VIII.3, IX.3). There is a partial
relaxation of EPR's and Bell's local realistic premises so that ``elements
of reality'' exist defined \emph{after} interaction with devices
(e.g. a polarizer or phase-shifter) that adjust measurement settings.
Based on the observation of ``branches'' in the simulation, there
is also consistency with \emph{weak macroscopic} \emph{realism} (wMR):
predetermined values $\lambda$ for outcomes emerge at a time $t_{m}$,
after sufficient amplification (Results IV.6-8, V.2, VII.10, VIII.2,
IX.2 concerning wMR, and Results VI.1-2). The simulation demonstrates
how the weak local realistic premises are consistent with Bell nonlocality
(Figs. \ref{fig:epr-bohm-premises} and \ref{fig:epr-bohm-premises-1-1}).
By analyzing the dynamics of the Q function and of the trajectories
(Secs. VII.B and IX), a mechanism for Bell nonlocality is given in
terms of ``hidden'' interference terms that contribute to measurable
probabilities only after a change of measurement setting $\theta$
and $\phi$ at both sites (Results VII.5, IX.4). For certain conditions
on $\theta$ and $\phi$, the contribution can lead to a failure of
Bell's local-hidden-variables assumptions (Results VII.6-8). The mechanism
explains how Bell nonlocality can be consistent with macroscopic realism
and no-signaling (Results IV.6-8, V.2, IX.2-4). The solutions of the
forward-backward equations for entangled systems show consistency
with the mechanism for Bell nonlocality, based on a model for projection
(Sec. IX). 

The theory presented in this paper explains why there is no genuine
retrocausality, which implies that a system is changed by a future
event. The realist variables $\lambda$ defined for a macroscopic
system are (according to the wLR and wLR premises) unchanged by a
future event, and unchanged by a spacelike-separated event (Results
IV.6-8, V.2, IX.2-3). The causal consistency as depicted in Figures
\ref{fig:causal-consistency-2-1} and \ref{fig:epr-causal-consistency-1}
explains the mechanism by which this is possible. The Q function $Q(\lambda,t)$
evolves causally in the forward-time direction, and yet is consistent
with the future boundary condition and counter-propagating trajectories
(Figs. \ref{fig:epr-bohm-premises} and \ref{fig:epr-bohm-premises-1-1}).
The ``paradox of the future boundary condition'' (Fig. \ref{fig:epr-bohm-premises-1})
is resolved by the nature of the causal (cause-and-effect) model based
on the simulation, which involves causal deterministic relations for
amplified variables (once settings are fixed) and a type of retrocausality
(meaning ``backward-in-time'' propagation) for random noise inputs
at the vacuum level (Secs. V, VII and IX).

\textbf{\emph{Interpretations:}}  An interesting question is whether
the Q-based model of reality favors any particular interpretation
of Bell nonlocality. There are real values $\lambda$ posited for
the outcome of the measurement that are defined for the system as
it exists at the time after the measurement settings have been adjusted
and after amplification, which supports a macroscopic-realist viewpoint.
There is no contradiction with the known violation of Leggett-Garg
inequalities because the derivation of these inequalities requires
the additional assumption of non-invasive measurability \citep{leggett1985quantum}.
There is also no contradiction with the existence of macroscopic superposition
states, because the Q-based reality model defines distributions for
variables (``states'') that are not defined in standard quantum
mechanics. Moreover, we have constructed (nonlocal) hidden-variable
(HV) models based on the simulation (Results IV.13, VII.9). The hidden
variables in these models are defined for the system at a time $t_{0}$
after preparation of the state of the system but before the choice
of settings which are considered independent, similar to Bell's model
(but apparently different to explanations of nonlocality based on
superdeterminism).

A possible explanation of Bell nonlocality is that the hidden-variable
distribution $\rho(\lambda)$ in the Bell model (\ref{eq:Bell-expression})
is not statistically independent of future settings. The causal consistency
provided by the Q model (refer Results V.1, IX.7, Appendix H) shows
that the distribution $\rho(\lambda)\equiv Q(\lambda,t_{0})$ where
$\lambda\equiv(x_{A},p_{A},x_{B},p_{B})$ is fixed at time $t_{0}$
(Figs. \ref{fig:epr-causal-consistency-1} and \ref{fig:epr-bohm-premises-1}),
contrary to this explanation. The probabilistic hidden-variable model
$\mathcal{P}_{Bell}$ (Result VII.9, Sec. VII.D) gives predictions
for the outcomes, conditioned on a set of hidden variables $\mathbf{\lambda}=(x_{A},p_{A},x_{B},p_{B})$
that describe the system at the time $t_{0}$ (prior to selection
of the measurement settings). We conclude that in the Q model, the
violation arises due to a failure of the Bell-locality assumption.

\textbf{\emph{Experiments: }}Finally, we consider experimental tests
that might be performed. The interaction $H_{amp}$ of Eq. (\ref{eq:ham-2-1})
is that of parametric amplification \citep{Yuen1976}, which has been
experimentally realized \citep{Wu1986generation,Yurke1989Observation}.
The states studied in this paper are superpositions of squeezed or
coherent states which have also been realized experimentally \citep{ourjoumtsev2007generation,frowis2018macroscopic,Brune1996,monroe1996schrodinger,leghtas2013deterministic,palomaki2013entangling,omran2019generation,ku2020experimental,wright1996collapses}.
Hence, the measurement model can be verified experimentally. A feature
would be to demonstrate the vanishing of the interference terms in
the Q function, as the system is amplified.

The prediction of causal consistency can be tested, as depicted in
Figures \ref{fig:causal-consistency-2-1} and \ref{fig:epr-causal-consistency-1}.
It is feasible to perform a simulation that matches boundary conditions
given by the experimental Q functions, which are measurable by tomography
\citep{Landon2018quantitativetomography}. The density of amplitudes
$x(t)$ and $p(t)$ in the simulation must match the distribution
given by an experimentally-measured $Q$ function $Q(x,p,t)$, regardless
of the choice of time $t_{f}$ of the future boundary condition.

A second experiment might illustrate a ``hidden loop'' and the cat
paradox (Fig. \ref{fig:scat-1-1}). A ``hidden loop'' is defined
so as to involve amplitudes that are not amplified (and hence not
observed) in the measurement, and a distribution $Q_{loop}$ that
cannot be described by a quantum wavefunction (Sec. IV.E). The $Q_{loop}$
for an individual branch of the superposition can be constructed from
the simulation that is based on experimental Q functions. If $Q_{loop}$
shows a reduction of variances below the Heisenberg uncertainty bound,
then, within the constraints of the model, this signifies a hidden
loop.

The simulations give an explanation of the EPR paradox, in terms of
the ``weak elements of reality'' that exist after sufficient amplification,
as calculated in Figure \ref{fig:epr-3} \citep{Reid2023Short,mcguigan2024resolving}.
 The simulations might be useful to analyze questions raised by Schrödinger
about the EPR argument, as in the recent experiment of Colciaghi et
al which reported EPR correlations for Bose-Einstein condensates \citep{colciaghi2023einstein}.
\begin{acknowledgments}
This publication was made possible through the support of Grant 62843
from the John Templeton Foundation. The opinions expressed in this
publication are those of the author(s) and do not necessarily reflect
the views of the John Templeton Foundation. This research has also
been supported by the Australian Research Council Grants schemes under
Grants DP180102470 and DP190101480, and the authors thank NTT Research
for their financial and technical support.
\end{acknowledgments}

\section*{Appendix A: stochastic path integral theorems\label{sec:AppendixA-Stochastic-path-integrals}}

\subsection{Definitions and notation}

We make use of methods that were first used for diffusive path integrals
\citep{wiener1930generalized,Graham1977Covariant}, here generalized
to forward-backward trajectories. For $M$ bosonic modes, the phase-space
vector $\bm{\lambda}$ is a $2M$-dimensional real vector, and $\bm{p},\bm{x}$
are $M$-dimensional real vectors with $\bm{\lambda}=\left(\bm{x},\bm{p}\right)$.
For the measurement interactions treated in this paper, the generalized
Fokker-Planck equation (GFPE) satisfied by a Q function is: 
\begin{equation}
\dot{Q}\left(\bm{\lambda},t\right)=\mathcal{L}\left(\bm{\lambda}\right)Q\left(\bm{\lambda},t\right).\label{eq:GFPE-1}
\end{equation}
The differential operator $\mathcal{L}\left(\bm{\lambda}\right)$
has first and second order differential terms $\mathcal{L}_{1}\left(\bm{\lambda}\right)$
and $\mathcal{L}_{2}\left(\bm{\lambda}\right)$, each with forward
and backward components $\mathcal{L}_{np}\left(\bm{p}\right)$, $\mathcal{L}_{jx}\left(\bm{x}\right)$:
so that: 
\begin{align}
\mathcal{L}\left(\bm{\lambda}\right) & =\mathcal{L}_{1}\left(\bm{\lambda}\right)+\mathcal{L}_{2}\left(\bm{\lambda}\right)\label{eq:Lxy-1-1}\\
 & =\sum_{n}\left[\mathcal{L}_{np}\left(\bm{p}\right)-\mathcal{L}_{nx}\left(\bm{x}\right)\right],
\end{align}
where $\mathcal{L}_{jp,x}$ are defined as: 
\begin{align}
\mathcal{L}_{1p} & =-\sum_{j}\partial_{j}^{p}a_{p}^{j}\left(\bm{p}\right)\\
\mathcal{L}_{2p} & =\frac{1}{2}\sum_{j}\partial_{j}^{p}\partial_{j}^{p}d^{j}\,.
\end{align}
Here $\mathcal{L}_{x}$ is identical to $\mathcal{L}_{p}$ except
for the substitution of $x$ for $p$, and we define $\partial_{j}^{p}\equiv\partial/\partial p^{j}$,
$\partial_{j}^{x}\equiv\partial/\partial x^{j}$. For the Hamiltonians
in this paper, we have $\bm{a}_{p}=\bm{a}_{p}\left(\bm{p}\right)$
and $\bm{a}_{x}=\bm{a}_{x}\left(\bm{x}\right)$, with diagonal diffusion
$\bm{d}$.

For use in the theorems proved below, we define a phase-space trajectory
and stochastic path integral as follows. Firstly, the trajectory is
a sequence $\left[\bm{\lambda}\left(t_{0}\right),\ldots\bm{\lambda}\left(t_{N}\right)\right]\equiv\left[\bm{\lambda}_{0},\ldots\bm{\lambda}_{N}\right]$,
with a total action of: 
\begin{equation}
S_{kn}=S_{kn}^{p}+S_{kn}^{x}=\sum_{m=k+1}^{n}S_{m}.
\end{equation}
The ``one-step'' action for the $n$-th step is given by $S_{n}=S_{n}^{p}+S_{n}^{x}$,
with $S_{n}^{p}$ having the same form as $S_{n}^{x}$ except for
the changed superscripts, where: 
\begin{align}
S_{n}^{p} & =S^{p}\left(\bm{\lambda}_{n-1},\bm{\lambda}_{n}\right)\nonumber \\
 & =\sum_{j}\left[\frac{\epsilon}{2d^{j}}\left|v_{n}^{j,p}\right|^{2}+\ln\left(\sqrt{\mathcal{N}^{j}}\right)\right].
\end{align}
This includes a path-integral normalization factor, $\mathcal{N}^{j}=\left(2\pi\epsilon d^{j}\right)$,
and relative velocity fields $\bm{v}\equiv\left(\bm{v}^{p},\bm{v}^{x}\right)$
which are functions of neighboring coordinates: 
\begin{align}
\bm{v}_{n}^{p} & \equiv\frac{1}{\epsilon}\left(\bm{p}_{n}-\bm{p}_{n-1}\right)-\bm{a}^{p}\left(\bm{p}_{n-1}\right)\nonumber \\
\bm{v}_{n}^{x} & \equiv\frac{1}{\epsilon}\left(\bm{x}_{n-1}-\bm{x}_{n}\right)-\bm{a}^{x}\left(\bm{x}_{n}\right).\label{eq:velocity-1}
\end{align}

The $N$-step total path probability $\mathcal{P}$ for a path $\underline{\bm{\lambda}}=\left(\bm{\lambda}_{0},\ldots\bm{\lambda}_{N}\right)$
at times $t_{k}=t_{0}+k\epsilon$, with $k=0,\ldots N$, is 
\begin{equation}
\mathcal{P}\left[\underline{\bm{\lambda}}\right]=e^{-S_{0N}}.
\end{equation}
The two-time propagator $G$ is then defined as: 
\begin{align}
G\left(\bm{\lambda},t_{k}\left|\tilde{\bm{\lambda}},\tilde{\bm{t}}\right.\right) & =\lim_{\epsilon\rightarrow0}\int\delta\left(\bm{\lambda}-\bm{\lambda}{}_{k}\right)\delta\left(\tilde{\bm{\lambda}}-\bm{\lambda}{}_{0N}\right)\mathcal{P}\left[\underline{\bm{\lambda}}\right]d\underline{\bm{\lambda}}.\label{eq:total-propagator-1}
\end{align}
Here, $d\underline{\bm{\lambda}}=\prod_{n=0}^{N}d\bm{\lambda}_{n}=\prod_{n=0}^{N}\prod_{k=1}^{M}dx_{n}^{k}dp_{n}^{k}$,
$\bm{\lambda}{}_{0N}=\left(\bm{x}_{N},\bm{p}_{0}\right)$, while $\tilde{\bm{\lambda}}=\left(\bm{x}_{f},\bm{p}_{i}\right)$
gives the initial momentum $\bm{p}_{i}$ and final position $\bm{x}_{f}$
at times $\tilde{\bm{t}}=\left(t_{i},t_{f}\right)=\left(t_{0},t_{N}\right)$.

Marginals have the notation $P\left(\bm{x},t\right)=\int Q\left(\bm{\lambda},t\right)d\bm{p}$,
with the variables that are integrated removed from the arguments.
Boundary conditions are imposed on the final marginal $P\left(\bm{x},t_{f}\right)$
and initial conditional, $P\left(\bm{p}\left|\bm{x},t_{i}\right.\right)=Q\left(\bm{\lambda},t_{i}\right)P^{-1}\left(\bm{x},t_{i}\right)$.
The remainder of this Section shows the equivalence of the differential
and path-integral approaches, both analytically and numerically.

\subsection{Path-integral equivalence theorems}

\subsection*{Theorem I:}

Any path-integral Q function $Q_{pi}$ obtained from multiplication
by the joint probabilities at the boundaries satisfies the Q function
dynamical equations, where: 
\begin{equation}
Q_{pi}\left(\bm{\lambda},t\right)=\int G\left(\bm{\lambda},t\left|\tilde{\bm{\lambda}},\tilde{\bm{t}}\right.\right)P\left(\tilde{\bm{\lambda}},\tilde{\bm{t}}\right)d\tilde{\bm{\lambda}}.\label{eq:propagator-form}
\end{equation}

The quantum Q function solution, $Q=Q_{pi}$ is found on solving simultaneous
equations for the joint distribution $P\left(\tilde{\bm{\lambda}},\tilde{\bm{t}}\right)$,
and hence for $Q_{pi}$, where: 
\begin{equation}
\int Q_{pi}\left(\bm{\lambda},t_{f}\right)d\bm{p}=P\left(\bm{x},t_{f}\right),
\end{equation}
and: 
\begin{equation}
\frac{Q_{pi}\left(\bm{\lambda},t_{i}\right)}{\int Q_{pi}\left(\bm{\lambda},t_{i}\right)d\bm{p}}=P\left(\bm{p}\left|\bm{x},t_{i}\right.\right).
\end{equation}

\subsection*{Proof:}

We wish to show that $Q_{pi}=Q$, by proving that $Q_{pi}$ satisfies
the GFPE (\ref{eq:GFPE-1}), and has an initial condition that corresponds
to the Q function for the initial quantum state. From linearity, proving
that $Q_{pi}\left(\bm{\lambda}\right)$ satisfies the GFPE \ref{eq:GFPE-1}
is achieved by verifying this for the propagator, $G\left(\bm{\lambda},t\left|\tilde{\bm{\lambda}},\tilde{\bm{t}}\right.\right)$.
To show this, define advanced and retarded propagators for $\bm{\lambda}$
at $t=t_{j}$ with $\tilde{\bm{p}}=\bm{p}_{0}$ at $t_{i}=t_{0}$
and $\tilde{\bm{x}}=\bm{x}_{N}$ at $t_{f}=t_{N}$, as 
\begin{align}
G^{r}\left(\bm{p}_{j},t\left|\tilde{\bm{p}},t_{i}\right.\right) & =\lim_{\epsilon\rightarrow0}\int e^{-S_{0j}^{p}}\prod_{n=1}^{n-1}d\bm{p}_{n},\nonumber \\
G^{a}\left(\bm{x}_{j},t\left|\tilde{\bm{x}},t_{f}\right.\right) & =\lim_{\epsilon\rightarrow0}\int e^{-S_{jN}^{x}}\prod_{n=j+1}^{N-1}d\bm{x}_{n}.
\end{align}

Due to its path-integral construction, $G^{r}$ satisfies a forward
Kolmogorov equation in $\bm{p}$ \citep{stratonovich1971probability,graham1977path}.
By reversing the sign of $t$, we find that $G^{a}$ also satisfies
a forward Kolmogorov equation in $\bm{x}$, but in the negative time
direction: 
\begin{align}
\dot{G}^{r}\left(\bm{p},t\left|\tilde{\bm{p}},t_{i}\right.\right) & =\mathcal{L}^{p}G^{r}\left(\bm{p},t\left|\tilde{\bm{p}},t_{i}\right.\right)\nonumber \\
\dot{G^{a}}\left(\bm{x},t\left|\tilde{\bm{x}},t_{f}\right.\right) & =-\mathcal{L}^{x}G^{a}\left(\bm{x},t\left|\tilde{\bm{x}},t_{f}\right.\right).\label{eq:Forward-Kolmogorov}
\end{align}

The normalization of the advanced Gaussian propagator terms in the
path integral for $\bm{x}_{n}$ with $n<j$ means that all these past
time factors integrate to unity. Similarly, the retarded propagator
is independent of $\bm{p}_{n}$ for future time points $n>j$, as
these also integrate to give unity. As a result, we can write that:

\begin{align}
G^{r}\left(\bm{p},t_{k}\left|\tilde{\bm{p}},t_{0}\right.\right) & =\lim_{\epsilon\rightarrow0}\int e^{-S_{0N}^{p}}\delta\left(\bm{p}-\bm{p}{}_{k}\right)\delta\left(\tilde{\bm{p}}-\bm{p}{}_{0}\right)d\underline{\bm{p}},\nonumber \\
G^{a}\left(\bm{x},t_{k}\left|\tilde{\bm{x}},t_{f}\right.\right) & =\lim_{\epsilon\rightarrow0}\int e^{-S_{0N}^{x}}\delta\left(\bm{x}-\bm{x}{}_{k}\right)\delta\left(\tilde{\bm{x}}-\bm{x}{}_{N}\right)d\underline{\bm{x}}.
\end{align}

From these results and the definitions above, it follows that the
total propagator factorizes as $G\left(\bm{\lambda},t\left|\tilde{\bm{\lambda}},\tilde{\bm{t}}\right.\right)=G^{r}\left(\bm{p},t\left|\tilde{\bm{p}},t_{0}\right.\right)G^{a}\left(\bm{x},t\left|\tilde{\bm{x}},t_{f}\right.\right)$.
Hence, using the chain rule for differentiation and Eq (\ref{eq:Forward-Kolmogorov}),
the required time-evolution can be written as: 
\begin{align}
\dot{G}\left(\bm{\lambda},t\left|\tilde{\bm{\lambda}}\right.\right) & =\dot{G^{r}}\left(\bm{p},t\left|\tilde{\bm{p}},t_{0}\right.\right)G^{a}\left(\bm{x},t\left|\tilde{\bm{x}},t_{f}\right.\right)\nonumber \\
 & +G^{r}\left(\bm{p},t\left|\tilde{\bm{p}},,t_{0}\right.\right)\dot{G^{a}}\left(\bm{x},t\left|\tilde{\bm{x}},t_{f}\right.\right)\nonumber \\
 & =\left(\mathcal{L}^{p}-\mathcal{L}^{x}\right)G\left(\bm{\lambda},t\left|\tilde{\bm{\lambda}},\tilde{\bm{t}}\right.\right).
\end{align}

Due to linearity, any integral of $G$ over its boundary values also
satisfies the GFPE. Therefore, the path integral construction must
obey the required GFPE, 
\begin{equation}
\dot{Q}_{pi}\left(\bm{\lambda},t\right)=\left(\mathcal{L}_{p}\left(\bm{p}\right)-\mathcal{L}_{x}\left(\bm{x}\right)\right)Q_{pi}\left(\bm{\lambda},t\right).
\end{equation}

Provided the joint probability $P\left(\tilde{\bm{\lambda}},\tilde{\bm{t}}\right)$
satisfies the boundary equations, one can verify that at the initial
time, $Q_{pi}\left(\bm{\lambda},t_{0}\right)=Q\left(\bm{\lambda},t_{0}\right)$.
Due to uniqueness of the solutions to a first order differential equations,
$Q_{pi}$ is equal to the quantum-mechanical Q function solution for
all times.

As a further check, in the short-time limit of $t_{0}=t_{f}$ we find
that 
\begin{equation}
\lim_{t_{0},t_{f}\rightarrow t}G\left(\bm{\lambda},t\left|\tilde{\bm{\lambda}}\right.\right)=\delta\left(\bm{\lambda}-\tilde{\bm{\lambda}}\right),
\end{equation}
hence, as $t_{0},t_{f}\rightarrow t$, one has that: 
\begin{align}
\lim_{t_{0},t_{f}\rightarrow t}Q_{pi}\left(\bm{\lambda},t\right) & =\int\delta\left(\bm{\lambda}-\tilde{\bm{\lambda}}\right)P\left(\tilde{\bm{\lambda}},t'\right)d\tilde{\bm{\lambda}},\nonumber \\
 & =P\left(\bm{p}\left|\bm{x},t'\right.\right)P\left(\bm{x},t'\right).
\end{align}
From the definition of the conditional probability, this is the initial
Q function.

\subsection*{Theorem II: }

The path-integral solution corresponds to a time-symmetric stochastic
differential equation (TSSDE) or forward-backward SDE, with a conditional
initial distribution in $\bm{p}$ and final marginal distribution
in $\bm{x}$ : 
\begin{align}
p^{j}\left(t\right) & =p^{j}\left(t_{0}\right)+\int_{0}^{t}a_{p}^{j}\left(t'\right)dt'+\int_{0}^{t}dw_{p}^{j}\nonumber \\
x^{j}\left(t\right) & =x^{j}\left(t_{f}\right)+\int_{t}^{t_{f}}a_{x}^{j}\left(t'\right)dt'+\int_{t}^{t_{f}}dw_{x}^{j}.
\end{align}
Here $\bm{x}\left(t_{f}\right)$ is distributed as $P\left(\bm{x},t_{f}\right)$,
while $\bm{p}\left(t_{0}\right)$ is distributed conditionally as
$C\left(\bm{p}\left|\bm{x}\left(t_{0}\right),t_{0}\right.\right)$,
and $\left[dw^{\mu}\right]=\left(d\bm{w}_{x},d\bm{w}_{p}\right)$
are independent real Gaussian noises correlated as 
\begin{equation}
\left\langle dw^{\mu}dw^{\nu}\right\rangle =\delta^{\mu\nu}d^{\mu}\epsilon,
\end{equation}
where $\mu=1,2M$ and $d^{j+M}=d^{j}$.

\subsection*{Proof:}

The TSSDE is obtained by discretizing the equation for times $t_{k}=t_{0}+k\epsilon$,
with $k=0,\ldots N$, and then taking the limit of $\epsilon\rightarrow0$.
We define ${\color{red}{\normalcolor \left[\bm{\lambda}\right]=\left[\bm{\lambda}_{0},\bm{\lambda}_{1},\dots\bm{\lambda}_{N}\right]}}$
as the stochastic path. The discretized solutions are then given as
the simultaneous solutions of the equations, for $k=1,\ldots N$ 
\begin{align}
\bm{p}_{k} & =\bm{p}_{k-1}+\bm{a}^{p}\left(\bm{p}_{k-1}\right)\epsilon+\bm{\Delta}_{k}^{p}\nonumber \\
\bm{x}_{k-1} & =\bm{x}_{k}+\bm{a}^{x}\left(\bm{x}_{k}\right)\epsilon+\bm{\Delta}_{k}^{x},\label{eq:Time-symmetricSDE-1-2-1}
\end{align}

To obtain an equivalent path integral to the TSSDE, we first obtain
the $N-$step trajectory probability density, conditioned on random
noises $\left[\bm{\Delta}\right]$. This is a product of Dirac delta
functions, 
\begin{align}
\mathcal{G}\left(\left[\bm{\lambda}\right]\left|\left[\Delta\right]\right.\right) & =\prod_{j=1}^{n}\delta^{2M}\left(\epsilon\bm{v}_{j}-\bm{\Delta}_{j}\right),\label{eq:Delta-function-path-1}
\end{align}
which gives a normalized probability conditioned on a specific noise
vector $\left[\bm{\Delta}\right]$. Solving for the resulting set
of trajectory values $\left[\bm{\lambda}\right]$ that satisfy the
delta-function constraints is straightforward in the parametric amplifier
case, due to the decoupling of forward and backward equations.

In a Fourier transform representation with $\bm{k}_{j}\equiv\left(k_{j}^{1},k_{j}^{2},\ldots k_{j}^{2M}\right)$,
one can expand the delta-functions as 
\begin{align}
\mathcal{G}_{n}\left(\left[\bm{\lambda}\right]\left|\left[\Delta\right]\right.\right) & =\prod_{j=1}^{n}\int\frac{d\bm{k}_{j}}{\left(2\pi\right)^{2M}}e^{-i\bm{k}_{j}\left(\bm{v}_{j}\epsilon-\bm{\Delta}_{j}\right)}.
\end{align}
The $2M$ real Gaussian noises $\bm{\Delta}_{k}$ at each step in
time, for $k>0$, are distributed as: 
\begin{align}
P\left(\bm{\Delta}_{k}\right) & =\frac{1}{\left(2\pi\epsilon d\right)^{M}}e^{-\left|\bm{\Delta}_{k}\right|^{2}/\left(2\epsilon d\right)}.
\end{align}
On integration over $\bm{k}_{j}$ one obtains the path probability
result as defined previously: 
\begin{equation}
\mathcal{G}_{n}\left(\left[\bm{\lambda}\right]\right)=\int\mathcal{G}_{n}\left(\left[\bm{\lambda}\right]\left|\left[\Delta\right]\right.\right)P\left(\left[\Delta\right]\right)d\left[\Delta\right].
\end{equation}

By construction, the initial values of $\bm{x}_{N}$ and $\bm{p}_{0}$
are sampled according to the required joint and conditional probabilities.
Hence the probability of a TSSDE solution is $Q_{pi}$, which is equal
to the quantum average Q from Theorem I. Our definitions are similar
to those in the mathematics literature \citep{ma1994solving}, except
for the use of a conditional probabilistic boundary.

\section*{Appendix B: Numerical $\chi^{2}$ tests\label{sec:Appendix-B:-Numerical}}

The analytic theorems obtained in Appendix A were verified numerically
using $\chi^{2}$ tests in several cases described here, with details
that are given below. The forward-backward stochastic equations were
integrated by first propagating $x$ backwards in time, generating
a conditional sample, then propagating $p$ forwards in time.

All trajectories plotted use $40$ sample trajectories, to provide
an intuitive demonstration of how they behave. In order to verify
the quantitative accuracy of the trajectory probability distributions,
large numbers of samples were generated and plotted using binning
methods.

These samples were statistically tested with $\chi^{2}$ methods \citep{pearson1900x,drummond2022simulating}
to compare them with analytic solutions. The test cases used $N_{s}=2\times10^{6}$
sample trajectories to obtain good statistics for the numerically
sampled distributions, 
\begin{equation}
p_{ijk}^{samp}=\frac{N_{ijk}}{N_{s}}.
\end{equation}
Here $N_{ijk}$ is the number of trajectories in the bin at $x_{i},p_{j}$,
and sampled time $t_{k}$. Such verification requires binning on a
three-dimensional $\left(x,p,t\right)$ grid, to obtain numerical
estimates of the integrated analytic probabilities $p_{ijk}$, where:
\begin{align}
p_{ijk} & =\int_{A}d\delta xd\delta pQ\left(x_{i}+\delta x,p_{j}+\delta p,t_{k}\right).
\end{align}
This was evaluated by numerical integration of the analytic $Q$ distribution,
using a two-dimensional Simpson's rule integrator in each bin.

In order to treat the dynamics of the Q-function, time-averages were
evaluated to give a definitive overall result. Due to the correlations
inside each trajectory, the \emph{range} of fluctuations of time-averaged
statistics are reduced compared to one-time tests, which exactly follow
the $\chi^{2}$ distribution. Individual tests at each time-point
are in agreement with $\chi^{2}$ statistics, and will be reported
elsewhere.

We therefore define: 
\begin{equation}
\bar{\chi}^{2}=\frac{1}{N_{t}}\sum_{i,j,k}\frac{\left\langle \left[p_{ijk}-p_{ijk}^{samp}\right]^{2}\right\rangle }{\left\langle \sigma_{ijk}^{2}\right\rangle },
\end{equation}
where, 
\begin{equation}
\sigma_{ijk}^{2}=p_{ijk}/N_{s}
\end{equation}
is the expected probability variance for $N_{s}$ total samples and
$N_{t}$ time points, with a phase-space bin area of $A$.

As an example, the simulations described in Figure 1 used $30$ time-steps
of $gdt=0.1$, combined with a midpoint stochastic integration method
for improved accuracy \citep{Drummond1990}. No significant discretization
error improvements were found with smaller time-steps. Comparisons
were made between the analytic and numerically sampled $Q\left(x,p,t\right)$
distributions with $dx=0.02$ and $dp=0.05$. This gave an average
of $\sim55,100$ comparison grid-points at each time step, after discarding
bins with non-significant sample populations of $N<10$ \citep{Rukhin2010}.

In a typical test with $2\times10^{6}$ sample trajectories and $A=10^{-3}$,
the time-averaged statistical error was $\bar{\chi}^{2}=55.2\times10^{3}$,
with $1.7\times10^{6}$ valid comparisons. There were an average of
$k=55.1\times10^{3}$ significant points per time-step. This shows
that $\bar{\chi}^{2}$ is within the expected range of $\left\langle \chi^{2}\right\rangle =k\pm\sqrt{2k}$.

Hence, as expected from the path-integral and stochastic theorems,
there is excellent agreement between the analytic $Q$-function probability
from quantum theory and the ensemble averaged stochastic trajectories.

\section*{Appendix C: Calculation of properties and Q functions\label{sec:Appendix-C:-Calculations}}

\emph{Q function of a squeezed state:} The calculation of the Q function
(Eq. (\ref{eq:q-sq-1})) of a squeezed state uses 
\[
S(r)|0\rangle=\frac{1}{\sqrt{\cosh r}}\sum_{n=0}^{\infty}(-\tanh r)^{n}\frac{\sqrt{(2n)!}}{2^{n}n!}|2n\rangle
\]
 and $\langle0|D(\beta-\alpha)=\langle\gamma|$ where $\gamma=\alpha-\beta$,
to deduce that
\begin{eqnarray*}
\langle\alpha|D(\beta)S(r)|0\rangle & = & \langle0|D(-\alpha)D(\beta)S(r)|0\rangle\\
 & = & e^{(-\alpha\beta^{*}+\alpha^{*}\beta)/2}\langle0|D(\beta-\alpha)S(r)|0\rangle\\
 & = & \frac{e^{(-\alpha\beta^{*}+\alpha^{*}\beta)/2}}{\sqrt{\cosh r}}e^{-|\gamma|^{2}/2}e^{[(-\tanh r)\gamma^{*2}/2]}
\end{eqnarray*}

\emph{Q function of the amplified state:} For the calculation of (\ref{eq:amp-sup-3}),
we use that $e^{-iH_{amp}t/\hbar}=S(-gt)$, implying squeezing in
$\hat{p}$.  Also $S(r)D(\beta)=D(\alpha)S(r)$ where $\alpha=\beta\cosh r-\beta^{*}\sinh r$.
For $\beta$ real, $\alpha=e^{-r}\beta$. Hence $S(-gt)D(\frac{x_{1}}{2})=D(\frac{Gx_{1}}{2})S(-gt)$
and $S(-gt)S(r)=S(-gt+r)$ where $G=e^{gt}$ is the amplification
factor.

\emph{Properties of the eigenstates of $\hat{x}$}: It is easy to
prove from the identities above that the eigenstates $|x_{j}\rangle$
defined by (\ref{eq:eigenstate-def}) as highly squeezed states in
$\hat{x}$ satisfy $|\langle x_{j}|x_{l}\rangle|^{2}=e^{-|x_{j}-x_{l}|^{2}e^{2r}}$,
becoming orthogonal (since $r\rightarrow\infty$) for $j\neq l$.\textcolor{blue}{
}Completeness is assumed in the limit where the eigenvalues $x_{j}$
form a continuous spectrum in analogy with the position representation
and noting that the overlap functions (e.g. $\langle x_{j}^{A}|\alpha\rangle$)
approach those (e.g. $\langle x|\alpha\rangle)$ of the position representation
as $r\rightarrow\infty$.

\emph{The Q function and quantum coefficients:} We note that the quantum
state $|\psi(t_{0})\rangle$ can be expanded in terms of the eigenstates
$|x_{j}\rangle$, as $|\psi(t_{0})\rangle=\sum_{j}c_{j}|x_{j}\rangle$,
where $c_{j}$ are probability amplitudes, and the Q function is derived
accordingly. The values $x_{j}$ are an important part of the description
of the model, since these are amplified by $H_{amp}$. The $|\psi(t_{0})\rangle$
and hence the $c_{j}$ are defined uniquely by $Q(\lambda,t_{0})$
and can in principle be determined by $Q(\lambda,t_{0})$. For example,
for a given $Q(\lambda,t_{0})$, the $|c_{j}|$ and $x_{j}$ can be
determined by applying $H_{amp}$ in the limit of $t$. From Sec.
III, we see that the Q function is of form $Q=\sum_{i}|c_{i}|^{2}\mathcal{G_{j}}+Int$,
where $\mathcal{G}_{j}$ are the known Gaussian functions involving
$x_{j}$, and $\mathcal{I}nt$ is an interference term, which can
hence be deduced from knowledge of $|c_{j}|$ and $x_{j}$. For a
simple two-state superposition as in Eq. (\ref{eq:sup-sq}), this
implies the phase for $c_{j}.$

\emph{Eigenstates of rotated quadratures:} The Q function of a rotated
squeezed state can also be calculated. We consider
\begin{equation}
|\beta_{j},z\rangle_{sq}=D(\beta_{j})S(z)|0\rangle\label{eq:sq-3}
\end{equation}
where $D(\beta_{j})=e^{\beta_{j}\hat{a}^{\dagger}-\beta_{j}^{*}\hat{a}}$
and $S(z)=e^{\frac{1}{2}(z^{*}\hat{a}^{2}-z\hat{a}^{\dagger2})}$
are the displacement and squeezing operators, where $z$ and $\beta_{j}$
are complex numbers. We define the rotated quadrature phase amplitudes
as $\hat{x}_{\theta}=\hat{a}e^{-i\theta}+\hat{a}^{\dagger}e^{i\theta}$
and $\hat{p}_{\theta}=(\hat{a}e^{i\theta}-\hat{a}^{\dagger}e^{-i\theta})/i$
i.e. $\hat{x}_{\theta}=\hat{x}\cos\theta+\hat{p}\sin\theta$ and $\hat{p}_{\theta}=-\hat{x}\sin\theta+\hat{p}\cos\theta$.
For the state with squeezed fluctuations in $\hat{x}_{\theta}=\hat{a}e^{-i\theta}+\hat{a}^{\dagger}e^{i\theta}$,
we note that $z=re^{2i\theta}$ where $r$ is a real, positive number.
Hence $(\Delta\hat{x}_{\theta})^{2}=e^{-2r}$ and $(\Delta\hat{p}_{\theta})^{2}=e^{2r}$.
 We choose the direction of $\beta_{j}$ to be along the direction
of squeezing, so that $\beta_{j}=x_{\theta j}e^{i\theta}/2$.The
eigenstate of $\hat{x}_{\theta}$ is denoted $|x_{\theta j}\rangle_{\theta}$
and is thus defined as the limiting squeezed state
\begin{equation}
|x_{\theta j}\rangle_{\theta}\equiv\lim_{r\rightarrow\infty}|\frac{x_{\theta j}}{2}e^{i\theta},z\rangle_{sq}\label{eq:eigenstate-def-3}
\end{equation}
Where the meaning is clear, we abbreviate the notation to $|x_{\theta j}\rangle$.
The Q function of $|x_{\theta_{j}}\rangle$ is $Q(\alpha)=\frac{1}{\pi}|\langle\alpha|x_{\theta_{j}}\rangle|^{2}$
where $\alpha=(x+ip)/2$ but we can calculate in the rotated axis,
with amplitude $\alpha_{new}=(x_{\theta}+ip_{\theta})/2$ and $x_{\theta}=x\cos\theta+p\sin\theta$,
$p_{\theta}=-x\sin\theta+p\cos\theta$. Hence, the Q function $Q_{x_{\theta j}}(x_{\theta},p_{\theta})$
of the eigenstate $|x_{\theta j}\rangle$ is
\[
Q_{x_{\theta j}}(x_{\theta},p_{\theta})=\frac{e^{-p_{\theta}^{2}/2\sigma_{p_{\theta}}^{2}}e^{-(x_{\theta}-x_{\theta j})^{2}/2\sigma_{x_{\theta}}^{2}}}{2\pi\sigma_{x_{\theta}}\sigma_{p_{\theta}}}
\]
where $\sigma_{p_{\theta}}^{2}=1+e^{2r}$ and $\sigma_{x_{\theta}}^{2}=1+e^{-2r}$.

\emph{Extensions to general superpositions:} The results of Eq. (\ref{eq:sup-sq})-
(\ref{eq:int-marg}) can be extended to the superposition $\sum_{j}c_{j}|x_{j}\rangle$
where the $|x_{j}\rangle$ are modeled as highly squeezed states.
The Q function is
\begin{eqnarray}
Q(x,p,t) & = & N\frac{e^{-p^{2}/2\sigma_{p}^{2}(t)}}{2\pi\sigma_{x}(t)\sigma_{p}(t)}\Bigr(\sum_{j}|c_{j}|^{2}e^{-(x-G(t)x_{j})^{2}/2}\nonumber \\
 &  & +\sum_{P\in\{j,k\}}2|c_{j}c_{k}|\mathcal{I}_{jk}(t)\Bigl)\label{eq:Q-ext}
\end{eqnarray}
where we sum over the pairs $P$ given by all sets of the non-ordered
pair $j$ and $k$. Here $\mathcal{I}_{jk}(t)=e^{-[(x-x_{k})^{2}+(x-x_{j})^{2}]/4}\mathcal{F}_{jk}(t)$
and ($c_{j}=|c_{j}|e^{i\theta_{j}}$)
\begin{eqnarray}
\mathcal{F}_{jk}(t) & = & \cos(\theta_{j}-\theta_{k})\cos[pG(t)(x_{j}-x_{k})/2]\nonumber \\
 &  & +\sin(\theta_{j}-\theta_{k})\sin[pG(t)(x_{j}-x_{k})/2]\nonumber \\
\end{eqnarray}
We see that the calculation of the marginal $Q(x,t)$ will involve
the integrals over $p$ as given by Eq. (\ref{eq:int}), so that the
interference terms will not contribute to $Q(x,0)$. Result III.1
follows. The extension to mixed states is straightforward.

Similarly, the proof of Result VII.5 in Sec. VII can be generalized
to hold for any arbitrary state $|\psi\rangle$ at time $t_{0}$,
by expanding in terms of the measurement basis, as $|\psi\rangle=\sum_{jk}c_{jk}|x_{j}^{A}\rangle|x_{k}^{B}\rangle$.
Here, we assume the $\{|x_{j}^{K}\rangle\}$ form a complete orthogonal
set for the states of system $K$ (proved in Appendix C). The Q function
of $|\psi\rangle$ is a sum of Gaussians and an interference term
$\mathcal{I}nt_{AB}$. Similar to Eq. (\ref{eq:Q-ext}), the $\mathcal{I}nt_{AB}$
is a sum of terms $\mathcal{I}nt_{AB,P}$ each involving a pair $P$
of product eigenstates, labelled by $P\in\Bigl\{\{j$, $k\}$, $\{j'$,
$k'\}\Bigr\}$ where either $k\neq k'$, $j\neq j'$ or both. The
calculation of the marginal $Q(x_{A},x_{B},t)$ involves integrals
over $p_{A}$ and $p_{B}$ as in (\ref{eq:int-ex}), for each pair.
The integral is zero, so that the interference terms do not contribute
to $Q(x_{A},x_{B},t)$. The same result applies when there is a rotation
of setting at $B$, as in Eq. (\ref{eq:q1-1-2}). With a rotation
of setting at both $A$ and $B$, each of the interference terms $\mathcal{I}nt_{AB,P}$
can contribute a nonzero value to the marginal $Q(x_{\theta A},x_{\phi B},t)$
because the integration over $p_{\theta A}$ and $p_{\phi B}$ is
not necessarily zero. This can only occur for the specified rotations,
as evident by Eq. (\ref{eq:bell-int-1-1-3}). The extension to mixed
states is straightforward.

\section*{Appendix D: Transformation in Q function due to measurement settings
\label{sec:Appendix-D-settings-Q}}

The Q function $Q(\lambda_{rot},t_{1})$ of the new state at time
$t_{2}$ after the interactions due to the $H_{\theta}$ and $H_{\phi}$
is that obtained by rotating the coordinates $x_{A},p_{A},x_{B},p_{B}$
in the original Q function, to new coordinates $x_{\theta A},p_{\theta A},x_{\phi B},p_{\phi B}$.
We see that 
\begin{equation}
\alpha e^{i\theta}=(x_{A}+ip_{A})e^{i\theta}=x_{A}'+ip_{A}'\label{eq:trans-1}
\end{equation}
where $x'_{A}=x_{A}\cos\theta-p_{A}\sin\theta$ and $p_{A}'=x_{A}\sin\theta+p_{A}\cos\theta$.
To obtain the Q function at time $t_{1}$, we hence transform an
initial Q function $Q(x_{A},p_{A})$ by substituting $\alpha$ for
$\alpha e^{i\theta}$, which means substituting $x_{A}$ and $p_{A}$
for $x_{A}'$ and $p_{A}'$, i.e. $x_{A}\rightarrow x_{A}\cos\theta-p_{A}\sin\theta$
and $p_{A}\rightarrow p'=x_{A}\cos\theta+p_{A}\sin\theta$ i.e. $x_{A}=x_{\theta A}\cos\theta-p_{\theta A}\sin\theta$
and $p_{A}=x_{\theta A}\sin\theta+p_{\theta A}\cos\theta$. The $x_{A}$
and $p_{A}$ in the rotated Q function are identified as $x_{\theta A}$
and $p_{\theta A}$, so that in terms of the original coordinates,
\begin{eqnarray}
x_{\theta A} & = & x_{A}\cos\theta+p_{A}\sin\theta\nonumber \\
p_{\theta A} & = & -x_{A}\sin\theta+p_{A}\cos\theta\label{eq:rot-6-a-1}
\end{eqnarray}
consistent with (\ref{eq:quad-1}). Similarly, the change of setting
at $B$ induced by the interaction $H_{\phi}$ amounts to a rotation
of coordinates 
\begin{eqnarray}
x_{\phi B} & = & x_{B}\cos\phi+p_{B}\sin\phi\nonumber \\
p_{\phi B} & = & -x_{B}\sin\phi+p_{B}\cos\phi\label{eq:rot-6-b-1}
\end{eqnarray}
consistent with (\ref{eq:quad-2}). Hence, the adjustment of measurement
settings is carried out in the simulation by changing the Q function
$Q(\lambda,t_{0})$ to obtain a new Q function $Q(\lambda_{rot},t_{2})$
where $\lambda_{rot}=(x_{\theta A},p_{\theta A},x_{\phi B},p_{\phi B})$.

\section*{Appendix E: transformation of coordinates in Q\label{sec:Appendix-G:-transformation-rot}}

The system interacts with the device so that the amplitudes are transformed
as above in Appendix D. The transformation proceeds by substituting
$x_{B}=x_{\phi B}\cos\phi-p_{\phi B}\sin\phi$ and $p_{B}=x_{\phi B}\sin\phi+p_{\phi B}\cos\phi$.
(Note the Jacobian of the transformation is unity.) To transform
the Gaussian $Q_{G}(x_{B},p_{B})=\frac{e^{-p_{B}^{2}/2\sigma_{p}^{2}}}{2\pi\sigma_{x}\sigma_{p}}e^{-(x_{B}-x_{k}^{B})^{2}/2\sigma_{x}^{2}}$
with variances $\sigma_{x}^{2}=1+e^{-2r}$ and $\sigma_{p}=1+e^{2r}$
where $r\rightarrow\infty$, we note that the term $\frac{1}{\sigma_{x}\sqrt{2\pi}}e^{-(x_{B}-x_{k}^{B})^{2}/2}$
dominates. Since $x_{k}^{B}=x_{k}^{B}(\cos^{2}\phi+\sin^{2}\phi)$,
this term becomes after some manipulation 
\begin{eqnarray*}
e^{-(x_{B}-x_{k}^{B})^{2}/2} & \rightarrow & e^{-\frac{(x_{\phi B}-x_{k}^{B}\cos\phi)^{2}}{2\sigma_{x_{\phi B}}^{2}}}e^{-\frac{(p_{\phi B}+x_{k}^{B}\sin\phi)^{2}}{2\sigma_{p_{\phi B}}^{2}}}\\
 &  & e^{-\frac{(x_{\phi B}-x_{k}^{B}\cos\phi)(p_{\phi B}+x_{k}^{B}\sin\phi)}{\sigma_{x_{\phi B}}\sigma_{p_{\phi B}}}}
\end{eqnarray*}
where $\sigma_{x_{\phi B}}^{2}=1/\cos^{2}\phi$ and $\sigma_{p_{\phi B}}^{2}=1/\sin^{2}\phi$.
The variances $\sigma_{x_{\phi B}}^{2}$ and $\sigma_{p_{\phi B}}^{2}$
of the transformed bivariate Gaussian are such that neither is now
necessarily infinite. Hence, provided $\phi\neq n\pi/2$ where $n$
is an integer, it is possible that neither variance of $\sigma_{p_{\phi B}}$
and $\sigma_{x_{\phi B}}$ will approach $\infty$. We note that we
have written the expressions (\ref{eq:q1-1-1}) as above, in the limit
of $r\rightarrow\infty$, which leads to functions that are not normalizable.
The full correct solutions take into account the term $\frac{1}{\sigma_{p}\sqrt{2\pi}}\exp(-p_{B}^{2}/2\sigma_{p}^{2})$,
allowing for the finite variance of $p$. To illustrate, we consider
the transformation $f_{x_{k}^{B}}(x_{\phi B},p_{\phi B})$ of the
full Gaussian $Q_{G}(x_{B},p_{B})=\frac{e^{-p_{B}^{2}/2\sigma_{p}^{2}}}{2\pi\sigma_{x}\sigma_{p}}e^{-(x_{B}-x_{k}^{B})^{2}/2\sigma_{x}^{2}}$
distribution, as it is passed through the phase-shifter $H_{\phi}^{B}$,
modeling the rotations from $x_{B}$, $p_{B}$ to $x_{\phi B}$, $p_{\phi B}$.
The evolution of the $x_{B}(t)$ and $p_{B}(t)$ is plotted in Figure
\ref{fig:Evolution-settingB}, where the equalization of the variances
in the $x_{B}(t)$ and $p_{B}(t)$ is evident for certain times $t$.
The joint probability density of the $x_{B}(t)$ and $p_{B}(t)$ in
Figure \ref{fig:Evolution-settingB} is given by the transformed function
$f_{x_{k}^{B}}(x_{\phi B},p_{\phi B})$.\textcolor{red}{{} }

The coefficients of transformation can be calculated. We rewrite the
eigenstate $|x_{k}^{A}\rangle$ is the new basis $\{|x_{\theta l}^{A}\rangle\}$
as $|x_{k}^{A}\rangle=\sum_{l}d_{kl}|x_{\theta l}^{A}\rangle$ where
$d_{kl}=\langle x_{\theta l}^{A}|x_{k}^{A}\rangle$ since the eigenstates
are orthogonal and complete. Then for any single-mode state $|\psi\rangle=\sum_{k}c_{k}|x_{k}^{A}\rangle$,
we find $|\psi\rangle=\sum_{l}f_{l}|x_{l}^{A}\rangle$ in the new
basis, where $f_{l}=\sum_{k}c_{k}d_{kl}$. The $\langle x_{\theta l}^{A}|x_{k}^{A}\rangle$
are readily calculated.

\section*{Appendix F: Result VII.3 and Asymmetric LHV model $\mathcal{P}_{asym,A}$\label{sec:Appendix-F:-Asymmetric}}

Following the text, we consider the system prepared with respect to
basis $\theta=\phi=0$ , and consider that there is then a change
of measurement setting at one site, $B$, only. We hence rotate the
basis at $B$, so that the system is prepared for final coupling to
a meter/ amplification to give the outcomes for a new measurement
$\hat{B}_{\phi}$ ($\phi\neq0$). Here, the predictions show no inconsistency
with Bell's local hidden variable condition. The simulation is consistent
with the premise that there is no impact on the observed value of
$\widetilde{\lambda}_{x}^{A}$ at $A$ by a change of setting at $B$.

We prove as follows. The initial system is expressed in terms of the
prepared basis:
\begin{eqnarray}
|\psi_{bell}\rangle & = & \sum_{ij}c_{ij}|i\rangle_{A}|j\rangle_{B}\label{eq:bell-1-1}\\
 & \rightarrow & c_{11}|+\rangle|+\rangle+c_{12}|+\rangle|-\rangle\nonumber \\
 &  & +c_{21}|-\rangle|+\rangle+c_{22}|-\rangle|-\rangle\nonumber 
\end{eqnarray}
Here $|i\rangle$ and $|j\rangle$ are eigenstates of measurements
$\hat{A}$ and $\hat{B}$, respectively, with eigenvalues $\lambda_{i}^{A}$
and $\lambda_{j}^{B}$, and the $c_{ij}$ are probability amplitudes.
We have generalized our notation to allow that the measurements may
be (for example) of spin, as in the standard Bell formulation. The
last line gives the example of the expansions for a spin Bell state
$=\frac{1}{\sqrt{2}}\{|+\rangle_{A}|+\rangle_{B}+|-\rangle_{A}|-\rangle_{B}\}$
written in the basis of Pauli spins $\sigma_{z}^{A}$ and $\hat{\sigma}_{z}^{B}$,
the $|\pm\rangle_{K}$ being the Pauli spin-$z$ eigenstates with
outcome $\pm1$ for system $K\in\{A,B\}$. For convenience, we denote
$|s_{A}\rangle_{A}|s_{B}\rangle_{B}$ where $s_{K}=\pm$ by $|s_{A}\rangle|s_{B}\rangle$,
dropping the subscripts where the meaning is clear. In the Result
VII.3, the eigenstates of $\hat{A}$ are $|i\rangle\equiv|x_{i}^{A}\rangle$.
We denote generally the eigenstates of the rotated measurement$\hat{B}_{\phi}$
to be $|k\rangle_{B}$ with eigenvalues $\lambda_{k}^{B}$.

With the rotation at $B$, we expand in the new basis: $|j\rangle=\sum_{k}d_{jk}|k\rangle$.
Thus, state becomes
\begin{eqnarray*}
|\psi_{bell}\rangle_{\phi} & = & \sum_{ij}c_{ij}|i\rangle_{A}[\sum_{k}d_{jk}|k\rangle_{B}]
\end{eqnarray*}
which simplifies to 
\begin{eqnarray}
|\psi_{bell}\rangle_{\phi} & = & \sum_{i}\sum_{k}\sum_{j}c_{ij}d_{jk}|i\rangle_{A}|k\rangle_{B}\nonumber \\
 & = & \sum_{i}\sum_{k}e_{ik}|i\rangle_{A}|k\rangle_{B}\label{eq:bell-ex2}
\end{eqnarray}
where $e_{ik}=\sum_{j}c_{ij}d_{jk}$. Looking at the example of the
Bell state, the Q function for this rotated state can be written as
a sum of the four new terms plus new interference terms $\mathcal{I}nt_{AB}(\phi)$.
The joint probability for outcomes $\lambda_{i}$ and $\lambda_{k}$
for the system in state $|\psi_{bell}\rangle$ is
\begin{eqnarray}
|e_{ik}|^{2} & = & |\sum_{j}c_{ij}d_{jk}|^{2}=|c_{11}d_{11}+c_{12}d_{21}|^{2}+..\nonumber \\
 & = & |c_{11}d_{11}+c_{12}d_{21}|^{2}+..\label{eq:bell-ex}
\end{eqnarray}

We compare this result for $|\psi\rangle$ with the system prepared
in a certain \emph{mixed} state $\rho_{mix,A}$, in which $A$ can
be viewed as being in one of the eigenstates $|i\rangle_{A}$ with
a definite probability $|f_{i}|^{2}=\sum_{j}|c_{ij}|^{2}$. This means
that $A$ can be viewed as having a definite outcome $\widetilde{\lambda}^{A}$
for measurement $\hat{A}.$ We define the joint probability for outcomes
of $\lambda_{i}$ and $\lambda_{k}$ to be $P_{ik}$. We will show
that after the change of setting at $B$, there is no difference between
the probabilities $P_{ik}$ as predicted by $|\psi_{bell}\rangle$
and $\rho_{mix,A}$. The system is consistent with the interpretation
of having a definite value $\widetilde{\lambda}^{A}$ for the outcome
of $\hat{A}$ throughout an interaction $H_{\phi}$ giving a change
of setting at $B$.

To show this, we rewrite (\ref{eq:bell-1-1}) as
\begin{equation}
|\psi_{bell}\rangle=\sum_{i}f_{i}|i\rangle_{A}|\psi_{i}^{B}\rangle\label{eq:bell-ex4}
\end{equation}
where $|\psi_{i}^{B}\rangle=\sum_{j}[c_{ij}/f_{i}]|j\rangle_{B}$.
We note that $\sum_{j}|c_{ij}|^{2}/|f_{i}|^{2}=1$. We consider the
mixed (non-entangled) state
\begin{equation}
\rho_{mix,A}=\sum_{i}|f_{i}|^{2}|i\rangle|\psi_{i}^{B}\rangle\langle\psi_{i}^{B}|\langle i|\label{eq:mixa}
\end{equation}
The rotation $\phi$ at $B$ for the state $|\psi_{i}^{B}\rangle$
gives
\begin{eqnarray}
|\psi_{i}^{B}\rangle_{\phi} & = & \sum_{j}\frac{c_{ij}}{f_{i}}\sum_{k}d_{jk}|k\rangle_{B}=\sum_{k}\sum_{j}\frac{c_{ij}}{f_{i}}d_{jk}|k\rangle_{B}\nonumber \\
 & \equiv & \sum_{k}g_{k}|k\rangle_{B}\label{eq:exp}
\end{eqnarray}
Hence
\begin{eqnarray}
\rho_{mix,A,\phi} & = & \sum_{i}|f_{i}|^{2}\sum_{k,k'}g_{k}g_{k'}^{*}|i\rangle|k\rangle\langle i|\langle k'|\label{eq:mix-ex4}
\end{eqnarray}
The joint probability of outcomes $\lambda_{i}^{A}$ and $\lambda_{k}^{B}$
for the mixed state $\rho_{mix,A}$ is
\[
|f_{i}|^{2}|g_{k}|^{2}=|f_{i}|^{2}|\sum_{j}c_{ij}d_{jk}|^{2}/|f_{i}|^{2}=|\sum_{j}c_{ij}d_{jk}|^{2}
\]
in agreement with (\ref{eq:bell-ex}), those of $|\psi_{bell}\rangle$.

Now we ask what happens to the Q function, as the change of setting
occurs at $B$? The Q function for $\rho_{mix,A,\phi}$ includes interference
terms $\mathcal{I}_{mix}(\phi)$ (arising from off-diagonal elements
in the new basis) but the $|\psi_{bell}\rangle$ has extra interference
terms, since its density matrix has extra terms arising from the fact
that the system $A$ is in a superposition (i.e. originally entangled
with $B$). Hence (even though the probabilities for outcomes $\lambda_{i}^{A}$
and $\lambda_{k}^{B}$ on rotation of basis at $B$ are the same for
$|\psi_{bell}\rangle$ and $\rho_{mixA,\phi}$), there is a difference
between the ``hidden'' interference terms\emph{ }$\mathcal{I}nt_{AB}$
of $|\psi_{bell}\rangle$ and $\rho_{mixA,\phi}$. However, because
the change of setting at $B$ induces only changes to the hidden terms
$\mathcal{I}nt_{AB}$ of $Q$ (which are not amplified by $H_{amp}$),
we see that the simulation is consistent with the premise that there
is no impact on the value of $\widetilde{\lambda}_{x}^{A}$ at $A$
by a change of setting at $B$ (no-signaling).

\textbf{\emph{LHV model: }}The result leads to a LHV model, which
we denote by $\mathcal{P}_{asym,A}$. We assume amplification $H_{amp}^{A}$
at $A$, so that there is the branch given by $\widetilde{\lambda}_{x}^{A}$.
Here, we designate the branch value as $\lambda_{x}^{A}\equiv\lambda_{i}^{A}$.
We can track the amplitudes back to the origin, as in Sec. IV.D. This
models projection as explained in Sec. VIII. The state for $B$ can
be inferred, as $Q(x_{B},p_{B}|\widetilde{\lambda}_{i}^{A}$) (Fig.
\ref{fig:projection-1}). The HV model is that of the mixed state
$\rho_{mix,A}$, in which the system is in a state with definite outcome
$\widetilde{\lambda}_{i}^{A}\equiv x_{i}^{A}$ with a fixed probability
$|f_{i}|^{2}$ (in the above $|i\rangle_{A}\equiv|x_{i}^{A}\rangle$
and $\hat{A}\equiv\hat{x}_{A}$). The state at $A$ with outcome $\widetilde{\lambda}_{i}^{A}\equiv x_{j}^{A}$
is coupled to the state $|\psi_{i}^{B}\rangle$ at $B$, which can
be a superposition state for $B$. In the Q-based model, this state
$|\psi_{i}^{B}\rangle$ is given by the variables $x_{B},p_{B}$ and
the Q distribution $Q(x_{B},p_{B}|\widetilde{\lambda}_{i}^{A})$.

\section*{Appendix G: Standard Bell example \label{sec:AppendixE-bell}}

Bell violations require adjustments of measurement settings $\theta$
and $\phi$. It is worth reviewing the role of the measurement settings
in standard Bell experiments. The choice of settings $\theta$ at
$A$ and $\phi$ at $B$ involves interacting the systems locally
with a physical device e.g. a Stern-Gerlach analyzer is orientated
so that the spin component $\hat{\sigma}_{\theta}$ is measured at
$A$ and $\hat{\sigma}_{\phi}$ is measured at $B$. After passing
through the analyzers, the systems $A$ and $B$ are prepared with
respect to the measurement basis, for measurements $\hat{\sigma}_{\theta}$
and $\hat{\sigma}_{\phi}$ respectively. The Bell state is
\begin{equation}
|\psi_{bell}\rangle=\frac{1}{\sqrt{2}}\{|+\rangle_{A}|-\rangle_{B}-|-\rangle_{A}|+\rangle_{B}\}\label{eq:bellspin-1}
\end{equation}
where $|\pm\rangle_{K}$, $K=A,B$, are the spin eigenstates of $\hat{\sigma}_{z}$
for the system denoted $K$. The probabilities for outcomes are calculated
from (\ref{eq:bellspin-1}), by rewriting $|\psi_{bell}\rangle$ in
terms of the measurement basis, corresponding to the eigenstates of
$\hat{\sigma}_{\theta,A}$ and $\hat{\sigma}_{\phi,B}$ (we take $\hat{\sigma}_{\theta}=\hat{\sigma}_{z}\cos\theta+\hat{\sigma}_{x}\sin\theta$).
For example, the state after a single rotation $\phi$ at $B$ is
\begin{eqnarray}
|\psi_{bell}\rangle_{\phi} & = & \frac{1}{\sqrt{2}}|+\rangle_{A}\{\cos\frac{\phi}{2}|-\rangle_{\phi B}+\sin\frac{\phi}{2}|+\rangle_{\phi B}\}\nonumber \\
 &  & -\frac{1}{\sqrt{2}}|-\rangle_{A}\{-\sin\frac{\phi}{2}|-\rangle_{\phi B}+\cos\frac{\phi}{2}|+\rangle_{\phi B}\}\nonumber \\
\label{eq:rot-bellphi-1}
\end{eqnarray}
where $|+\rangle_{\phi B}=\cos\frac{\phi}{2}|+\rangle+\sin\frac{\phi}{2}|-\rangle$
and $|-\rangle_{\phi B}=-\sin\frac{\phi}{2}|+\rangle+\cos\frac{\phi}{2}|-\rangle$.
The probabilities for outcomes $+-$, $++$, $--$, $-+$ are $\frac{1}{2}\cos^{2}\frac{\phi}{2}$,
$\frac{1}{2}\sin^{2}\frac{\phi}{2}$, $\frac{1}{2}\sin^{2}\frac{\phi}{2}$
and $\frac{1}{2}\cos^{2}\frac{\phi}{2}$ respectively. We may compare
with predictions for the non-entangled mixed states
\begin{eqnarray}
\rho_{mix}^{(AB)} & = & \frac{1}{2}\Bigl\{|+\rangle_{A}|-\rangle_{B}\langle-|_{B}\langle+|_{A}\nonumber \\
 &  & \ \ \ \ \ \ +|-\rangle_{A}|+\rangle_{B}\langle+|_{B}\langle-|_{A}\Bigr\}\label{eq:mix-spin-1}
\end{eqnarray}
The state after the rotation $\phi$ at $B$ is
\begin{eqnarray}
\rho_{mix,\phi} & = & \frac{1}{2}\{\rho_{+}^{(A)}\rho_{-\phi}^{(B)}+\rho_{-}^{(A)}\rho_{+\phi}^{(B)}\}\label{eq:mix-rotphi-1}
\end{eqnarray}
where $\rho_{\pm}^{(A)}=|\pm\rangle_{A}\langle\pm|_{A}$ and $\rho_{\pm\phi}^{(B)}=|\pm\rangle_{\phi B}\langle\pm|_{B\phi}$,
which gives the same predictions for the probabilities as $|\psi_{bell}\rangle_{\phi}$.

However, after a \emph{further} change of setting (basis) to $\theta$
at $A$, the predictions between the entangled and non-entangled states
diverge. For the Bell state, the state in the measurement basis becomes
\begin{eqnarray}
|\psi_{bell}\rangle_{\phi,\theta} & = & \cos(\frac{\theta-\phi}{2})\frac{1}{\sqrt{2}}\{|+\rangle_{\theta A}|-\rangle_{\phi B}-|-\rangle_{\theta A}|+\rangle_{\phi B}\}\nonumber \\
 &  & -\sin(\frac{\theta-\phi}{2})\frac{1}{\sqrt{2}}\{|+\rangle_{\theta A}|+\rangle_{\phi B}+|-\rangle_{\theta A}|-\rangle_{\phi B}\}\nonumber \\
\label{eq:bell-rot-2-1-1}
\end{eqnarray}
which gives probabilities $\cos^{2}[(\theta-\phi)/2]$ (and $\sin^{2}[(\theta-\phi)/2]$)
for outcomes $++$ and $--$ (and $+-$ and $-+$) respectively, which
are \emph{different} to those of the mixed state $\rho_{mix,AB}$.
This is seen by writing both $|\psi_{bell}\rangle_{\phi}$ and $\rho_{mixA,\phi}$
in the new basis for $A$, and is evident by the fact that $|\psi_{bell}\rangle$
violates a Bell inequality, whereas $\rho_{mix,AB}$ does not. The
interference terms that are present in the density operator (when
written in the original basis) for the Bell state $|\psi_{bell}\rangle$,
but not for the mixed state $\rho_{mix}^{(AB)}$, do not manifest
as a difference in predictions between $|\psi_{bell}\rangle$ and
$\rho_{mix}^{(AB)}$ when there are no changes of settings, or only
one change of setting. The violation of the Bell inequality requires
a change of settings at both sites, $A$ and $B$.

\section*{Appendix H: causal consistency (bipartite case) and projection\label{sec:Appendix-H:-causal-cons}}

\subsection*{Causal consistency}

\emph{Result AH:1} We give the proof of causal consistency as stated
in Result IX.9 and depicted in Figure \ref{fig:epr-causal-consistency-1}
(Secs. I and IX).

The simulation for the general superposition state $|\psi\rangle=\sum_{j,k}c_{jk}|x_{j}^{A}\rangle|x_{k}^{B}\rangle$
begins at the future boundary condition, with values $x_{A}(t)$ and
$x_{B}(t)$ emanating from particular branches $x_{j}^{A}$, $x_{k}^{B}$
with probability $|c_{jk}|^{2}$. For each branch-pair $x_{j}^{A}$,
$x_{k}^{B}$, the distribution of $x(t_{f})$ is given by the Gaussian
$\frac{e^{-(x_{A}-Gx_{j}^{A})^{2}/2}}{\sqrt{2\pi}}\frac{e^{-(x_{B}-Gx_{j}^{B})^{2}/2}}{\sqrt{2\pi}}$
where $G=e^{gt_{f}}$. We first consider just one branch-pair, $x_{j}^{A}$
and $x_{k}^{B}$. At time $t$, the distribution is $\frac{e^{-(x_{A}-e^{gt}x_{j}^{A})^{2}/2}}{\sqrt{2\pi}}\frac{e^{-(x_{B}-e^{gt}x_{j}^{B})^{2}/2}}{\sqrt{2\pi}}$
(where $t_{0}<t<t_{f}$). The distribution at time $t_{0}$ is $Q_{jk}(x_{A},x_{B})=\frac{e^{-(x_{A}-x_{j}^{A})^{2}/2}}{\sqrt{2\pi}}\frac{e^{-(x_{B}-x_{j}^{B})^{2}/2}}{\sqrt{2\pi}}$,
a set of values for $x_{A}$ and $x_{B}$ that we denote by $\mathcal{B}_{jk}$.
At time $t_{0}$, there is a connection with a set of values for $p_{A}$,
$p_{B}$ determined by $Q_{0}(p_{A},p_{B}|x_{A},x_{B})$ (Eq. (\ref{eq:cond-1})).
The joint distribution associated with the branch-pair $\widetilde{\lambda}_{x}^{A}\equiv x_{j}^{A}$,
$\widetilde{\lambda}_{x}^{B}\equiv x_{k}^{B}$ is (refer Sec. IV.D)
\begin{eqnarray}
Q_{loop}(\mathbf{x},\mathbf{p},t_{0}|x\in\mathcal{B}_{jk}) & \equiv & Q_{0}(\mathbf{x,p}|x\in\mathcal{B}_{j})\label{eq:loopq}
\end{eqnarray}
where $\mathbf{x}=(x_{A},x_{B})$ and $\mathbf{p}=(p_{A},p_{B})$.

We take the simple case of the superposition of $|x_{j}^{A}\rangle|x_{k}^{B}\rangle$
and $|x_{l}^{A}\rangle|x_{m}^{B}\rangle$ (Eq. \ref{eq:entbasis})
where there are just two branch-pairs. We take $|c_{1}|=|c_{2}|$
and $x_{j}^{A}=x_{1}^{A}$, $x_{k}^{B}=x_{1}^{B}$, $x_{l}^{A}=-x_{j}^{A}$,
$x_{m}^{B}=-x_{k}^{B}$. Hence, it is convenient to denote $\mathcal{B}_{jk}$
and $\mathcal{B}_{lm}$ by $\mathcal{B}_{+}$ and $\mathcal{B}_{-}$
respectively. Denoting the distribution of the branch $\mathcal{B}_{\pm}$
at $t_{0}$ by $Q_{\pm}(\mathbf{x})=\frac{e^{-(x_{A}\mp x_{1}^{A})^{2}/2}}{\sqrt{2\pi}}\frac{e^{-(x_{B}\mp x_{1}^{B})^{2}/2}}{\sqrt{2\pi}}$,
we consider $\mathcal{B}_{+}$ and find 
\begin{eqnarray}
Q_{0}(\mathbf{x},\mathbf{p}|x\in\mathcal{B}_{+}) & = & Q_{+}(\mathbf{x})Q_{0}(\mathbf{p}|\mathbf{x})\label{eq:loopcons}
\end{eqnarray}
We use $Q_{0}(\mathbf{p}|\mathbf{x})=Q(\mathbf{x},\mathbf{p},t_{0})/Q_{0}(\mathbf{x})$
where $Q_{0}(\mathbf{x})=\sum_{\pm}Q_{\pm}(\mathbf{x})$. Here $Q(\mathbf{x,p},t_{0})$
is given by (\ref{eq:qex}) which here simplifies to 
\begin{eqnarray}
Q(\mathbf{x,p},t_{0}) & = & N\frac{e^{-p_{A}^{2}/2\sigma_{p}^{2}}}{\sqrt{2\pi}\sigma_{p}}\frac{e^{-p_{B}^{2}/2\sigma_{p}^{2}}}{\sqrt{2\pi}\sigma_{p}}\Bigl\{ Q_{+}(\mathbf{x})+Q_{-}(\mathbf{x})\nonumber \\
 &  & \ \ +2g(\mathbf{x})\mathcal{F}_{AB}\Bigl\}\label{eq:qcc}
\end{eqnarray}
where $g(\mathbf{x})=e^{-(x_{1}^{A})^{2}/2}e^{-(x_{1}^{B})^{2}/2}e^{-x_{A}^{2}/2}e^{-x_{B}^{2}/2}$
and $\mathcal{F}_{AB}=\cos[p_{A}\Delta_{A}/\sigma_{x_{A}}+p_{B}\Delta_{B}/\sigma_{x_{B}}]$,
$\Delta_{A}=x_{1}^{A}/\sigma_{x_{A}}$ and $\Delta_{B}=x_{1}^{B}/\sigma_{x_{B}}$.
A similar expression can be obtained for $\mathcal{B}_{-}$. 
We note that (on extending the calculation of Eq. (\ref{eq:Q-sup-1})
to two modes, refer Appendix I below), the Q function $Q(\mathbf{x,p},t)$
of the amplified system is given by $Q(\mathbf{x,p},t_{0})$ on substituting
$\sigma_{p_{A}}\rightarrow\sigma_{p_{A}}(t)$, $\sigma_{p_{B}}\rightarrow\sigma_{p_{B}}(t)$,
$x_{1}^{A}\rightarrow G(t)x_{1}^{A}$, $x_{1}^{B}\rightarrow G(t)x_{1}^{B}$.

The densities for $x_{A}(t)$, $x_{B}(t)$ and $p_{A}(t)$, $p_{B}(t)$
at a given time $t_{0}\leq t\leq t_{f}$ can be evaluated, for each
branch $\mathcal{B}_{+}$ and $\mathcal{B}_{-}$, by examining $Q_{loop}(\mathbf{x,p},t|x\in\mathcal{B}_{\pm})$.
The value of $x_{j}$ deterministically amplified to $G(t)x_{j}$,
so that 
\begin{equation}
Q_{\pm}(\mathbf{x})\rightarrow Q_{G\pm}(\mathbf{x})=\frac{e^{-(x_{A}\mp G(t)x_{1}^{A})^{2}/2}}{\sqrt{2\pi}}\frac{e^{-(x_{B}\mp G(t)x_{1}^{B})^{2}/2}}{\sqrt{2\pi}}\label{eq:Q-pmG}
\end{equation}
and similarly in $\mathcal{I}nt_{AB}$, we replace $x_{1}\rightarrow Gx_{1}$,
and $\sigma_{p}\rightarrow\sigma_{p}(t)$. Hence, denoting $Q_{loop}(\mathbf{x,p},t|x\in\mathcal{B}_{\pm})$
by $Q_{loop\pm}(\mathbf{x,p},t)$ for simplicity, we find 
\begin{eqnarray}
Q_{loop\pm}(\mathbf{x,p},t) & = & N_{t}Q_{G\pm}(\mathbf{x})\frac{e^{-p_{A}^{2}/2\sigma_{p}^{2}(t)}}{\sqrt{2\pi}\sigma_{p}(t)}\frac{e^{-p_{B}^{2}/2\sigma_{p}^{2}(t)}}{\sqrt{2\pi}\sigma_{p}(t)}\nonumber \\
 &  & \times\Bigl(1+2g_{G}(\mathbf{x})\frac{\mathcal{F}_{AB}(t)}{\sum_{\pm}Q_{G\pm}(\mathbf{x})}\Bigr)\label{eq:qllop-cc}\\
\nonumber 
\end{eqnarray}
where $g_{G}(\mathbf{x})=e^{-(G(t)x_{1}^{A})^{2}/2}e^{-(G(t)x_{1}^{B})^{2}/2}e^{-x_{A}^{2}/2}e^{-x_{B}^{2}/2}$,
$G(t)=e^{gt}$, $\mathcal{F}_{AB}(t)=\cos[p_{A}\Delta_{A}/\sigma_{x_{A}}+p_{B}\Delta_{B}/\sigma_{x_{B}}]$
where $\Delta_{A}=G(t)x_{1}^{A}/\sigma_{x_{A}}$ and $\Delta_{B}=G(t)x_{1}^{B}/\sigma_{x_{B}}$
and $N_{t}$ is a normalization factor. The Q function $Q(x,p,t)$
for the state $|\psi(t)\rangle=e^{-iH_{amp}^{A}t/\hbar}e^{-iH_{amp}^{B}t/\hbar}|\psi\rangle$
at time $t$ corresponds to the joint density given as the average
of that for the branches. Hence
\begin{eqnarray}
Q(\mathbf{x,p},t) & = & \frac{1}{2}\sum_{\pm}Q_{loop\pm}(\mathbf{x,p},t)\label{eq:avbranch}
\end{eqnarray}
Evaluating,
\begin{eqnarray}
Q(\mathbf{x,p},t) & = & N_{t}\frac{e^{-p_{A}^{2}/2\sigma_{p}^{2}(t)}}{2\sqrt{2\pi}\sigma_{p}(t)}\frac{e^{-p_{B}^{2}/2\sigma_{p}^{2}(t)}}{\sqrt{2\pi}\sigma_{p}(t)}\nonumber \\
 &  & \times\Bigl(Q_{G+}(\mathbf{x})+Q_{G-}(\mathbf{x})+2g_{G}(\mathbf{x})\mathcal{F}_{AB}(t)\Bigr)\nonumber \\
\label{eq:fullq-1}
\end{eqnarray}
reduces to the Q function $Q(\mathbf{x,p},t)$ defined above
(refer Appendix I) as required. The extension to the more general
superposition $|\psi\rangle=\sum_{j,k}c_{jk}|x_{j}^{A}\rangle|x_{k}^{B}\rangle$
follows along the same lines. Assuming the eigenstates $\{|x_{j}^{K}\rangle\}$
form a complete set for a single mode system $K\in\{A,B\}$, this
completes the first part of the proof.

In the second part of the proof, we consider the dynamics of the measurement
settings. The Figure \ref{fig:epr-causal-consistency-1} illustrates
the operations that adjust the measurement settings $\theta$ and
$\phi$, prior to amplification. Here, causal consistency follows
because the associated dynamics is deterministic and reversible, given
by equations (\ref{eq:rot-6-a-1}) and (\ref{eq:rot-6-b-1}). To give
detail, we consider two scenarios (refer Fig. 5).

\emph{Case I: }First, we consider that the changes of settings occur
at each site between the times $t_{0}$ and $t_{2}$, as in Figure
\ref{fig:sim}, where amplification takes place after time $t_{2}$.
We wish to prove causal consistency between these times. The dynamics
in terms of the amplitudes is given by the transformations (\ref{eq:rot-6-a-2})
and (\ref{eq:rot-6-b-3}) (Fig. \ref{fig:Evolution-settingB}) as
explained in Appendix E. Different sequences for the setting-changes
are possible, but the amplitudes in each case are defined by the deterministic
transformations, or as a sequence of them. For successive deterministic
transformations, it is assumed that the trajectories flow continuously,
so that output of one transformation is the input of the other. This
defines amplitudes $x_{A}(t),p_{A}(t),x_{B}(t),p_{B}(t)$ at a time
$t_{0}\leq t\leq t_{2}$. For each transformation, the amplitudes
are the variables of a probability density function. The probability
density function for such transformed variables is known (from ordinary
calculus) to be found by expressing the $x_{A}$, $p_{A}$, $x_{B}$,
$p_{B}$ in terms of the $x_{A}(t),p_{A}(t),x_{B}(t),p_{B}(t)$, and
substituting in the initial density function, which is given by $Q(\mathbf{x,p},t_{0})$
(the Jacobian is unity, refer Sec. VII and Appendix E). Since this
corresponds to the procedure used to evaluate the evolved Q function
(Sec. VII), we confirm causal consistency.

This concludes the proof for the system depicted in Figure \ref{fig:sim}.
We note that we prove the equivalence at any one time $t$, where
we consider two time intervals, from $t_{0}$ to $t_{2}$, and then
from $t_{2}$ to $t_{f}$. We place the boundary condition at time
$t_{2}$, just prior to amplification, and hence define the hidden
loop at the time $t_{2}$ ($Q_{loop}(x,p,t_{2})$). We note that a
distribution $Q(x_{\theta A},p_{\theta A},x_{\phi B},p_{\phi B})$
at a time $t_{2}$ defines a unique distribution $Q(x_{A},p_{A},x_{B},p_{B})$
at time $t_{0}$ (and vice versa) in accordance with the transformations
(\ref{eq:rot-6-a-2}) and (\ref{eq:rot-6-b-3}). We could also give
a proof where the hidden loop is at the time $t_{0}$, prior to the
setting dynamics.

\emph{Case II:} The second scenario is to consider where a setting
change occurs at one site ($B$ say), while amplification occurs simultaneously
at $A$ from time $t_{0}$ to $t_{f}$. The system at time $t_{0}$
is prepared in the state $Q(\lambda,t_{0})$. The interactions given
by Hamiltonians $H_{\phi}^{B}$ and $H_{amp}^{A}$ are then applied,
up to a time $t_{f}$. We consider a time $t$, such that $t_{0}<t<t_{f}$.
The Q function is given by $Q(\mathbf{x,p},t)$ defined above but
we put $G=1$ in the expressions relating to $B$, and the amplitudes
of $B$ are rotated (refer Sec. VII and Appendix E). We find
\begin{eqnarray}
Q(\mathbf{x,p},t) & \rightarrow & N_{t}\frac{e^{-p_{A}^{2}/2\sigma_{p_{A}}^{2}(t)}}{2\sqrt{2\pi}\sigma_{p_{A}}(t)}\frac{e^{-p_{B}(t_{0})^{2}/2\sigma_{p_{B}}^{2}(t_{0})}}{\sqrt{2\pi}\sigma_{p_{B}}(t_{0})}\nonumber \\
 &  & \times\Bigl(Q_{G_{A}+}(\mathbf{x})+Q_{G_{A}-}(\mathbf{x})+2g_{G_{A}}(\mathbf{x})\mathcal{F}_{AB_{0}}(t)\Bigr)\nonumber \\
\label{eq:fullq-1-2}
\end{eqnarray}
where ($G(t)=e^{gt}$)
\[
Q_{G_{A}\pm}(\mathbf{x})=\frac{e^{-(x_{A}\mp G(t)x_{1}^{A})^{2}/2}}{\sqrt{2\pi}}\frac{e^{-(x_{B}(t_{0})\mp x_{1}^{B})^{2}/2\sigma_{x_{B}}^{2}(t_{0})}}{\sqrt{2\pi}}
\]

\begin{eqnarray*}
g_{G_{A}}(\mathbf{x}) & = & e^{-(G(t)x_{1}^{A})^{2}/2}e^{-(x_{1}^{B})^{2}/2\sigma_{x_{B}}^{2}(t_{0})}\\
 &  & \times e^{-x_{A}^{2}/2}e^{-x_{B}(t_{0})^{2}/2\sigma_{x_{B}}^{2}(t_{0})}
\end{eqnarray*}
and $\mathcal{F}_{AB_{0}}(t)=\cos[p_{A}\Delta_{A}+p_{B}(t_{0})\Delta_{B_{0}}/\sigma_{x_{B}}]$
where $\Delta_{A}=G(t)x_{1}^{A}$ and $\Delta_{B_{0}}=x_{1}^{B}/\sigma_{x_{B}}(t_{0})$.
The variances $\sigma_{x_{A}}^{2}(t)$, $\sigma_{p_{A}}^{2}(t)$,
$\sigma_{x_{B}}^{2}(t_{0})$ and $\sigma_{p_{B}}^{2}(t_{0})$ are
given below in Appendix I ($\sigma_{x_{A}}^{2}(t)=1$). In the expression
above for $Q(\mathbf{x,p},t)$, we use 
\begin{eqnarray}
x_{B}(t_{0}) & = & x_{B}(t)\cos\phi-p_{B}(t)\sin\phi\nonumber \\
p_{B}(t_{0}) & = & x_{B}(t)\cos\phi+p_{B}(t)\sin\phi\label{eq:det-2}
\end{eqnarray}
(noting that we simplify to write $x_{B}(t)\equiv x_{B}$, $p_{B}(t)\equiv p_{B}$
in the final Q function $Q(\mathbf{x,p},t)$). We consider the branches
emanating from $t_{f}$, as in Eqs. (\ref{eq:loopq})-(\ref{eq:fullq-1})
above, but here, for a branch-pair $x_{j}^{A}$, $x_{k}^{B}$, the
distribution of $x(t_{f})$ is given by $\frac{e^{-(x_{A}-G(t)x_{j}^{A})^{2}/2}}{\sqrt{2\pi}}\frac{e^{-(x_{B}(t_{0})-x_{j}^{B})^{2}/2}}{\sqrt{2\pi}}$.
The solution for $Q_{loop\pm}(\mathbf{x,p},t_{0})$ is as above, where
we note $\sigma_{p}\equiv\sigma_{p_{A}}(t_{0})=\sigma_{p_{B}}(t_{0})$.
Using the deterministic relations $x_{1}^{A}\rightarrow G(t)x_{1}^{A}$,
$\sigma_{p_{A}}\rightarrow\sigma_{p_{A}}(t)$, and (\ref{eq:det-2})
as above, the distribution for the $Q_{loop\pm}(\mathbf{x,p},t)$
becomes
\begin{eqnarray}
Q_{loop\pm}(\mathbf{x,p},t) & = & N_{t}Q_{G_{A}\pm}(\mathbf{x})\frac{e^{-p_{A}^{2}/2\sigma_{p_{A}}^{2}(t)}}{\sqrt{2\pi}\sigma_{p_{A}}(t)}\nonumber \\
 &  & \times\frac{e^{-p_{B}(t_{0})^{2}/2\sigma_{p_{B}}^{2}(t_{0})}}{\sqrt{2\pi}\sigma_{p_{B}}(t_{0})}\nonumber \\
 &  & \Bigl(1+2g_{G_{A}}(\mathbf{x})\frac{\mathcal{F}_{AB_{0}}(t)}{\sum_{\pm}Q_{G_{A}\pm}(\mathbf{x})}\Bigr)\nonumber \\
\label{eq:qloop2}
\end{eqnarray}
Following the proof above, averaging over both branches, we deduce
equivalence to the probability density $Q(\mathbf{x,p},t)$. $\square$

\subsection*{Projection}

Figure 5 also depicts certain results about the projected states.
We explain and prove these results.

First, it is noted that for the bipartite system with Q function $Q(\lambda,t_{0})$,
the Q function for the system $A$ is given by the marginal $Q(x_{A},p_{A},t_{0})=\int Q(\lambda,t_{0})dx_{B}dp_{B}$.
To illustrate, the bipartite state can be expanded as $|\psi\rangle=\sum_{ij}c_{ij}|x_{i}^{A}\rangle|x_{j}^{B}\rangle$,
where $c_{ij}$ are probability amplitudes, where the $|x_{j}^{B}\rangle$
form an orthogonal set (Appendix C). The reduced state for system
$A$ is $|\psi_{A}\rangle=\sum_{i}d_{i}|x_{i}^{A}\rangle$ where
$d_{i}=\sum_{k}c_{ik}$, for which the Q function is 
\begin{equation}
Q(x_{A},p_{A},t)=\frac{1}{\pi}\sum_{i}\sum_{k}d_{i}d_{k}^{*}\langle\alpha|x_{i}^{A}\rangle\langle x_{k}^{A}|\alpha\rangle\label{eq:qproj}
\end{equation}
The Q function of the bipartite state is $Q(\lambda,t)=|\langle\alpha|\langle\beta|\psi\rangle|^{2}/\pi^{2}$
which simplifies to

\begin{eqnarray}
Q(\lambda,t) & = & \frac{1}{\pi^{2}}\sum_{i}\sum_{k}\langle\alpha|x_{i}^{A}\rangle\langle x_{k}^{A}|\alpha\rangle\nonumber \\
 &  & \times\sum_{j}c_{ij}\langle\beta|x_{j}^{B}\rangle\sum_{l}c_{kl}^{*}\langle x_{l}^{B}|\beta\rangle\label{eq:qprojbipa}
\end{eqnarray}
Integrating over $x_{B}$ and $p_{B}$, we find $\int Q(\lambda,t)dx_{B}dp_{B}\rightarrow Q(x_{A},p_{A},t)$
as required.

\emph{Result AH 2:} The postselected and projected distributions defined
before and after settings are fixed are related by a simple transformation
of coordinates.

\emph{Proof: }We examine Figure \ref{fig:sim} where the system is
prepared in state $Q(\lambda,t_{0})$ at time $t_{0}$. Let us assume
the measurement setting is changed to $\phi$ at $B$ first, and then
at time $t_{1}$, is changed to $\theta$ at $A$. At time $t_{2}$,
both settings are fixed at $\phi$ and $\theta$ respectively. A postselected
distribution $Q_{loop}(\{x_{\phi k}^{B}\},\lambda_{rot2},t_{2}|x_{\theta j}^{A})$
can be defined at time $t_{2}$, from a branch $\mathcal{B}_{j}^{A}$
based on the outcome $x_{\theta j}^{A}$ for system $A$. Here, $\lambda_{rot2}\equiv(x_{\theta A},p_{\theta A},x_{\phi B},p_{\phi B})$
are the variables defined after the transformations that determine
the settings. We recall that the postselected distribution must be
consistent with the deterministic relations $\mathcal{D}$ that determine
the correlation between outcomes $x_{\phi k}^{B}$ for system $B$
and the outcome $x_{\theta j}^{A}$ at $A$, which are defined in
the causal model at the time $t_{2}$, given the fixed settings $\theta$
and $\phi$ (refer Sec. VIII). With this specified, we simplify the
notation, denoting $Q(\{x_{\phi k}^{B}\},\lambda_{rot2},t_{2}|x_{\theta j}^{A})\equiv Q(\lambda_{rot2},t_{2}|x_{\theta j}^{A})$.
From $Q_{loop}(\lambda_{rot2},t_{2}|x_{\theta j}^{A})$, a projected
distribution $Q_{\phi}(x_{\phi B},p_{\phi B}|x_{j}^{A})$ can be determined
for system $B$ at time $t_{2}$, where 
\begin{equation}
Q_{t_{2}}(x_{\phi B},p_{\phi B}|x_{\theta j}^{A})=\int dx_{\theta A}dp_{\theta A}Q_{loop}(\lambda_{rot2},t_{2}|x_{\theta j}^{A})\label{eq:qa5}
\end{equation}
Mathematically, it is always possible to transform to different coordinates
$x_{A},p_{A}$ for system $A$, and to $x_{B},p_{B}$ for system $B$,
as defined by (\ref{eq:rot-6-a-1}) and (\ref{eq:rot-6-b-1}) (Jacobian
factors are unity).  A coordinate transformation for $A$ corresponds
to the amplitudes $x_{A}(t_{1})$ being combined with $p_{A}(t_{1})$
according to the inverse deterministic transformation (\ref{eq:rot-6-a-1})
which physically is due to the setting interaction $H_{\theta}^{A}$.
Hence, the trajectories for the branch $\mathcal{B}_{j}^{A}$ can
be traced back to the time $t_{1}$, prior to the setting change at
$A$. This defines  a probability density function $Q_{loop}(\lambda_{rot1},t_{1}|x_{\theta j}^{A})$
in terms of amplitudes $\lambda$ at time $t_{0}$, so that $\lambda_{rot1}=(x_{A},p_{A},x_{\phi B},p_{\phi B})$.
The projected distribution for system $B$ is (refer Result VIII.1e)
\begin{equation}
Q_{t_{1}}(x_{\phi B},p_{\phi B}|x_{\theta j}^{A})=\int dx_{A}dp_{A}Q_{loop}(\lambda_{rot1},t_{1}|x_{\theta j}^{A})\label{eq:qa3-1}
\end{equation}
Yet, we can mathematically transform the coordinates for $B$ to define
\begin{equation}
Q_{t_{0}}(x_{B},p_{B}|x_{\theta j}^{A})=\int dx_{A}dp_{A}Q_{loop}(\lambda_{rot1},t_{1}|x_{\theta j}^{A})\label{eq:qa3}
\end{equation}
hence making a variable change from $(x_{\phi B},p_{\phi B}$) to
$(x_{B},p_{B})$. Hence $Q_{t_{1}}(x_{\phi B},p_{\phi B}|x_{j}^{A})$
can be transformed to a density function $Q_{t_{0}}(x_{B},p_{B}|x_{j}^{A})$
for coordinates $(x_{B},p_{B})$. Since the transformation corresponds
to the dynamics of the amplitudes as given by $H_{\phi}^{B}$, given
by the inverse of (\ref{eq:rot-6-b-1}), this corresponds to the projected
distribution for amplitudes at time $t_{0}$ prior to the change of
setting from $\phi=0$ to $\phi$ at $B$.

This result is used in the diagrams of Figures 5 and in Section IX,
where the projected distributions $Q(x_{A},p_{A}|\widetilde{\lambda}_{\phi}^{B})$
and $Q(x_{B},p_{B}|\widetilde{\lambda}_{\theta}^{A})$ are drawn in
at an unspecified time $t$, but with reference to system $A$ and
$B$ respectively. The time $t$ is unspecified because the distribution
e.g. $Q(x_{A},p_{A}|\widetilde{\lambda}_{\phi}^{B})$ that applies
at time $t_{0}$ can be readily transformed to apply to the amplitudes
at a later time e.g. $t_{2}$, by transforming the variables $x_{A},p_{A}$
in the density function.

\emph{Result AH:3} The projected distribution, for example $Q(x_{A},p_{A}|\widetilde{\lambda}_{\phi}^{B})$,
defines the predictions for outcomes at $A$ for any future setting
adjustment $\theta$ at $A$. This follows from the proof above, since
one can make a transformation to coordinates $x_{\theta A},p_{\theta A}$
based on knowledge of $Q(x_{A},p_{A}|\widetilde{\lambda}_{\phi}^{B})$,
and this defines the projected distribution for outcomes of $\hat{x}_{\theta A}$
given the outcome defined by $\widetilde{\lambda}_{\phi}^{B}$.

\emph{Mutual Consistency:} Result IX.3 is stated in Sec. IX.

\emph{Proof:} Following the proof for Result IX.3, the evaluation
of $Q(x_{A},p_{A}|\widetilde{\lambda}_{\phi}^{B})$ requires the knowledge
of the setting $\phi$ at $B$ and of the initial Q function $Q(\lambda,t_{0})$.
The evaluation does not require knowledge of $\theta$, the setting
at $A$. The distribution $Q(x_{A},p_{A}|\widetilde{\lambda}_{\phi}^{B})$
gives predictions of probabilities for all choices of setting $\theta$
at $A$, as explained in Result AH.2 above. The proof of Result AH.2
shows how the projected distributions defined from $Q_{loop}(\lambda_{rot2},t_{2}|x_{\theta j}^{A})$
and $Q_{loop}(\lambda,t_{0}|x_{\theta j}^{A})$ are evaluated consistently
with each other. The proof of causal consistency (Result AH.1) shows
how $Q_{loop}(\lambda_{rot2},t_{2}|x_{\theta j}^{A})$ and $Q_{loop}(\lambda,t_{0}|x_{\theta j}^{A})$
are consistent with $Q(\lambda,t_{0})$. $\square$

\section*{Appendix I: Calculation of Q function: amplified bipartite state\label{sec:Appendix-I:-CalculationQ-ampbipartite}}

Consider the state
\begin{eqnarray}
|\psi\rangle & = & |x_{j}^{A}\rangle|x_{k}^{B}\rangle+|x_{l}^{A}\rangle|x_{m}^{B}\rangle\nonumber \\
 & \rightarrow & |\frac{x_{j}^{A}}{2},r\rangle_{sq}|\frac{x_{k}^{B}}{2},r\rangle_{sq}+|\frac{x_{l}^{A}}{2},r\rangle_{sq}|\frac{x_{m}^{B}}{2},r\rangle_{sq}\nonumber \\
\label{eq:sqappendix}
\end{eqnarray}
Here, $|x_{j}^{A}\rangle$ and $|x_{k}^{B}\rangle$ are eigenstates
of $\hat{x}_{A}$ and $\hat{x}_{B}$ respectively, which we approximate
by highly squeezed states in $\hat{x}_{A}$ and $\hat{x}_{B}$ (Eq.
(\ref{eq:eigenstate-def})). With amplification at each site, the
state becomes $|\psi(t)\rangle=e^{-iH_{amp}^{A}t/\hbar}e^{-iH_{amp}^{B}t/\hbar}|\psi\rangle$.
Applying the techniques of Appendix C, we find
\begin{eqnarray}
|\psi(t)\rangle & = & N\Bigl(|G(t)\frac{x_{j}^{A}}{2},r'\rangle_{sq}|G(t)\frac{x_{k}^{B}}{2},r'\rangle_{sq}\nonumber \\
 &  & +|G(t)\frac{x_{l}^{A}}{2},r'\rangle_{sq}|G(t)\frac{x_{m}^{B}}{2},r'\rangle_{sq}\Bigr)\label{eq:ampbipartite}
\end{eqnarray}
($N$ is a normalization constant). Hence, the calculation of the
Q function is as for that of (\ref{eq:sqappendix}), on substituting
in the final result $x_{j}^{A}\rightarrow G(t)x_{j}^{A}$, $x_{l}^{A}\rightarrow G(t)x_{l}^{A}$,
$x_{k}^{B}\rightarrow G(t)x_{k}^{B}$, $x_{m}^{B}\rightarrow G(t)x_{m}^{B}$,
and $\sigma_{x}\rightarrow\sigma_{x}(t)$, $\sigma_{p}\rightarrow\sigma_{p}(t)$.
The variances are given by (\ref{eq:amp-var}): $\sigma_{x_{K}}^{2}(t)=1+e^{-2r+2gt}=1+G(t)^{2}\left[\sigma_{x_{K}}^{2}\left(0\right)-1\right]$
and $\sigma_{p_{k}}^{2}(t)=1+e^{2r-2gt}=1+[\sigma_{p_{k}}^{2}(0)-1]/G(t)^{2}$
($K\in\{A,B\}$), from which we see that $\sigma_{x_{K}}^{2}(t)=1$
given the system is initially in the eigenstate of $\hat{x}_{K}$.

\section*{Appendix J: Fine-tuning\label{sec:Appendix-H:-Fine-tuning}}

The simulations of this paper give a model in which no-signaling is
explained consistently with observation of Bell nonlocality. The result
does not however conflict with results that exclude classical causal
models that are not fine-tuned: In Ref. \citep{wood2015lesson}, it
is stated that an ``\emph{observed statistical independence between
variables should not be explained by fine-tuning of the causal parameters}''.....
\emph{``In other words, all conditional independences (CIs) should
be a consequence of the causal structure alone, not a result of the
causal-- statistical parameters taking some particular set of values.''}
The authors of \citep{wood2015lesson} showed that in order for a
causal model to explain Bell violations consistently with no-signaling,
it is necessary that there be a fine-tuning of the causal parameters.
The causal parameters include the measurement settings $\theta$ and
$\phi$. 

We have seen that the simulations of this paper support Results IV.9\textbf{\emph{
}}and\textbf{\emph{ }}IX.3, which imply no-signaling. An interaction
$H_{\phi}^{B}$ (as in a change of setting) at $B$ \emph{can change
unobservable hidden terms} in the Q function, \emph{without changing
the outcome} $\widetilde{\lambda}_{x}^{A}$ for the setting $\theta\equiv0$
(Fig. \ref{fig:feedback-settingB}). This implies no-signaling.

However, with an interaction $H_{\theta}^{A}$ at $A$ so that $\theta\neq0$
(i.e. over two rotations), Bell nonlocality can emerge (Fig. \ref{fig:settings-AB}).
Thus, no-signaling does not rule out Bell nonlocality. We see from
the analysis of interference terms for the Bell state that specific
values of $\theta$ and $\phi$ contribute to the observable probabilities
that violate the Bell inequality. The value of $\theta$ needs to
be adjusted from $0$ in order to observe the nonlocality, consistent
with the conclusions of Ref. \citep{wood2015lesson}. It is not the
case that there can be assumed to be no impact on $A$ due to the
change of setting at $B$, in the sense that there is a change to
the interference term in the Q function. We summarize with the following
result.

\textbf{\emph{Result A: Fine-tuning: }}There is a fine-tuning of the
causal parameters (the measurement settings) in order to observe the
violation of the Bell inequality consistently with the observation
of no-signaling.

\emph{Proof:} No-signaling is defined as the condition that there
is no change to the outcome of $A$ due to a change of setting at
$B$, with the setting at $A$ fixed \citep{wood2015lesson}. This
corresponds to fixing the setting of $\theta$ at $A$, which we take
without loss of generality to be $\theta=0$ so that we assume $\hat{x}$
is measured at $A$ (as in Figs. \ref{fig:epr-sim-xx-1} and \ref{fig:feedback-settingB}).
Result IX.4\textbf{\emph{ }}states that the outcome given by $\widetilde{\lambda}_{x}^{A}$
in the model is unchanged as the setting $\phi$ at $B$ is changed.
In order to detect Bell nonlocality, the simulations \emph{(}Results
VII.4-7) show that we require to adjust to a nonzero setting $\theta\neq0$
at $A$ (Fig. \ref{fig:settings-AB}). Hence, a fine-tuning of the
causal parameters is required. $\square$

\bibliographystyle{apsrev4-2}

\end{document}